\documentclass[12pt,preprint,numberedappendix,twocolappendix]{emulateapj}

\newcommand\bibinc{n}		% set to y if bib pasted in .tex, set to n to use bibtex

\usepackage{subeqnarray}
\usepackage{amsmath}
\usepackage{amssymb}
\usepackage{hyperref}
\usepackage{xcolor}
\usepackage{ulem}
\usepackage{stmaryrd}

\hypersetup{colorlinks=true,linkcolor=black,citecolor=blue,urlcolor=black}

\usepackage{natbib}
\def\tt{\bf   }
\newcommand{\wpm}{$\rm{W\;m^{-2}}$}
\newcommand{\cpunit}{$\rm J\;kg^{-1}\;K^{-1}$}
\newcommand{\mps}{$\rm m\;s^{-1}$\;}
\newcommand{\kms}{$\rm km\;s^{-1}$\;}

\newcommand{\trad} {$\tau_{\rm rad  }$\;}
\newcommand{\tadv} {$\tau_{\rm adv }$\;}
\newcommand{\twave} {$\tau_{\rm wave }$\;}
\newcommand{\mpstwo}{$\rm m\;s^{-2}$\;}

\setcitestyle{citesep={,}}
\begin{document}

\slugcomment{Space Science Reviews special issue on ``Understanding the Diversity of Planetary Atmospheres"}

\shorttitle{Atmospheric Dynamics of Hot Giant Planets and Brown Dwarfs}
\shortauthors{Showman, Tan \& Parmentier}

\title{ATMOSPHERIC DYNAMICS OF HOT GIANT PLANETS AND BROWN DWARFS}
\author{Adam P. Showman\altaffilmark{1,$\dagger$}, Xianyu Tan\altaffilmark{2,$\ast$} and Vivien Parmentier\altaffilmark{2,$\star$}}
\affil{$^1$Lunar and Planetary Laboratory, University of Arizona, 1629 University Boulevard, Tucson, AZ 85721, USA \\ \url{showman@lpl.arizona.edu} \\
$^2$Atmospheric Oceanic  and Planetary Physics, Department of Physics, University of Oxford, OX1 3PU, UK \\}
\altaffiltext{$\dagger$}{\rm This may be the last published work led by Adam Showman, whose sudden death during the revision process of this manuscript deprived the field of one of his giant. He will be missed by all. }
\altaffiltext{$\ast$}{\url{xianyu.tan@physics.ox.ac.uk}}
\altaffiltext{$\star$}{\url{vivien.parmentier@physics.ox.ac.uk}}

\begin{abstract}
 Groundbased and spacecraft telescopic observations, combined with an intensive modeling effort, have greatly enhanced our understanding of hot giant planets and brown dwarfs over the past ten years. Although these objects are all fluid, hydrogen worlds with stratified atmospheres overlying convective interiors, they exhibit an impressive diversity of atmospheric behavior. Hot Jupiters are strongly irradiated, and a wealth of observations constrain the day-night temperature differences, circulation, and cloudiness. The intense stellar irradiation, presumed tidal locking and modest rotation leads to a novel regime of strong day-night radiative forcing. Circulation models predict large day- night temperature differences, global-scale eddies, patchy clouds, and, in most cases, a fast eastward jet at the equator---equatorial superrotation. The warm Jupiters lie farther from their stars and are not generally tidally locked, so they may exhibit a wide range of rotation rates, obliquities, and orbital eccentricities, which, along with the weaker irradiation, leads to circulation patterns and observable signatures predicted to differ substantially from hot Jupiters. Brown dwarfs are typically isolated, rapidly rotating worlds; they radiate enormous energy fluxes into space and convect vigorously in their interiors. Their atmospheres exhibit patchiness in clouds and temperature on regional to global scales---the result of modulation by large-scale atmospheric circulation. Despite the lack of irradiation, such circulations can be driven by interaction of the interior convection with the overlying atmosphere, as well as self-organization of patchiness due to cloud-dynamical-radiative feedbacks. Finally, irradiated brown dwarfs help to bridge the gap between these classes of objects, experiencing intense external irradiation as well as vigorous interior convection. Collectively, these diverse objects span over six orders of magnitude in intrinsic heat flux and incident stellar flux, and two orders of magnitude in rotation rate---thereby placing strong constraints on how the circulation of giant planets (broadly defined) depend on these parameters. A hierarchy of modeling approaches have yielded major new insights into the dynamics governing these phenomena. 
\end{abstract}
%\keywords{hydrodynamics -- methods: numerical -- planets and satellites: atmospheres -- planets and satellites: gaseous planets -- stars: low-mass, brown dwarfs}

\section{introduction}
\label{ch.intro}

Giant planets, broadly defined, span an enormous range of objects. Limiting ourselves to substellar bodies comprised primarily of hydrogen\footnote{Thus we exclude not only terrestrial planets, but Uranus and Neptune-like planets, which have primarily fluid interiors com-
prised of denser materials, such as water (e.g., \citealp{hubbard1991, fortney2010interior}).}, such bodies nevertheless encompass an impressive diversity. Jupiter and Saturn represent the canonical prototypes, and of course are the best observed due to their proximity to Earth. Outside our solar system, hundreds of extrasolar giant planets (EGPs) have been discovered. Hot Jupiters are the most easily observationally characterized; they orbit extremely close to their stars, at distances of typically $\sim0.03-0.1$ AU, receive thousands of times more starlight than Jupiter, and thereby achieve temperatures of 1000 K or more
~\citep{showman2002}. {At dayside temperatures exceeding $\sim2200$ K, the molecular constituents of their atmospheres start to dissociate and they are called ultra hot Jupiters
~\citep{bell2018,parmentier2018}}. Also amenable to atmospheric characterization are the directly imaged planets---that is, planets that are sufficiently hot and distant from their host stars to be imaged as distinct entities. They are hot not because they are strongly irradiated, but because they are massive and young, so that they still glow from their heat of formation---and therefore also have temperatures of typically $\sim$1000 K
~\cite{bowler2016}. Intermediate between these extremes are a large population of EGPs that are irradiated, but less so than hot Jupiters, and also which are old enough to have lost much of their internal heat of formation~\citep{guillot1996}; as a result, they exhibit cooler temperatures and are harder to observe. One might usefully define ``warm Jupiters" to be those objects with temperatures of 300 to 1000K (corresponding to distances from a sunlike star of approximately 1 to 0.1 AU) and ``cool Jupiters" to be those objects with temperatures less than 300 K (corresponding to orbital distances from a sunlike star exceeding 1 AU). Although the warm and cool Jupiters are harder to observe than hot Jupiters and directly imaged planets, far more have been discovered, and they will be increasingly amenable to observational characterization in the future { when more sensitive instruments will become available.}.

Brown dwarfs are objects thought to have formed like stars but which have insufficient mass to fuse hydrogen~\citep{chabrier2000,burrows2001}; they are typically defined as objects of $\sim$10 to 80 Jupiter masses (the stellar mass limit). Lacking strong internal thermonuclear heat generation, they cool off over time, but their super-Jovian mass implies that even after billions of years they may still exhibit atmospheric temperatures of 1000 K or more (e.g., \citealp{burrows2001}). Typically, they are isolated objects, far from any star, which makes them easier to observe than exoplanets. Although their formation mechanisms differ from EGPs, brown dwarfs share many physical similarities to the currently known directly imaged planets; from an atmospheric dynamics point of view, the former may be considered high-mass, high-gravity versions of the latter.

Although far less numerous than known EGPs and field brown dwarfs, a population of brown-dwarf companions to stars has also been discovered. Some such objects orbit sunlike stars in tight, several-day orbits and resemble high-mass, high-gravity versions of hot Jupiters (see a summary in \citealp{bayliss2016}). Other brown dwarfs orbit white dwarf stars so closely that the two objects nearly touch, and have orbital periods of just several hours (see a summary in \citealp{casewell2015}). In some such systems---the cataclysmic variables---the brown dwarf lies so close to the white dwarf that it continually sheds mass onto the white dwarf, while in other systems, the orbital separation (while still tight) is great enough to prevent mass exchange.

Despite this broad diversity, there is merit in considering these objects together as a class. They share in common the fact that they are all fluid, hydrogen-dominated objects; they have radii similar to Jupiter to within a factor of two; their atmospheres { merges} continuously into their interiors; and because of the large opacity and low viscosity of hydrogen in the conditions of their interiors, they are generally expected to lose their internal heat by convection, implying convective well-mixed interiors.\footnote{Whether the interior is indeed convective and well mixed depends on whether the internal heat flux is strong enough to overcome the barrier from molecular weight gradients created during the object formation
~\citep[e.g.][]{Leconte2017,Mankovich2016}.} Like all planetary atmospheres, giant planets and brown dwarfs should share in common many of the fundamental dynamical, physical, and chemical processes that shape the structure and circulation of atmospheres generally. Considering these objects together therefore provides a unique opportunity to better understand the physical and dynamical processes that operate in atmospheres over a wide range, and to understand how these common processes leads to diverse outcomes for different objects. This ``grand challenge" is far easier to complete for giant planets than for terrestrial planets because the atmospheres of giant planets are currently amenable to observational characterization over a wide range, whereas for terrestrial planets, such observational characterization over a comparably wide range is still decades away.

\begin{figure}      % use "figure*" instead of "figure" if you want your figure to span both columns
\epsscale{1.15}      % adjust this number to change the size of your figure
\includegraphics[scale=0.32]{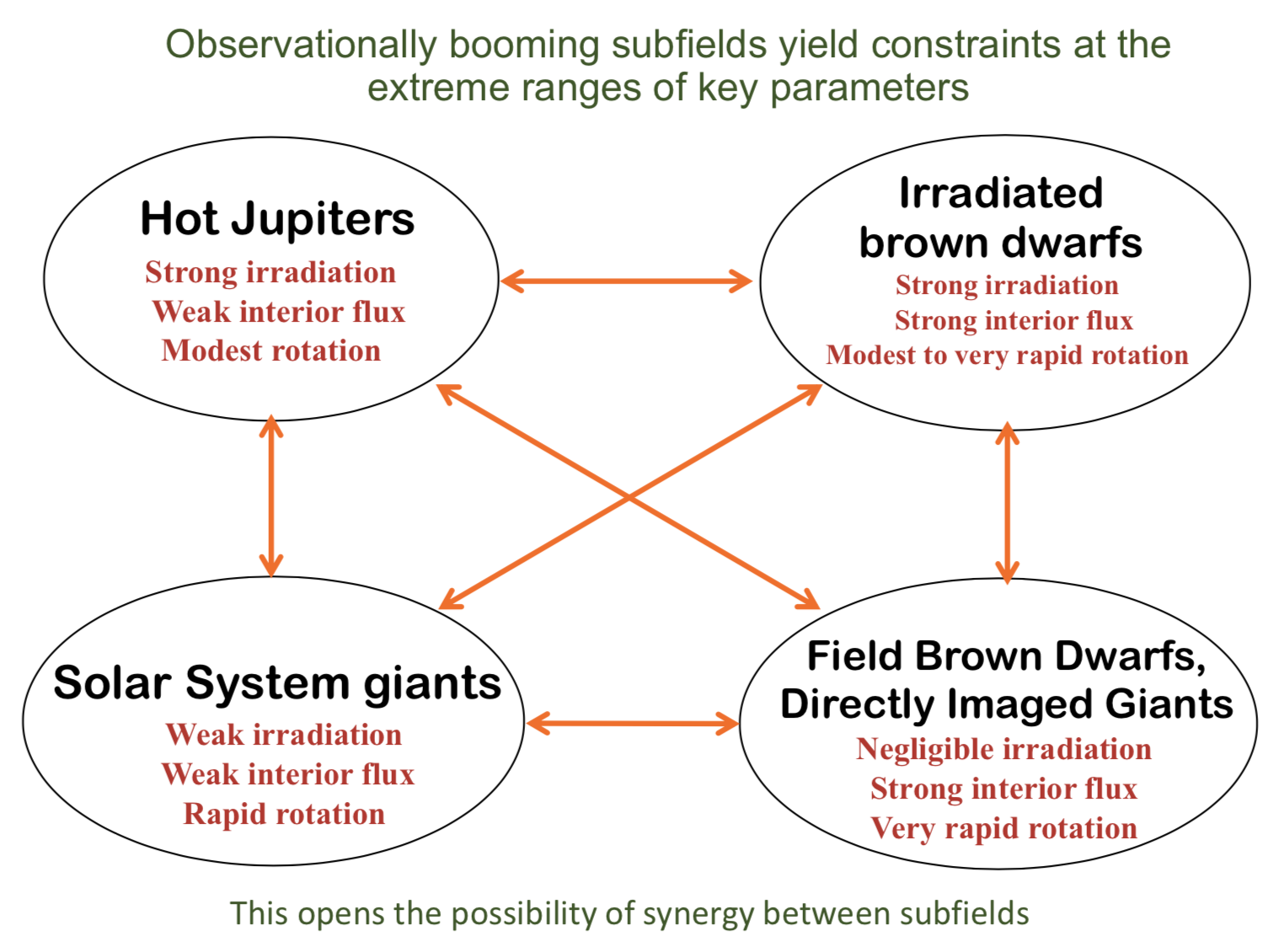}
\caption{Four distinct subfields of astronomy and planetary science are providing important constraints on atmospheric structure and circulation of giant planets---hot Jupiters; solar system giant planets; brown dwarfs and directly imaged giant planets; and irradiated brown dwarfs. They together span a wide range of physical properties. }
\label{fig.1}
\end{figure}

The atmospheric circulation and structure of giant planets is shaped by a variety of factors, including the external irradiation (that is, the radiative flux the atmosphere receives from a nearby star), the internal convective heat flux, the planetary mass (therefore gravity), the rotation rate, and the atmospheric composition, including the overall bulk metallicity, as well as specific elemental ratios such as C/O. Even among known objects, these factors vary over ranges of $\sim10^7$ , $10^6$ , $10^2$ , $10^2$, and $10^2$, respectively. Thus an enormous diversity is represented. Figure \ref{fig.1} summarizes the subpopulations where key observations constraining the atmospheric circulation have been obtained. Fortuitously, these populations span a wide range in several of the above parameters. Jupiter and Saturn exhibit weak irradiation, weak interior heat flux, and rotate rapidly (a factor that strongly influences the dynamical regime). Hot Jupiters are strongly irradiated, but likely have small interior heat fluxes (perhaps within an order of magnitude of Jupiter), and due to tidal locking, they are expected to have modest rotation rates. Field brown dwarfs and directly imaged giant planets receive negligible irradiation from their stars, but have enormous interior heat fluxes and are rapidly rotating—in many cases faster than Jupiter. \\

\begin{figure*}      % use "figure*" instead of "figure" if you want your figure to span both columns
\epsscale{1.}      % adjust this number to change the size of your figure
\includegraphics[scale=0.4]{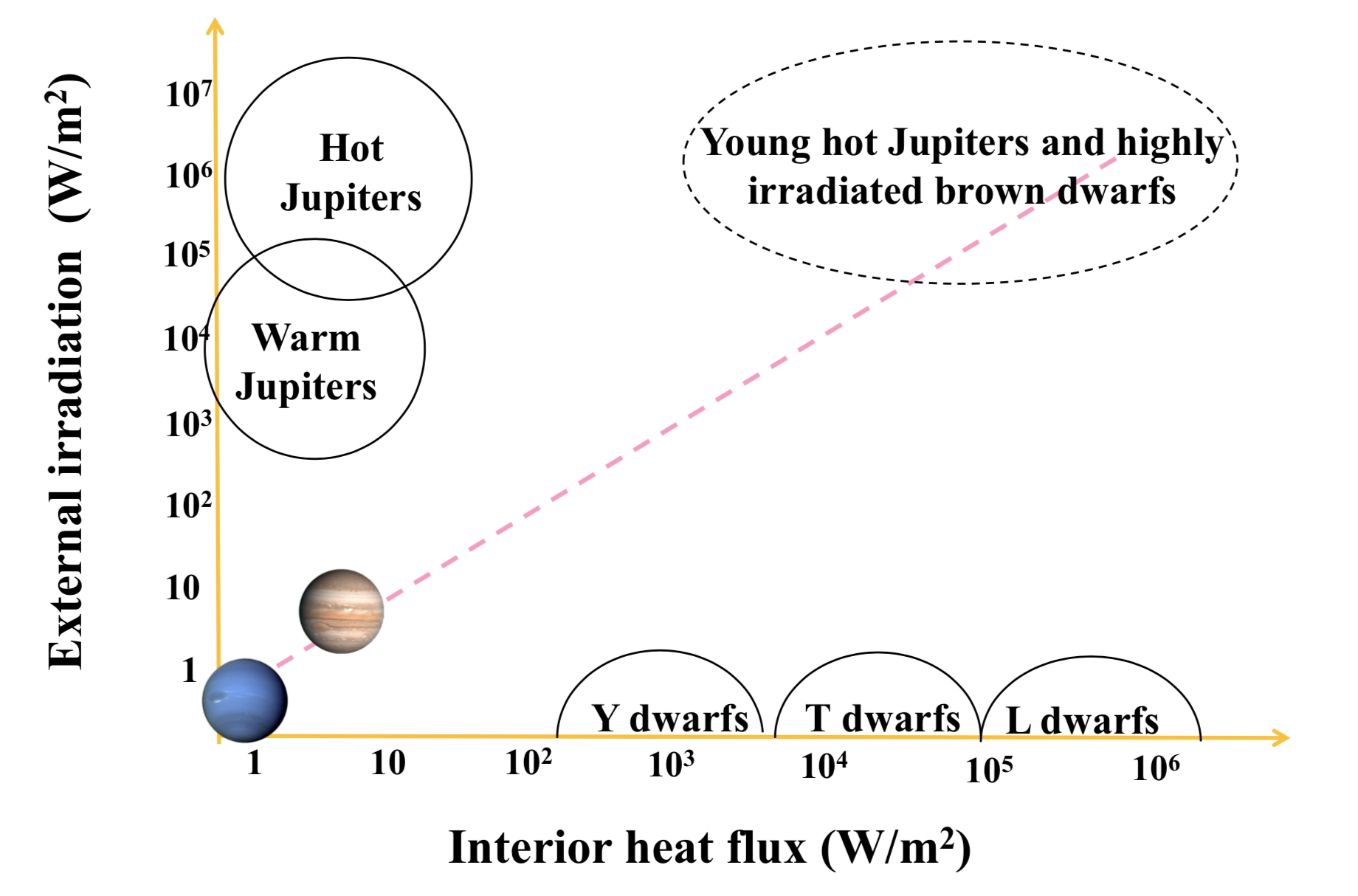}
\centering
\caption{Regime diagram of various giant planet populations as a function of interior heat flux (abscissa) and global-mean energy flux received from the star (ordinate). Solar system giant planets lie in the lower left corner, hot Jupiters in the upper left corner, and brown dwarfs in the lower right corner. A handful of highly irradiated brown dwarfs are also providing information and reside in the upper right corner. Observational constraints from these diverse populations place tight constraints on how fundamental theories for the atmospheric circulation scale with these heat fluxes over many orders of magnitude. This paper will provide a tour across this parameter space.}
\label{fig.2}
\end{figure*}

To highlight the wide parameter space involved, Figure \ref{fig.2} shows these diverse populations on a parameter space of external irradiation and internal heat heat flux, which are two of the factors that matter most for driving an atmospheric circulation, and which vary over the widest range across the known population. Jupiter, Saturn, Uranus, and Neptune occupy the ``weak forcing" regime near the lower left corner, with external and interior heat fluxes that are small and comparable. Jupiter exhibits internal and irradiation fluxes of 7.5 \wpm and 6.6 \wpm \citep{li2018} absorbed and internal fluxes are 0.27 and 0.70 W m$^{-2}$ \citep{pearl1991}. In contrast, hot Jupiters reside in the upper left corner, with external fluxes of $10^4$ to $10^7$ \wpm, and internal fluxes that are expected to be far smaller. The hot Jupiters show us how giant planets behave when external forcing dominates. Brown dwarfs embody the opposite extreme, with enormous interior fluxes of $10^3$ to $10^6$ \wpm and typically negligible external irradiation. Lying in the lower right corner of Figure \ref{fig.2}, brown dwarfs yield insights on giant-planet behavior under extreme internal fluxes when external forcing is zero. Finally, the irradiated brown dwarfs represent the ``strong forcing" regime in both internal and external fluxes; they occupy the upper right corner of Figure \ref{fig.2}.

Our grand challenge is to understand the atmosphere and interior circulation and structure on giant planets and brown dwarfs, broadly defined. Key questions include the following:
\begin{itemize}
    \item What is the nature of the atmospheric circulation--- including the distribution and importance of zonal jets, vortices, storms, waves, and turbulence? What are the characteristic wind speeds, temperature variations, length scales, and time variability? How do they depend on parameters? These questions could be viewed as one of {\it characterization}, both from observations and careful numerical experiments.
    
    \item How does the circulation work? What are the dynamical mechanisms controlling it? How does the interplay of these mechanisms lead to various outcomes in the behavior of the global circulation across the wide parameter space occupied by giant planets? What is the link between these mechanisms and those mechanisms well-known from study of more familiar atmospheres like Earth and Jupiter? This could be viewed as a question of {\it understanding} the behavior characterized in the first point. 
    
    \item What is the role of condensation and clouds? How critical are they to driving (or influencing) the circulation and climate, via radiative feedbacks, latent heating, re-distribution of condensable chemical species, or other mechanisms?
    
    \item What is the role of coupling between atmospheric dynamics, { radiative transfer} and chemistry?
    
   \item { How do the magnetic field and the atmospheric circulation interact with each other, particularly in the hottest atmospheres?}
    
    \item Can we achieve a unified theory of giant planet atmospheric circulation that explains observations of hot Jupiters, brown dwarfs, and solar system giant planets? 

    \item Does this knowledge provide insights into the circulation and climate of (less easily observed) smaller planets, including habitable terrestrial planets?
\end{itemize}

Our aim is to broadly survey the atmospheric circulation across these diverse classes of giant planets and brown dwarfs. We place particular attention on dynamical mechanisms and the way they vary among these populations. This review can be viewed as a guided tour through Figure \ref{fig.2}, starting with hot Jupiters in the upper left corner (Section \ref{section.2}), moving downward to the warm Jupiters (Section \ref{section.3}), hopping across to the brown dwarfs in the lower-right corner (Section \ref{brown-dwarfs}), and then finishing with irradiated brown dwarfs in the upper-right corner (Section \ref{irradiated-brown-dwarfs}). Given the existence of many prior reviews of atmospheric dynamics on Jupiter and Saturn, we touch on the dynamics of solar system planets only briefly, as points of reference with hotter giant planets. Our review updates prior reviews on the atmospheric dynamics of exoplanets (\citealp{showman2008b,showman2010,showman2013,heng2015})  and complements the many excellent reviews covering observations and radiative structure of EGPs (e.g. \citealp{deming2009,deming2017,seager2010,burrows2014,madhusudhan2014,madhusudhan2019,zhang2020}) and brown dwarfs (\citealp{stevenson1991,burrows2001,helling2014,marley2015,biller2017,zhang2020}). 

\section{Hot  Jupiters}
\label{section.2}

EGPs orbiting extremely close to their stars---hot Jupiters---comprise the most easily characterizable type of exoplanet, and therefore our understanding of these exoplanets is best developed. As little as 0.03–0.05 AU from their parent star, hot Jupiters have orbital periods of just a few Earth days. The immense stellar irradiation heats their atmospheres to temperatures of $\sim$1000– 3000K, and they therefore radiate enormous IR heat fluxes to space, promoting direct detection of their thermal emission. Close-in planets such as hot Jupiters are also more likely to transit their stars—the transit probability for a planet on a 0.05 AU, randomly inclined orbit around a sunlike star is $\sim$10\%, versus 0.1\% for a planet at Jupiter’s distance---and when such a transiting planet is detected, it enables the determination of the planet's radius and allows atmospheric characterization through a wide suite of observation methods.

\begin{figure*}      % use "figure*" instead of "figure" if you want your figure to span both columns
\epsscale{1.}      % adjust this number to change the size of your figure
\includegraphics[scale=0.4]{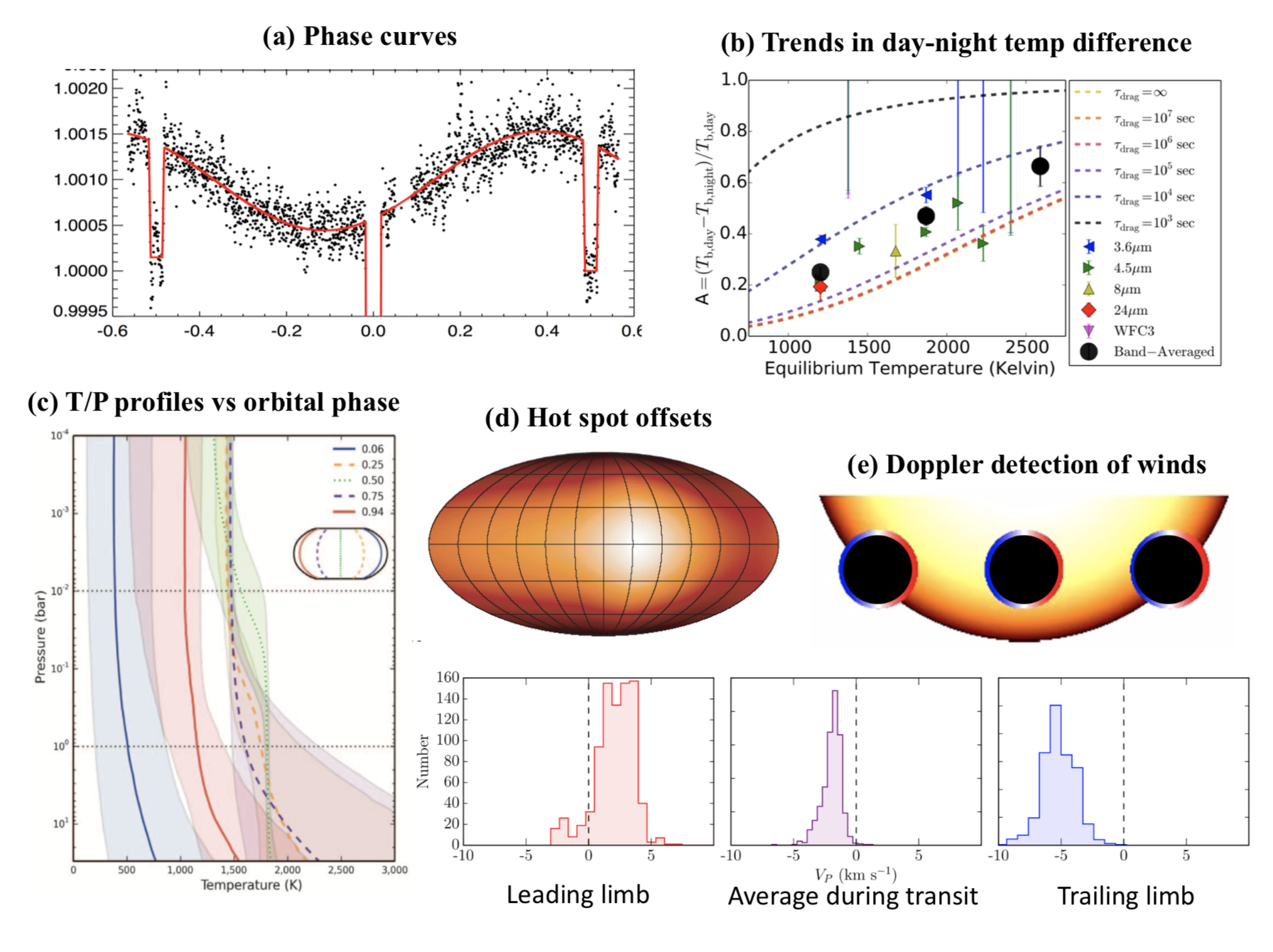}
\centering
\caption{{ Hot Jupiter observations and dynamical insights that can be gather from them: {\bf (a)} Infrared phase curves of the hot Jupiter HD209458b observed with Spitzer at 4.5 $\mu m$. {\bf (b)} Day-night temperature contrast estimated from phase curve observations~\citep{komacek2017}. {\bf (c)} Thermal structure at different phases estimated based on the spectral phase curve of WASP-43b observed by HST/WFC3~\citep{stevenson2016}. {\bf (d)} Brightness map of HD189733b reconstructed from the $8\mu m$ phase curve observed by Spitzer
~\citep{knutson2007}. {\bf (e)} Doppler measurements at the planetary limb during transit allow direct detection of atmospheric winds on the different limbs for the hot Jupiter HD189733b~\citep{louden2015}.}}
\label{fig.3}
\end{figure*}

Hot Jupiters are too close to their stars to be distinctly resolvable from their stars in images---what we observe is the combined light from the planet-star system---and indirect methods are therefore needed to tease apart the planetary light from the starlight. When the planet passes behind its star---an event known as secondary eclipse---only the star is visible. Subtracting the total system flux received during secondary eclipse from that received immediately before and afterward (when both the planet and the star contribute to the combined light) yields a spectrum of the planet, and in particular of the planet's dayside. Moreover, observing the planet throughout its orbit---as its dayside and nightside rotate in and out of view---allows the measurement of the phase variations of the planet's outgoing IR flux, and thereby provides inferences on the longitudinal variation of temperature near the photosphere (Figure \ref{fig.3}). In particular, such light curve observations yield day-night temperature differences and offsets of the hot spot from the substellar point. And observations during transit---when the planet lies in front of its star---probe the atmosphere in transmission. Starlight passing through the planet's atmosphere on its way to Earth is preferentially blocked at wavelengths where the atmosphere is more absorbing; therefore, the planet essentially appears bigger at wavelengths corresponding to atmospheric absorption, and a spectrum of the planet's size is essentially a transmission spectrum of the planet's atmosphere at its terminator. For recent reviews of these and other methods for characterizing EGP atmospheres, see \cite{deming2017} and \cite{madhusudhan2019}.

The atmospheric regime of hot Jupiters differs radically from that of any planet in the solar system. Hot Jupiters are blasted by starlight---they typically receive stellar fluxes of $10^5–10^6$ \wpm (Figure \ref{fig.2}), which is $\gtrsim10^4$ times the flux received by Jupiter and hundreds to thousands of times that received by Earth. Moreover, the close-in orbital distances lead to strong tidal forces that slow their spin; for example, a typical hot Jupiter orbiting at 0.05 AU around a Sun-like star has a synchronization timescale of $\sim$$10^6$ years, which is orders of magnitude shorter than the typical multi-Gyr system ages \citep{guillot1996}. Therefore, hot Jupiters are generally presumed to be synchronously rotating, with permanent daysides and nightsides. This property, coupled with the intense irradiation, leads to a unique climate regime of permanent day-night forcing, which is not experienced by any planet in the solar system. 

In what follows, we first sketch out the basic dynamical regime of hot Jupiters before proceeding to a detailed discussion from simulations and theory of the atmospheric circulation and the processes that maintain it. We also cover various topics of current interest including coupling of the dynamics to clouds and chemistry.

\subsection{Basic dynamical arguments}

The relative slowness of rotation (compared to Earth, Jupiter, and brown dwarfs) { causes} several important effects. First, it implies that the dynamical length scale on hot Jupiters will be relatively large, approaching the global scale \citep{showman2002,menou2003,cho2003}. One measure of this effect is the equatorial deformation radius, given by
\begin{equation}
L_{\rm eq}=\left(\frac{NH}{\beta}\right)^{1/2}
\label{eq.1}
\end{equation}
where $N$ is the Brunt-Vaisala frequency, $H$ is the pressure scale height, and $\beta$ is the gradient of Coriolis parameter $f$ with latitude, that is $\beta = df/dy$, where $f = 2\Omega\sin{\phi}$, $\Omega$ is the planetary rotation rate ($2\pi$ over the rotation period), $\phi$ is latitude, and $y$ is northward distance. On a sphere, $\beta = 2\Omega/a$ at low latitudes, where $a$ is the planetary radius. The quantity $NH$ can be thought of as the phase speed of horizontally propagating gravity waves, which for a vertically isothermal atmosphere is just $R\sqrt{T/c_p}$, where $R$ and $c_p$ are the specific gas constant and specific heat at constant pressure, and $T$ is temperature. Under typical hot Jupiter conditions, these expressions yield an equatorial deformation radius $L_{\rm eq} \sim 5 \times 10^4$ km---more than half a planetary radius.\footnote{For vertically isothermal conditions, $R = 3700$ \cpunit and $c_p = 1.3 \times 10^4$ \cpunit relevant to a $\rm{H_2}$-He atmosphere, and $T \approx 1500$ K appropriate for a typical hot Jupiter, we obtain $NH \approx 1250$ \mps. Adopting a rotation period of 3 Earth days yields $\Omega = 2.4\times10^{-5} \rm{s^{-1}}$ and, with a radius of $8\times10^7$ m, implies that $\beta = 6\times10^{-13} \rm{m^{-1}\;s^{-1}}$. Together these imply $L_{\rm eq} \approx 5\times10^7$ m.
\citep{showman2011}. }
 In contrast, for Earth and Jupiter, respectively, the equatorial deformation radius is about 30\% and 11\% of the planetary radius, respectively.

Thus one expects that the dynamically ``tropical" conditions { (i.e.,  where the Coriolis force is not dominant in the horizontal force balance)} that prevail within a deformation radius of the equator---including equatorially trapped waves and the effect they exert in adjusting the planet's atmosphere--- will extend to significantly higher latitudes than they do on Earth and especially Jupiter. Indeed, the equatorial regions comprise a waveguide, of meridional half-width $L_{\rm eq}$, for a wide population of tropical baroclinic waves \citep{andrews1987,holton2012}. These waves are confined to very low latitudes on Jupiter and brown dwarfs but can extend meridionally to mid-latitudes or farther on typical hot Jupiters.

Outside the equatorial waveguide, the deformation radius is given by
\begin{equation}
L_D=\frac{NH}{f}.
\label{eq.2}
\end{equation}
Inserting typical numbers for a hot Jupiter yields $L_D \approx 4 \times 10^7$ m, again about half the planetary radius.

An alternate measure of the role of rotation comes from directly assessing the amplitude of Coriolis forces relative to other forces in the equation of motion. This can be accomplished by using the Rossby number, which represents the ratio of the characteristic amplitude of advection forces per mass, $U^2/L$, to Coriolis accelerations, $fU$, in the horizontal equation of motion, where $U$ and $L$ are the characteristic horizontal wind speed and length scale. Taking the ratio of these forces, we have that the Rossby number is $Ro = U/fL$. Typical wind speeds in hot Jupiters are expected to be $\sim$$1–3$ \kms, and adopting a rotation period of 3 days and a length scale of half a planetary radius, we obtain $Ro \approx 0.6-2$ at midlatitudes. With Rossby numbers of order unity, hot Jupiters are thus transitional between regimes where the rotation plays little role ($Ro \gg 1$) and where it dominates the dynamics ($Ro \ll 1$).  As we will discuss later, the large-scale circulation away from the equator on Jupiter, Earth's atmosphere and oceans, and probably brown dwarfs are in the latter regime, which will lead to differing dynamics between hot Jupiters and these other planets. Note that because the Coriolis parameter goes to zero at the equator, the Rossby number always tends to become large near the equator, and in fact one useful dynamical measure of the ``tropics" corresponds to the range of latitudes within which $Ro \gtrsim 1$ (e.g. \citealp{showman2013}). According to this measure, some hot Jupiters--- particularly those with faster wind speeds and/or slower rotation rates—will be ``all tropics" worlds where the Rossby number always exceeds one even at the poles; in contrast, other hot Jupiters with faster rotation and/or slower wind speeds may exhibit a transition where the low and midlatitudes comprise the tropics but the poles exhibit a more extratropical ($Ro \lesssim 1$) behavior.
 
The high temperatures of hot Jupiters imply that, in the observable atmosphere, they will experience short radiative time constants. Near the photosphere, the radiative time constant is approximately \citep{showman2002}
\begin{equation}
    \tau_{\rm rad} = \frac{pc_p}{4g\sigma  T^3} \sim 10^5\left(\frac{p}{0.3 ~\rm{bar}}    \right)\left(\frac{1200~ \rm{K}}{T}\right)^3 \rm{s},
    \label{eq.3}
\end{equation}
where $\sigma$
$\sigma$ is the Stefan-Boltzmann constant, and $p$ should be interpreted as a pressure near the IR photosphere. Note that, because of the $T^3$ dependence, the radiative time constant varies significantly across the hot Jupiter population, from $\sim$$10^4$ s for ultra hot Jupiters, to $10^5$ s for intermediate-temperature planets like HD 189733b, to $10^6$ s for warm Jupiters.
 
 What are the expected wind speeds, to order of magnitude? A simple estimate can be obtained by balancing the pressure-gradient force that drives the flow with the greater of the Coriolis force, advective forces, or drag force (if present) in the horizontal momentum equation. Generally, if the frictional drag is weak, and if $Ro \ll 1$, then the balance is between the pressure-gradient and Coriolis force. This leads to the well-known thermal-wind equation \citep{vallis2006,holton2012}:
 \begin{equation}
     f\frac{\partial u}{\partial \ln p} = R\frac{\partial T}{\partial y},
     \label{eq.4}
 \end{equation}
 where $u$ is zonal wind, $R$ is the specific gas constant, and $y$ is northward distance on the sphere. To order of magnitude, this equation yields
 \begin{equation}
     U\sim \frac{R\Delta T_{\rm horiz} \delta \ln p}{fL},
     \label{eq.5}
 \end{equation}
 where $\Delta T_{\rm horiz}$ is the characteristic horizontal temperature difference, assumed to extend vertically over a number of scale heights $\delta \ln p$, $L$ is the characteristic horizontal length scale, and $U$ is the difference in characteristic horizontal wind speed between some upper level of interest (e.g., the photosphere) and deeper levels. This should be interpreted as the characteristic eddy speed (e.g., associated with the wind flow from day to night) and not the speed of equatorial superrotation. Alternately, if $Ro \gtrsim 1$, then the pressure gradient forces are typically balanced by horizontal advection forces\footnote{One can also have cyclostrophic balance, which on the sphere is between the pressure-gradient force and a metric term, but the scaling is the same.}. To order of magnitude, the former is $R\Delta T_{\rm horiz} \delta \ln p/L$, and the latter is $U^2/L$, so their balance implies a wind speed
 \begin{equation}
     U\sim(R\Delta T_{\rm horiz} \delta \ln p)^{1/2}. 
     \label{eq.6}
 \end{equation}
 Since $Ro \sim 1$ on typical hot Jupiters, the Coriolis and advection forces are comparable, and we would expect that these two expressions would yield similar estimates. Indeed, when we plug in typical numbers (e.g., $R$ = 3700 \cpunit, $\Delta T_{\rm horiz} \approx 400$ K, $\delta \ln p \approx 3$, $f \approx 3 \times 10^{-5}$, and adopting $L$ of a Jupiter radius), we obtain $U \approx 2$ \kms  from both estimates.
 
 \subsection{GCM experiments and comparison to observations}
 
 General circulation models (GCMs) for hot Jupiters have been developed using a variety of different codes and numerical approaches. Most commonly, these solve the primitive equations of atmospheric dynamics over the full globe, assuming that the circulation is driven by the intense stellar irradiation gradient, under conditions of synchronous rotation. Consistent with the expectations described above, these models typically assume a thermal structure that is deeply stratified throughout the atmosphere. The vertical domain typically extends over many pressure scale heights from a pressure of $\lesssim 1$ mbar at the top, to commonly $\sim$100 bars at the bottom.
 
\begin{figure*}      % use "figure*" instead of "figure" if you want your figure to span both columns
\epsscale{1.}      % adjust this number to change the size of your figure
\includegraphics[scale=0.7]{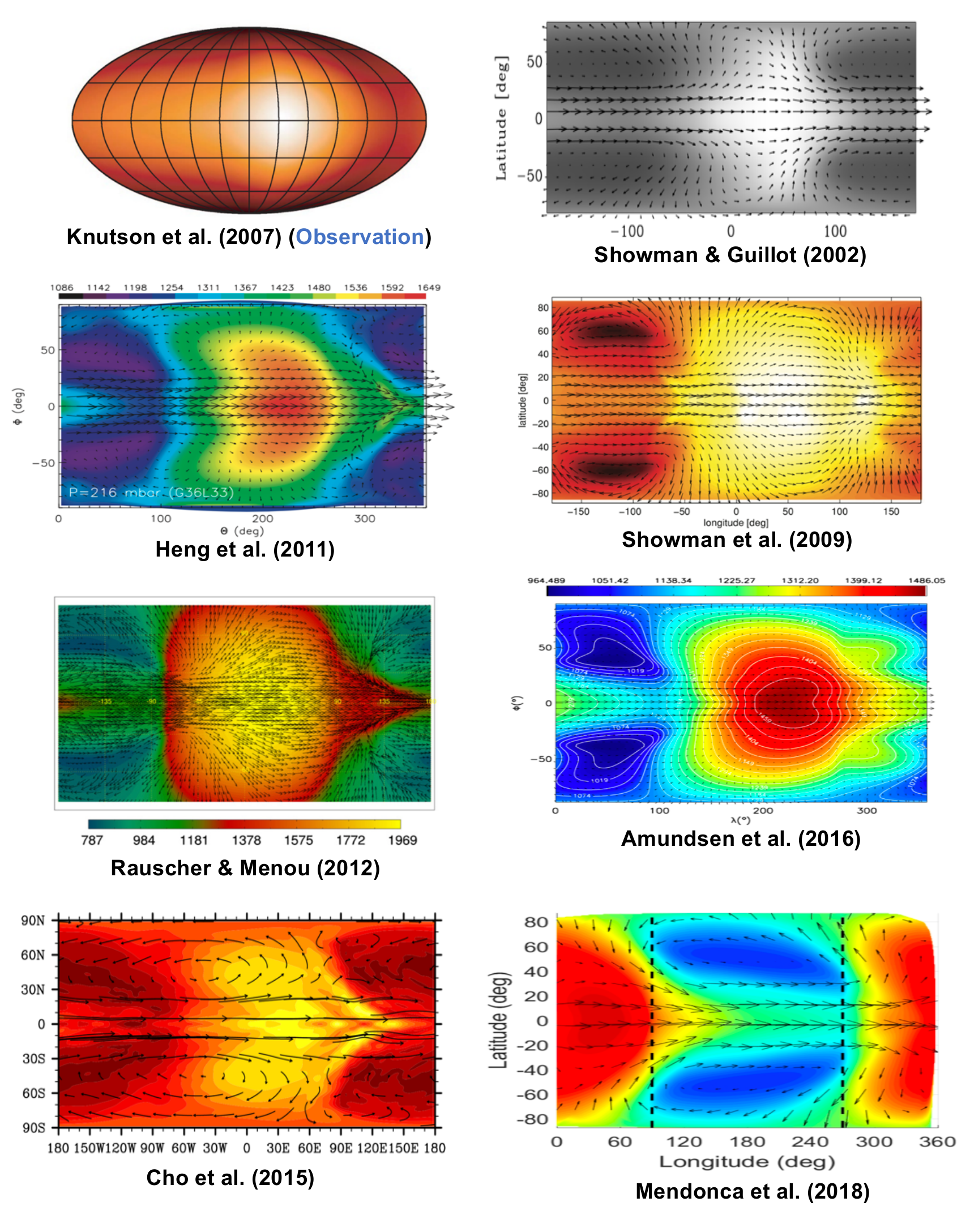}
\centering
\caption{Example GCM simulations of hot Jupiters from a variety of groups in the field. Despite differences in forcing setup, numerics, and other factors, all the models exhibit similar circulation regimes, comprising significant day-night temperature differences, a fast eastward equatorial jet, and an eastward shifted dayside hot spot. All models assume synchronously rotation and conditions appropriate for typical hot Jupiters. Top left shows observations of HD 189733b from \cite{knutson2007}. Simulations of HD 209458b are shown from \cite{showman2002}, \cite{heng2011}, \cite{rauscher2012b}, \cite{amundsen2016}, \cite{cho2015}. Simulations of HD 189733b from \cite{showman2009}. Simulations of WASP-43b  from \cite{mendonca2016}. These seven simulations were performed with totally distinct numerical codes, involving varying approximation of radiative forcing, and using seven independent dynamical cores. For each image, the substellar point is in the center of the panel, except for \cite{mendonca2018}), where the antistellar point is in the center.}
\label{fig.4}
\end{figure*}

 These GCMs predict atmospheric flows comprising several key features, including (1) eddy and jet structures of near-global scale; (2) large day-night temperature differences reaching hundreds of K; and, (3) most interestingly, a wide, fast eastward equatorial jet---so-called equatorial superrotation---which straddles the equator, extends to latitudes $\sim$$30^{\circ}$, and achieves zonal-mean zonal wind speeds of typically 2-4 \kms (Figure \ref{fig.4}). Despite the synchronous rotation---which in radiative equilibrium would lead to a temperature field comprising a simple day-night temperature difference with the hottest regions at the substellar point---the dynamics distorts the temperature structure in a complex manner, most prominently by inducing an eastward displacement of the day-side hot spot by tens of degrees longitude from the sub-stellar point. The earliest GCMs capturing these general features preceded observations \citep{showman2002,cooper2005}, and predicted that the large day-night temperature differences and eastward offsets would be detectable in IR lightcurves of these planets. This helped motivate searches for these features in IR light curves. Observations from the Spitzer Space Telescope first confirmed this prediction for the hot Jupiter HD 189733b \citep{knutson2007}, and subsequent full-orbit IR light curve observations from the Spitzer and Hubble Space Telescopes have detected such an eastward offset on the majority of hot Jupiters that have been observed (Figure \ref{fig.3}; for reviews, see \citealp{heng2015,parmentier2015,parmentier2018review}).

Over the last 15 years, many additional atmospheric dynamics models have been employed to investigate the global circulation of hot Jupiters; interestingly, most of these models agree reasonably well in their qualitative predictions, including the presence of large day-night temperature differences, equatorial superrotation, and---under appropriate conditions---eastward shifted hot spots (e.g., \citealp{menou2009,rauscher2010,rauscher2012,rauscher2013,DobbsDixon2008,dobbsdixon2013,heng2011,heng2011b,mayne2014,mayne2017,showman2008,showman2009,showman2013b,showman2015,parmentier2013,parmentier2016,kataria2015,kataria2016,lewis2010,lewis2013,cho2015,zhang2017,menou2019,tan2019,mendonca2020}). Figure \ref{fig.4} presents representative snapshots from several different groups, which, despite the many differences in model setup, highlight the overall similarity in qualitative regime across these diverse models. The earliest hot-Jupiter GCMs adopted extremely idealized schemes to represent the day-night thermal forcing; more recent work has implemented radiative transfer schemes of varying levels of realism that represent the absorption of incoming starlight and radiation of thermal IR.
 
Interestingly, the qualitative properties of the circulation in these models---including the emergence of equatorial superrotation---seem to be robust to model assumptions, numerics, and the detailed formulation of the day-night thermal forcing. This contrasts with Jupiter and Saturn, for which circulation models can readily predict either eastward or westward\footnote{``Eastward" refers to the same direction as the solid-body rotation, whereas ``westward" refers to the opposite.}  equatorial jets depending on the detailed model assumptions (for reviews, see \citealp{vasavada2006,delGenio2009,showman2018review}). This suggests that the mechanism that causes the superrotation on hot Jupiters is extremely robust. 

Observed IR light curves of hot Jupiters are now sufficiently accurate that detailed comparison to GCM experiments is a fruitful exercise. Performing such comparisons self-consistently requires a model that represents the heating/cooling with an explicit radiative transfer scheme.\footnote{The simplest approaches to forcing a day-night temperature difference, such as a Newtonian cooling scheme, are ideal for understanding dynamical mechanisms; however, they do not include a formal radiative energy budget, so it is difficult to formally compare their output to those of observed IR lightcurves.} Most GCMs to date represent the angular dependence of radiation using the two-stream approximation. Approaches to treating the opacities come in several flavors. The simplest is a dual-band (double grey) scheme, with one opacity band in the IR and one in the visible (e.g., \citealp{heng2011b,rauscher2012,rauscher2013,rauscher2014,perna2012,komacek2017,tan2019,mendonca2018,flowers2019}). The advantage of this approach is its simplicity---it is ideal for wide parameter explorations, understanding dynamical processes, and capturing the bulk radiative budget, which is sufficient for many applications.
 
 On the other hand, the gaseous radiative transfer is inherently non-grey, with opacities that vary by orders of magnitude from wavelength to wavelength. For the price of increasing the model complexity, capturing the non-grey behavior affords two advantages. First, it allows a more realistic representation of the radiative heating/cooling, so that the overall thermal structure is accurately simulated. Second, IR spectra and lightcurves at different wavebands probe different pressures, where the temperature differ. As a result, IR spectra and lightcurves indicate that hot Jupiters are inherently non-grey bodies. Accurately capturing the wavelength-dependence of IR spectra and lightcurves requires non-grey radiative transfer. Several hot-Jupiter GCMs have been developed that treat the opacities using the well-known correlated-k method: the SPARC/MITgcm \citep{showman2009,lewis2010,kataria2013}, and the Exeter group’s hot-Jupiter implementation of the UK Met Office GCM \citep{amundsen2014,amundsen2016}. In this approach, opacities are treated by dividing the spectra into typically 10-40 wavelength bins, and then statistically representing the opacity information from $\sim$$10^4-10^5$ wavelength points within each bin. This allows accuracy of typically 1\% or better in heating rates \citep{showman2009,kataria2013,amundsen2014} while retaining much greater computational efficiency than a line-by-line radiative transfer calculation, which is computationally prohibitive in GCMs. Intermediate approaches include bin methods, which divide the spectrum into several dozen wavelength bins, but treat the opacity as essentially grey within each bin \citep{dobbsdixon2012,dobbsdixon2013}. A key point is that none of these differing approaches are inherently better or worse; they should be viewed as distinct tools useful for different applications.
 
 Detailed comparisons between IR lightcurve observations (obtained with Spitzer and Hubble) and GCM simulations including radiative transfer have been performed for a wide range of hot Jupiters, including HD 189733b \citep{showman2009,knutson2012,dobbsdixon2013,drummond2018,steinrueck2019,flowers2019}, HD 209458b \citep{zellem2014,amundsen2016,drummond2018b}, WASP-43b \citep{kataria2015,stevenson2017,mendonca2018}, WASP-18b \citep{Arcangeli2019}, WASP-19b \citep{wong2016}, HAT-P-7b \citep{wong2016}, WASP-103b \citep{kreidberg2018}, WASP-121b \citep{parmentier2018}, HAT-P-2b \citep{lewis2014}, and HD 80606b \citep{lewis2017}. Most of these studies compute the radiative transfer driving the GCM assuming a solar-composition, cloud-free hydrogen atmosphere that is in local chemical equilibrium, and perform straight-up comparisons of the observed lightcurves to a nominal model, without any special attempts to search for an exact match via model tuning. Nevertheless, a few of the models explore the influence of supersolar metallicities, nonsynchronous rotation, specified atmospheric drag (perhaps representing the effect of Lorentz forces), disequilibrium chemistry, or specified haze distributions on the lightcurves.
 
\begin{figure*}      % use "figure*" instead of "figure" if you want your figure to span both columns
\epsscale{1.}      % adjust this number to change the size of your figure
\includegraphics[scale=0.9]{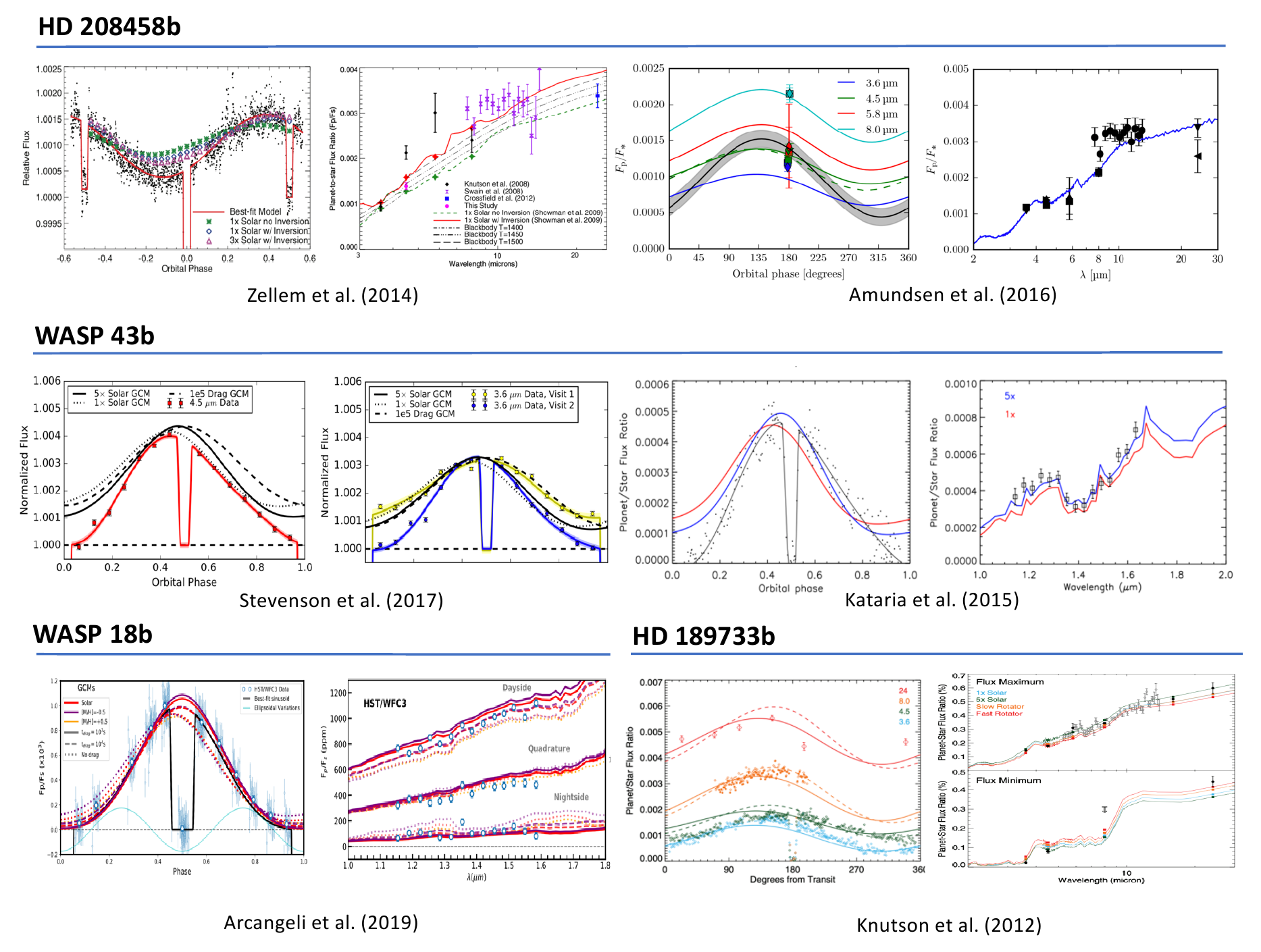}
\centering
\caption{{ Observed thermal phase curves of benchmark hot Jupiters exoplanets compared to 3D global circulation models outputs. For HD209458b (top) we show the comparison of the~\citet{zellem2014} Spitzer observation at $4.5\mu m$ with the SPARC/MITgcm output (left) and with the UK Met office Unified Model (right). For WASP-43b (middle row) we show the comparison between the SPARC/MITgcm and, from left to right, the Spitzer $4.5\mu m$ phase curve, the Spitzer $3.6\mu m$ phase curve, the HST/WFC3 phase curve and the HST/WFC3 dayside spectrum. For WASP-18b (bottom row left) we show the comparison between the HST/WFC3 spectral phase curve and spectrum at several phases and the SPARC/MITgcm output. Finally for HD189733b we show the four Spitzer phase curve and the related day and nightside spectrum and compare them to the SPARC/MITgcm outputs.}}
\label{fig.5}
\end{figure*}

 Generally speaking, the lightcurves predicted from these models capture the qualitative features of the observations---including large day-night flux differences and phase offsets wherein the fluxes peak before secondary eclipse, as expected due to the eastward offset of the dayside hotspot from the substellar point\footnote{We note that in one case, CoRoT-2b, the offset was claimed to be westward
~\citep{Dang2018}, but this remains an outlier.}. Given the lack of tuning, these aspects of qualitative agreement could be viewed as major successes suggesting the models are in approximately the correct regime. However, the GCM lightcurves also exhibit significant discrepancies from the observations---they tend to overpredict the hotspot offset, overpredict the nightside flux, underpredict the dayside flux, and therefore underpredict the day-night flux difference (i.e., the phase curve amplitude). Figure \ref{fig.5} shows examples for the benchmark planets HD 189733b, HD 209458b, WASP-43b and WASP-18b that highlight these successes and failures~\citep[see also][for a qualitative comaprison]{parmentier2018review}. Although the degree of discrepancy varies from planet to planet, the persistence of these aspects of discrepancy across the hot-Jupiter sample suggests that the models may consistently be missing one or more crucial ingredients.

As yet, there exists no consensus on what these missing factors may be, although several authors have explored the influence of various factors in an attempt to better match the observed lightcurves. Super-solar metallicities (e.g., 5 or 10 times solar) tend to increase the day-night flux differences and decrease the hotspot offset \citep{showman2009,lewis2010,kataria2015}, which improves the agreement for certain planets, but generally do not solve the overall problem. For some planets, the addition of atmospheric drag improves the agreement (e.g., \citealp{kreidberg2018} for WASP-103b, and \citep{Arcangeli2019} for WASP-18b); however, the Lorentz forces such drag parameterizes are only relevant at hot temperatures ($\gtrsim$ 2000 K), and so this effect cannot provide a solution for the cooler planets in Figure \ref{fig.5}. Chemical disequilibrium between CH4 and CO was suggested as a possible culprit for HD 189733b and HD 209458b \citep{knutson2012,zellem2014}, but recent GCM experiments rule out this explanation \citep{steinrueck2019,drummond2018,drummond2020}. More likely is that the discrepancies result from clouds and/or hazes, although this possibility remains to be quantified in detail.

Measurements of the Doppler shift of the atmospheric winds on the planet's terminator, as detected during transits, has been used to characterize the wind regime for several hot Jupiters. Using terminator-averaged measurements, \cite{snellen2010} showed that the high-altitude winds on HD 209458b exhibit a preferential blueshift of 2 \kms , indicating that the atmospheric circulation comprises net flow toward Earth at high altitude (with the return flow presumably occurring at a deeper, unobserved level). \cite{louden2015,louden2019ESS} used transit ingress and egress measurements of HD 189733b and WASP-49b to discriminate between the leading and trailing limbs, and they showed that both planets exhibit equatorial superrotation (corresponding to a redshift on the leading limb, as the air flows from the planet's nightside to dayside, and a blueshift on the trailing limb, carrying the return flow from the dayside back to the nightside). For HD 189733b, the observations indicate eastward velocities on the leading and trailing limbs of about 3-4 \kms and 4-6 \kms, respectively. For WASP-49b, their study likewise identified atmospheric superrotation, and they were further able to determine that the circulation over both poles flows from day to night.

Motivated by these measurements, several GCMs investigations have explicitly determined the Doppler signatures that would result from the atmospheric circulation, and made predictions for various planets \citep{kempton2012,showman2013b,rauscher2014,flowers2019}. These models show that at high altitude (pressures of $\sim$0.01-1 mbar, the approximate levels sensed by the Doppler measurements), the circulation tends to comprise a day-night flow at high latitudes combined with a superrotation at lower latitudes, yielding a net signature where the windflow is redshifted at low latitudes on the leading terminator and blue shifted everywhere else along the terminator. These studies show that the relative contributions of the superrotation and the day-night flow to the limb signature depend on the stellar irradiation, planetary rotation rate, and the exact altitude sensed (which depends on the atmospheric composition), among other factors. The addition of sufficiently strong frictional drag or Lorentz forces can damp the jet, causing the day-to-night flow regime to predominate, although the regimes under which this can occur need further study. Generally, the GCM results agree reasonably well with the observed signatures, lending confidence that superrotation exists on hot Jupiters and that the GCMs are in approximately the correct regime.

\subsection{Dynamical mechanisms: superrotation}
We next turn to the dynamical mechanisms responsible for maintaining the flow, starting with the eastward equatorial jets prominent in hot Jupiter simulations (Figure \ref{fig.4}), as well as the mechanisms responsible for controlling the characteristic day-night temperature differences, wind speeds, and eastward offsets of the hot spots. Significant progress has been made on these topics in the past ten years.
 
The eastward equatorial jets emerging in circulation models of hot Jupiters are important for several reasons. Being perhaps the dominant aspect of the atmospheric flow, they are critical in explaining observations, because they help to cause the eastward shift of the dayside hot spot, influence the day-night temperature distribution, and cause differing conditions on the leading and trailing terminators. Such eastward equatorial jets are also dynamically interesting---they are an example of what is termed {\it superrotation}.\footnote{The term superrotation is typically defined to correspond to regions of atmospheric flow where the angular momentum per unit mass about the rotation axis, $m= (\Omega a \cos{\phi} + u)a\cos{\phi}$ exceeds that corresponding to zero wind at the equatorial surface, $\Omega a^2$. In this sense, zonal jets at the equator naturally correspond to superrotation if they exhibit eastward wind of essentially any speed. In contrast, zonal jets at higher latitudes are superrotating only if their zonal wind exceeds the value $u = \Omega a \sin^2{\phi}/ \cos{\phi}$. Generally, eastward zonal jets that are away from the equator in planetary atmospheres (such as those on Jupiter and Earth) exhibit angular momentum less than that of the equatorial surface and are therefore not superrotating: despite being local maxima of zonal wind as a function of latitude, they are not local maxima of angular momentum. In this sense, eastward equatorial jets are a very different phenomenon from either eastward or westward jets at any other latitude.} The equator is the region of the planet farthest from the rotation axis, and therefore, an eastward equatorial jet represents a local maximum of angular momentum per unit mass as a function of latitude and height---air at higher latitudes or deeper levels will tend to have smaller values of angular momentum per unit mass due to its closer proximity to the rotation axis. Maintaining such a superrotating jet against friction or other processes requires that the atmospheric circulation transport angular momentum up-gradient from regions where it is small (outside the jet) to where it is large (inside the jet). This is the opposite of diffusion, which causes down-gradient transport. Hide's theorem \citep{hide1969} states that such superrotating flows cannot result from axisymmetric circulations such as angular-momentum-conserving Hadley cells, but rather that the necessary up-gradient momentum transport can only be accomplished by waves and/or turbulence. Indeed, in many branches of fluid dynamics, up-gradient momentum transport is a fairly common outcome of turbulence and wave propagation (indeed, wave propagation is non-local in the sense that the wave can cause a flux of angular momentum that does not depend on the local gradients of the flow, as would occur in a diffusive problem). The puzzle in the present context is to understand the specific dynamical mechanism that causes the superrotation on hot Jupiters and to understand why it appears to be so robust. 
 
 The dynamical mechanism for equatorial superrotation on tidally locked planets was first investigated by \cite{showman2010b,showman2011}, who showed that the intense day-night heating pattern on synchronously rotating planets induces a standing pattern of large-scale atmospheric waves, and that these waves have a structure that naturally leads to a flux of angular momentum towards the equator, causing the superrotation.
 
 To see how this works, it is useful to first consider the behavior of freely propagating tropical waves. Importantly, tropical waves tend to be confined to an equatorial waveguide whose half-width is equal to the equatorial Rossby deformation radius (Equation \ref{eq.1}). Two classes of tropical wave are particularly relevant here--- the Kelvin waves and the Rossby waves (see \citealp{holton2012,andrews1987}). The {\it Kelvin waves} are essentially gravity (buoyancy) waves in the east-west direction, with broad-scale pressure anomalies that are centered on, and symmetric about, the equator. In the north-south direction, Kelvin waves are geostrophically balanced—meaning the wave-induced pressure gradients are supported by Coriolis forces associated with the wave-induced wind anomalies. This balance implies (1) that isolated Kelvin waves exhibit very weak meridional wind relative to the zonal wind, and (2) that the zonal wind exhibits eastward phase in the regions of pressure maxima, and westward phase in the region of pressure minima. It turns out that trait (2) implies that only eastward-propagating waves are possible, and therefore Kelvin waves exhibit phase and group propagation to the east.
 
 In contrast, {\it Rossby waves} are vortical waves whose restoring force relies on the planetary rotation, or more specifically, the fact that the Coriolis parameter is a function of latitude. This effect is referred to as the ``$\beta$ effect," where $\beta = df/dy$ is the gradient of the Coriolis parameter with northward distance $y$. Because of the tendency of fluid parcels to conserve their potential vorticity following the flow\footnote{For a stratified atmosphere, the potential vorticity is generally defined as $(\zeta + f)/h$, where $\zeta =  \mathbf{k}\cdot \mathbf{\nabla}\times \mathbf{v}$ is the relative vorticity and $h$ is some measure of the vertical thickness of a fluid column (e.g., the vertical spacing between constant-entropy surfaces). In the adiabatic, frictionless limit, the dynamical equations conserve potential vorticity following the flow, which means that changes in $f$ caused by meridional deflections of a fluid parcel tend to be counteracted by changes in $\zeta$. Thus, meridional motions of a fluid parcel tend to ``spin up" a fluid parcel, generating relative vorticity.}, the latitude variation of the planetary vorticity implies that fluid parcels deflected poleward tend to generate anticyclonic vorticity, whereas those deflected equatorward tend to generate cyclonic vorticity. This vorticity comprises both zonal and meridional motions, and it turns out that the meridional motions associated with this wave-generated vorticity tend to deflect the fluid parcel back to its original latitude. This chain of events thus acts as a restoring force that allows wave propagation. These properties imply that Rossby waves exhibit phase propagation to the west \citep{holton2012}. When equatorially trapped, Rossby waves exhibit pressure patterns that are symmetric about the equator, with maxima that peak off the equator, and velocities skirting the pressure maxima in a vortical pattern. For low-order Rossby waves of long wavelength, this structure tend to fill the equatorial waveguide.
 
 Time variable, stochastic forcing can trigger freely propagating Rossby and Kelvin waves, which in a system with weak damping are able to propagate zonally for long distances within the equatorial waveguide. For example, in Earth's tropics, cumulus convection tends to trigger both of these wave classes, which can be observed because of their influence on the cloud patterns \citep{wheeler1999,kiladis2009}. Models of stochastic forcing in the tropics likewise lead to these and other wave modes (e.g., \citealp{salby1987}). A large heating pulse applied at the equator, with a size comparable to a deformation radius, for example, might naturally produce a Kelvin-wave packet that propagates to the east, and a Rossby-wave packet that propagates to the west.
 
\begin{figure}      % use "figure*" instead of "figure" if you want your figure to span both columns
\epsscale{1.}      % adjust this number to change the size of your figure
\includegraphics[scale=0.45]{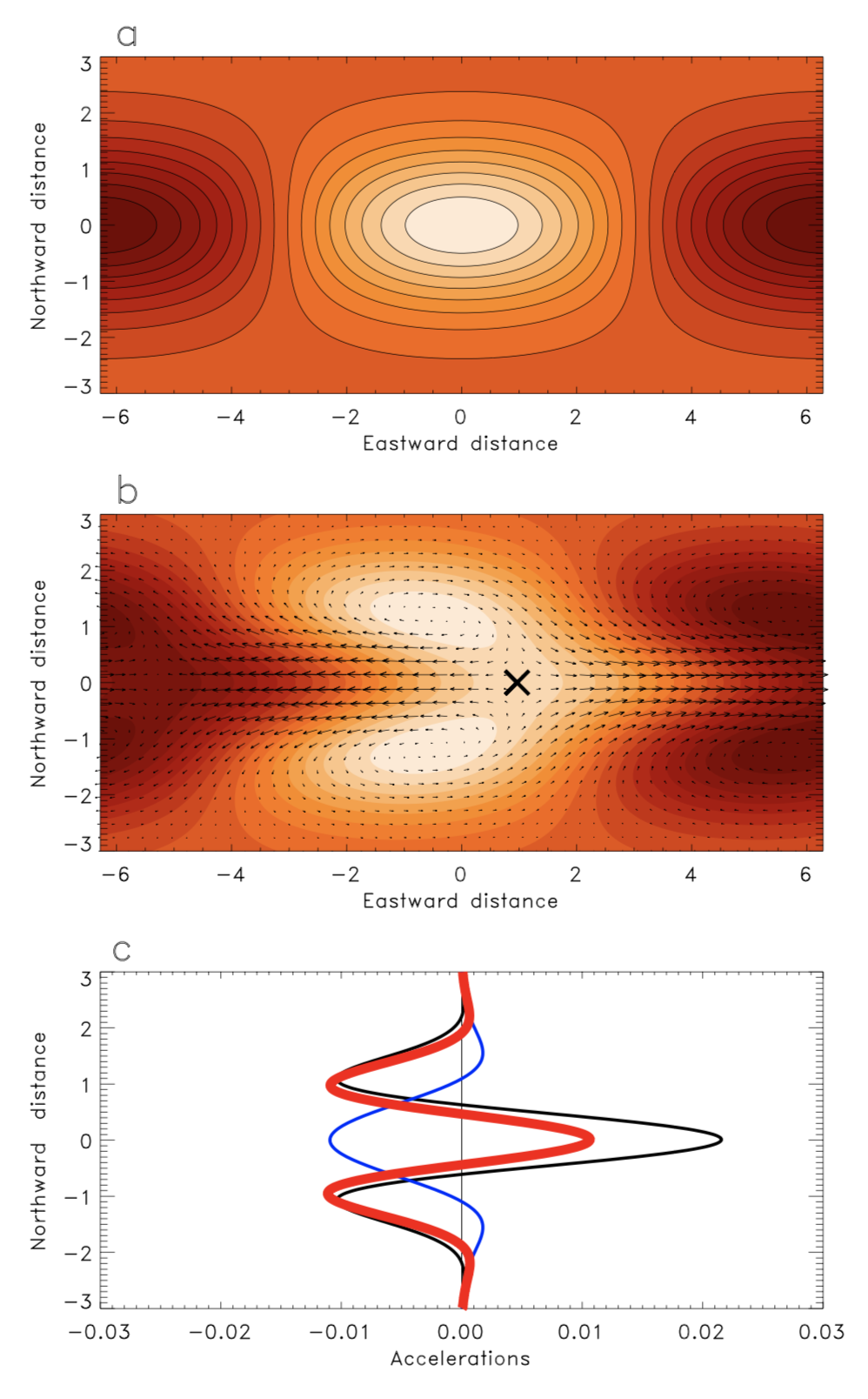}
\centering
\caption{Linear, analytical solutions of the shallow-water equations demonstrating how day-night heating can induce equatorial superrotation on tidally locked exoplanets, from \cite{showman2011}. Panel (a) shows the day-night heating pattern; specifically, the colors and contours represent the radiative equilibrium state as a function of eastward distance and northward distance (zero northward distance represents the equator and the center of the plot represents the substellar point; distance is plotted in units of the equatorial deformation radius, which for a hot Jupiter is a significant fraction of a planetary radius). The dayside is thick (hot) and the nightside is thin (cold). Panel (b) depicts the steady-state, linear solution that emerges in response to this heating pattern when no background flow is present. The solution can be interpreted as standing Kelvin and equatorially trapped Rossby waves. Notice that the predominant phase tilts of the eddy velocities are from northwest-southeast in the northern hemisphere and southwest-northeast in the southern hemisphere; this pattern causes transport of angular momentum from midlatitudes to the equator. Panel (c) shows the resulting acceleration of the zonal-mean zonal wind. The acceleration from meridional momentum convergence is in black, the acceleration from vertical momentum exchange is in blue, and the net acceleration is shown in red. In the net, eastward acceleration occurs at the equator, which will cause superrotation to emerge.}
\label{fig.6}
\end{figure}

 \cite{showman2010b,showman2011} demonstrated that the intense day-night thermal heating pattern on sychronously rotating exoplanets triggers a global-scale, standing wave pattern, with dynamics analogous to the tropical wave dynamics just described, and which explains the emergence of equatorial superrotation. These solutions, which built on the classic solutions of \cite{matsuno1966} and \cite{gill1980}, adopted the 1.5-layer shallow-water equations, which represent the flow of an active, constant-density atmosphere (here representing the photosphere levels of a giant planet) that overlies a denser interior (representing the deep atmosphere and interior) that is assumed quiescent. Figure \ref{fig.6} depicts the linear, analytic, $\beta$-plane solutions for typical hot Jupiter conditions, and adopting radiative and drag time constants that are equal. In the solutions, the steady, day-night heating pattern (Figure \ref{fig.6}a) leads to standing eddy pattern known as a Matsuno-Gill pattern (Figure \ref{fig.6}b). At low latitudes, it comprises predominantly east-west flows that diverge from a point (marked by an X) lying east of the substellar point. This structure can be interpreted as a standing Kelvin wave; the eastward displacement of the thermal structure at low latitudes results from the fact that Kelvin waves propagate to the east. At higher latitudes it comprises vortical flow, with off-equatorial pressure maxima (anticyclones) in the northern and southern hemispheres of the dayside, and pressure minima (cyclones) in the northern and southern hemispheres of the nightside. This structure can be understood as an equatorially trapped Rossby wave; the westward displacement results from the fact that long-wavelength Rossby waves propagate to the west. Importantly, by distorting the thermal field eastward at low latitude (due to the Kelvin wave component) and westward at higher latitudes (due to the Rossby wave component), this latitudinally varying zonal-phase shift induces a global-scale chevron-shaped eddy pattern in which the pressure contours and eddy velocities are tilted northwest-southeast in the northern hemisphere and southwest-northeast in the southern hemisphere. These phase tilts are clearly visible in Figure \ref{fig.6}b. See \cite{heng2014} for a wider range of Matsuno-Gill-type solutions, and \cite{pierrehumbert2019} for a tutorial on the Matsuno-Gill model.

 This tilted pattern of eddy velocities causes an equatorward flux of momentum and is therefore key in driving the equatorial superrotation. Let $u'$ and $v'$ be the eddy zonal and meridional velocities, respectively, defined as the deviations of the zonal and meridional winds from their zonal-means. Examination of Figure \ref{fig.6} shows that, in the northern midlatitudes, fluid parcels moving east ($u' > 0$) tend to be moving south ($v' < 0$), whereas those moving west ($u' < 0$) tend to be moving north ($v' > 0$). Thus, the velocities are correlated such that $\overline{u'v'} < 0$, where the overbar denotes a zonal average. In the southern midlatitudes, the correlation is reversed—fluid parcels moving east tend to be moving north, whereas fluid parcels moving west tend to be moving south, such that $u'v' > 0$. The quantity $u'v'$ represents a meridional flux of eastward momentum per unit mass---a positive value implies northward transport of eastward momentum, whereas a negative value implies southward transport. Thus, the Matsuno-Gill pattern (Figure \ref{fig.6}) naturally causes a convergence of momentum onto the equator, which leads to an eastward acceleration that drives equatorial superrotation. In this theory, the meridional width of these Matsuno-Gill solutions---and the equatorial jet that they drive—is comparable to the equatorial deformation radius.
 
 The solutions also show that, at the equator, the standing waves cause a net downward transport of zonal momentum, which plays a critical role in the momentum balance. Because of the eastward phase shift of the Kelvin-wave structure, the equatorial eddy velocity on the nightside is preferentially eastward while that on the dayside is preferentially westward (see Figure 6b). Nightside cooling causes downward transport of air, which takes its eastward eddy momentum with it. Dayside heating brings up air from the deeper atmosphere with only weak zonal velocity. Thus, there is a net downward transport of eastward momentum out of the atmosphere into the interior. This loss of momentum from the atmosphere causes a westward acceleration at the equator that partially---but not completely---cancels out the eastward acceleration caused by the convergence of momentum onto the equator (In Figure \ref{fig.6}c, the acceleration due to this vertical exchange is shown in blue, that due to meridional momentum convergence is shown in black, and the net acceleration is shown in red.)
 
 The solution shown in Figure \ref{fig.6} adopts equal radiative and drag timescales, but the linear behavior differs when these timescales are unequal. In particular, \cite{showman2011} and \cite{heng2014} explored a wide range of differing $\tau_{\rm rad}$ and $\tau_{\rm drag}$; they showed analytically that, as the drag becomes weak (i.e., the drag timescale becomes long), the equatorial waveguide corresponding to these forced-damped solutions becomes confined closer and closer to the equator. Their solutions show that, in the linear limit of $\tau_{\rm drag} \rightarrow \infty$, the meridional width of the region with prograde phase tilts shrinks to zero\footnote{By ``prograde" phase tilts, we mean northwest-southeast (southwest-northeast) in the northern (southern) hemisphere.}, and a pattern of reverse phase tilts instead emerges—northeast-southwest in the northern hemisphere and southeast-northwest in the southern hemisphere. Such a pattern would not be expected to converge eddy momentum onto the equator or drive strong superrotation. The presence of these reverse phase tilts are well-understood (\citealp{showman2011}, Appendix C), but this leads to a puzzle: how can superrotation develop in a 3D hot Jupiter GCM with no explicit frictional drag, when the meridional width of the Matsuno-Gill pattern (and its prograde phase tilts) should shink to zero?
 
 The answer is provided by nonlinearity. At the high forcing amplitudes relevant to typical hot Jupiters, the dynamics becomes nonlinear, which modifies the details of the wave-mean-flow interactions, both due to the nonlinearity of the eddies themselves, and because a strong zonal flow develops when the forcing amplitude is large, and the zonal flow modifies the eddy structures. \cite{showman2011} showed numerically that when nonlinearity is progressively introduced back into the dynamics, the meridional width of the Matsuno-Gill-type pattern broadens accordingly, even in the limit of weak drag, and at very large amplitudes exhibits a meridional width comparable to the equatorial deformation radius. This effect of nonlinearity on the eddies can be understood by appreciating that nonlinear momentum advection can qualitatively play the role of drag in the multi-way force balance that controls eddy behavior, allowing prograde phase tilts to persist to high latitudes, and preventing the Matsuno-Gill-type behavior from collapsing to the equator (\citealp{showman2011}; for a visual explanation of how this effect can lead to prograde phase tilts, see \citealp{showman2013b}). Mathematically, this effect can be appreciated from the momentum equation governing the 1.5-layer shallow-water system (\citealp{showman2011}, Equation 12):
 \begin{equation}
     \frac{d\mathbf{v}}{dt}+g\nabla h+f\mathbf{k}\times\mathbf{v}=-\mathbf{v}\left[\frac{1}{\tau_{\rm drag}} + \frac{1}{\tau_{\rm rad}}\left(\frac{h_{\rm eq}-h}{h}\right)\mathcal{H}(h_{\rm eq}-h)\right]
     \label{eq.7}
\end{equation}
 where $\mathbf{v}$ is the horizontal velocity vector, $d/dt$ is the material derivative, $h$ is layer thickness, $\mathbf{k}$ is the local vertical unit vector, $h_{\rm eq}$ is the local radiative equilibrium thickness, and $\mathcal{H}$ is the Heaviside step function, which equals one when its argument is positive and zero otherwise. The entire quantity in square brackets plays the role of one over an ``effective drag" time constant. At low amplitude, the quantity $(h_{\rm eq} - h)/h \ll 1$, and thus the second term in square brackets is not generally important. At high amplitude, however, deviations from radiative equilibrium tend to be substantial,\footnote{An exception occurs when the radiative time constant is much shorter than all the dynamical timescales, in which case the solution is close to radiative equilibrium even at high nonlinearity.} implying that $(h_{\rm eq} - h)/h \sim 1$. This implies that, on the dayside, the second term in square brackets tends to have a magnitude of order $1/$\trad. Thus, under typical hot-Jupiter conditions, the ``effective" drag time constant caused by the nonlinearity has comparable magnitude to the radiative time constant---even when the actual drag time constant $\tau{\rm drag}$ is infinite. The fact that the highly nonlinear, zero-drag regime tends to naturally exhibit radiative and effective drag timescales that are comparable to each other can explain why these solutions have a Matsuno-Gill-like pattern that extends to high latitudes, just like linear solutions with equal radiative and drag timescales (Figure \ref{fig.6}). This behavior also explains more generally why the standard definition of equatorial deformation radius (Equation \ref{eq.1}) provides a reasonable fit to the actual jet width in most 3D hot-Jupiter GCMs, despite the fact that these GCMs generally do not include explicit frictional drag in the region where the superrotation forms. { \cite{debras2020} further confirmed the above picture by performing  full 3D linear wave calculations  using the numerical package developed by \cite{debras2019}.}
 
 In sum, high-amplitude, fully nonlinear shallow-water solutions on the sphere show that essentially the same mechanism holds even in the nonlinear regime: although the details of the wave-mean-flow interactions are sensitive to nonlinearity, the qualitative mechanism still occurs even at high amplitude. As mentioned above, the eddy nonlinearity is critical for preventing the Matsuno-Gill solutions from collapsing to the equator in the weak-drag limit \citep{showman2011}. Once the jet builds up, these planetary-scale waves can exhibit finite meridional width even in the linear, weak-drag limit \citep{hammond2018,debras2020}, presumably because they are then propagating relative to the moving airflow (even though they are standing in the reference frame of the planet). As the jet builds up, the concomitant changes in the eddy structure lead to changes in the amplitudes of the horizontal and vertical eddy momentum convergences. Steady state is achieved when they balance (along with drag, if any).
 
\begin{figure*}      % use "figure*" instead of "figure" if you want your figure to span both columns
\epsscale{1.}      % adjust this number to change the size of your figure
\includegraphics[scale=0.4]{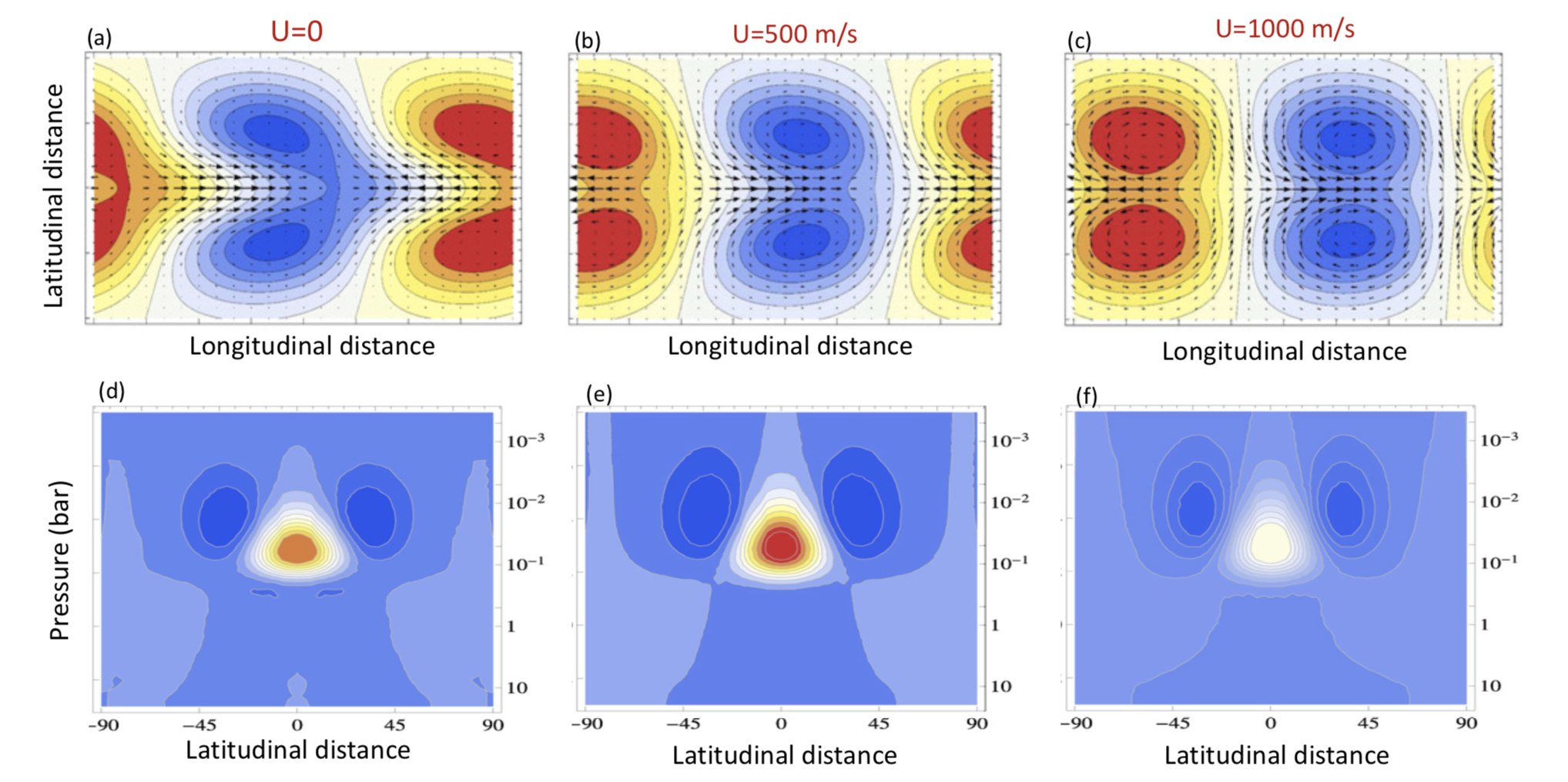}
\centering
\caption{Linear, analytic, steady $\beta$-plane solutions of the Matsuno-Gill problem in 3D primitive equations, assuming equal radiative and frictional time constants that are constant with depth, from \cite{tsai2014}. Left, middle, and right columns adopt a uniform background wind of $U=0$, 500, and 1000 \mps, respectively. Top row shows eddy geopotential and eddy wind structure; antistellar point lies in the middle of each plot. Bottom row shows the total wave-induced acceleration;  exact values are arbitrary but for plausible forcing amplitudes range from $-10^{-6}$\mpstwo (westward) in dark blue to $3\times10^{-6}$\mpstwo (eastward) in  red. }
\label{fig.7}
\end{figure*}

 In a major study, \cite{tsai2014} showed that a deeply stratified, 3D atmosphere exhibits global-scale, Matsuno-Gill-type wave solutions that are very similar to those in shallow water. They adopted radiative and frictional damping times that are equal and invariant with depth. The day-night heating was assumed to occur over a broad vertical layer peaking around 0.1 bar. Tsai et al.'s solutions in the absence of a background flow, shown in the left column of Figure \ref{fig.7}, yield a Matsuno-Gill pattern very similar to the shallow-water solutions identified by Showman \& Polvani, comprising superposed equatorial Kelvin and Rossby waves whose differential zonal propagation produces eddies tilting predominantly northwest-southeast (southwest-northeast) in the northern (southern) hemisphere. Just as in the shallow-water system, the eddy structure induces a meridional momentum convergence onto the equator in the forcing layer, as well as a downward momentum transport at the equator from the forcing levels to deeper levels. Within the forcing layer, the net accelerations in the meridional plane are eastward at the equator and westward at higher latitudes (Figure \ref{fig.7}d), qualitatively similar to the pattern of accelerations predicted by the shallow-water solutions (Figure \ref{fig.6}c). The striking similarity between the shallow-water and 3D solutions is not surprising, because, as is standard in tidal theory and many other wave problems \citep{chapman1970}, this 3D problem is mathematically separable and determined by solving a shallow-water equation for the horizontal structure, coupled to a 1D vertical-structure equation for the vertical eigenfunctions. 
 
 In \cite{tsai2014}'s solutions, below the $\sim$1-bar level, the waves propagate downward away from the forcing layer. This propagation causes the Rossby gyres to shift westward at greater depth, while the Kelvin wave component shifts eastward with depth. The overall longitudinal/vertical tilts of the solution depend on the relative amplitudes of the Kelvin and Rossby components, which vary with the adopted damping time constant. In particular, the Kelvin wave component dominates when the damping timescales are short, while the Rossby wave component dominates when the damping timescales are long. Despite these complications, the visual appearance of their solutions is that the vertical tilts appear to be modest under damping conditions relevant to hot Jupiters.
 
 Still, analysis of fully nonlinear GCMs shows that the build-up of the equatorial jet strongly modifies the standing waves and disrupts the coherency of the prograde phase tilts, leading to a wave structure in the fully equilibrated state that differs substantially from the classical Matsuno-Gill pattern \citep{showman2011,tsai2014,showman2015,mayne2017,mendonca2020}. Because of the disruption of the phase tilts, the net meridional momentum transports---while still equatorward|are weaker than would be predicted from the classic Matsuno-Gill model at the same wave amplitude. Quantitatively understanding how the equatorial jet equilibrates therefore requires an understanding of how the jet modifies the eddies. Several authors have addressed this question.

The addition of a uniform background zonal wind to the Matsuno-Gill model has long been known to modify the standing wave modes \citep{phlips1987} and the momentum fluxes they induce \citep{arnold2012}. To investigate this effect in the context of hot Jupiters, \cite{tsai2014} included a uniform background zonal wind $U$ in their analytic solutions (Figure \ref{fig.7}, middle and right columns). The background  zonal flow influences the amplitudes  and relative phase offset between the Kelvin and Rossby wave components, with crucial implications for the wave-induced accelerations. Because the group velocity for long Rossby waves is about one-third that of the Kelvin wave, the phase offset of the Rossby-wave component is much more easily influenced by the background winds. In the solution in Figure \ref{fig.7} background winds ranging from zero to 1 \kms hardly change the offset of the Kelvin-wave component, but they cause the Rossby wave gyres to shift from $\sim$$20^{\circ}$ west of the substellar longitude to $50^{\circ}$ east of it. (Compare Figure \ref{fig.7}a and c; see \citealp{tsai2014},  Figure 8, for a decomposition of the Kelvin and Rossby-wave components of the solutions.) At zonal winds of $\sim$1 \kms, the pressure maxima of the Rossby and Kelvin components are nearly in phase, and as can be seen from Figure \ref{fig.7}, this destroys the strong polewared-westward to equatorward-eastward phase tilts. After a transient increase in the strength  of the acceleration at modest $U$ (Figure \ref{fig.7}e), the zonal acceleration then plummets as the zonal flow increases (Figure \ref{fig.7}f).  Still, the net acceleration remains eastward throughout the entire range of $U$ shown  in Figure \ref{fig.7},  suggesting  that the jet will continually accelerate to speeds exceeding 1 \kms. \cite{tsai2014} showed that, for the particular model parameters of Figure \ref{fig.7}, the net, wave-induced zonal acceleration at the equator reaches zero only when the background wind reaches about 3 \kms. This intriguing result could help to explain why the equilibrated zonal jet speeds in GCMs are typically 3-4 \kms (Figure \ref{fig.4}), although the meridional and vertical shear of the zonal wind are also likely important. 

\cite{hammond2018} explored the effect of meridional shear of the equatorial jet on the wave structure and momentum fluxes in a linear shallow-water model. They assumed a Gaussian-shaped equatorial jet whose zonal-mean wind is $\overline{U}=U_0e^{-y^2/L^2_{\rm eq}}$, qualitatively similar to those in GCM solutions. Their results indicate that, even in the presence of the sheared equatoral jet, the eddies still transport momentum from midlatitudes onto the equator, leading to an eastward acceleration at the equator that can help maintain the superrotation. {Furthermore, they showed that the inclusion of the meridionally shearing  jet and its associated geopotential anomaly  triggers more forced wave  modes, which results in  a better  match between the overall surface height from the linear shallow-water model  and  the GCM temperature field than those without a jet. }

Several authors have further investigated the dynamics of the equatorial superrotation in full 3D GCM experiments. As mentioned above, the equilibrated, standing  eddy structure shows significant distortion from the classic Matsuno-Gill pattern. Despite this complication, diagnostics from these GCMs still generally support a picture wherein the day-night forcing leads to a large-scale, standing wave pattern that causes a meridional convergence of eddy angular momentum onto the equator, driving the superrotation, and a downward eddy momentum transport at the equator from above the photosphere to deeper levels \citep{showman2011, tsai2014,showman2015,mayne2017,mendonca2020,debras2020}. This downward eddy momentum transport helps cause the equatorial jet to slowly penetrate more deeply over time at pressures of a few bars and greater. At mid-to-high latitudes, all the studies agree that the steady state zonal-momentum balance involves not only the eddy flux convergences (both meridional and vertical), but also the Coriolis accelerations and momentum advection associated with the mean-meridional circulation. At the equator, the zonal-momentum balance at and above the photosphere tends to comprise a balance primarily between the (eastward) acceleration due to meridional momentum transport and the (westward) acceleration due to the vertical momentum transport (e.g., \citealp{showman2011}). Nevertheless, several authors have highlighted the role of vertical momentum transport due to the zonal-mean circulation as well \citep{mayne2017,mendonca2020}). Interestingly, although the primary day-night forcing would appear to comprise a zonal wavenumber-one (diurnal) mode, \cite{mendonca2020}  argues that the wavenumber-two (semidiurnal) mode may be at least as important. The details of all these issues are likely to be sensitive to planetary rotation rate, strength and implementation of radiative forcing, and other modeling aspects, so it would be worth exploring these issues over a wider range of conditions.

 To summarize, the mechanism predicts that the equatorial jet has a meridional half-width approximately equal to the Rossby deformation radius. For conditions typical of hot Jupiters, this length scale is comparable to a planetary radius, explaining the broad structure of the equatorial jet in hot-Jupiter circulation models (Figure \ref{fig.4}). The process that drives the jet is, in its essence, a direct wave-mean-flow interaction between the eddies and the planetary-scale mean flow. Turbulent cascades or other eddy-eddy interactions are possible, and would affect the details, but are not essential to the basic mechanism.
 
 \subsection{Day-night temperature differences and recirculation efficiency}
 
 Synchronously rotating exoplanets exhibit permanent daysides and nightsides and are therefore subject to a fascinating climate regime of day-night thermal forcing that has no analogy in the solar system. Lacking any direct irradiation, the only possible means for the nightside to exhibit a temperature exceeding tens of K is for heat to be transported from the dayside to the nightside by dynamical motions in the atmosphere or interior. This raises the intriguing question of what processes control this day-night heat transport, what the expected day-night temperature difference is, and how it should vary with the stellar irradiation, planetary rotation rate, atmospheric composition, and other parameters. This problem is analogous to the important question of what controls the meridional temperature gradients and equator-to-pole temperature differences on planets like Earth, but in the case of hot Jupiters, the heat transport occurs not only meridionally but zonally as well. Because the $\beta$ effect inhibits heat transport meridionally, whereas no such constraint exists in the zonal direction, the dynamics are therefore likely to be quite different.
 
 This problem is intriguing not only because it touches on basic issues in geophysical fluid dynamics (GFD) but because it is directly amenable to observational characterization: on a synchronously rotating planet, energy balance implies that the energy transported from the dayside to the nightside by the atmosphere/interior circulation is radiated to space on the nightside. Therefore, measurements of the total IR flux radiated to space on the nightside---which can be directly estimated from IR lightcurves---provides an observed measure of the day-night dynamical heat transport.
 
 Many authors have posited that the problem of whether the day-night temperature difference is large or small can be cast as a comparison between two timescales, the radiative timescale (Equation 3) and an advection timescale for air to travel across a hemisphere from day to night:
 \begin{equation}
     \tau_{\rm adv} \sim \frac{a}{U},
     \label{eq.8}
 \end{equation}
 where $a$ is the planetary radius and $U$ is a characteristic horizontal wind speed. \cite{showman2002} first suggested that hot Jupiters will exhibit large day-night differences when \trad $ll \tau_{\rm adv}$  and small day-night temperature differences when \trad $gg \tau_{\rm adv}$. Many authors have suggested that this timescale comparison helps to explain the dependence of the day-night temperature difference on pressure, opacity, and incident stellar irradiation \citep{cooper2005,showman2008,fortney2008,rauscher2010,heng2011,cowan2011,cowan2011b,perna2012}.
 
 The radiative timescale scales as $T^{-3}$ (Equation \ref{eq.3}), implying that it becomes very short for ultra hot Jupiters. In contrast, order-of-magnitude expressions for the typical horizontal velocities (Equation \ref{eq.5}–\ref{eq.6}) suggest that the horizontal velocity should also be greater on strongly irradiated hot Jupiters, but that the dependence is weaker. Thus, one might expect that for hot Jupiters whose photospheres exceed some critical temperature---perhaps of order 2000 K---the atmosphere is close to radiative equilibrium and the day-night temperature difference is large. For hot Jupiters cooler than this value, the day-night temperature difference should become smaller.
 
\begin{figure}      % use "figure*" instead of "figure" if you want your figure to span both columns
\epsscale{1.}      % adjust this number to change the size of your figure
\includegraphics[width=\linewidth]{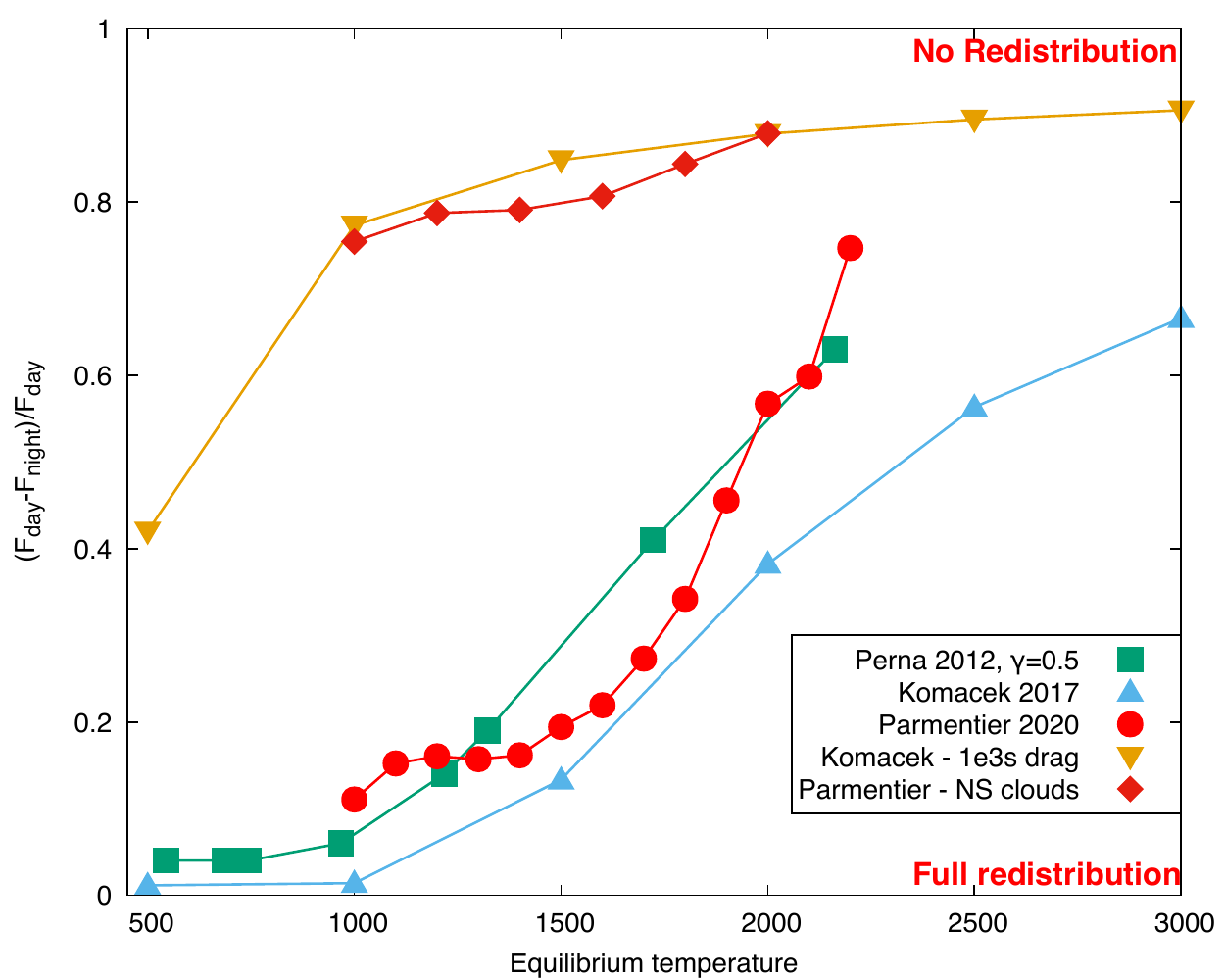}
\centering
\caption{{ Fractional day-night IR flux differences in GCM simulations of synchronously rotating hot Jupiters as a function of their equilibrium temperature for models with different assumptions. \citet{perna2012} uses the FMS model with semi-grey radiative transfer.~\citet{komacek2017} use the MITgcm model with semi-grey radiative transfer but slightly different choices of opacities and rotation period than~\citet{perna2012}. ~\citet{Parmentier2020} uses the SPARC/MITgcm with non-grey, temperature dependent opacities (assuming chemical equilibrium without TiO/VO) and rotation periods similar to~\citet{perna2012}. Despite their differences, the three models provide similar results when used in similar conditions, showing that the decrease of heat redistribution efficiency with increasing temperature is robust. However, when either drag (orange triangles from
~\citealp{komacek2017}) or the radiative feedback of nightside clouds (diamonds from~\citealp{Parmentier2020}) are included  in the models, the heat redistribution becomes much smaller. The difference between physical ingredients is much larger than the difference between modelling frameworks.  
}}\label{fig.8}
\end{figure}

 \cite{perna2012} were the first authors to systematically explore the influence of stellar irradiation on day-night temperature structure in GCM simulations. Their GCM adopted a dual-band radiative transfer approach with one broad opacity band in the visible and one in the IR, neglecting scattering, with opacities chosen to yield a range of plausible temperature profiles. In a series of numerical experiments, they varied the stellar irradiation over nearly three orders of magnitude—corresponding to planetary effective temperatures of about 600 to 2500K---and examined how the circulation responds. Their simulations show that, as qualitatively expected, the day-night temperature differences are large for effective temperatures $\gtrsim 2000$ K but becomes small for $T \lesssim 1500$ K (Figure \ref{fig.8}). They showed that the transition is consistent with the expectation based on the timescale argument presented above: for planets with effective temperatures less than $\sim$1500K, the advective and radiative timescales (as post-processed from the simulation results) are similar, but at effective temperatures exceeding 1500 K, the radiative timescale becomes shorter than the advective timescale, and the day-night flux differences become large (Figure \ref{fig.8}).

\begin{figure*}      % use "figure*" instead of "figure" if you want your figure to span both columns
\epsscale{1.}      % adjust this number to change the size of your figure
\includegraphics[scale=0.45]{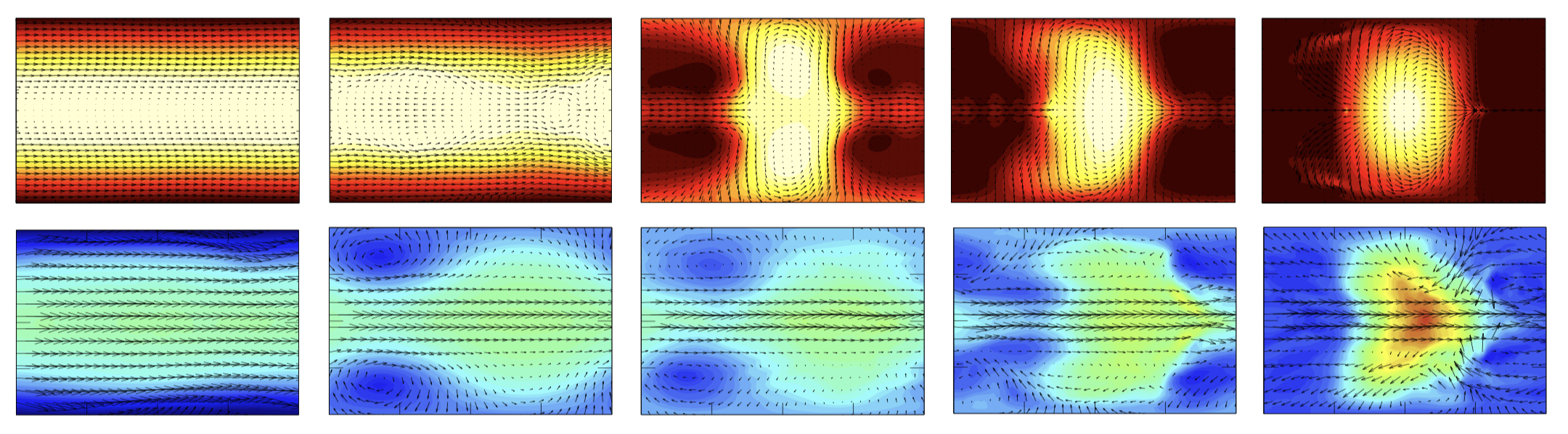}
\centering
\caption{Dependence of circulation in synchronously rotating hot Jupiter models over a wide range of radiative time constant. The radiative heating is implemented with an idealized Newtonian cooling scheme that relaxes the circulation toward a specified radiative equilibrium structure. Top row presents shallow-water simulations from \cite{showman2013b} and \cite{perezbecker2013}; from left to right panels, radiative time constant is 100, 10, 1, 0.1, and 0.01 Earth days, respectively. Rotation period is 2.2 days in all models. Plot shows thickness (colorscale) and flow velocity (arrows). Bottom row: 3D primitive equation simulations from \cite{komacek2016} showing temperature (colorscale) and flow velocity (arrows) structure at 80 mbar; from left to right panels, radiative time constant is $10^7$, $10^6$, $10^5$, $10^4$, and $10^3$ s, respectively. Rotation period is 3.5 days in all models. Both sets of models exhibit a transition from small day-night thermal contrast at \trad $\gtrsim 10^5$ s to large day-night contrast at \trad $\lesssim 10^5$ s. }
\label{fig.9}
\end{figure*} 

There are several issues with the timescale comparison between \trad and \tadv, however \citep{perezbecker2013,komacek2016}. First, it was not derived in any self-consistent manner from the governing dynamical equations, but rather represents an ad hoc, albeit attractive, hypothesis. Second, it is not predictive--- the advection timescale can only be evaluated once the wind speeds are known, yet the wind speeds themselves depend on the temperature differences one is trying to understand. Third, the criterion obscures any role for other key timescales, including rotational timescale, frictional timescale (if any), and timescale for horizontal and vertical wave propagation. These timescales surely affect the circulation, including the day-night temperature differences, and therefore one wound expect a timescale criterion for day-night temperature differences to depend on  them. 

\cite{perezbecker2013} and \cite{komacek2016} performed idealized numerical experiments and constructed analytic scaling theories for the day-night temperature differences and characteristic wind speeds, with the aim of obtaining a self-consistent understanding of {\it both} the day-night temperature differences and characteristic wind speeds (hence advection timescale). The former study adopted the 1.5-layer shallow water model, while the latter adopted the 3D primitive equations, but the setup was otherwise quite similar across both studies, involving an idealized day-night forcing where the thermal structure was relaxed towards a prescribed radiative equilibrium (hot on the dayside and cold on the nightside) over a prescribed radiative timescale \trad. In agreement with the results of \cite{perna2012}, the simulations show that the thermal structure exhibits essentially no day-night variation when \trad is long, but that the day-night contrast becomes large when \trad  is short (Figure \ref{fig.9}).  The transition occurs near  \trad  $\sim10^5$ s. Frictional drag was also included in some models to parameterize the possible effects of Lorentz forces, and  these authors found that  sufficiently strong drag  could also lead to large day-night temperature differences.

The analytic theory in \cite{perezbecker2013} and \cite{komacek2016} is constructed by balancing dominant terms in the dynamical equations. \cite{ginzburg2016} also present a related analysis. In the momentum equation, the day-night pressure-gradient force, which emerges from the day-night temperature difference and drives the circulation, is balanced by the greater of the Coriolis, frictional drag, horizontal momentum advection, or vertical momentum advection forces, respectively; this leads to  different expressions for the day-night contrast in each regime, along with criteria for the transition between them. For example, when the dominant force balance is between the day-night pressure-gradient  force and the Coriolis force, which holds when   drag is weak and  the rotation is sufficiently fast, the theory predicts that the fractional day-night contrast is 
\begin{equation}
    A=\left(1+\frac{\tau_{\rm rad}}{f\tau^2_{\rm wave}}\right)^{-1},
\end{equation}
where here $A$ is the fractional day-night thermal contrast (i.e., the fractional day-night thickness contrast in shallow water and fractional day-night temperature difference in 3D). { $f=2\Omega sin\phi$ is the Coriolis parameter, $\Omega$ the planetary rotation rate and $\phi$ the latitude}. The quantity \twave$=a/NH$ is the characteristic timescale for gravity (or Kelvin) waves of long vertical wavelength to propagate over a zonal distance equal to a planetary radius. The typical horizontal wave propagation speed can be estimated by $NH$, which is about 1 \kms for a hot Jupiter (see Section 2.1), which leads to a timescale \twave $\sim 10^5$ s, where we have adopted a Jupiter radius $a = 7 \times 10^7$ m. In the limit of large \trad, this equation predicts $A = 0$, corresponding to no variation from day-to-night; in the limit of small $\tau_{\rm rad}$, it predicts $A=1$, corresponding to a temperature structure in radiative equilibrium, exhibiting large day-night temperature differences. The transition between these limits occurs when
\begin{equation}
    \tau_{\rm rad} \sim f\tau^2_{\rm wave},
\end{equation}
where $f$ is the Coriolis parameter.
Evaluating the expression for a typical hot Jupiter value $f=3\times 10^{-5} \rm{s^{-1}}$ yields a critical radiative time constant \trad $\sim 3\times 10^5$ s. Thus, under typical hot Jupiter conditions, this theory predicts that day-to-night temperature differences will be small when \trad $\lesssim 3\times 10^5$ s and large when \trad $\gtrsim 3\times 10^5$ s. It can be seen that this explains the transition occurring in Figure  \ref{fig.9}. More generally,  the theory  yields self-consistent predictions of the day-night thermal contrast and wind speeds over the full parameter space of \trad, \trad, forcing amplitude (i.e., the day-night difference in radiative-equilibrium temperature) and other parameters, with no free parameters and no tuning. \cite{perezbecker2013} and \cite{komacek2016} showed that the theory matches the simulations reasonably well over this wide range. In contrast, the comparison between radiative and horizontal advection timescales does not, particularly when the forcing amplitude is weak.

\cite{perezbecker2013} and \cite{komacek2016} showed that the dynamical mechanism for regulating the day-night temperature differences can be interpreted in terms of a wave-adjustment mechanism: the day-night heating contrast triggers planetary-scale waves---including the Kelvin and Rossby wave discussed previously---which propagate longitudinally around the planet. The horizontal convergence and divergence associated with these waves changes the vertical thickness of fluid columns, inducing vertical motion that pushes isentropes up or down. One can make an analogy to throwing a rock into a pond: the perturbation to the water surface caused by throwing the rock into the pond leads to the propagation of gravity waves whose horizontal convergence or divergence change the thickness; once the waves radiate away, the end result is a flat surface. In the atmosphere, a similar process acts to flatten surfaces of constant entropy (for a review, see \citealp{showman2013}). If the damping is weak (i.e. the damping timescales are long), the waves can easily propagate from dayside to nightside and this adjustment process is efficient, leading to small day-night temperature difference. If the damping is strong (i.e., the damping timescales are short), however, then the waves become damped before they can propagate from day to night. This suppresses the wave regulation mechanism, allowing the temperature difference to approach radiative equilibrium.
 
 What is the relationship of this criterion to the comparison between radiative and advective timescales? \cite{perezbecker2013} and \cite{komacek2016} showed that the timescale criterion for transition between large and small fractional day-night temperature differences  can be expressed as a comparison between the radiative timescale and an appropriately defined {\it vertical} advection timescale $\tau_{\rm vert}$. This comparison emerges naturally from the dynamical equations because, on global scales, the vertical entropy advection term generally dominates over the horizontal entropy advection term in the thermodynamic energy equation (see \citealp{komacek2016} for a discussion). This is essentially the ``weak temperature gradient" (WTG) balance well-known from Earth tropical meteorology (for reviews, see \citealp{showman2013,pierrehumbert2019}). The timescale comparison between \trad and $\tau_{\rm vert}$ subsumes the classic comparison between \trad and \tadv: they are essentially the same in the highly forced limit where the day-night temperature difference is order unity, but they differ greatly in the weak forcing limit, and in that limit, the former comparison explains the simulation results far better than the latter one.
 
 \cite{zhang2017} combined these regimes into a single compact expression for the temperature difference, valid across the entire parameter space:
 \begin{equation}
     \frac{\Delta T}{\Delta T_{\rm eq}} = 1-\frac{2}{\alpha+\sqrt{\alpha^2+4\gamma^2}}
 \end{equation}
 where the non-dimensional, pressure-dependent constants $\alpha$ and $\gamma$ are defined as
\begin{equation}
    \alpha=1+\frac{\left(\Omega+\frac{1}{\tau_{\rm drag}}\right) \tau^2_{\rm wave}}{\tau_{\rm rad}\delta \ln p}
\end{equation}
 and 
 \begin{equation}
     \gamma=\frac{\tau^2_{\rm wave}}{\tau_{\rm rad}\tau_{\rm adv,eq}\delta \ln p}.
 \end{equation}
 In these expressions, $\tau_{\rm adv,eq} = a/U_{\rm eq}$ is the timescale for advection by a reference ``equilibrium cyclostrophic wind", defined as $U_{\rm eq} = \sqrt{\Delta T_{\rm eq} \delta ln p/2}$. The quantity $\delta \ln p$ is the difference in log-pressure between some deep pressure where the flow is assumed to decay to zero (assumed to be 10 bars by \citealp{komacek2016})
and some lower pressure of interest, for example at the IR photosphere.

\begin{figure}      % use "figure*" instead of "figure" if you want your figure to span both columns
\epsscale{1.}      % adjust this number to change the size of your figure
\includegraphics[width=\linewidth]{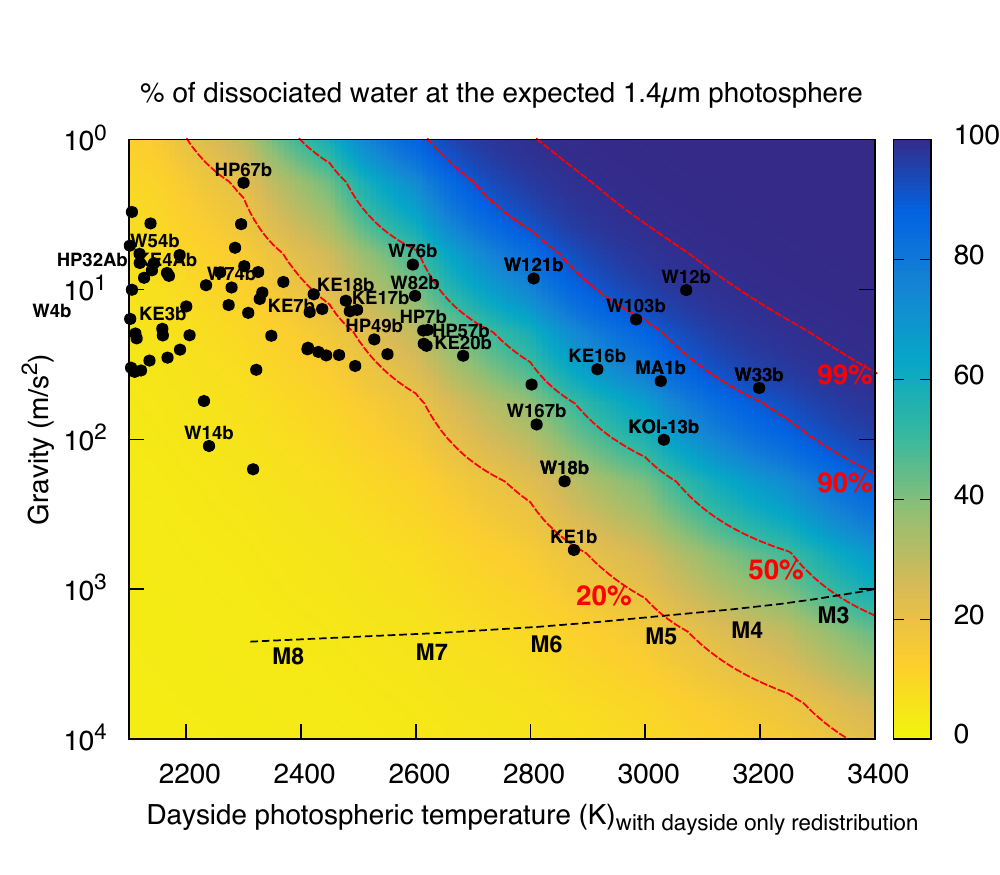}
\centering
\caption{{ Percentage of water that is dissociated at the infrared photosphere of irradiated planets as a function of their dayside temperatures and gravity. Planets on the right hand side of the $20\%$ of water dissociation iso-contour can be considered as ultra hot Jupiters. The efficiency of thermal dissociation is pressure dependent, hence why planets with higher gravity, tend to be less affected by it. A stellar isochrone for an age of 4 Gyr from~\citet{baraffe2008} is shown as a dashed black line. }}
\label{fig.9a}
\end{figure} 
 
\begin{figure}      % use "figure*" instead of "figure" if you want your figure to span both columns
\epsscale{1.}      % adjust this number to change the size of your figure
\includegraphics[scale=0.45]{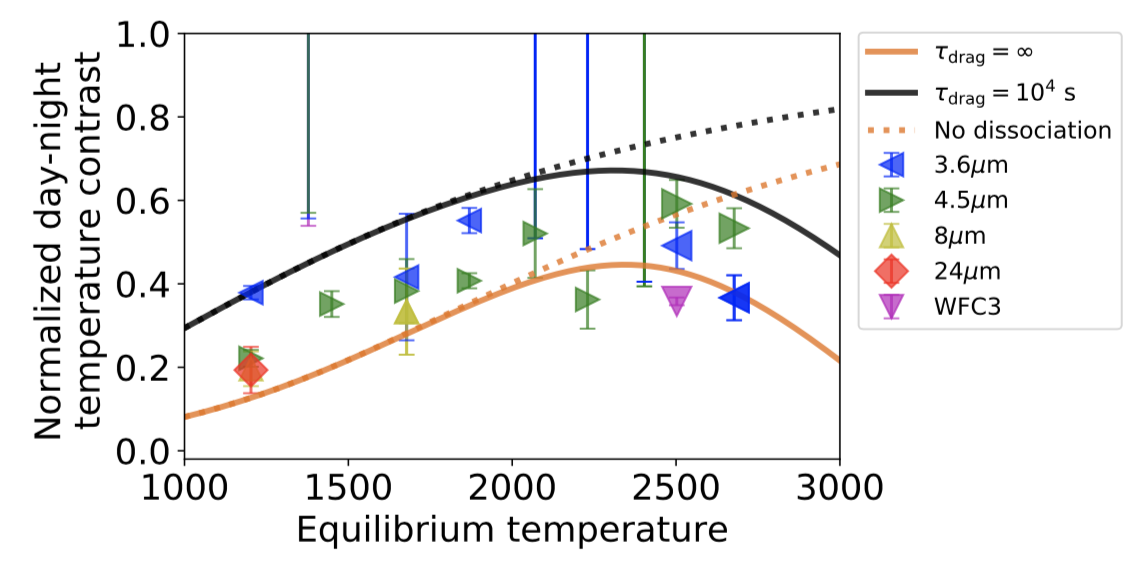}
\centering
\caption{Predictions for day-night temperature differences from analytic theories and comparison to measurements from lightcurves of various hot Jupiters, from \cite{komacek2018rnaas}. Dotted lines show the predictions from \cite{komacek2016}, and solid lines show the predictions from \cite{komacek2018rnaas} accounting for the latent heat caused by dissociation and recombination of hydrogen at high temperature. Orange curve assume there is no explicit frictional drag in the observable atmosphere, while the black curves adopt a spacially constant  drag coefficient with a timescale of $10^4$ s.}
\label{fig.10}
\end{figure}

 Atmospheres of ultra-hot Jupiters (here loosely defined as having equilibrium temperature $\gtrsim2000$ K) are sufficiently irradiated such that all molecules, including molecular hydrogen and water are partially dissociated in their dayside atmospheres (see Figure~\ref{fig.9a} and ~\citet{parmentier2018}).
 Heat is absorbed on the dayside to break the strong $\rm{H_2}$ bond, and then the atomic hydrogen is transported by atmospheric flows towards cooler regions of the atmosphere, where hydrogen atoms recombine to molecules and release heat. The energy per unit mass associated with the molecule-atom transition of hydrogen ($2.18 \times 10^8 \; \rm{J\; kg^{-1}} $) is two orders of magnitude higher than that associated with water phase change, and together with the fact that hydrogen is the main atmospheric constituent, suggests that heat transport by hydrogen dissociation and recombination can be enormous in ultra-hot atmospheres. \cite{bell2018} first studied this effect using an semi-analytic energy transport model, and they found that the heating from hydrogen recombination could significantly reduce the day-night phase-curve amplitude and increase phase curve offsets. Building upon the theory of \cite{komacek2016}, \cite{komacek2018rnaas} analytically investigated this effect on the day-night heat transport of ultra-hot Jupiters. They found that the day-night temperature difference increases with increasing temperature until reaching an equilibrium temperature of about 2300K, after which the day-night temperature difference decreases with increasing temperature. This finding can qualitatively explain the relatively small phase curve amplitudes observed for several ultra-hot Jupiters (Figure \ref{fig.10}). Next, \cite{tan2019} implemented the effects of heating and cooling, change of mean molecular weight and gas thermodynamic properties caused by hydrogen recombination and dissociation into an idealized GCM similar to that in \cite{komacek2017}, and they systematically investigated the influence on the atmospheric circulation of ultra-hot Jupiters. They likewise confirmed that the fractional day-night temperature difference as a function of equilibrium temperature peaks at $\sim$2300 K, in good agreement with the prediction of \cite{komacek2018rnaas}. Additionally, wind speeds and direction of the equatorial flows at relatively low pressures are strongly affected by including the hydrogen dissociation and recombination, and this influence is more prominent with higher equilibrium temperature.

\subsection{Clouds, hazes, and chemistry on hot Jupiters}
There now exist several lines of evidence for clouds and hazes in the atmospheres of hot Jupiters. First, many hot Jupiters exhibit transit (transmission) spectra that are considerably flatter than predicted by cloud-free models; by causing broadband scattering, high-altitude clouds/hazes can flatten spectral features in the observed way. Moreover, in the visible, these transmission spectra commonly exhibit a Rayleigh-like slope attributed to scattering by small aerosol particles (e.g., \citealp{sing2011,sing2016,pont2013,stevenson2016,iyer2016,heng2016,barstow2017,wakeford2017}). Second, measurements of secondary eclipse at visible wavelengths imply that several hot Jupiters exhibit sufficiently high albedo to require clouds (e.g., \citealp{evans2013, heng2013,esteves2013,esteves2015,angerhausen2015}). Energy-balance arguments for planets with full-orbit IR lightcurves also imply significant Bond albedos for several hot Jupiters \citep{schwartz2015}. Most spectacularly, Kepler visible lightcurves of several hot Jupiters peak after secondary eclipse, indicating that the dayside bright spot in visible wavelengths is west, rather than east, of the substellar point \citep{demory2013,angerhausen2015,esteves2015, shporer2015}. This is interpreted as an inhomogeneous cloud distribution, with a cloud-dominated region west of the substellar point. 

Atmospheric circulation modulates the 3D distribution of condensate clouds, but in turn the cloud distribution influences the radiative heating/cooling and therefore the circulation. We consider these issues in turn, followed by brief discussions of cloud-induced variability, hazes, and coupling between chemistry and dynamics.

\subsubsection{How does the atmospheric circulation affect the cloud distribution?}
Two main processes can shape the cloud distribution in hot Jupiter
atmospheres. First, the temperature affects the cloud distribution,
leading to a day/night contrast in the cloud abundance. Second, the
meridional atmospheric circulation leads to an equator to pole cloud
variation that superimposes on the day/night variation.

 \begin{figure}      % use "figure*" instead of "figure" if you want your figure to span both columns
\epsscale{1.}      % adjust this number to change the size of your figure
\includegraphics[width=\linewidth]{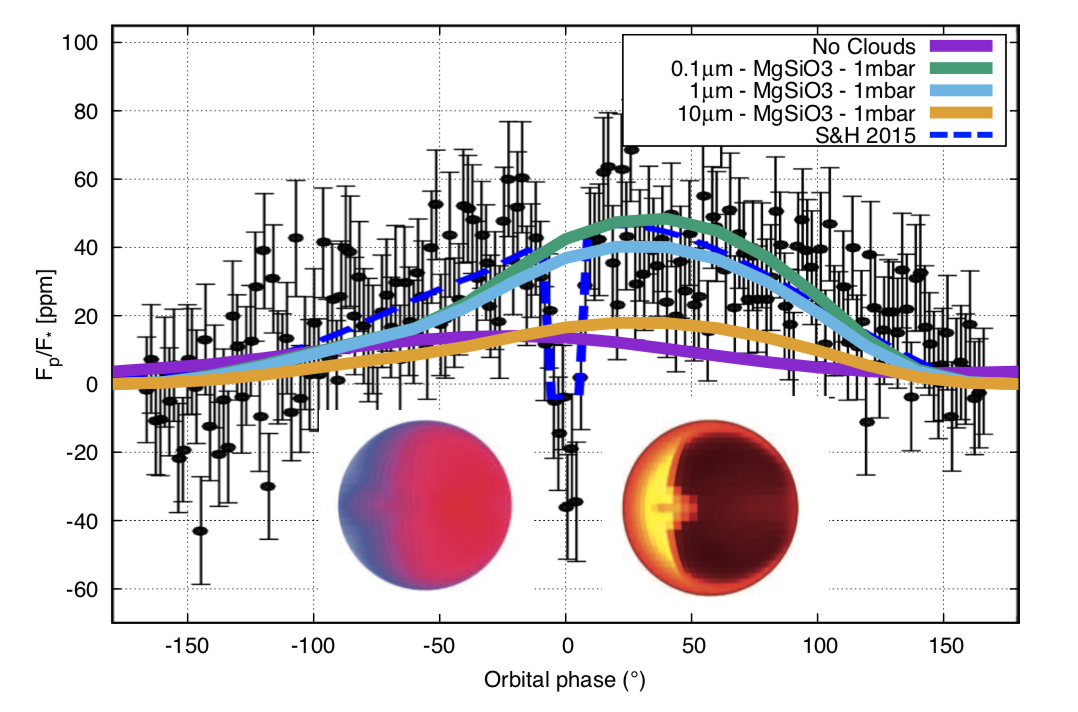}
\centering
\caption{The optical phase curve of Kepler-41b observed with the
  Kepler space telescope~\citep{shporer2015} compared with the
  output from cloudless global circulation and models including the
  presence of silicate clouds of different particle sizes. The left
  inset shows the temperature distribution on the dayside of a global
  circulation model with similar equilibrium
  temperature~\citep{parmentier2016}. The right inset shows the
  brightness distribution on the dayside of the same model integrated
  over the Kepler bandpass. The bright crescent on the left is due to
  the reflection by clouds and creates the positive shift of the
  optical phase curves.}
\label{fig.11}
\end{figure} 

The temperature effect has been highlighted by the observation of optical phase curves \citep{demory2013,esteves2013,esteves2015} from which the distribution of clouds on the dayside atmosphere can be mapped \citep{shporer2017}. As shown
by~\citet{shporer2015}, all the best targets for cloud mapping showed
a partial cloud coverage on their dayside, indicating that most hot
Jupiters are likely partially cloudy, with clouds present on the
western part of the dayside and absent from the substellar point and
the hot spot. As seen in Fig.~\ref{fig.11}, this cloud
distribution is anticorrelated with the temperature distribution
derived from infrared phase curves which points toward an eastward
shift of the hottest part of the atmosphere~\citep[see][for a review]{parmentier2018}. This indicates that clouds are present
where the atmosphere is cold and absent where it is hot. For the ultra
hot Jupiters this can be easily explained by the fact that no known
species can condense at the high temperatures of their
dayside~\citep{helling2019}. For cooler planets, however, something
must prevent the clouds to form in the dayside. One possibility,
proposed by~\citet{parmentier2016}, is that the most refractory
elements that could form clouds on the dayside of the cooler hot
Jupiters are trapped in the deep layers of the planet. If that is
correct, hotter planets would be expected to have silicate
(MgSiO$_{\rm 3}$ and Mg$_{\rm 2}$SiO$_{\rm 4}$), corundum (Al$_{\rm
  2}$O$_{\rm 3}$), Iron or perovskite (CaTiO$_{\rm 3}$) clouds whereas
planets with an equilibrium temperature cooler than $1500\,\rm K$
would have an atmosphere dominated by sulphide clouds such as
manganese sulphide (MnS), zinc sulphide (ZnS) or sodium sulphide
(Na$_{\rm 2}$S). Such a transition would be very similar to the
transition between L and T brown dwarfs~\citet{morley2012}. However,
this mechanism operates only if cloud particles can be sequestered
below the photosphere for cooler planets, which is a balance between
between settling and vertical mixing. The presence of such a deep
sequestration of condensates has been simulating for silicate clouds
by~\citet{powell2018}. The study shows that part, but not all, of the
cloud material was sequestered in the deep layers of the
atmosphere. The formation of larger particle sizes, through the
inclusion of a larger number of species such as iron could potentially
help sequester a larger amount of silicates and explain the trends
seen in the Kepler optical phase curves. Overall, determining more
precisely the cloud composition of hot Jupiters would shed light into
the deep atmosphere, including the rate of vertical mixing and the
deep thermal structure, both related to the deep atmospheric
circulation.

 \begin{figure*}      % use "figure*" instead of "figure" if you want your figure to span both columns
\epsscale{1.2}      % adjust this number to change the size of your figure
\includegraphics[width=\linewidth]{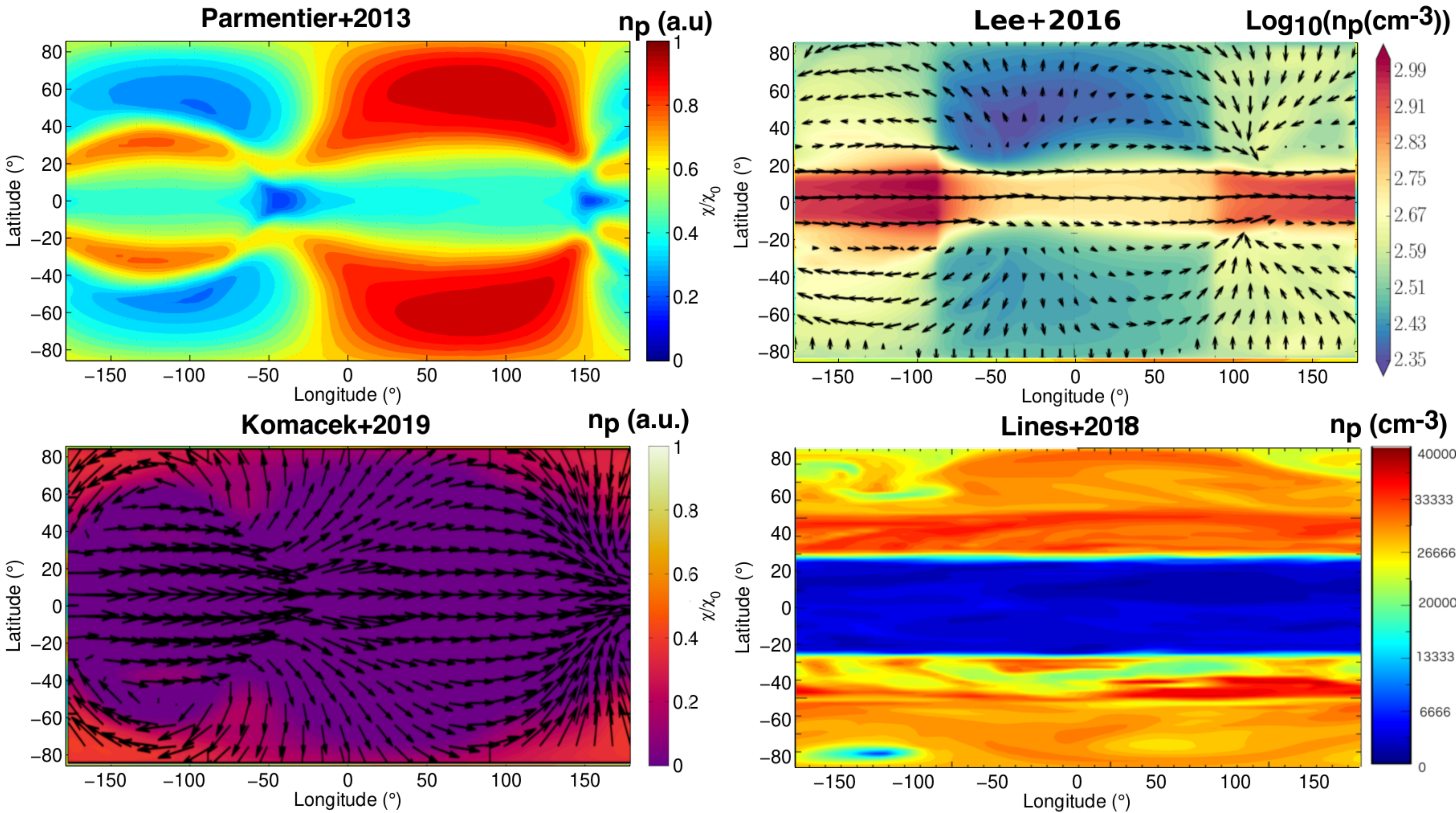}
\centering
\caption{Cloud particle number density at 1 mbar of four different
  models using different assumptions.~\citet{parmentier2013}
  and~\citet{komacek2019} model HD209458b using passive tracers
  advected by the flow with a settling velocity corresponding to
  $2.5\mu m$ particles. Whereas the former uses a non-grey radiative
  transfer scheme including TiO/VO, the second one uses a semi-grey
  radiative transfer.\citet{lee2016} and~\citet{lines2018}
  use the same microphysical cloud model. The former is coupled to the
  GCM of~\citet{dobbsdixon2013} with the parameters of
  HD189733b whereas the latter is coupled to the Metoffice ENDgame
  global circulation model of~\citet{mayne2014} with the
  parameters of HD209458b without TiO. Both include the radiative
  feedback of the clouds. All models show that the particle number
  density is different between the equator and the mid-latitudes,
  showing that banded clouds can be formed by the circulation. Models
  disagree, however, as to whether the equator should be depleted or
  enhanced in cloud particles. The dynamical mechanism responsible for
  this pattern is not yet fully understood.}
\label{fig.12}
\end{figure*}

The atmospheric circulation can also affect the cloud coverage by
advecting cloud material from one part of the planet to another
one. The level of coupling between flow and particles depends on the
pressure. For a given particle, the settling timescale increases with
decreasing pressure and, below some pressure, always becomes smaller
than any circulation timescale. At 1 mbar in HD209458b,
~\citet{parmentier2013} show that particles $1\mu m$ or smaller should
be coupled to the flow whereas larger particles should be more
strongly affected by the vertical settling. As a consequence,
simulations often show that the particles horizontal distribution is
more homogeneous for small than for large particles. Particularly, for
large particles, the interaction between the atmospheric circulation
and the settling particles often lead to a peculiar behaviour at the
equator. As seen in Fig.~\ref{fig.12} in three out of
four models the equator gets depleted in clouds whereas in the
~\citet{lee2016} models the equator gets enhanced in clouds. Overall,
more investigation is needed to understand the banding of clouds in
hot Jupiters.

\subsubsection{How does the cloud distribution affect the atmospheric circulation?}
 
In hot Jupiter atmospheres, clouds affect the atmospheric circulation
mainly through their optical properties, the latent heat release
during their formation being negligible. When present on the dayside,
they can cool the planet by reflecting some of the incoming stellar
radiation but can also warm the atmosphere by increasing the
greenhouse effect. When on the nightside, clouds always tend to warm
up the atmosphere. The final effect of clouds on the atmospheric
circulation therefore depends on their precise
location.~\citet{roman2017} show that nightside clouds tend to warm
the planet nightside, leading to smaller day/night contrast on
isobars. A global cloud, however, produces a temperature structure
similar, but cooler than the one of a clear atmosphere. Although the
cloud radiative feedback can strongly affect the day/night contrast
and the hot spot offset on isobars, all models found that they do not
change qualitatively the general behaviour of the atmospheric
circulation such as the presence of a superotating jet (see
~\citet{parmentier2016},~\citet{lee2016},~\citet{lines2018} or
~\citet{roman2019}).

{ Recently,~\citet{Parmentier2020} carried a systematic study of the effect of non-grey nightside clouds on hot Jupiter atmospheres and their observational consequences using the SPARC/MITgcm. They confirm that nightside clouds dramatically reduce the overall day/night heat transport (see Figure~\ref{fig.8}). Furthermore, they highlight an apparent contradiction: whereas on isobars nightside clouds decrease the day/night temperature contrast and increase the hot spot shift on isobars, they usually increase the phase curve amplitude and decrease the phase curve offset. This apparent contradiction is due to the fact that phase curves probe hemispherically averaged flux maps. When sharp variations in flux with longitude are present due to sharp variation of the atmospheric opacities, the center of the brightest hemisphere does not necessarily correspond to the longitude of the brightest point. Overall, this shows that phase curve amplitude and phase curve offsets do not necessarily probe the day/night temperature contrast and hot spot offsets on isobars. This should be kept in mind when interpreting the results from analytical theories that often assume isobaric heat transport.}

Importantly, for a given planet, the effect of the clouds on the
atmospheric circulation can easily be over or
underestimated. Neglecting the wavelength dependence of the cloud
opacity (such as~\citet{roman2019}), or the scattering properties of
the clouds ~\citep{lee2016} can strongly enhanced the heating rates
and affect the thermal structure and modelled observation in a much
stronger way than in reality~\citep{harada2019}. Similarly,
considering a species that is not present in the atmosphere can lead
to a strong overestimate of the cloud radiative feedback. As shown by
~\citet{lines2019}, using two different cloud parametrisation can lead
to very different radiative feedback, thermal structure and
observational consequences.

\subsubsection{Clouds and atmospheric variability}

Atmospheric variability has been long sought in hot Jupiter infrared
observations. Clear sky models of hot Jupiters are predicted to vary
by less than $1\%$ in their global
temperature~\citep{showman2009, rauscher2012b,
  komacek2020}, which would not lead to observational
evidence given current observational facilities. Optical phase curves,
however, have been shown to vary significantly for two
planets~\citep{armstrong2016, jackson2019}. Although the
source of the time variation is still under debate, and possibly
includes magnetic coupling between the winds and the planetary magnetic
field~\citep{rogers2014komacek, rogers2017} it has been postulated
that the presence of clouds could significantly enhance the observable
variability~\citep{lines2018}. Indeed, a planet dayside can rapidly evolve from barely
cloudy to fully cloudy when the global temperature changes by one to
two hundreds degrees~\citep{parmentier2016}. Moreover, on a given
bandpass, the relative importance of the reflected light and thermal
emission can be extremely sensitive to the cloud abundance and the
temperature.~\citet{armstrong2016} postulated that the atmosphere of
HAT-P-7b could oscillate between two states, one hot and relatively
clear dominated by thermal emission and having a bright spot at the
temperature maximum, east of the substellar point and a colder,
cloudier state dominated by reflected light and having a bright spot
at the cloud maximum, west of the substellar point. Such an
oscillation would lead to large variation from positive to negative in
the phase curve offset while keeping the total dayside luminosity
relatively constant.

\subsubsection{Haze}

Haze modelling, assuming that methane is a main haze precursor, have
shown that hazes can form and explain some features in the
transmission spectrum of hot
Jupiters~\citet{lavvas2017}. However, since all the observed
planets do not have large methane concentrations, other pathways might
be more important. Particularly, recent experiments at $\approx
1500\,K$ by~\citep{fleury2019} have shown that photochemical
hazes can form in a pure H2-CO gas at high temperature. Haze and
condensation clouds are expected to be affected differently by the
atmospheric circulation. Whereas cloud particles form and evaporate at
similar temperature and pressures, hazes can have a much higher
evaporation temperature than the temperature they formed. As a
consequence, whereas clouds are expected to track the temperature maps
of hot Jupiters, hazes might provide a more homogeneous aerosol
cover. The two could potentially be distinguished by measuring the
difference between the east and west limb of the
planet~\citep{line2016, powell2019}. Hazes would
cover both the east and west limbs whereas condensation clouds would
only be present on the cooler western limb~\citep{kempton2017}. Recent GCM investigations are starting to test these ideas \citep{steinrueck2019}.

\subsection{Chemistry}

If the atmospheres of hot Jupiters were in local chemical equilibrium,
the large day/night temperature contrast would naturally lead to large
horizontal variation in their chemical composition. However, the
consensus is that hot Jupiter atmospheres are not in chemical
equilibrium. The reason is that the chemical timescale is an
exponential function of both temperature and pressure. As an example,
the time it takes to convert carbon monoxide to methane at 0.1 bar is
of order of 10 days at 2500K but of the order of millions of years at
1000 Kelvin~\citep{visscher2012}. For comparison, the horizontal
advection timescale is on the order of days. As first shown
by~\citet{cooper2006}, atmospheric mixing is expected to
homogenise the atmospheric abundances horizontally extremely
easily. The exact value of the resulting abundances is, however,
subject to the details of the atmospheric circulation. To first
order,~\citet{agundez2014} show that the dayside photospheric
abundances set the nightside abundances of the atmosphere, with the
dayside abundances being themselves set by the vertical quenching. As
a consequence the final atmospheric abundances are determined by the
vertical mixing strength on the dayside atmosphere and and the
pressure and temperature at the quench point, where the vertical
mixing timescale equals the chemical
timescale.~\citet{drummond2018} { and~\citet{drummond2020} additionally showed that the meridional
circulation could also mix the chemical composition latitudinally, meaning that the 3D nature of the atmospheric mixing by the circulation should not be ignored.} Finally,
differences between chemical schemes~\citep[e.g.][]{moses2011,
  venot2019} and the simplifications needed to couple the scheme
to the atmospheric circulation~\citep{tsai2018} can also change
the expected quenched abundances~\citep[see Fig. 13 of][]{drummond2018b}

Changing the chemical abundances of the atmosphere through chemical
quenching changes the optical properties of the atmosphere, hence the
energy balance and the atmospheric circulation. However, as shown by
~\citet{steinrueck2019, drummond2018,
  drummond2018b}, for the methane/carbon monoxide reaction,
quenching affects the thermal structure of the atmosphere by
$\pm100\,K$, which is small compared to the day/night contrast of the
atmosphere. Although important when comparing model outputs to
observations, chemical quenching does not fundamentally alter the
atmospheric circulation.

\begin{figure}      % use "figure*" instead of "figure" if you want your figure to span both columns
\epsscale{1.2}      % adjust this number to change the size of your figure
\includegraphics[scale=0.65]{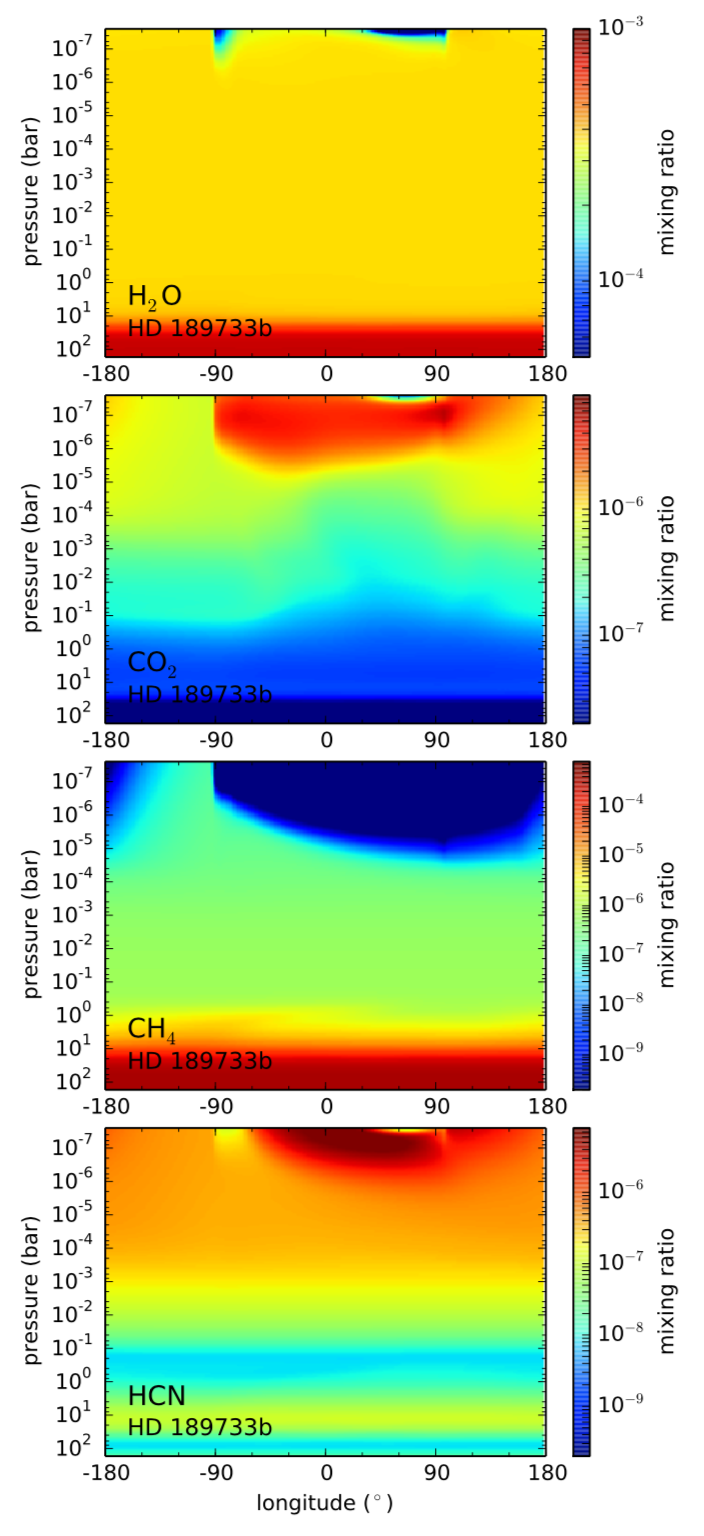}
\centering
\caption{Abundances of several molecules at the equator of HD189733b
  calculated by~\citet{agundez2014} using a 2D model
  (longitude/pressure). The abundances are determined by a combination
  of diffusive vertical mixing, advective horizontal mixing and
  photochemistry. At pressures larger than 1-10 bars, the abundances
  are in chemical equilibrium, between 1-10 bars and $100-10\,\mu\rm
  bar$ the abundances are determined by a combination of vertical and
  horizontal quenching. For pressures lower than $100-10\,\mu\rm bar$
  photochemistry starts being important.}
\label{fig.13}
\end{figure}

There are two situations where the chemical abundances of CO, $\rm{CH}_4$, and $\rm{H_2O}$ are expected to depart from being vertically and horizontally constant.  First, at pressures lower than
$\approx 10\mu\,bar$, photochemistry is expected to significantly
deplete the dayside atmosphere from molecules such as methane, water
or ammonia while increasing the abundances of other species such as
HCN or CO2. Horizontal advection is expected to extend the effect of
photochemistry to the limb and part of the nightside atmosphere (see
Fig.~\ref{fig.13}). Importantly, photochemistry can also affect
the higher pressures through vertical mixing, resulting on an inverted
quenching as seen for the case of HCN in Fig.~\ref{fig.13}.

The second situation where abundances are not constant through the atmosphere is extremely high temperatures, such as occur on ultra-hot Jupiters, when molecular dissociation becomes important.  The timescale to thermally dissociate
and reform molecules is much faster than the timescale for other
chemical reactions~\citep{bell2018, kitzmann2018}. As a consequence,
ultra-hot Jupiters are expected to be in close to chemical equilibrium
with respect to the dissociation reactions. The molecular abundances
of almost all species, including $H_{\rm 2}$, are expected to vary
both vertically and horizontally, CO being the only molecule to not
dissociate at these temperatures due to its higher binding
energy~\citep{parmentier2018, kitzmann2018}. In the hottest planets the
dissociation of $H_{\rm 2}$ on the dayside and its recombination on
the nightside is expected to significantly enhance the day to night
heat transport through latent heat
transport~\citep{bell2018, komacek2018rnaas, tan2019, 
mansfield2020, wong2019}.

\subsubsection{Vertical mixing}

Vertical mixing is a fundamental outcome of atmospheric
circulation. Understanding mixing is critical to understanding how the atmosphere interacts with clouds and chemistry.  It affects the
abundance of clouds, determines their particle size distribution and
whether particles can sequester important chemical species deep in the
atmosphere. However, understanding vertical mixing in hot Jupiter is a
challenge. Despite being locally stable to convection, the atmospheres
of these planets can transport and mix material with large scale
atmospheric dynamics. Using passive tracers in a global circulation
model of HD209458b,~\citet{parmentier2013} show that the vertical
mixing in hot Jupiters is due to the combined effect of large scale
upwelling on the dayside, downwelling on the nightside and localised
updraft and downdraft close to the terminators. These mixing patterns
are very different from known solar system and brown dwarfs
equivalent, and, for example, vertical mixing calculations based on
mixing length theory cannot be easily applied to the hot Jupiter
case. As an example, for HD209458b~\citet{parmentier2013} estimated
the vertical mixing by measuring the vertical flux of passive tracers
along isobars of their global circulation model and found a vertical
mixing coefficient that is 100 times smaller than the prediction from
mixing length theory. Nonetheless, the resulting vertical mixing at
the photosphere is much larger  than
at the photosphere of known solar-system planets (e.g., eddy diffusivity of $\approx 10^5m^2/s$ at 100 mbar).

Earth stratosphere is probably the best solar-system analogue for
vertical mixing in hot Jupiters. As shown by~\citet{holton1986},
mixing in the stratosphere is determined by large scale atmospheric
motions. The formulation of~\citet{holton1986} served as the base of
the analytical models developed by~\citet{zhang2018a,
  zhang2018b} and ~\citet{komacek2019}. They show that
the vertical mixing is tightly tied to the correlation between the
abundance on isobars of the species being mixed and the vertical
motions. { When a chemical species has an equilibrium background abundance
varying with pressure (for example, decreasing with decreasing pressure), the atmospheric circulation naturally causes a
correlation between vertical motion and abundances. Where the winds
are upward they carry parcel of gas with high chemical abundance, { while where winds are downward they carry less chemical abundance. The net effect leads to a net chemical flux from regions with high chemical abundance to regions with low abundance.  }} Efficiency of chemical transport by this correlation can be modulated by two different phenomena. The first one
is horizontal mixing: the larger the isobaric winds, the harder it is
to maintain a horizontal perturbation due to the vertical winds. The
second one is relaxation timescale. Given enough time a chemical
species abundance will tend towards local chemical equilibrium. If
this equilibrium abundance is only pressure dependent, then the
fastest the chemical timescale, the smaller the horizontal variations
and thus the smaller the mixing. Following these ideas,
~\citet{komacek2019} derived that the vertical mixing coefficient
should scale as:
\begin{equation}
K_{\rm zz}=\frac{w^2}{\frac{1}{\tau_{\rm chem}}+\frac{1}{\tau_{\rm adv}}},
\end{equation}
where $w$ is the vertical velocity, $\tau_{\rm chem}$ is the chemical
timescale and $\tau_{\rm adv}$ the horizontal advective
timescale. This formula shows that a vertical mixing coefficient,
$K_{\rm zz}$, cannot be extracted from an atmospheric flow
independently from the chemical species. Otherwise said, different
chemical species are advected following different paths in the 3D
atmosphere which are equivalent to different 1D vertical mixing
coefficients. Finally, the formula is valid for species such as
chemical species that are non-conservative. Cloud particles cannot be
well represented by this model since they do not disappear but fall
down. Intriguingly, despite having investigated settling timescales
spanning several orders of magnitude, ~\citet{parmentier2016},
~\citet{zhang2018b} and ~\citet{komacek2019} all found that the
vertical mixing coefficient derived from the global circulation models
was rather independent of the particle size. More work is needed to
understand this behaviour and derive a relationship estimating
vertical mixing for settling particles.

Finally, because the vertical mixing timescale inherently depends on the horizontal mixing, any estimate of the vertical mixing rely on a
global averaging on isobars (e.g.~\citet{parmentier2013}). If one is
lucky, the horizontally averaged circulation on isobars might act like
a one-dimensional diffusive column. However, on {\it local} scales, mixing is not generally diffusive.  As such,  it is  generally not correct to calculate local $K_{\rm zz}$ profiles from 3D circulation. Doing so will likely produce spurious variations of mixing with height (e.g., the local vertical mixing coefficient would go to zero when the local vertical velocity changes sign) and the sensitivity of the outputs of the model to the method to calculate the vertical mixing coefficient should be thoroughly tested to understand which conclusions are robust and which ones are not (see e.g. \citealp{helling2019}, for a discussion).

 \subsection{Other aspects}
 \subsubsection{Choice of equations}
 \begin{figure*}      % use "figure*" instead of "figure" if you want your figure to span both columns
\epsscale{1.2}      % adjust this number to change the size of your figure
\includegraphics[scale=0.65]{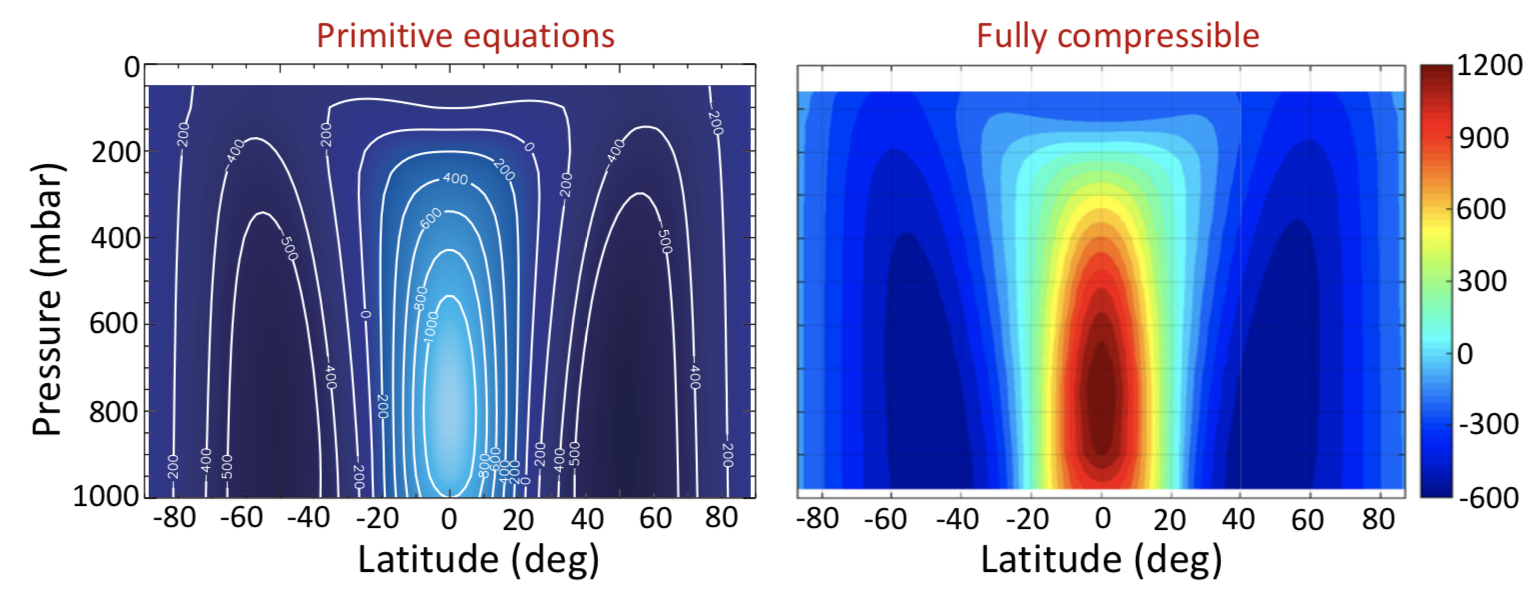}
\centering
\caption{Comparisons of solutions from the primitive equations (left)
  versus full Navier-Stokes equations (right) in the equilibrated
  ``shallow hot Jupiter'' test case defined by
  \citet{menou2009}, in which strong day-night heating
  (implemented with a Newtonian cooling scheme) occurs in an
  atmosphere with a surface pressure of 1~bar.  Shown are zonal-mean
  zonal winds versus latitude and pressure.  Left panel shows the
  primitive equation implementation of this test case from
  \citet{heng2011}.  Right panel shows the fully compressible,
  Euler-equation implementation of the test case, from
  \citet{mendonca2016}. }
\label{fig.14}
\end{figure*} 

All the global circulation models described in the studies above do not solve the same set of dynamical equations. The primitive equations (used in e.g. the SPARC/MITgcm) are the standard dynamical equations appropriate for
circulations in stratified atmospheres having horizontal length scales
greatly exceeding their vertical scales (see \citealt{vallis2006} or
\citealt{showman2010} for a discussion of equation sets in
atmospheric dynamics).  These conditions of stratification and large
aspect ratio are generally met for the large-scale flow in planetary
atmospheres, including hot Jupiters, where the typical horizontal
length scales are $10^4$--$10^5\rm\,km$ and atmospheric scale heights
are $\sim$200--$500\rm\,km$, leading to aspect ratios of
$\sim$20--500.  Large aspect ratio and stable stratification generally
allow the vertical momentum equation to be replaced by local
hydrostatic balance, i.e., $\partial p/\partial z=-\rho g$, where $p$
is pressure, $\rho$ is density, $z$ is height, and $g$ is gravity;
this balance means that spatially and temporally varying
meteorological perturbations to the density are generally in
hydrostatic balance with meteorological perturbations to the vertical
pressure gradient.  \citet{showman2008, showman2008b}
performed a scaling analysis of the vertical momentum equation for the
large-scale flow on hot Jupiters, which suggests that, for large-scale
flows, hydrostatic balance is valid locally to typically $\sim$1\% or
better.\footnote{\label{eq-sets} In addition to hydrostatic balance,
  the primitive equations also generally adopt the ``traditional
  approximation,'' in which Coriolis and metric terms involving
  vertical velocity are dropped from the horizontal momentum equations
  and the Coriolis term is dropped from the vertical momentum
  equation, and a ``thin-shell'' or ``shallow-atmosphere''
  approximation, in which distance from the planetary center is
  replaced with a reference planetary radius when it does not occur
  inside a derivative.  Conservation of angular momentum requires that
  the thin-shell and traditional approximations be taken (or not)
  together.  Under conditions when the thin-shell approximation is
  valid, distance from the planetary center varies only slightly
  across the vertical extent of the atmosphere, and so gravity is also
  typically assumed to be constant rather than varying radially.  See
  \citet{vallis2006} for a general treatment of these
  issues. \citet{mayne2014} presents a nice summary of the
  relationship among these approximations in the context of
  hot-Jupiter models.}  Importantly, the primitive equations make no
assumptions that density variations are small, nor do they set any
explicit limits on the wind speeds, in contrast to some other reduced
equation sets in atmospheric dynamics.\footnote{For example, the
  non-hydrostatic Boussinesq and anelastic equation sets, which are
  often used to study convection in planetary interiors, explicitly
  assume that dynamical perturbations to the density are small and
  that the wind speeds are small compared to the speed of sound.}
However, the fastest wind speeds in hot-Jupiter circulation models
commonly exceed the speed of sound, at least in local regions (the
speed of sound in an ideal-gas hydrogen atmosphere is
$2.2\rm\,km\,s^{-1}$ at $1000\rm\,K$, rising to nearly
$4\rm\,km\,s^{-1}$ at $3000\rm\,K$).  This raises the question of
whether acoustic shocks may form in the atmospheres of hot Jupiters.
Shocks, by definition, are sharp features with small horizontal scales
across the shock, which would necessarily imply a local violation of
hydrostatic balance.  Thus, an important question is the extent to
which shocks are important and whether the primitive equations break
down in a significant way.
{ Furthermore, in fluid planets, the smooth connection between the deep interior and the observable atmosphere can lead to flows spanning a large vertical extent that can get close to violate the traditional approximation used in the primitive equations~\citep{mayne2019}.}

To address this issue, several groups have performed 3D simulations of
hot Jupiters using the fully compressible (Euler or Navier-Stokes)
equations.  Dobbs-Dixon and collaborators were the first to tackle
this problem
%first with a model using idealized day-night thermal forcing, and
%progressing to more realistic, multi-band treatments of radiative
%transfer
\citep[e.g.,][]{DobbsDixon2008, DobbsDixon2010, dobbsdixon2012,dobbsdixon2013}.  Their simulations produce circulations that
are qualitatively very similar to those of primitive-equation GCMs,
although differences in radiative forcing, implementation of friction,
and other modeling details prevent a direct, head-to-head benchmark
comparison.

 \begin{figure*}      % use "figure*" instead of "figure" if you want your figure to span both columns
\epsscale{1.2}      % adjust this number to change the size of your figure
\includegraphics[scale=0.65]{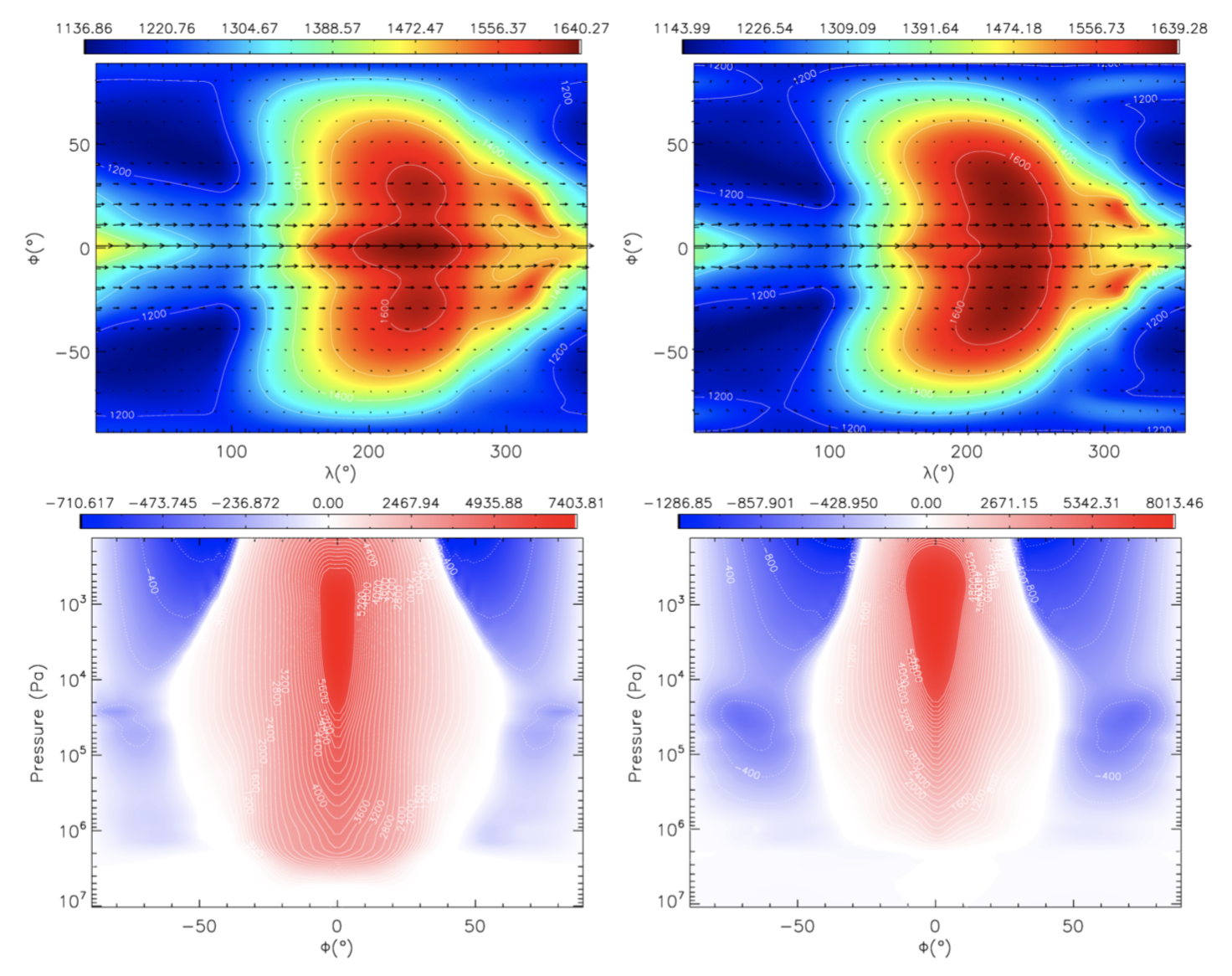}
\centering
\caption{Comparisons of primitive equations (left) versus full
  Navier-Stokes equations (right) in spin-up models of hot Jupiters
  after 10,000~days of integration, from
  \citet{mayne2017}. Planetary parameters of HD~209458b are
  adopted.  Day-night forcing is implemented with Newtonian
  cooling. Top row shows temperature (colorscale, K) and winds
  (arrows) at $0.2\rm\,bar$; bottom row shows zonal-mean zonal winds
  ($\rm m\,s^{-1}$) versus latitude and pressure.}
\label{fig.15}
\end{figure*}

\citet{mayne2014} and \citet{mendonca2016} presented new,
non-hydrostatic hot Jupiter GCM codes, the former based on the UK Met
Office dynamical core, and the latter a custom code called THOR, which
both allow several levels of approximation to be made in the dynamical
equations.  The most sophisticated system is the fully compressible,
non-hydrostatic Euler equations with radially varying gravity and no
thin-shell approximation (``full'' in the notation of
\citealt{mayne2014}); these are followed by equation sets
that successively introduce various simplifications, including
constant gravity (``deep''), thin-shell and traditional approximations
(``shallow''), followed by the addition of hydrostatic balance (the
primitive equations)$^{\ref{eq-sets}}$.
%\citet{mayne-etal-2014, mayne-etal-2017} presented a new
%model based on the UK Met Office dynamical core, which allows several
%levels of approximation to be made, including the fully compressible,
%non-hydrostatic Euler equations with radially varying gravity and no
%thin-shell approximation (``full''); fully compressible,
%non-hydrostatic equations that adopt constant gravity but do not make
%a thin-shell approximation (``deep''); fully compressible,
%non-hydrostatic equations that adopt constant gravity and the
%thin-shell approximation (``shallow''), and the primitive equations,
%which also adopt hydrostatic balance and the traditional approximation
%(see footnote~\ref{eq-sets}). 
The inclusion of multiple choices of approximation in a single code is
extremely useful, because it allows head-to-head comparisons between
the results of different equation sets, where the radiative forcing,
domain, and other aspects of the numerics are held fixed.  As a
benchmark comparison, these studies both evaluated the ``shallow hot
Jupiter'' test case defined in \citet{menou2009}.
\citet{mendonca2016} showed that, when integrated with the
``full'' equations, this shallow hot-Jupiter benchmark yields very
similar behavior to its primitive-equation implementation
\citep{menou2009, heng2011, bending2013}, as
shown in the top row of Figure~\ref{fig.15}.  \citet{mayne2014}'s
``full'' implementation of this test case did not show such good
agreement, although subsequent work \citep{mayne2017} suggested
that the discrepancies may have resulted from their model's
implementation of numerical diffusion rather than inherent differences
between the primitive and fully compressible equations.

\citet{mayne2017} presented further comparisons between
primitive and fully compressible equations sets, for a hot-Jupiter
setup extending deeper into the atmosphere.  Day-night forcing was
implemented with Newtonian cooling.  No frictional drag was included
near the base of the model, which implies that the winds spin up over
time in these models.  Comparisons between the primitive equations and
the ``full'' set after 10,000 days of integration are shown in
Figure~\ref{fig.15}.  Qualitatively, the two implementations produce
very similar circulations, with roughly the same day-night temperature
differences, qualitatively similar spatial temperature patterns, and
maximum zonal-mean zonal winds that differ by only $\sim$10\%.  Still,
a number of important quantitative differences occur, most notably the
equatorial jet in the primitive equation variant (Figure~\ref{fig.15},
left) extends considerably deeper than in the ``full'' variant
(Figure~\ref{fig.15}, right).  Because these are not equilibriated---the
zonal jet is continuing to spin up over time---it is unclear if these
differences reflect true differences in the equilibrium state that
would be achieved between the two equation sets, or rather simply
reflect differences in the {\it rates} at which the zonal jet spins up
in the two equation sets.  Further comparisons and diagnostics of this
type are needed to clarify the situation.

\subsubsection{ Hot Jupiter inflation mechanisms }
{
 Many hot Jupiters have observed radius larger than expectations from standard planetary evolution models (e.g., see Figure 1 in \citealp{komacek2017structure}). Strong stellar incident irradiation together with the assumption of efficient global homogenization of the incoming energy significantly reduce the interior cooling and help to sustain the large radius (e.g., \citealp{guillot1996,fortney2007}), but is still not sufficient  to explain the  observed   radius. Various mechanisms involving external energy injection to the planetary interior or further reducing the interior cooling   have been proposed. For thorough reviews of this topic, readers are referred to, for example, \cite{fortney2010}, \cite{baraffe2014} and \cite{dawson2018}. Here we only briefly survey several hydrodynamic mechanisms related to the understanding of atmospheric dynamics of hot  Jupiters. 

{ If a small fraction (on the order of 1\%) of the stellar irradiated energy is deposited near or below the radiative-convective boundary (RCB), the planetary contraction can be significantly slowed down or even halted  \citep{Guillot2002,komacek2017structure}.  \cite{showman2002} and \cite{Guillot2002} proposed that if significant vertical wind shear develops in the deep layers approaching the RCB where the stratification becomes weak, hydrodynamic instability such as the Kelvin-Helmholtz instability could occur and dissipate the kinetic energy associated with the mean flow into thermal energy via turbulent cascade near the RCB. A ``Mechanical Greenhouse"  mechanism has been proposed by \citet{Youdin2010}, in which forced turbulence (which may come from shear-wind instabilities) mix higher entropy in the outer radiative zone towards the interior, pushing the RCB  to depth and significantly reducing the cooling. These  mechanisms are qualitatively attractive,  but the quantitative details involving the development and organization of small-scale turbulence and their interactions with the large-scale fields remain inconclusive.  So far only very few studies exist in this direction (e.g., \citealp{fromang2016, menou2019}), and future investigations unifying our understanding of the global-scale flow and small-scale (over lengthscales much smaller than grid sizes typically used in global models) turbulence are desired. Additional uncertainty of the above mechanisms also arises from our lack of understanding on the deep circulation of hot Jupiters (see section 2.7.4).}

Another mechanism driven by the atmospheric dynamics was proposed by~\citet{tremblin2017}. They developed a 2D radiative-advective non-grey model of hot Jupiters equator with a parametrized meridional and vertical mass transport. The simplicity of the model allows one to solve directly for the steady-state solution rather than  integrating forward in time, which impedes most global circulation models to solve for a converged steady state. The model shows that for non-zero vertical transport, the deep atmospheric temperature pressure profile was converging towards a hot adiabat rather than an isotherm as in a one dimensional radiative-convective equilibrium
~\citep[e.g.][]{fortney2008,guillot2010,Parmentier2014}. When large enough vertical transport is assumed, the downward transport of energy is large enough to explain the inflated radii of hot Jupiters. This behavior was further explored with a 3D GCM by ~\citet{Sainsbury-Martinez2019} who used a Newtonian cooling scheme and integrated their models for several thousands of Earth years. They found that for large enough relaxation timescale for the temperature (e.g. larger then 3000years at 200 bars) the model converges towards a hot adiabat as in the 2D model.

The large-scale circulation models proposed by \citet{tremblin2017} and \cite{Sainsbury-Martinez2019} is attractive in the sense that no (yet unknown) small-scale turbulence is required, but it has not been fully demonstrated. First, the mechanism behind the results is still unclear. To mechanically transport thermal energy from low to high pressure, one requires a net downward heat flux  ($\overline{\omega T'}>0$, where $\omega$  is the vertical velocity in  pressure coordinates, $T'$ is the isobaric temperature variation relative to the global mean,  and  the overbar denotes a global mean at isobar surface)  through a level near the upper deep layer given that the lower boundary is impermeable. In the deep layer, this  is a thermally  {\it indirect} circulation, meaning that the circulation works against buoyancy\footnote{Circulation of the  whole system, including the photosphere which is driven by the large day-night temperature contrast,  is still thermally {\it direct}---otherwise no motions will be  maintained against dissipation.}  (e.g., \citealp{holton2012}). It cannot occur spontaneously within the deep layer by its own, and  driving forces to the deep layers from the  upper photosphere have to be involved. It is not yet clear what is driving the thermally indirect  circulation (i.e., where $\overline{\omega T'}>0$) throughout the deep layer. 

Secondly, both the 2D and the 3D models parametrize a fundamental part of the problem. The former parametrizes the vertical flow whereas the latter one parametrizes the heat transfer through a Newtonian relaxation scheme. Whereas both studies argue that their models are efficient to transport energy downward in most of the parameter space they explored, it is worth noting that the radiation and the circulation are intimately linked. Whether the actual physical parameters of hot Jupiters allow for the aforementioned thermally indirect circulation to develop is not clear. Particularly, a more recent study by ~\citet{mendonca2020} using a 3D GCM with semi-grey radiative transfer predicts a much smaller warming of the deep atmosphere then~\citet{Sainsbury-Martinez2019}, which would not be able to explain the inflated radius of hot Jupiters. 

%Therefore, the existence of a meridional  circulation throughout the deep layer does not automatically indicate a mechanism (in other words, it does not tell how and why  $\overline{\omega T'}>0$ is present in the deep layers). More careful diagnoses is required to understand the detailed dynamical mechanism.
%Indeed, when the radiative timescales are much longer than the vertical transport timescale, any atmospheric circulation would transport energy adiabaticaly and lead towards and adiabatic thermal gradient. 
Finally, one often needs to be cautious when applying the simulated results which are in a closed domain to real hot Jupiters. The current simulations are often dependent on the choice of deep  boundary conditions (see section 2.7.4). For example, considering a more realistic situation wherein the GCM is connected to a cold planetary interior (equivalent to the assumed cold initial deep layers in \citealp{Sainsbury-Martinez2019}), the downward heat transport forced by the photospheric circulation could eventually trigger a sharp thermal inversion near the interface between the upper hot adiabat and the lower-entropy interior. Hence a strong and sharp stratification may suppress the circulation further down and stop the heat transport. Additionally, hot Jupiters likely have a convective interior and it is not yet clear how the presence of a radiative-convective boundary would affect the deep atmospheric circulation
~\citep[see e.g.][]{rauscher2014influence}.}

\subsubsection{Magnetic coupling}
{
In the atmospheres of hot Jupiters, sodium, potassium, calcium and aluminium become ionized when dayside temperatures become higher than $2000\,K$~\citep{Batygin2013,Helling2018,helling2019}, rendering the atmosphere significantly conductive. Interactions between the magnetic field and the flow are therefore expected.

Magnetic fields are expected to interact with the atmospheric circulation in two ways. First they dissipate kinetic energy through Ohmic dissipation and hence slow down the winds. Second they affect the atmospheric circulation differentially in different directions, leading to a change in the  circulation patterns. The dissipation effect was recognized early on and implemented in several GCMs by simply adopting frictional drag terms in the hydrodynamic equation sets~\citep[e.g.][]{perna2012}. Further studies proposed that observational measurements, such as smaller than expected hot spot offsets, smaller than expected wind speeds and and higher than expected day/night contrast could be caused by the presence of additional magnetic drag
~\citep{komacek2017,kreidberg2018,Arcangeli2019,koll2018}. 

In all the aforementioned studies, the additional drag terms tend to drive the atmospheric circulation from a jet-dominated towards a day-to-night flow~\citet{komacek2016}. However, as pointed out by~\citet{Batygin2013} zonal flows should be much more stable than  meridional flows when the magnetic field is aligned with the spin axis of the planet\footnote{For misaligned magnetic dipole we would expect a misaligned equatorial jet
~\citep{Batygin2014}}.  

More self-consistent magnetohydrodynamic models were performed by~\citep{rogers2014komacek}. They confirmed that the presence of magnetic coupling leads to a reduced meridional flow. Furthermore, when the conductivity is allow to change due to large day-night temperature difference, ~\cite{rogers2017} show that the direction of the zonal jet can start to oscillate, leading to a variation in the hot spot offset of up to 20 degrees. Such a large variation in the hot spot offset may be responsible for the time variability observed in the Kepler phase curves of two hot Jupiters
~\citep{armstrong2016,jackson2019}.

The coupling between magnetic field and the circulation leads to the downward transport of stellar irradiated energy to planetary interior through the Ohmic dissipation. Stellar energy is deposited at the photosphere which fuels a vigorous atmospheric circulation. This circulation produces currents that can connect to deeper atmospheric layers and dissipate energy~\citep{Batygin2010}. If more than $\sim1\%$ of the stellar energy can be deposited near the RCB, this could explain that many hot Jupiters have an inflated radius
~\citep[e.g.][]{Guillot2002,Menou2012,Thorngren2019}. However, as shown by~\citet{rogers2014} and ~\citet{rogers2014komacek}, the simulations show that magnetic drag slows the deep winds significantly more than predicted by~\citet{Menou2012b}, leading to Ohmic heating $\sim$100 times too small to explain the inflated radii. 

Additional magnetic effects in the atmospheres of hot Jupiters that have been explored theoretically but have not been included in GCMs include the presence of thermal instabilities above the photospheric layers triggered by local heat deposition by Ohmic dissipation~\citet{Menou2012c} or the presence of an atmospheric dynamo triggered by the large day/night variation in conductivity~\citep{Rogers2017McElwaine}, which could change the magnetic field topography. 

Finally, it is worth pointing out that no fully satisfactory model currently exists to compare to the observations of ultra hot Jupiters. The only MHD model that has been applied to hot Jupiter originates from stellar interior modeling from the Rogers group. The anelastic approximation used systematically leads to smaller wind speed than predicted by models solving the primitive or fully compressible equations. In addition, no radiative transfer scheme was used (only Newtonian cooling schemes) and therefore the model outputs cannot be directly compared to observations. 
On the other hand, models including non-grey radiative transfer are pure hydrodynamic models that only consider the magnetic effect through the inclusion of simple drag terms, which  could lead to an incorrect atmospheric circulation pattern in very hot atmospheres. In the coming decades, coupling the two approaches will be a necessary step to interpret the observations of ultra hot Jupiters. }

 \subsubsection{Deep boundary conditions and integration times}

 { Gaseous planets lack a distinctive boundary between the atmosphere and the interior, and  the choices of bottom boundary conditions in  GCMs  could affect the results both at the photospheric levels and in the deep layers. Most hot Jupiter GCMs adopted a slip-free, non-permeable bottom boundary at a pressure of around 100 or 200 bars (e.g., \citealp{showman2009, heng2011b, rauscher2012, mendonca2016,mayne2017}).  Compared to the shallow hot Jupiter simulations with a bottom pressure of about 1 bar (e.g., \citealp{heng2011,mayne2014}), flows in the ``deep" simulations are typically more time-invariant. Recently,  \cite{carone2020} proposed that for rapidly rotating hot Jupiters like WASP-43b, the simulated domain should extend to a deeper pressure ($\sim$700 bars) to properly capture dynamics in the observable layers. }
 
 { The radiative timescale increases rapidly with increasing pressure due to the increased opacity and atmospheric mass. Convergence of  the whole simulated domain is therefore bottlenecked by the extremely long radiative timescale in the deep layers. Recent GCM experiments by \cite{wang2020} showed that several hundred simulated years are needed for the deep flow to converge if assuming an 80-bar bottom boundary pressure. \cite{mendonca2020} showed that several tens of simulated years are required for the equatorial jet to equilibrate. Such long integration time is yet challenging for GCMs coupled with non-grey radiative transfer, chemistry and cloud microphysics, and is also demanding on the conservation properties of dynamical cores. Proper simulating strategies should be investigated in  the near future to cop requirements of both convergence and comprehensive physics. }
 
{ In GCMs utilizing radiative transfer schemes,  a horizontally isotropic net heat flux is typically applied at the bottom boundary. In this case the deep thermal structure is unconstrained and can evolve according to the dynamics driven by the upper atmosphere. In the case of inflated hot Jupiters, the deep model layer may have reached the convective zone given the likely high interior entropy (e.g., \citealp{Thorngren2019}). Scaling analysis and global convection models of planetary interior with uniform surface cooling  predict small ($\sim$a few Kelvin) isobaric temperature variation  in the convective zone (e.g., \citealp{stevenson1991, kaspi2009,showman2011scaling,showman&kaspi2013}). In the absence of strong molecular gradient, the interior is expected to be nearly adiabatic. This means that in the presence of convection, the zeroth-order  temperature structure in the upper convective zone could be  constrained by the specific interior entropy of the planet, which does not evolve over timescales relevant for GCM integration time. In addition, interactions between the stratified layer and the interior convection may trigger a wealth of turbulence and waves that propagate  upward and interact with the mean flow (see Section \ref{brown-dwarfs}).    The extent of which the interior convection can affect  the atmospheric  circulation and deep thermal structure remains unexplored. In the other way around, it remains an open question whether and how interior convection will differ from those forced by uniform surface cooling when the hot-Jupiter-like atmospheric circulation is applied to the surface condition of interior convection models.  }

{ The interior of giant planets is expected to be electronically conducting, and interaction of the magnetic field with the flow leads to Ohmic dissipation and retards the flow (e.g., \citealp{liu2008}). The large-scale flow in the interior is expected to be significantly slower than that near the photosphere. Convection leads to a nearly barotropic state of the interior, witch together with the slower winds (which implies a small  global Rossby number) may result in a Taylor–Proudman effects \citep{pedlosky_book} that  tends to drag down winds in the deep GCM layers (e.g., \citealp{schneider2009}). To crudely represent this effect,  some GCMs adopted a frictional drag near the bottom boundary that relax winds towards zero over characteristic drag timescales  (e.g., \citealp{liu2013, komacek2017,tan2019,carone2020}). Although easy to implement, the choice of drag timescale and the exact form of the drag are rather loosely chosen.  In addition, recent gravity measurements on Jupiter by the {\it Juno} spacecraft have revealed that the zonal jets of Jupiter could penetrate down to $\sim$3000 km below the cloud deck \citep{kaspi2018,guillot2018}. Similarly, the zonal jets of Saturn may extend to $\sim$9000 km below the cloud deck which is constrained by gravity measurements from  the {\it Cassini} spacecraft \citep{iess2019}. The implication is that, for cooler planets in which the conducting  region is far below the GCM domain,  whether a deep drag that relaxes winds toward zero is questionable in general conditions.   }

{ Understanding the interactions and coupling between the photospheric level dynamics and interior dynamics is in a pressing need given the above issues. A self-consistent coupling between them is numerically challenging partly because traditionally two different sets of equations are  used in different parts---primitive or fully compressible equations are used in the upper atmosphere and anelastic approximation is used in the interior. But more critically, in order to achieve a statistically steady state with realistic computational cost, global interior models are overly forced with non-dimensional dynamical parameters that are many orders of magnitude different to realistic values (e.g., \citealp{showman2011scaling}), while models of the upper atmosphere do not suffer from this issue. Proper modeling strategies are likely needed to circumvent this challenge. For example, without a fully coupling between the two parts, modeling of the  individual part can be used as idealized boundary condition of the other part.    }

\section{Warm Jupiters}
\label{section.3}

A bit farther from their stars than hot Jupiters lie the warm
Jupiters, which we define to be approximately Jupiter-mass planets
with effective temperatures of $\sim$300 to $1000\rm\,K$.  They lie
just below hot Jupiters in the upper left corner of
Figure~\ref{fig.2}, with global-mean incident stellar 
and radiated IR fluxes ranging from a few hundred to a $\rm few
\times10^4\rm\,W\,m^{-2}$.  As with hot Jupiters, their interior
fluxes are expected to be relatively weak, perhaps $10$ to $1000$
times less than the incident stellar flux they receive (depending on
the planet's effective temperature, history, and other factors),
implying that, like hot Jupiters, the warm Jupiters should have
atmospheres that in the time-mean are nearly in radiative equilibrium
with their parent star.

As yet, there are far fewer observational constraints on the
atmospheres of warm Jupiters than hot Jupiters, but this situation may
change in the future.  Over 40 transiting warm Jupiters have been
discovered.  Although most of these are Kepler detections that are
difficult to follow up due to their great distance from Earth, over a
dozen have been discovered by groundbased surveys and the Transiting
Exoplanet Survey Satellite (TESS) mission around brighter, closer
stars amenable to further observational
characterization.\footnote{Prominent examples include WASP-69b and 84b
  \citep{anderson2014}, HAT-P-17b \citep{howard2012},
  HATS-17b \citep{brahm2016}, HATS-71b \citep{bakos2018},
  TOI-813b \citep{Eisner2020}, and TOI-216b and c
  \citep{kipping2019}.}

The greater orbital separations of warm Jupiters from their stars will
lead to key differences in behavior relative to hot Jupiters.  
%The smaller incident stellar flux and cooler temperatures imply that
%the radiative time constant is longer, weaking the ability of the
%day-night radiative contrast to drive the circulation.
In particular, the tidal effects that drive hot Jupiters into a
(presumed) state of synchronization and that circularize their orbits
are less dominant at the greater orbital distances of warm Jupiters.
Therefore, unlike typical hot Jupiters---where it is generally a
reasonable assumption that the rotation rate is equal to the
synchronous value, and the obliquity and orbital eccentricity are
zero---warm Jupiters should be expected to exhibit a range of rotation
rates, obliquities, and orbital eccentricities.  This may lead to a
wider range of possible behavior.

Let us quantify the distances at which these transitions occur.  The
tidal synchronization timescale from a primordial rotation rate
$\Omega_p$ is \citep{guillot1996}
\begin{equation}
\tau \sim Q\left({R_p^3\over G M_p}\right)\Omega_p \left({M\over
  M_*}\right)^2 \left({a_{\rm orb}\over R_p}\right)^6 ,
%\sim1\times10^6\left({Q\over 10^5}\right) \left({a_{\rm orb}\over
%  0.05\rm\,AU}\right)^6\rm \,yr
\label{spindown}
\end{equation}
where $G$ is the gravitational constant, $M_*$ is the stellar mass,
$a_{\rm orb}$ is the orbital semi-major axis, and $Q$, $R_p$ and $M_p$
are the planet's tidal dissipation factor, radius, and mass,
respectively.  Evaluating this expression using values appropriate
for a typical hot Jupiter yields
\begin{equation}
\tau \sim 1\times10^6\left({Q\over 10^5}\right) \left({a_{\rm
    orb}\over 0.05\rm\,AU}\right)^6\rm \,yr ,
\label{spindown2}
\end{equation}
where we have used a solar mass for the star, a Jupiter mass and
rotation rate for the planet, along with a planetary radius of 1.2
Jupiter radii, which is typical of hot Jupiters.  For a canonical hot
Jupiter 0.05~AU from its star and adopting $Q\sim 10^5$ (an
appropriate time-averaged value for the giant planets in the solar
system), the equation yields a spindown time of
$\sim$$10^6\rm\,yr$---which underlies the common expectation that hot
Jupiters should be close to a state of synchronous rotation.  However,
the sychronization timescale increases greatly with only modest increases in
semi-major axis: for a planet 0.2~AU from its star, the spindown
timescale increases to 4~Byr, comparable to system ages.  Thus, EGPs
beyond $\sim$0.15~AU around sunlike stars will generally not be
synchronized.  They may thus exhibit a range of rotation rates and
obliquities.

We now discuss, in turn, the effect of non-synchronous rotation,
non-zero obliquity, and non-zero eccentricity on the atmospheric
circulation of warm Jupiters, as currently understood.  All of the
research done so far assumes the incident stellar flux greatly exceeds
the interior flux, which should be valid in relatively old systems for
planets inward of $\sim$1~AU.  We close this section by offering a few
remarks about the way these regimes may be modified by an interior
convective flux.

\subsection{Non-synchronous rotation}

Several GCM investigations have explored the effect of non-synchronous
rotation under conditions of zero obliquity and eccentricity
\citep{showman2009, rauscher2014, showman2015, penn2017}.  Non-synchronous rotation exerts two
effects---first the differing rotation rate changes the strength of
the Coriolis parameter and its gradient $\beta$, thereby influencing
the Rossby number, the Rossby deformation radius, and other factors
that depend on rotation.  Second, non-synchronous rotation implies
that the planet no longer has permanent daysides and nightsides, but
rather that the dayside heating pattern sweeps in longitude across the
planet over time.  This effect might at first seem trivial, but in
fact it can significantly alter the planetary wave modes that arise
from the day-night heating pattern, with consequent implications for
the superrotation, hot-spot offset, and other aspects of the
circulation. Indeed, GCM experiments demonstrate that both effects
critically influence the dynamical behavior.

\citet{showman2009} performed simulations of HD~189733b at
rotation rates 0.5, 1, 1.5, and 2~times the synchronous value, and
\citet{rauscher2014} performed a similar study for both
HD~189733b and HD~209458b, with the goal of understanding how the two
effects listed above influence the circulation regime and IR
lightcurves. For HD~189733b, the rapidly rotating models in both
studies showed the emergence of high-latitude eastward jets in
addition to the primary superrotating equatorial jet.  The slower
rotating model exhibited a robust equatorial jet flanked by strong
westward flow.  In this case, the equatorial jet was fast
and---despite the slower rotation---narrower than the equatorial jet
in the synchronously rotating model.  These changes do not follow the
trends that occur in sychronously rotating GCM experiments when
rotation rate {\it alone} is varied
\citep[e.g.,][]{showman2008}, which suggests that the
longitudinal migration of the dayside heating pattern is critical in
setting the detailed jet properties.

Strikingly, however, the slowly rotating HD~209458b experiment in
\citet{rauscher2014}---corresponding to a rotation period of
6.6~days---developed a qualitatively different circulation pattern
comprising a strong {\it westward} equatorial jet, a dayside hotspot
shifted westward of the substellar point, and day-night flow across
most of the terminator.  They report that equally slowly rotating
models of HD 189733b (i.e., using a rotation period of 6.6~days, three
times the synchronous value) also exhibit a similar circulation, with
a fast westward equatorial jet rather than superrotation.  The
dynamical mechanism causing this transition remains unclear, but it
appears that the slow rotation and migrating dayside heating patterns
are key factors. Interestingly, \cite{mendonca2020} found in synchronously rotating hot-Jupiter models that rotation periods longer than 5 days allowed equilibrated states containing either eastward or westward equatorial jets, which may be a related phenomenon.   As expected, synthetic IR lightcurves for the models
with eastward jets reach maximum flux before the secondary eclipse,
but the slowly rotating HD 209458b model exhibits an IR flux peak {\it
  after} secondary eclipse.  Thus, this phenomenon presents a clear
prediction for future lightcurve observations, as well as having a distinct Doppler-shift signature during transit \citep{rauscher2014}.

 \begin{figure*}      % use "figure*" instead of "figure" if you want your figure to span both columns
\epsscale{1.2}      % adjust this number to change the size of your figure
\includegraphics[scale=0.5]{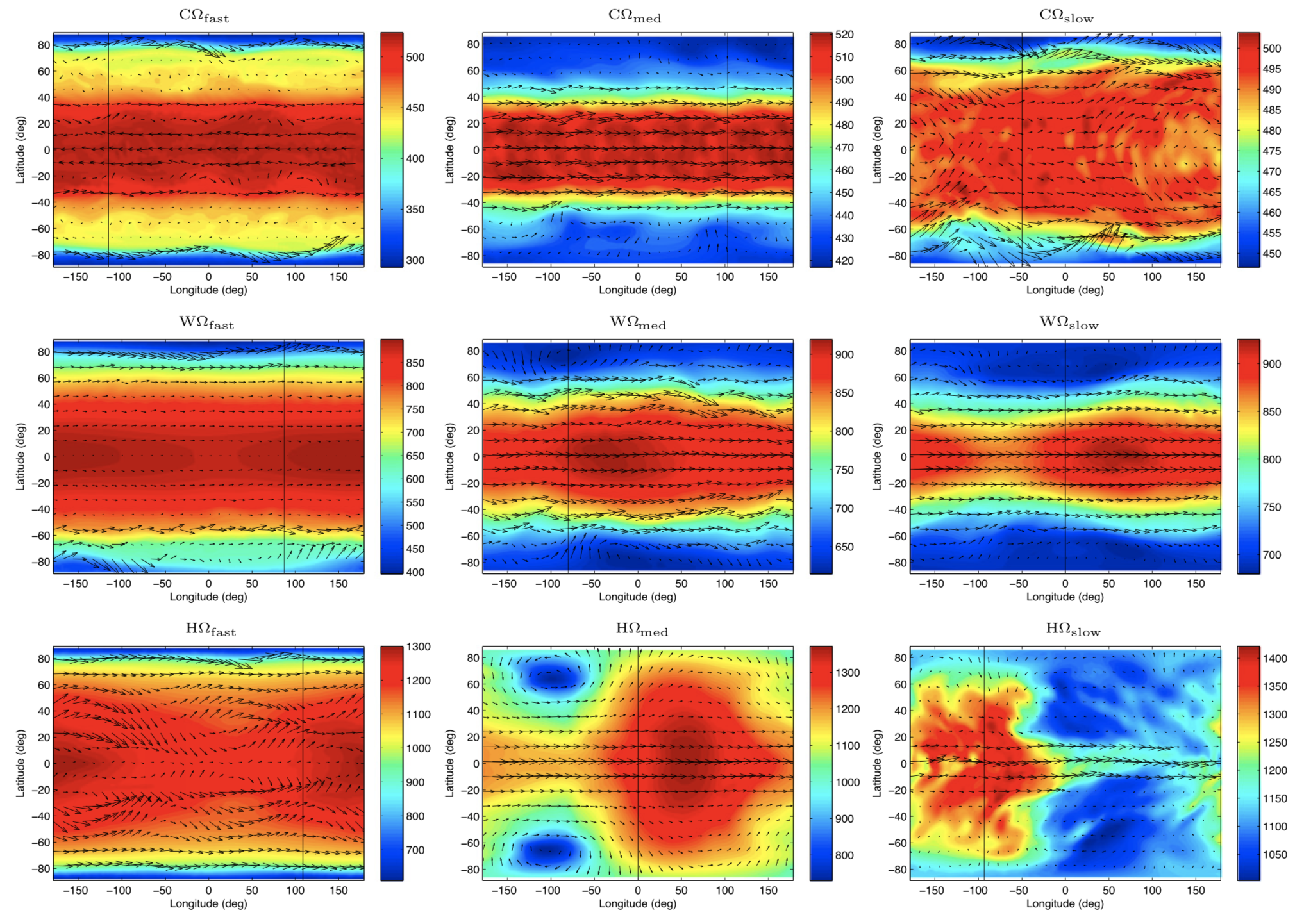}
\centering
\caption{Temperature pattern over the globe (colors) and winds
  (arrows) for a set of 9 models straddling the transition between hot
  and warm Jupiters, illustrating a dynamical transition in the
  behavior, from \citet{showman2015}.  Models in the left,
  right, and middle columns adopt rotation periods of 0.55~days,
  2.2~days, and 8.8~days, respectively; the top, middle, and bottom
  columns adopt orbital semimajor axes of 0.2, 0.08, and 0.03~AU,
  respectively, for planets orbiting a K0 star with the properties of
  HD~189733.  Models in the lower right have radiative time constants
  shorter than their solar days, and exhibit canonical hot-Jupiter
  circulation patterns with large day-night temperature differences
  and equatorial superrotation.  Models in the upper right have
  radiative time constants longer than their solar days, and instead
  exhibit little day-night temperature variation; baroclinic
  instabilities transport heat toward the poles, and the fastest wind
  speeds tend to occur in midlatitudes rather than the equator.  Model
  H$\Omega_{\rm med}$ and W$\Omega_{\rm slow}$ are synchronously
  rotating; the former has parameters identical to HD~189733b. The solid
  vertical bars in each panel show the substellar longitude at the 
  time of these snapshots.}
\label{fig.16}
\end{figure*}

\citet{showman2015} suggested that the circulation of
non-sychronously rotating hot and warm Jupiters systematically splits
into two dynamical regimes, depending on whether the radiative time
constant is shorter or longer than the solar day.  If the radiative
time constant is less than the solar day, the day-night heating
pattern (diurnal cycle) is strong, and the circulation resembles the
canonical hot-Jupiter regime already discussed: the day-night
temperature differences are large, and the global wave modes triggered
by the strong day-night forcing drive equatorial superrotation through
the mechanism described in Section~2.

If instead the radiative time constant is greater than the solar day,
the amplitude of day-night heating (the diurnal cycle) is expected to
be weak.  Longitudinal variations of temperature are therefore small,
and the circulation is driven by the equator-to-pole gradient in
zonal-mean heating, and primarily plays the role of transporting heat
meridionally.  At low obliquities, the sun- light predominantly illuminates the low latitudes, and the greatest temperature gradients tend to occur in mid- latitudes, with zonal-mean temperature decreasing poleward. Through thermal-wind balance (Equation 4), this implies that the strongest winds occur at midlatitudes rather than the equator, and furthermore that those winds are eastward at photosphere levels. The midlatitudes become baroclinically unstable, leading to baroclinic eddies that transport heat poleward; this triggers the formation of Rossby waves, which propagate meridionally away from their latitude of generation. The meridional
propagation of the Rossby waves arising from these baroclinic eddies
causes the convergence of angular momentum into the instability
latitudes (for a review, see \citealp{vallis2006, showman2013}), promoting the generation of midlatitude, eddy-driven jets
analogous to those observed on the Earth.

 \begin{figure*}      % use "figure*" instead of "figure" if you want your figure to span both columns
\epsscale{1.2}      % adjust this number to change the size of your figure
\includegraphics[scale=0.5]{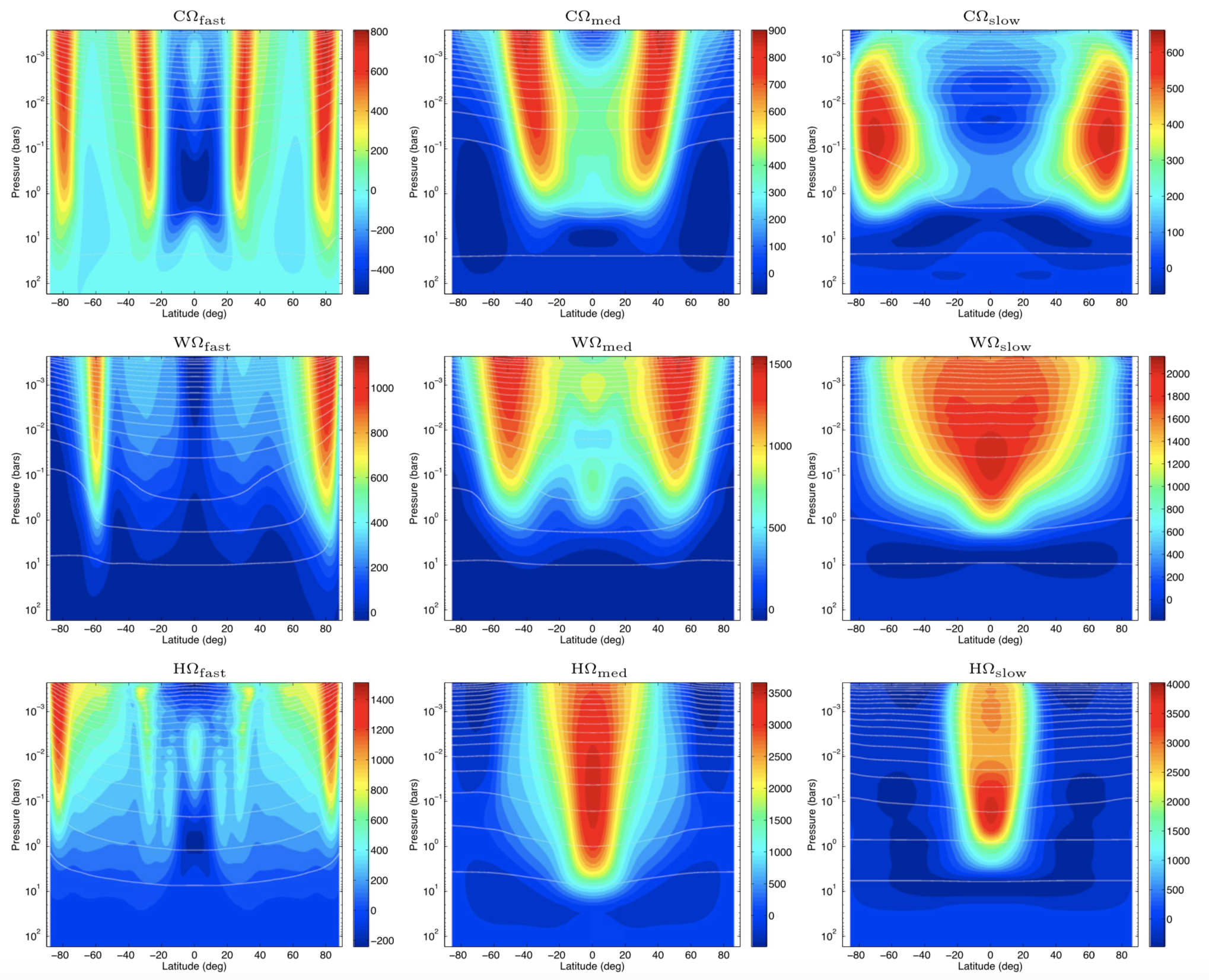}
\centering
\caption{Zonal-mean zonal winds versus latitude and pressure for the
  same 9 models shown in Figure~\ref{fig.16}.  There is
  a striking transition from strong equatorial superrotation for
  models in the lower right to mid-latitude eastward jet streams for
  models in the upper left.  The fastest rotating, least irradiated
  model (upper left panel) exhibits multiple zonal jets in each
  hemisphere, and illuminates the dynamical continuum between the
  warm Jupiters and Jupiter and Saturn themselves. 
  From \citet{showman2015}.}
\label{fig.17}
\end{figure*}

This transition tends to correspond approximately to a
particular orbital distance inside of which the diurnal cycle is
critical and beyond which it becomes less important or 
irrelevant.  To calculate this distance, note that the effective
temperature of a planet in energy balance with its star (and assuming
zero albedo for simplicity) is
\begin{equation}
T_{\rm e} ={1\over \sqrt{2}}\left({R_* \over a_{\rm orb}}\right)^{1/2} T_*,
\end{equation}
where $R_*$  and $T_*$ are the stellar radius and effective temperature,
respectively.  Inserting this expression into Equation~(3)
for the radiative time constant yields
\begin{equation}
\tau_{\rm rad} \sim {p c_p\over g\sigma T_*^3}\left({a_{\rm orb}\over R_*}
\right)^{3/2}
\end{equation}
which implies that, for a given stellar type, the radiative time
constant at the IR photosphere increases approximately as semi-major
axis to the $3/2$ power.\footnote{Of course, the photosphere pressure
  $p$ can depend on atmospheric composition and temperature, which
  would complicate this trend. These effects seem to be relatively
  modest, however, due to the fact that a wide variety of gas-phase
  species become optically thick at pressures a bit greater than
  $0.1\rm\,$bar \citep[e.g.,][]{robinson2014}.}  The solar
day is $P_{\rm solar} = 1/|P_{\rm rot}^{-1} - P_{\rm orb}^{-1}|$,
where $P_{\rm rot}$ is the rotation period and $P_{\rm orb}$ is the
orbital period. Equating the solar day to the radiative time constant
and solving for the semi-major axis, we obtain
\begin{equation}
%Here's the version where I ignore the ``1/k'' Kepler's constant, which
% is equivalent to just equating solar day to rotation period:
%a_{\rm orb} \sim \left( {C P_{\rm rot}g \sigma\over p c_p}\right)^{2/3} T_*^2 R_*   
a_{\rm orb} \sim P_{\rm rot}^{2/3} \left[{1\over k} + {g\sigma T_*^3\over
p c_p}R_*^{3/2}\right]^{2/3}
\label{diurnal-transition}
\end{equation}
where we have used Kepler's law, $P_{\rm orb}^2 = k^2 a_{\rm orb}^3$
to relate the orbital period and semi-major axis, where the constant
$k=2\pi/\sqrt{G(M_* + M_p)}$.  For a sunlike star, and adopting
typical planetary parameters  ($g\approx 20\rm\,m\,s^{-2}$, $p=0.25\rm\,bar$,
and $c_p=1.3\times10^4\rm\,J\,kg^{-1}\,K^{-1}$), we find that the
radiative time constant becomes longer than the solar day for
orbital semimajor axes of 0.045~AU, 0.08~AU, 0.13~AU, and 0.2~AU for
rotation periods of 0.5, 1, 2, and 4 days, respectively.  
%For parameters relevant for HD~189733b, the radiative time constant
%becomes longer than a solar day for semimajor axes of 0.03~AU, 0.04, 0.08~AU,
%and 0.13~AU for rotation periods of 0.5, 1, 2, and 4 days.

Thus, these arguments suggest that, around a sunlike star, planets
inward of $\sim$0.05--0.1~AU will exhibit strong diurnal cycles, but
outward of $\sim$0.15--0.2~AU, the diurnal cycle will be weak, and the
circulation will be driven by the zonal-mean stellar heating.

Figure~\ref{fig.16} and \ref{fig.17}
provide a test of this prediction from \citet{showman2015} for
planets around an HD189733-like star, which is slightly dimmer than
the sun.  Simulations were performed on a regular grid in rotation
rate and orbital semimajor axis.  The condition
(19) is just met for the lower left, middle, and
upper right panels, respectively, of the two figures.  The results
indeed show the expected regime transition: models in the lower right
exhibit essentially canonical hot-Jupiter-like behavior, while models
in the upper left are in a regime where zonal temperature variations
are minimal, baroclinic instabilities transport heat poleward in the
midlatitudes, and the fastest jets occur in midlatitudes.  The
transition is broad and, not surprisingly, models near the predicted
transition exhibit hybrid behavior with fast midlatitude zonal jets
superposed on weak equatorial superrotation.

Interestingly, the fastest rotating, least-irradiated simulation of
this set---in the upper left panel of Figures
\ref{fig.16} and \ref{fig.17}---exhibits
multiple eastward jets in each hemisphere, reminiscent of the pattern
of multiple zonal jets on Jupiter and Saturn themselves.  The zonal
jets in this simulation exhibit meridional spacings that are
qualitatively consistent with Rhines scaling, which implies that 
if the jet speeds were slower (as might be expected under even weaker
stellar irradiation), then the jets would have smaller meridional
spacings and they would be more numerous, becoming even more similar
to Jupiter and Saturn.  {  The transition of the jet structure along with the change in the rotation rate  has been studied for conditions appropriate for terrestrial planets and qualitatively similar results to those in Figure \ref{fig.17} were obtained (e.g., \citealp{williams1988}).}

In the appropriate temperature range, a similar transition can occur among planets of differing atmospheric composition receiving a given flux of starlight. \cite{lewis2010} showed that the atmospheric circulation regime of the hot Neptune GJ 436b, which has an effective temperature of 650K, depends on the metallicity. \cite{Menou2012}, \cite{kataria2014} and \cite{charnay2015} found just the same phenomenon for the super Earth GJ 1214b, whose effective temperature is 550K. For these planets, atmospheres with metallicities $\gtrsim$ 30 times solar have high opacities, and therefore low photosphere pressures, leading to short radiative time constants at the photosphere. This leads to strong day-night forcing at the photosphere, causing substantial day-night temperature contrasts and a fast equatorial jet whose maximum eastward winds are at the equator. On the other hand, at metallicities of a few times solar or less, the gas opacities are smaller, so the photosphere levels are deeper, leading to larger radiative time constants at the photosphere. These models have comparatively weaker day-night forcing, leading to temperatures that are nearly constant with longitude, with the fastest zonal winds occurring in midlatitude jets rather than at the equator. These two regimes are directly analogous to those occurring in Figures 16---17, as well as to that occurring in Figure 9, and they occur for essentially the same reasons. The difference is that the change in radiative time constant that causes the transition is brought about not by a difference in the incident stellar flux but rather by a change in the photosphere pressure, via the metallicity.

\subsection{Non-zero obliquity}

We next turn to consider the effects of non-zero obliquity.  Only a
few papers have so far addressed the influence on obliquity on hot and
warm Jupiters \citep{langton2007, rauscher2017,
  ohno2019a, ohno2019b}, using both shallow-water and 3D
models.

Let's first consider some basic aspects.  When obliquity is zero, of
course, there is no seasonal cycle and the starlight predominantly
irradiates low latitudes.  When the obliquity is non-zero, the
sunlight received by each hemisphere varies throughout the year,
leading to seasonal effects, but additionally, the {\it annual-mean}
irradiation changes, shifting annual-mean sunlight away from low
latitudes toward high latitudes.  For small obliquities, the strongest
instantaneous (daily mean) starlight received remains at low
latitudes, even near the time of summer solstice.  However, when the
obliquity reaches approximately $18^\circ$, the daily mean insolation
quickly shifts so that, at summer solstice, the maximum starlight is
received at the summer pole---even though the {\it annual-mean}
insolation still maximizes at low latitudes \citep{ohno2019a}.
This implies that, if the radiative time constant is less than the
planet's year, the seasonal effects will exhibit a large increase in
amplitudes for obliquities crossing a threshold of $\sim$$18^\circ$.
A second transition occurs when the obliquity exceeds $54^\circ$,
above which even the {\it annual mean} irradiation is greater at the
poles than the equator.  If planets with such large obliquities have
short radiative time constants (relative to the planet's year), they
will of course have extremely strong seasons.  If their radiative time
constants are long compared to the year, they will lack strong
seasons, but they will exhibit the unusual situation of exhibiting
warm poles and a cold equator, with the atmospheric circulation
transporting heat from high latitudes toward low latitudes.

These insolation changes can trigger rich transitions in dynamics, but
basic aspects can be anticipated by simple arguments.  At small
obliquities, low latitudes tend to be hotter than high latitudes, and
through thermal-wind balance (Equation~4), this
poleward-decreasing temperature pattern implies that the midlatitude
zonal winds should become more {\it eastward} with altitude.  This is
precisely the way Earth behaves, and is likewise consistent with the
warm-Jupiter models shown in the upper left corner of
Figures~\ref{fig.16}--\ref{fig.17}.  On
the other hand, if the poles are hotter than the equator, as expected
at high obliquities and long radiative time constants, then
thermal-wind balance implies that the midlatitude winds become more
{\it westward} with altitude.  If, during a given season, strong
seasonal effects imply that one pole is warmer than the equator while
the other pole is colder than the equator, then thermal-wind balance
implies that the warmer (colder) hemisphere will exhibit midlatitude
winds becoming more westward (eastward) with altitude.

 \begin{figure*}      % use "figure*" instead of "figure" if you want your figure to span both columns
\epsscale{1.2}      % adjust this number to change the size of your figure
\includegraphics[scale=0.5]{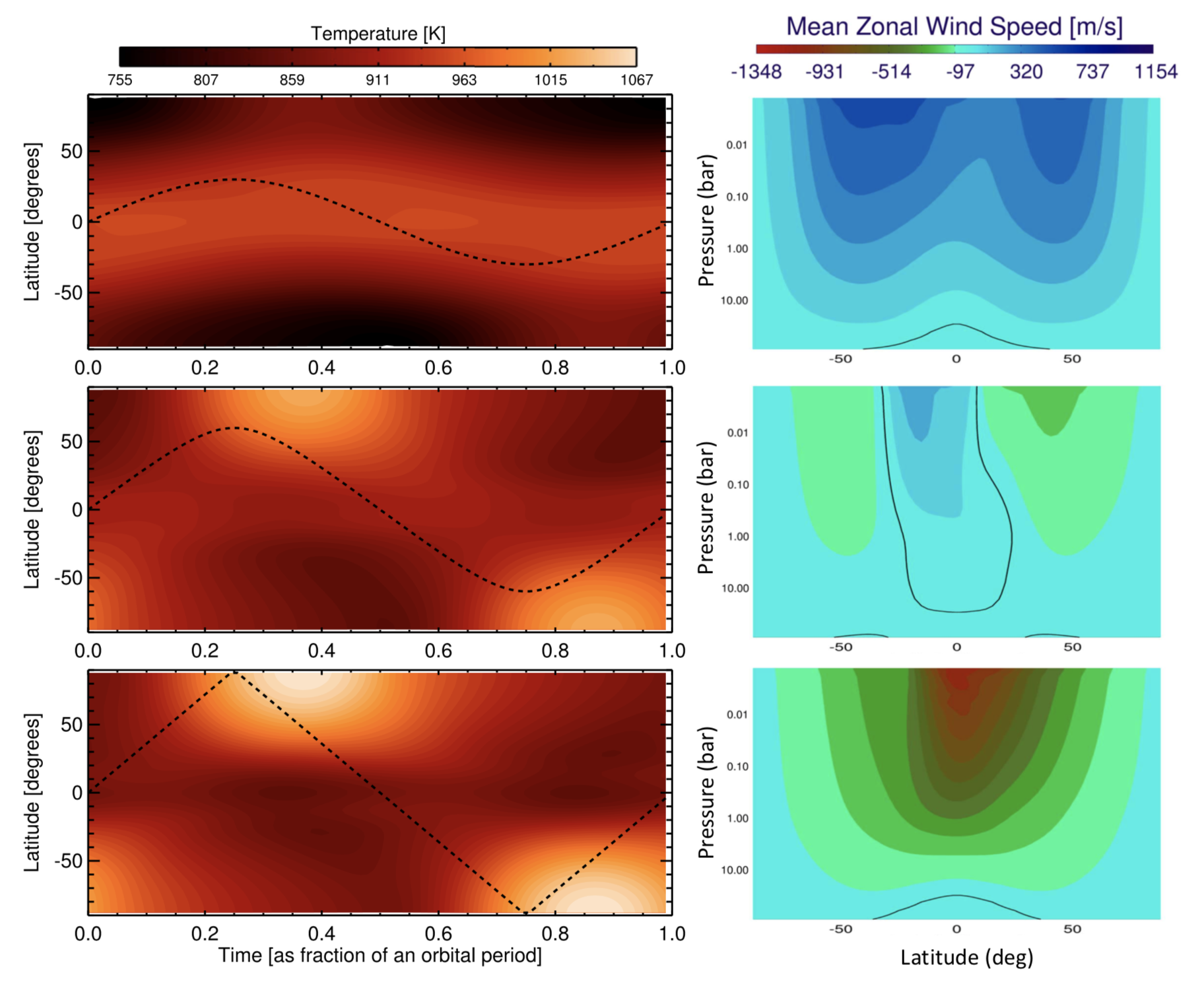}
\centering
\caption{GCM simulations from \citet{rauscher2017} showing the effect of
obliquity on rapidly rotating warm Jupiters.  Top, middle, and bottom
rows show simulations with obliquities of $30^\circ$, $60^\circ$, and $90^\circ$,
respectively.  Left column shows the zonal-mean temperature at the IR
photosphere throughout one orbital period.  Right column shows the
zonal-mean zonal wind at the norther summer solstice. }
\label{fig.18}
\end{figure*}

\citet{rauscher2017} presented the first 3D study of how obliquity
affects the circulation and observables of EGPs.  They considered a
rapidly rotating (10-hour period) giant planet on a 10-day period
orbiting a sunlike star.  Under these conditions, the planet's
effective temperature is $880\rm\,K$; the rotation rate and
irradiation level of their models are thus very similar to the
``W$\Omega_{\rm fast}$'' models of \citet{showman2015} (see
Figures~\ref{fig.16} and \ref{fig.17}).
Given the expectation that the diurnal cycle would likely be
unimportant, \citet{rauscher2017} applied diurnally averaged (i.e.,
longitudinally axisymmetric) heating and investigated obliquities of
$0^\circ$, $3^\circ$, $10^\circ$, $30^\circ$, and $90^\circ$.  Their
low-obliquity simulations ($\psi\lesssim 30^\circ$) exhibit a warm
equator, cold poles, and eastward zonal jets maximizing in the
mid-to-high latitudes (Figure~\ref{fig.18}), consistent with
expectations from thermal-wind balance, and qualitatively in agreement
with the rapidly rotating, weakly irradiated models of
\cite{showman2015}.

 \begin{figure}      % use "figure*" instead of "figure" if you want your figure to span both columns
\epsscale{1.2}      % adjust this number to change the size of your figure
\includegraphics[scale=0.4]{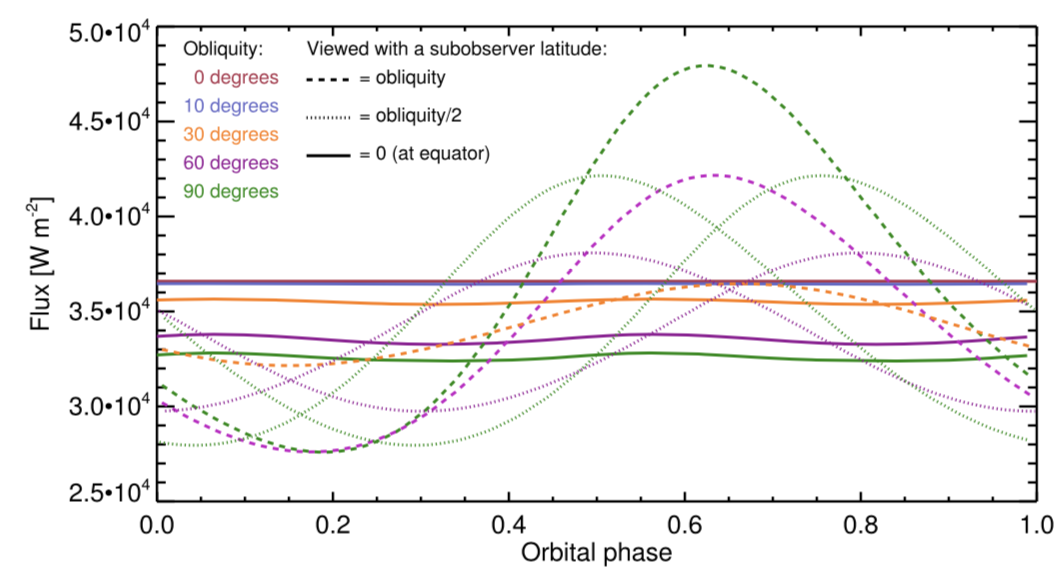}
\centering
\caption{IR lightcurves of oblique warm Jupiters from the GCM simulations of \cite{rauscher2017}.  These lightcurves are calculated for transiting planets, such that the sub-Earth latitude can lie be-tween zero and the obliquity.  Different colors represent GCM simulations performed at different obliquities.  For a given obliquity,different line styles represent different possible orientations of the planet's  rotation  axis  relative  to  the  line-of-sight  to  Earth.   Orbital  phase  is  defined  such  that  transit  occurs  at  0  and  1,  while secondary eclipse occurs at 0.5 }
\label{fig.19}
\end{figure}

\citeauthor{rauscher2017}'s high-obliquity ($60^\circ$ and
$90^\circ$) cases, however, exhibit a strong seasonal cycle and a very
different circulation pattern.  Near and after solstice, the summer
hemisphere is much warmer than the winter hemisphere, and the hottest
location on the planet is the summer pole. As expected from thermal
wind balance, this sign reversal in the meridional temperature
gradient---with high latitudes warmer than the low latitudes---leads
to a sign reversal in the thermal-wind shear, and therefore the zonal
flow is to the {\it west}.  This effect is strongest at $90^\circ$
obliquity, where the global flow is dominated by a single, broad
westward zonal jet with speeds of $\sim$$1\rm\,km\,s^{-1}$.
Figure~\ref{fig.18} clearly shows that, over the majority
of the seasonal cycle---including much of the winter---both poles are
warmer than the equator.  This behavior is consistent with the fact
that, over an annual average, the poles receive more starlight than
the equator for obliquities greater than $54^\circ$, an effect which
is modest for \citeauthor{rauscher2017}'s $60^\circ$ case but severe
for their $90^\circ$ case.

Figure~\ref{fig.19} presents IR lightcurves resulting
from \citet{rauscher2017}'s models for the particular situation of
transiting planets (where, depending on the orientation of the
planetary rotation vector relative to Earth, the sub-Earth latitude
can lie anywhere between zero and a latitude equal to the obliquity).
Generally, when the obliquity is small, or when the obliquity vector
is large but in the sky plane (such that neither pole preferentially
tilts toward Earth), the lightcurves are flat.  However, if the
rotation is oriented such that one pole aims toward Earth, the
lightcurve exhibits significant variations, as the Earth-facing pole
alternates between colder winters and warmer summers.  If the planet's
rotation axis is tilted toward Earth (i.e., if the projection of the
rotation vector into the planet's equatorial plane is aimed directly
at Earth), then the flux maxima occur after secondary eclipse, which
is caused by the thermal lag whereby the summer hemisphere is hottest
$\sim$1/8 orbit {\it after} the solstice (see
Figure~\ref{fig.18}).  If the rotation vector tilts in
other directions, the situation is complicated, and the flux maxima of
the lightcurve can occur either before or after secondary eclipse. The
degeneracies in Figure~\ref{fig.19} imply that it will be
hard to disentangle circulation behavior and obliquity for planets
with unknown obliquity, but \citeauthor{rauscher2017} showed that the
acquisition of dayside flux maps from ingress/egress eclipse mapping
can help break this degeneracy.  A key takaway point is that the flux
peaks in Figure~\ref{fig.18} result from a {\it completely
  different} physical mechanism than those for close-in hot Jupiters.

\begin{figure*}      % use "figure*" instead of "figure" if you want your figure to span both columns
\epsscale{1.2}      % adjust this number to change the size of your figure
\includegraphics[scale=0.45]{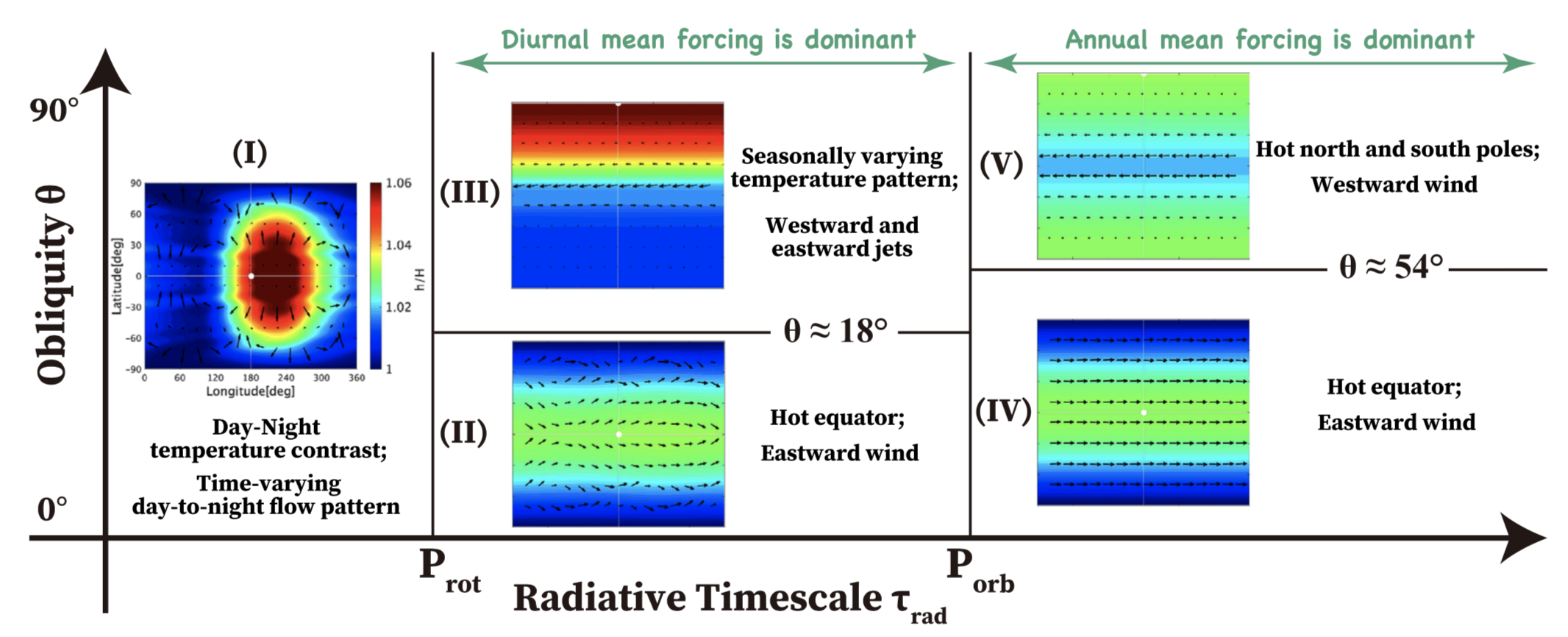}
\centering
\caption{Regimes of behavior for warm Jupiters as a function of
  radiative time constant, length of day, and orbital period, from
  \citet{ohno2019a}.  When the radiative time constant is less
  than the stellar day, the day-night contrast is large (regime I). At
  $P_{\rm rot} < \tau_{\rm rad}< P_{\rm orb}$, the diurnal cycle is
  weak; when obliquity is less than $\sim$$18^\circ$, the seasonal
  cycle is weak, whereas when obliquity is greater than $18^\circ$, it
  is strong (regimes II and III, respectively).  When $\tau_{\rm
    rad}>P_{\rm orb}$, the seasonal cycle is weak and the {\it
    annual-mean} insolation drives the circulation; for obliquities
  less than $54^\circ$, the height is thicker at the equator than the
  poles and the winds are predominantly eastward, whereas obliquities
  greater than $54^\circ$, the reverse is true, and the winds are
  predominantly westward. }
\label{fig.20}
\end{figure*}

\citet{ohno2019a} presented shallow-water models of warm
Jupiters that explore the diverse range of regimes as a function of
obliquity and radiative time constant, using an idealized model that
represents the day-night forcing with a Newtonian cooling scheme in
which the radiative time constant is an external parameter
(Figure~\ref{fig.20}).  Unlike \citet{rauscher2017}, they
allowed a diurnal cycle, rather than adopting a diurnally averaged,
axisymmetric forcing.
%They adopted rotation and orbital periods of
%0.5 and 30~days, respectively, and explored several values of
%radiative time constant of 0.1, 5, and 100~days.  
They found five classes of behavior.  As expected, when the radiative
time constant is shorter than the stellar day ($\tau_{\rm
  rad}=0.1\rm\,day$), the diurnal cycle leads to a strong day-night
thermal (height) contrast.  When the radiative time constant is longer
than the stellar day but shorter than the orbital period ($\tau_{\rm
  rad}=5\rm\,days$), the diurnal cycle is weak and the dynamical
structure is more zonally symmetric; as expected, the seasonal cycle
is weak when the obliquity is less than $\sim$$18^\circ$ but strong
when it is greater than $18^\circ$.  When the radiative time constant
is longer than the orbital period ($\tau_{\rm rad}=100\rm\,days$),
then the seasonal cycle is weak at any obliquity; if the obliquity is
less than $54^\circ$, the height is thickest at the equator and
thinnest at the poles, and the predominant winds are eastward.  When
the obliquity is greater than $54^\circ$, the height is thicker at the
poles than the equator, and the predominant winds are westward.

\subsection{Orbital eccentricity}

EGPs with semimajor axes greater than $\sim$$0.05\rm\,AU$ exhibit a
range of orbital eccentricities from zero to nearly one.  In contrast
to non-synchronous rotation and non-zero obliquities---which alter the
distribution of starlight across the globe but not the total starlight
hitting the planet---a non-zero eccentricity means that the total
stellar power intercepted by the planet varies throughout its orbit.
Many eccentric planets are transitional objects: the incident stellar
flux received at periapse puts them in the hot-Jupiter regime, whereas
the orbit-mean flux is more typical of warm Jupiters.  Prominent
examples amenable to observational characterization include the
transiting planets HAT-P-2b ($e=0.5$), HD~17156b ($e=0.67$), HD~80606b
($e=0.93$), which have stellar fluxes that vary throughout their orbit
by factors of 9, 27, and 828, respectively.  The time variable
forcing places eccentric EGPs in a novel regime that is both
inherently interesting and may yield insights into atmospheric
processes not easily obtainable from planets in circular orbits.

Generally, eccentric planets are not expected to exhibit rotation
periods equal to their orbital periods.  At large eccentricities,
tidal effects are orders of magnitude stronger near periapse than
throughout the rest of the orbit when the planet is farther from the
star.  Therefore, a common expectation is that for EGPs whose
periapses lie very close to the star, tides act to drive eccentric
planets toward ``pseudo-synchronization,'' a state in which the
planet's rotation rate equals the instantaneous rate of orbital motion
{\it when the planet is near periapse}.  Debate exists about the
details, but for large eccentricities, this state corresponds to
rotation periods much shorter than the orbital period.  The rotation
period is of course constant throughout the orbit, and given the
varying rate of orbital motion, a pseudo-synchronous rotation state
implies that, for high eccentricities, the stellar day will be much
shorter near apoapse than near periapse.

Eccentric planets can be subject to several possible thermal regimes,
due to the fact that their mean temperature---and therefore
characteristic radiative time constant at the photosphere---vary
throughout the orbit.  The first regime corresponds to the situation
where the radiative time constant is shorter than the stellar day
throughout the orbit, so that the diurnal cycle (day-night radiative
forcing) is important at all orbital phases.  A second regime occurs
when the peripase is small enough that, near periapse, the radiative
time constant is shorter than the stellar day, whereas near apoapse it
is longer than the stellar day.  In this regime, the diurnal cycle is
important near periapse, leading to an intense burst of day-night
radiative forcing, but the diurnal cycle is not important throughout
the rest of the orbit.  A third regime occurs when the entire
orbit---including the periapse---is sufficiently far from the star
that the diurnal cycle is irrelevant throughout the orbit.  So far,
most eccentric EGPs whose atmospheres are being characterized lie in
the first or second regime.

Several investigations of the atmospheric circulation of eccentric
EGPs have been performed \citep{langton2008,
  kataria2013, lewis2010, lewis2014,
  lewis2017}.  \citet{langton2008} adopted an
idealized 2D model, whereas the other studies adopted a 3D GCM with
sophisticated non-grey radiative transfer.  For eccentricities that
are small---less than perhaps $0.2$---the effects of eccentricity on
the qualitative circulation regime are modest \citep{lewis2010,
  kataria2013}.  But for larger eccentricities, the
investigations performed to date all agree that the intense stellar
irradiation near periapse passage plays a dominant role in driving the
circulation.  This ``flash heating'' causes transient formation of a
Matsuno-Gill-like pattern of standing equatorial waves that can drive
equatorial superrotation, which can persist throughout the orbit even
though it is not strongly forced throughout the more distant parts of
the orbit \citep{kataria2013, lewis2014, lewis2017}.
The flash heating event can also trigger shocks and other wave fronts
that propagate from day to night.  Still, the 3D studies performed to
date indicate that this time-variable forcing appears to produce a
time-variable version of the circulation regime already familiar from
hot Jupiters on circular orbits, as opposed to a qualitatively
different, totally new circulation regime.  As mentioned, these
studies tend to be in the first or second regimes described above, and
one could imagine that eccentric EGPs in the third regime would lack
Matuno-Gill patterns and would not have strong equatorial jets (at
least not via the mechanism described in Section~2).
Future work is needed to explore this issue.

For large orbital eccentricities and small periapse distances, the
pseudo-synchronous rotation rate is fast, and this leads to an
equatorial jet much narrower than occurs in GCM experiments of hot
Jupiters on circular orbits.  \citet{kataria2013} performed a
systematic investigation across a wide range of semi-major axes and
eccentricities, exploring both pseudo-synchronous and synchronous
rotation rates, and she showed that the meridional half-widths of the
equatorial jet are reasonably well matched by the equatorial
deformation radius, as predicted by the analytical theory (see Section 2.3).

These studies have important implications for observations of
eccentric EGPs.  The increase in temperature caused by periapse
passage tends to lag the pulse of stellar insolation, such that the
maximum temperatures occur after the planet is already past periapse
and receding farther from its star.  The overall timescale for this
thermal pulse to damp out provides an observational measure of the
atmosphere's radiative time constant, a quantity than is difficult to
infer for planets on circular orbits.  The large day-night temperature
difference can persist for times considerably past periapse passage, and
if the rotation period is short, this can lead to a ``ringing''
phenomenon where the flux received at Earth oscillates up and down as
the hotter hemisphere rotates in and out of view
\citep{langton2008, cowan2011b, kataria2013,
  lewis2017}.

Synthetic lightcurves show that the observational signatures of
transiting eccentric EGPs can be subtle and depend on the orientation
of the planet's orbit relative to Earth.  Temporal flux variations in
IR lightcurves reflect both spatial effects---in which hot and cold
regions on the EGP move in and out of view as the planet
rotates---and temporal effects in which the entire planet heats and
then cools as it passes through periapse.  Disentangling these effects
can be difficult.  Depending on the orbital alignment relative to
Earth, the secondary eclipse can occur either before or after periapse
passage, and, partly as a result, the peak IR flux can either lead or
lag the secondary eclipse \citep{kataria2013}.  In most cases,
the peak IR flux lags the periapse passage, although the timing
depends on the orbital configuration, since the orbital alignment
controls the time at which the newly flash-heated hot hemisphere first
rotates into view as seen from Earth \citep{kataria2013}.  
%For specific planets, the orbital geometry is fixed and some of these
%degeneracies disappear, but in general, the range of possible
%lightcurve signatures is diverse.
As a result of these issues, the range of possible lightcurve
signatures is diverse.

Spitzer IR lightcurves have been obtained for HAT-P-2b over its full
5-day orbit \citep{lewis2013} and for HD~80606b near its
periapse passage \citep{laughlin2009, dewit2017}.  These
observations show a large spike in the IR flux due to flash heating,
which peaks around the time of periapse passage and then slowly
decays.  In general, this flux spike is reproduced in GCM simulations
of these two planets \citep{lewis2014, lewis2017}, but the
detailed way the flux decays in the simulations exhibits discrepancies
with the observations, especially at times $\gtrsim 1\rm\,day$ after
periapse, which suggests possible influences of clouds, chemistry, or
interior heat flux that are not yet self-consistently included in the simulations~\citep{lewis2017}.

\section{Brown dwarfs and Directly Imaged Planets}
\label{brown-dwarfs}

Brown dwarfs are fluid hydrogen objects thought to form like stars but
that contain insufficient mass to fuse hydrogen.  They range in mass
from approximately several Jupiter masses ($M_J$) up to the stellar
mass limit of $\sim$$80 M_J$.\footnote{The mode of origin of
  substellar objects is difficult to ascertain for individual objects,
  so for convenience brown dwarfs are sometimes defined as hydrogen
  objects with masses between $\sim$10 and $80 M_J$.  The mode of
  formation is not important for the present discussion of atmospheric
  dynamics.}  Despite their large mass range, the hydrogen equation of
state implies that these objects have radii close to Jupiter's radius
over nearly two orders of magnitude in mass, from less than $1 M_J$ to
$80M_J$ \citep[e.g.,][]{guillot1999,chabrier2000}.  Their lack of an internal heat
source\footnote{Sufficiently massive brown dwarfs can fuse deuterium,
  but its abundance is sufficiently small that this process generally
  causes only a modest change to the evolutionary trajectory of a
  brown dwarf \citep{burrows2001}.} implies that, like Jupiter,
brown dwarfs cool down over time.  Nevertheless, massive brown dwarfs
contain so much internal energy that they cool down very slowly and,
even after billions of years, can exhibit temperatures at the IR
photosphere exceeding $1000\rm\,K$.  Brown dwarfs were first
definitively discovered in the mid 1990s \citep{nakajima1995,
  oppenheimer1995}, around the same time as the first EGPs were
discovered around main sequence stars \citep{mayor1995}, and
over the past 25 years the field of brown dwarf discovery and
characterization has developed in parallel with the field of
exoplanets.  Because brown dwarfs are typically isolated---with no
nearby star to drown out their light---they are generally much easier
to observationally characterize than exoplanets.  The quality and
quantity of IR emission spectra for brown dwarfs is therefore
exquisite by the standards of exoplanets.

Like stars, brown dwarfs have been subdivided into spectral types, the L, T, and Y types, based on their spectral characteristics \citep{kirkpatrick2005,cushing2011,kirkpatrick2012,leggett2017}. L dwarfs are the hottest, with effective temperatures $\sim$2100--1300K, as well as reddish colors in the near-IR and rather shallow spectral features indicating the prevalence of cloud decks of silicates and other refractory condensates. T dwarfs are cooler, with effective temperatures of 1300 to $\sim$600K. Their IR spectra exhibit deep spectral features of water and methane, with blue near-IR colors, indicating comparatively cloud-free atmospheres. Y dwarfs are the coldest class, with effective temperatures less than $\sim$600 K and deep spectral features of water, methane, and other species. The disappearance of silicate clouds in T and Y dwarf spectra makes sense, because at their relatively cool temperatures, refractory materials like silicates condense only very deep in their atmospheres, below the optical-depth unity level, where the cloud layers cannot strongly affect their outgoing IR spectra. Since brown dwarfs cool down over time, this spectral sequence is also an evolutionary sequence, meaning a given object can transition from L to T to Y over billions of years. This process happens more quickly for lower mass objects, such that for objects of a given age, a massive brown dwarf may remain an L dwarf while a lower-mass object may have already transitioned to a T or Y dwarf.

A key point is that, starting in the L dwarfs, and proceeding through the T and Y dwarfs, the atmospheric temperatures of brown dwarfs are sufficiently cool that they should magnetically decouple from any background dipole magnetic field they may possess (e.g., \citealp{gelino2002}), making them less like stars, and strengthening the analogy between the atmospheric dynamics of brown dwarfs and those of solar system planets like Jupiter and Saturn.

In addition to brown dwarfs, a variety of young, hot EGPs are being
directly imaged and characterized.  These include the planets around
HR 8799, $\beta$ Pic, and 2M1207, among other systems.  Because of the
difficulty of discerning the planet from the much-brighter star,
planets that can be directly imaged tend to be distant from their
stars (10~AU or more) and have photospheric temperatures of
$\sim$500~K or more.  This implies that they receive negligible stellar
flux compared to the interior heat flux they radiate to space; therefore,
from a meteorology standpoint, directly imaged EGPs resemble low-mass,
low-gravity versions of brown dwarfs, and they will likely exhibit
analogous dynamical processes.

 \begin{figure*}      % use "figure*" instead of "figure" if you want your figure to span both columns
\epsscale{1.2}      % adjust this number to change the size of your figure
\includegraphics[scale=0.5]{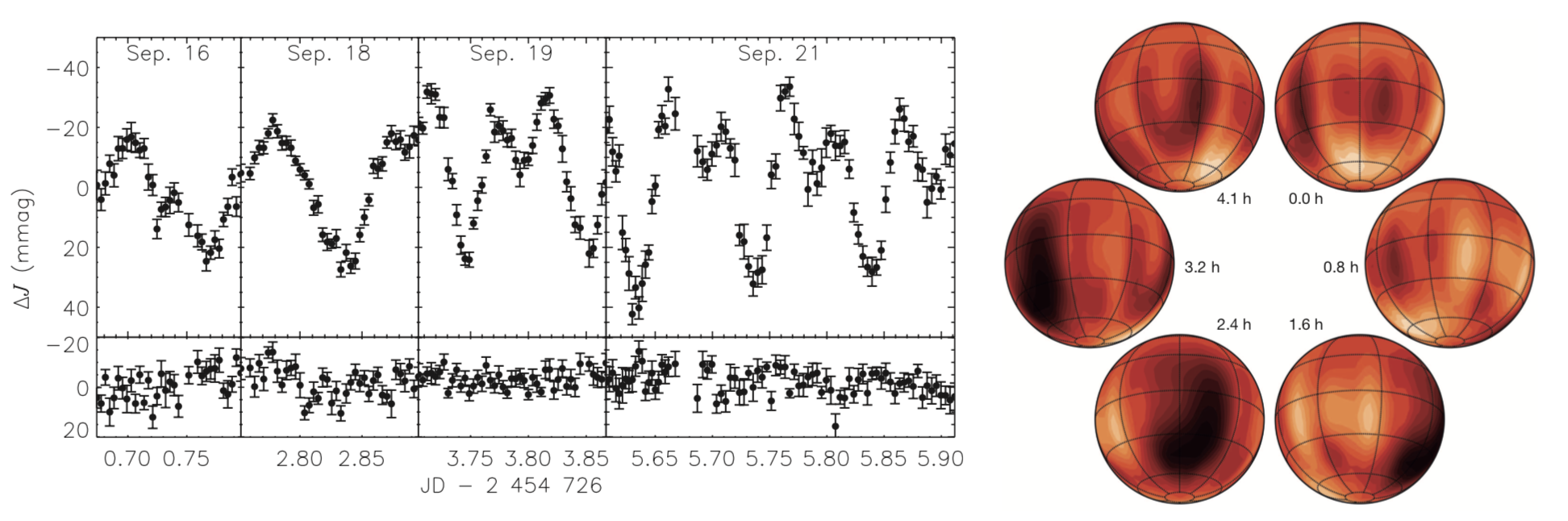}
\centering
\caption{Observations indicating surface patchiness on brown dwarfs.
  {\it Left panel} shows the IR flux versus time measured in J-band
  ($\sim$$1.2\rm\,\mu$m) for brown dwarf SIMP0136 by
  \citet{artigau2009}.  The J-band flux varies by about 5\% with
  a period of 2.4~hours, indicating that regional patchiness in clouds
  and temperature move in and out of view as the brown dwarf rotates
  every 2.4~hours, causing significant variability in a disk average.
  As expected, the period is constant throughout the observing
  sequence, but the {\it shape} of the lightcurve varies substantially
  on intervals of several days, which implies that the global pattern
  of cloud patchiness evolves on this timescale.  {\it Right panel}
  shows global maps throughout a rotation period of the brown dwarf
  Luhman 16B obtained by Doppler imaging from
  \citet{crossfield2014}.  The maps indicate regional patchiness
  on a scale of tens of thousands of km.  Note that although the
  patchiness is real, the preferential north-south elongation of
  features seen in the maps, as well as some of the lower-amplitude
  features, are probably analysis artefacts; moreover, at the
  signal-to-noise ratio of these maps, the method may not be
  sufficiently reliable to retrieve a zonally banded structure, if any
  \citep{crossfield2014}. }
\label{fig.21}
\end{figure*}

\begin{figure}      % use "figure*" instead of "figure" if you want your figure to span both columns
\epsscale{1.2}      % adjust this number to change the size of your figure
\includegraphics[scale=0.35]{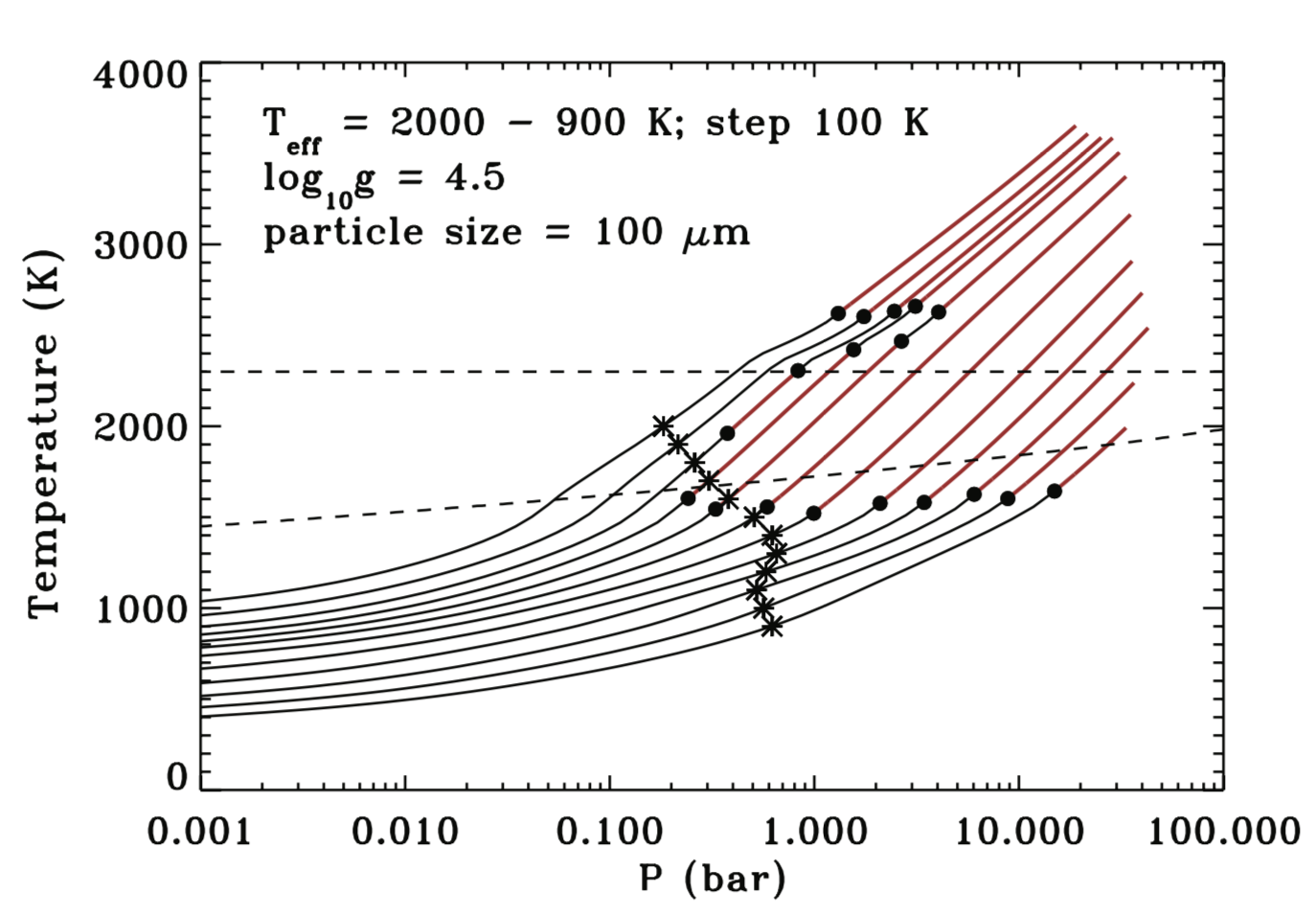}
\centering
\caption{Modeled 1D temperature-pressure profiles of brown dwarfs in radiative-convective equilibrium, from \cite{burrows2006}. Twelve models are shown, with effective temperatures ranging from 900 to 2000K in steps of 100K (solid curves). On each curve, red represents the convection zones and black represents stably stratified layers in radiative equilibrium. Black dots show the radiative-convective boundaries. Stars show the approximate IR photosphere level where the bulk of the radiation escapes to space. Note that in most cases, the IR photosphere occurs in the stably stratified atmosphere. Understanding observations of IR spectra, variability, surface patchiness, and chemical disequilibrium therefore requires an understanding of the atmospheric structure, including the ability of atmospheric dynamics to mix clouds and chemical tracers in this stratified region. }
\label{fig.22}
\end{figure}

Despite exhibiting similar temperature ranges, the regime of brown
dwarf atmospheres differs significantly from that of hot Jupiters.
Most known brown dwarfs are isolated objects that receive no
starlight; they are hot not due to irradiation, but because they are
massive and still losing their formidable heat of formation.  This
implies that, unlike typical hot Jupiters, they transport an enormous
internal heat flux through their interiors and into their atmospheres
to be radiated to space.  For example, a 1500-K brown dwarf has an IR
heat flux of $290,000\rm\,W\,m^{-2}$, while even a 1000-K object
radiates an energy flux of $60,000\rm\,W\,m^{-2}$, hundreds of times
greater than Earth's IR flux to space of $240\rm\,W\,m^{-2}$.  In the
interiors, this prodigious heat flux is expected to be transported by
convection, but the transition to optically thin radiative transfer
near $\sim$1~bar pressure decreases the radiative equilibrium
temperature gradient, ensuring that the temperature structure
transitions from nearly adiabatic in the deep interior to a stably
stratified atmosphere at low pressure, with the radiative-convective
boundary expected to occur typically at a few bars pressure
(Figure \ref{fig.22}, \citealp{stevenson1991, burrows2001, baraffe2014}).

Several lines of evidence indicate that brown dwarfs exhibit vigorous
atmospheric circulations.  First, the IR spectra and colors of many
brown dwarfs---particularly L dwarfs---indicate the presence of
silicate clouds (dust) in the visible atmospheres
\citep[e.g.,][]{chabrier2000, tsuji2002, knapp2004, kirkpatrick2005, cushing2006}.  In the absence of dynamics,
cloud particles would settle gravitationally, so the existence of
clouds implies the presence of atmospheric motions to loft the
particles vertically.  Second, the transition between L and T dwarfs
is a puzzling phenomenon that seemingly requires a role for
atmospheric dynamics, either in generating cloud patchiness or in
influencing how cloud microphysics depends on spectral type
\citep{ackerman2001, burgasser2002, knapp2004, burrows2006, marley2010}.  Third, many brown dwarfs
exhibit chemical disquilibrium of CO, CH$_4$ and NH$_3$, which seems
to require vertical mixing of air between deep levels and the
atmosphere
%where the
%chemical equilibrium abundances differ from those in the atmosphere,
%followed by the chemical quenching of those deep abundances due to
%long chemical kinetics timescales in the low-pressure, low-temperature
%conditions of the observable atmosphere 
\citep{fegley1996, saumon2000, saumon2006,leggett2007,stephens2009,miles2020}.  Fourth, many brown dwarfs
exhibit IR variability over rotational timescales, which probably
results from regional patches of differing temperature and cloud
opacity---and therefore differing IR fluxes to space---moving in and
out of view as the brown dwarf rotates (e.g.,
\citealt{artigau2009}; \citealt{radigan2012};
\citealt{apai2013}; \citealt{wilson2014}; \citealt{buenzli2014};\citealt{buenzli2015};
\citealt{metchev2015}; \citealt{yang2016}; \citealt{lew2016}; \citealt{vos2019}; \citealt{vos2020}; \citealt{zhou2020};  \citealt{bowler2020}; for reviews, see \citealt{biller2017}
and \citealt{artigau2018}).  Interestingly, the shape of the
lightcurves changes on short timescales, indicating that the {\it
  spatial pattern} of patchiness changes over time---presumably due to
time-evolving organization of turbulence, vortices, or other
atmospheric structures over the globe.  In some situations, the lightcurves
can be used to place constraints on the size and spatial distribution
of atmospheric spots on the globe \citep{karalidi2016,
  apai2017}.  Fifth, Doppler imaging of the closest brown dwarf
to Earth, Luhman 16B, provides direct confirmation of a patchy surface
structure \citep{crossfield2014}, a technique that may be
extended to other brown dwarfs in the future.

Several authors have attempted to constrain atmospheric wind speeds
from observations.  \citet{allers2020} recently presented the
first true atmospheric speed measurement for a brown dwarf.  They independently measured the rotation period of IR variability and the
period from radio emission for the brown dwarf 2MASS J1047, showing
that the former is about one minute shorter than the latter on this
$\sim$1.75-hour-period object---a 1\% difference.  The IR period
senses atmospheric cloud patchiness, while the radio period presumably
senses the magnetic field rotation rate, which is rooted in the deep
interior.  Thus these measurements imply that the atmospheric features
move eastward relative to the deep interior at a velocity of $600\pm
300\rm\,m \,s^{-1}$.  This phenomenon is most easily explained by
advection of atmospheric features by a fast eastward zonal jet, though
a contribution due to zonal wave propagation is not ruled out.  In
fact, this method is essentially the same approach used to infer the
atmospheric wind speeds on Jupiter relative to those in the Jovian
interior.

The number of brown dwarfs for which radio and IR periods can both be
measured is limited, however, and so several authors have attempted to
infer wind speeds solely from IR lightcurves.  This requires
plausibility arguments about the nature of the atmospheric features
causing the variability.  \citet{artigau2009} and
\citet{radigan2012} found that time evolution of their
lightcurves could be explained by the assumption that discrete,
coherent atmospheric features change in longitudinal separation over
time due to advection by an assumed zonal wind; the derived wind
speeds are $\sim$300--500 and 30--$50\rm\,m\,s^{-1}$, respectively.
Similarly, \citet{apai2017} inferred a speed of
$\sim$$600\rm\,m\,s^{-1}$ but suggested this may result from
differential propagation of atmospheric waves, rather than advection
by zonal jets.  
%Still, a rapidly evolving pattern of cloud
%patches---rather than advection of quasi-steady features by a zonal
%flow---could perhaps also explain these lightcurves, in which case no
%constraint on zonal wind would emerge.  
\citet{burgasser2014}
and \citet{karalidi2016} used lightcurves to place constraints
on the diameter of atmospheric spots; under the assumption that the
spots may coexist with jets of equal width and that the jets obey
Rhines scaling, the inferred wind speed is $\sim$600$\rm\,m\,s^{-2}$
or faster.  
%These arguments are plausible, but as we will describe,
%large-scale coherent structures can exist without zonal jets, and the
These approaches are worth exploring, but---as the above authors are
careful to admit---the assumptions they invoke are non-unique and
uncertain, so these estimates should be taken as plausibility
inferences rather than tight observational constraints.

Here, we survey the atmospheric dynamics of brown dwarfs, first using
basic arguments to sketch the expected dynamical regime, and then
summarizing current atmospheric circulation models of brown dwarfs,
emphasizing physical insights, basic processes, and observational
implications derived from these studies.

\subsection{Basic dynamical regime}

Isolated brown dwarfs rotate rapidly, and this exerts a strong control
over their atmospheric dynamics.  Doppler broadening of atmospheric
spectral emission lines \citep{reiners2008}, and rotation
periods directly inferred from IR variability
\citep[e.g.,][]{artigau2009, radigan2012,
  metchev2015}, indicate rotation periods typically ranging
from 1.5 to 12 hours.  Directly imaged planets likewise rotate rapidly,
with periods of order several to 11~hours \citep{snellen2014, zhou2016}.

Because of the fast planetary rotation, the regional-to-global circulation on
brown dwarfs and directly imaged planets exhibits Rossby number $Ro
\ll 1$, implying that the large-scale dynamics are rotationally
dominated \citep{showman&kaspi2013}.
%Consider wind speeds of $50$--$1000\rm\,m\,s^{-1}$ and rotation
%periods of 1--$10\rm\,hr$.
Adopting wind speeds of $50$--$1000\rm\,m\,s^{-1}$ and rotation
periods of 1--$10\rm\,hr$ consistent with the observationally inferred
ranges, global-scale flows ($L\sim10^8\rm\,s$) yield Rossby numbers of
0.0002 to 0.04.\footnote{For these estimates we adopted a value of the
  Coriolis parameter relevant to midlatitudes, implying that $f\sim
  2\times 10^{-3}\rm\,s^{-1}$ to $2\times 10^{-4}\rm\,s^{-1}$ for
  rotation periods of 1 to 10 hours.}  If the dominant flow length
scale is instead a regional scale ($L\sim10^7\rm\,m$), similar to the
dominant length scales on Jupiter and Saturn, then the Rossby numbers
range from 0.002 to 0.4.  Over almost the entire range considered,
these values are much less than one, implying that the dominant
horizontal force balance is between Coriolis and pressure-gradient
forces, that is, geostrophic balance \citep{vallis2006, holton2012}.  The large-scale circulation on Earth, Jupiter,
Saturn, Uranus and Neptune are also geostrophically balanced, and
therefore a substantial body of literature in the dynamics of rapidly
rotating turbulence developed over the past 40 years will likely have
important insights for understanding brown dwarfs.  Note for a typical
brown dwarf with a rotation period of a few hours, the ``tropical
regime'' corresponding to $Ro\gtrsim1$ obtains only within a few
degrees of the equator.

A second influence of rotation is that it leads to small Rossby
deformation radii.  Under brown dwarf conditions, the deformation
radius in midlatitudes is $L_D = NH/f \sim 500$ to
$5000\rm\,km$---just a few percent of the planetary radius---for brown
dwarfs with rotation periods of 1 to 10 hours.  The Rossby deformation
radius is a typical length scale that emerges from a variety of
processes involving an interplay between buoyancy and rotation.  For
example, baroclinic instability and convection tend to produce eddies
with sizes comparable to the deformation radius.  On Jupiter and
Saturn, this fact helps to explain the large population of vortices
with sizes of $\sim$$1000$--$2000\rm\,km$
\citep[e.g.][]{maclow1986, li2004,
  vasavada2006, choi2009}: they are probably a natural
outcome of convection interacting with the overlying, stratified
atmosphere.  This argument suggests that brown dwarfs will likewise
exhibit a large population of eddies with sizes of order $1000\rm\,km$.

Still, the processes that drive a circulation in brown dwarf
atmospheres differ significantly from those on typical irradiated
planets.  On hot Jupiters, Earth, and other irradiated planets, the
large-scale atmospheric circulation results primarily from the
gradient in stellar irradiation between the dayside and nightside (or
between the equator and pole on planets where the diurnal cycle is
weak).   But most brown dwarfs lack external source of
irradiation, and so this familiar way of generating an atmospheric
circulation cannot occur on a brown dwarf.   Convective mixing should lead to a nearly
constant entropy in their interiors, and if so, the temperature on an
isobar near the radiative-convective boundary should vary little with
latitude.  As a result, one might expect the upward IR flux emitted to
space---and therefore the atmospheric temperature-pressure profiles in
radiative equilibrium with that upwelling flux---to be nearly
independent of longitude or latitude.  In such a situation, the
stratified atmosphere would exhibit essentially no horizontal
temperature contrasts at all.

Still, as we will see, several mechanisms exist for generating a circulation in the stratified atmosphere of brown dwarfs. Because the IR radiation that escapes to space generally originates from within the stratified atmosphere rather than the convection zone (see Figure \ref{fig.22}), understanding the dynamics in this stratified region is critical to explaining observations of lightcurve variability, patchy cloud structure, and chemical disequilibrium.

 \begin{figure}      % use "figure*" instead of "figure" if you want your figure to span both columns
\epsscale{1.2}      % adjust this number to change the size of your figure
\includegraphics[scale=0.5]{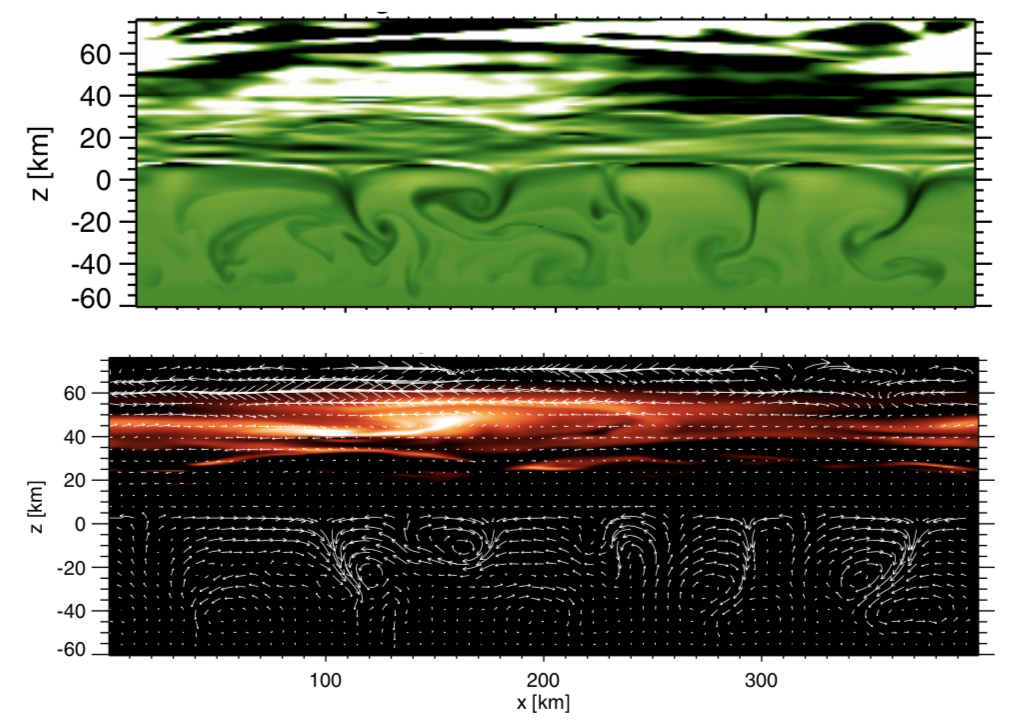}
\centering
\caption{Entropy anomalies (top) and velocity and dust distribution
  (bottom) versus horizontal distance and height in a local 2D box
  model of convection from \citet{freytag2010}.  This simulation
  corresponds to a brown dwarf with effective temperature $1858\rm\,K$
  and $g=1000\rm\,m\,s^{-2}$.  The convective zone roughly corresponds
  to $z<0$, and convective plumes can be readily seen there.  They
  interact with the overlying stratified layer at $z>0$ to generate
  small-scale gravity waves. }
\label{fig.23}
\end{figure} 

\subsection{Circulation models of brown dwarfs}
There exist at least three independent mechanisms for generating an atmospheric circulation in the stably stratified atmosphere of an isolated brown dwarf: (1) interaction of the interior convection with the stratified atmosphere; (2) 1D cloud feedbacks between radiation, vertical mixing, and cloud physics; and (3) multi-D cloud-radiative-dynamical feedbacks that drive an overturning circulation that maintains cloud patchiness. We summarize recent models investigating each of these three mechanisms.

\subsubsection{Atmospheric circulation I: interaction of convection with the stably stratified atmosphere}

In a pioneering study, \citet{freytag2010} presented 2D
numerical models of convection in a local box, representing a small
patch of a brown dwarf, aimed at determining how convection interacts
with an overlying, stably stratified atmosphere.  Rotation was
neglected, and the typical domain size was 300--$400\rm\,km$, about
0.5\% of the radius of a brown dwarf.  They coupled the 2D dynamics to
a radiative transfer scheme and a dust cycle that computes the
transport of a condensable gas species and its condensate---silicate
dust---as well as the interconversion between them.  In the
simulations, convective plumes drip off the top of the convection zone
and descend into the interior.  The convection interacts with the
overlying stratified layer, generating a wide population of
small-scale gravity waves (Figure~\ref{fig.23}).  These waves break,
triggering localized regions of vertical mixing in the stratified
atmosphere, which can transport the clouds and condensable vapor
upward, and generate cloud patchiness on a scale of tens of km.
\citet{freytag2010} analyzed the characteristic velocities in
the convection zone and stratified layer, characterized how they
depend on effective temperature, and derived effective vertical eddy
diffusivities for upward mixing of tracer due to the breaking gravity
waves.  This helps to explain the cloud decks on typical L dwarfs, and
may be relevant for understanding the L/T transition, but its
lengthscale is too small to generate the lightcurve variability shown
in Figure~\ref{fig.21}.  The 2D geometry and neglect of rotation also
implies that the convection in their models cannot generate Rossby
waves and has nothing to say about whether zonal jets, large-scale
coherent vortices, or large-scale cloud patchiness are expected on
brown dwarfs.

\citet{showman&kaspi2013} presented the first global-scale models of
the atmospheric circulation on brown dwarfs.  They demonstrated that
rotation plays a key role in the dynamics, leading to behavior on
large scales ($\gtrsim1000\rm\,km$) greatly differing from the
small-scale convection simulated by \citet{freytag2010}.
\citet{showman&ingersoll1998} presented 3D spherical-shell, global models
of convection in the interior to understand the global-scale
convective organization.  They solved the Navier-Stokes equations
subject to the anelastic approximation, which are valid when the
dynamical perturbations to the background density are small and the
wind speed is much less than the speed of sound, as expected to be
appropriate in the interiors of brown dwarfs.  They demonstrated that
the rotation exerts a strong control over the structure of the
interior circulation, and they characterized how the characteristic
convective velocities vary with heat flux, rotation rate, and radius
in the interior.  Convection occurs more efficiently near the poles
than the equator, leading to equator-to-pole temperature differences
that may reach a few~K in the interior.

 \begin{figure}      % use "figure*" instead of "figure" if you want your figure to span both columns
\epsscale{1.2}      % adjust this number to change the size of your figure
\includegraphics[scale=0.5]{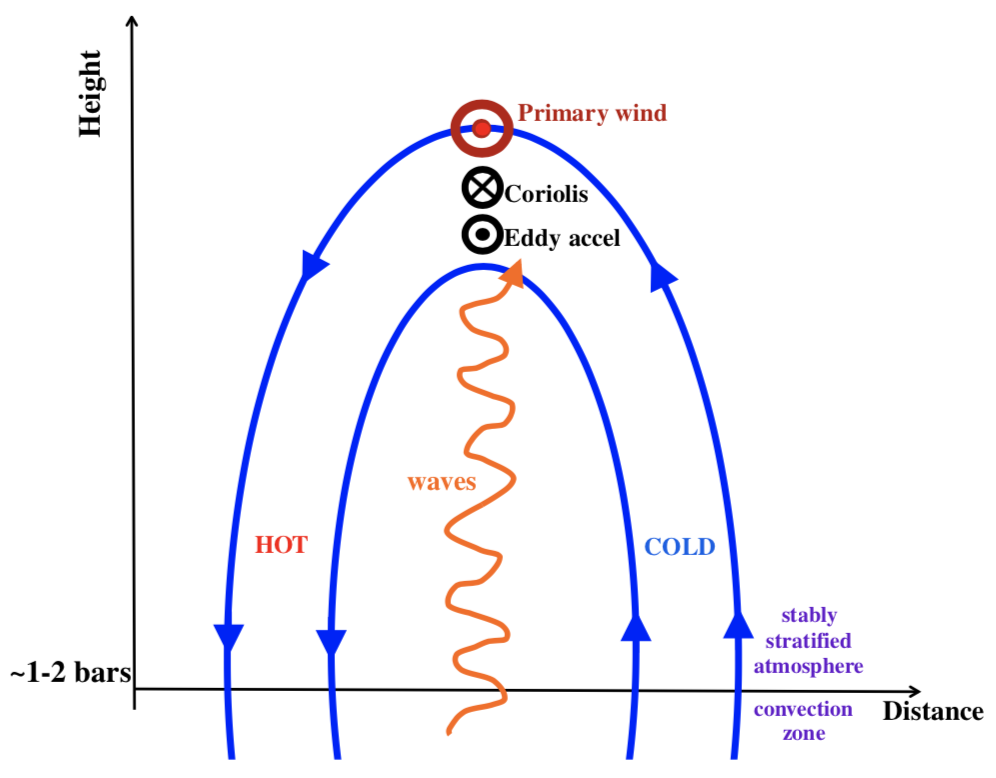}
\centering
\caption{Illustration of a large-scale, wave-driven atmospheric
  circulation, as occurs in the stratospheres of solar system planets
  (including Earth, Jupiter, and Saturn) and as
  \citet{showman&kaspi2013} proposed also occurs in the stratified
  atmospheres of brown dwarfs.  Vertically propagating gravity and
  Rossby waves are generated by convection near the
  radiative-convective boundary and propagate upward into the
  atmosphere (orange wavy arrows), where they are absorbed or break.
  Their damping generates a coherent eddy acceleration (black $\odot$
  symbol, representing a vector coming toward the viewer), which in
  turn drives a mean wind (black and red $\odot$ symbols,
  respectively, representing vectors coming toward the viewer).  In
  steady state, the eddy acceleration is balanced by a Coriolis
  acceleration (black $\otimes$ symbol, representing a vector pointing
  away from the viewer) associated with a slow secondary circulation
  (blue curve). The ascending and descending branches of this circulation
  advect entropy vertically, leading to horizontal temperature contrasts.}
\label{fig.24}
\end{figure}

Next, \citet{showman&kaspi2013} constructed analytic models for the
circulation in the stratified atmosphere.  As mentioned previously,
the absence of external stellar irradiation impinging on brown dwarfs
suggests that they lack the day-night or equator-pole radiative
forcing that drives the circulation in the tropospheres of most solar
system planets.  \citet{showman&kaspi2013} suggested that the
convection will generate a wide population of atmospheric waves,
including both gravity waves \citep{freytag2010} and Rossby
waves, and that these will propagate vertically and drive a mean flow
comprising zonal jets, turbulence, and coherent structures.
Figure~\ref{fig.24} illustrates the mechanism.
Gravity and Rossby waves are generated by convection near the
radiative-convective boundary and propagate vertically into the
atmosphere, where they break or dissipate, causing a zonal
acceleration, which in steady state is balanced by an equal and
opposite zonal acceleration due to a meridional flow (blue curves).
The ascending and descending motion associated with this circulation
advects entropy vertically; because entropy increases with height in a
stably stratified atmosphere, this implies that, on an isobar, low
entropy---hence cold---air will accompany regions of ascent, whereas
high entropy---hence warm---air will accompany regions of descent.
Such a circulation is thermally indirect; ascending air is denser than
descending air, and so the circulation increases rather than decreases
the potential energy of the circulation.  In steady state, this
potential energy is destroyed by radiation, which causes cooling in
the warm regions and warming in the cool regions.  Of course,
such a circulation cannot occur in isolation; it is driven by
the  absorption of the waves propagating upward from below,
and is therefore referred to as a wave-driven circulation.   Such
circulations are well known in the solar system community; they
occur in the stratospheres of Earth, Jupiter, Saturn, Uranus, and 
Neptune \citep[e.g.][]{andrews1987, conrath1990,
west1992, moreno1997}.

\citet{showman&kaspi2013} constructed an analytic theory for such a
circulation with the aim of predicting the characteristic horizontal
temperature contrasts and wind speeds in brown dwarf atmospheres.  The
efficiency with which wave absorption drives the mean flow is difficult
to predict from first principles, so \citet{showman&kaspi2013} considered
it a free parameter, described by $\eta$, which is a
dimensionless number giving the ratio of the power per area used to
drive the circulation to the total IR heat flux radiated to space.
No fundamental theory for $\eta$ in planetary stratospheres
yet exists, but observations of Earth and Jupiter indicates $\eta\sim10^{-3}$. { This number is generally yet unknown in the context of brown dwarfs, and here we apply a  similar number $\eta\sim10^{-3}$ in the following analysis. } \citet{showman&kaspi2013}
showed that, to within numerical factors of order unity, 
the characteristic horizontal temperature differences $\Delta T_{\rm horiz}$,
horizontal wind speeds $\Delta u$, and vertical wind speeds $w$ are
given by
\begin{equation}
{\Delta T_{\rm horiz}\over T} \sim \eta^{1/2} {NH\over c_s}
\end{equation}

\begin{equation}
{\Delta u \over c_s} \sim \eta^{1/2} \, l L_D
\end{equation}

\begin{equation}
w \sim \eta^{1/2} {H \over \tau_{\rm rad}} {c_s \over NH}
\end{equation}
where, as before, $NH$ is the characteristic horizontal phase speed of
long-wavelength gravity waves, $c_s$ is the speed of sound, $L_D$ is
the deformation radius, $l$ is the horizontal wavenumber of the flow,
corresponding to a characteristic horizontal length scale $L\approx
\pi/l$, and $\tau_{\rm rad}$ is the radiative time constant.  For a
vertically isothermal, ideal-gas H$_2$ atmosphere, $NH/c_s \approx
0.4$, which is order unity.  The length scale of the flow is unknown,
but values several times the deformation radius are plausible, which
suggests that $l L_D$ likewise is order unity.  For small wave-driving
efficiencies ($\eta\ll 1$), these equations therefore predict that the
fractional temperature differences are small, the wind speeds are much
less than the speed of sound, and the timescale for air to advect
vertically over a scale height, $H/w$, is much longer than the
radiative time constant.  For $\eta\sim 10^{-3}$ and typical brown
dwarf conditions, these equations predict that $\Delta T_{\rm
  horiz}\sim 30\rm\,K$, $\Delta u\sim 10^2 \rm\,m\,s^{-1}$, and
$H/w\sim 10\tau_{\rm rad}\sim 10^6\rm\,s$, corresponding to vertical
velocities of order $10^{-2}\rm\,m\,s^{-1}$, given a typical brown
dwarf scale height of 5 to 10 km.  Note that these velocities
correspond to vertical velocities in the stably stratified atmosphere
associated with the large-scale meridional, wave-driven overturning
circulation; convective velocities in the underlying convection zone
are of course greater.

 \begin{figure}      % use "figure*" instead of "figure" if you want your figure to span both columns
\epsscale{1.2}      % adjust this number to change the size of your figure
\includegraphics[scale=0.55]{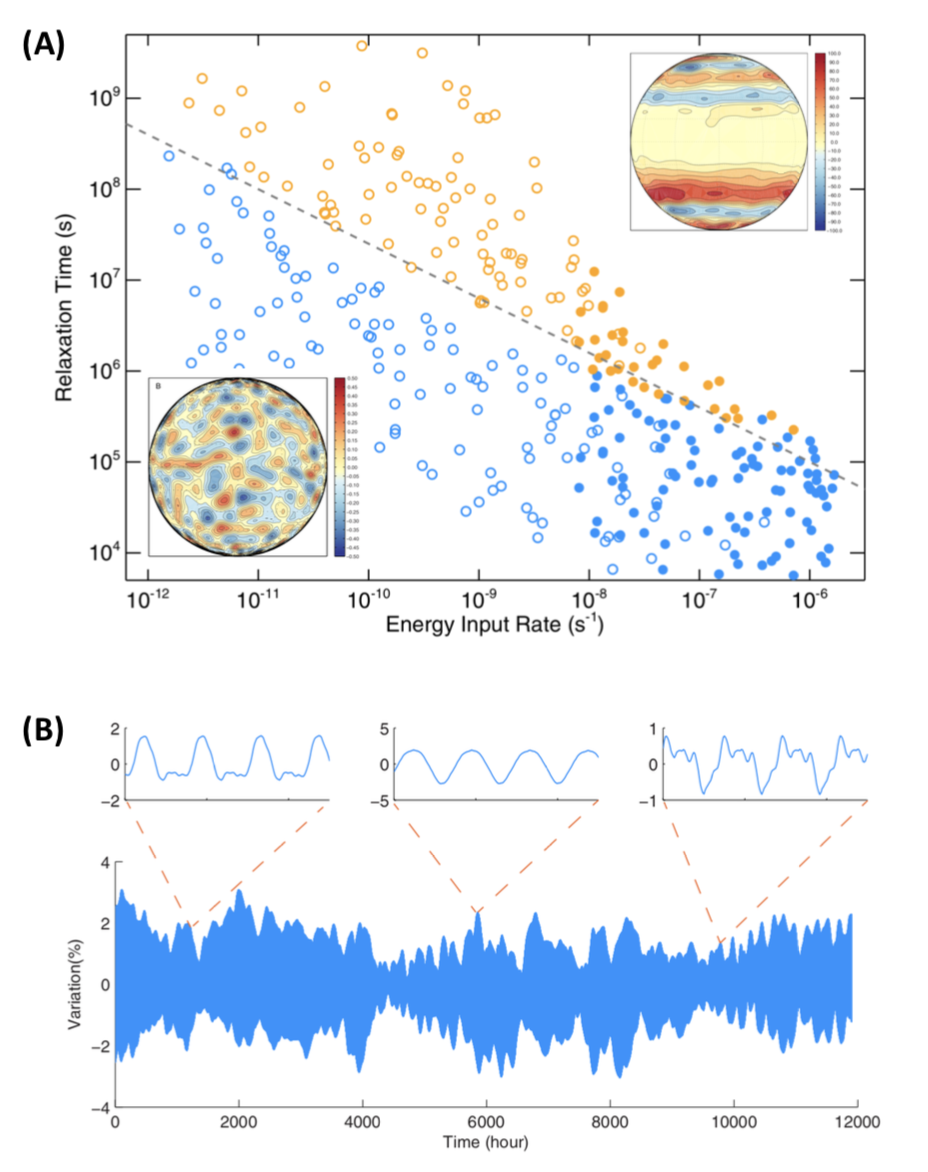}
\centering
\caption{Shallow-water simulations of the global circulation on
  isolated brown dwarfs, from \citet{zhang&showman2014}.  {\bf (A)}
  Dynamical behavior as a function of the convective forcing rate
  (abscissa) and strength of damping (ordinate), here represented by
  the radiative time constant.  Note that short radiative time
  constant implies strong damping of the flow, and vice versa.  Orange
  circles represent models exhibiting a banded flow with prominent
  zonal jets, which occurs with strong forcing and/or weak damping.
  Blue circles depict models dominated by horizontally isotropic
  turbulence, which occurs with weak forcing and/or strong damping.
  { An example of the turbulent regime is shown as the lower left inset and the jet-dominated regime is shown in the upper right inset. Colors in the insets represent geopotential anomaly (in unit of $10^5 \;\rm{m^2s^{-2}}$).} {\bf (B)} A
  typical synthetic lightcurve from these simulations, calculated by
  integrating the height (thickness) field over the Earth-facing
  hemisphere.  The solid blue envelope in the main panel shows the
  time evolution of the overall variability amplitude, which ranges
  over almost an order of magnitude on timescales of thousands of
  hours.  The insets show the actual lightcurves over short timescales
  (four rotation periods long each) at three different intervals.
  These insets illustrate how the lightcurve shapes vary with time, at
  some times exhibiting complex multi-peaked structures (left and
  right insets) and at other times exhibiting nearly sinusoidal
  behavior (middle inset).}
\label{fig.25}
\end{figure}

 \begin{figure*}      % use "figure*" instead of "figure" if you want your figure to span both columns
\epsscale{1.2}      % adjust this number to change the size of your figure
\includegraphics[scale=0.5]{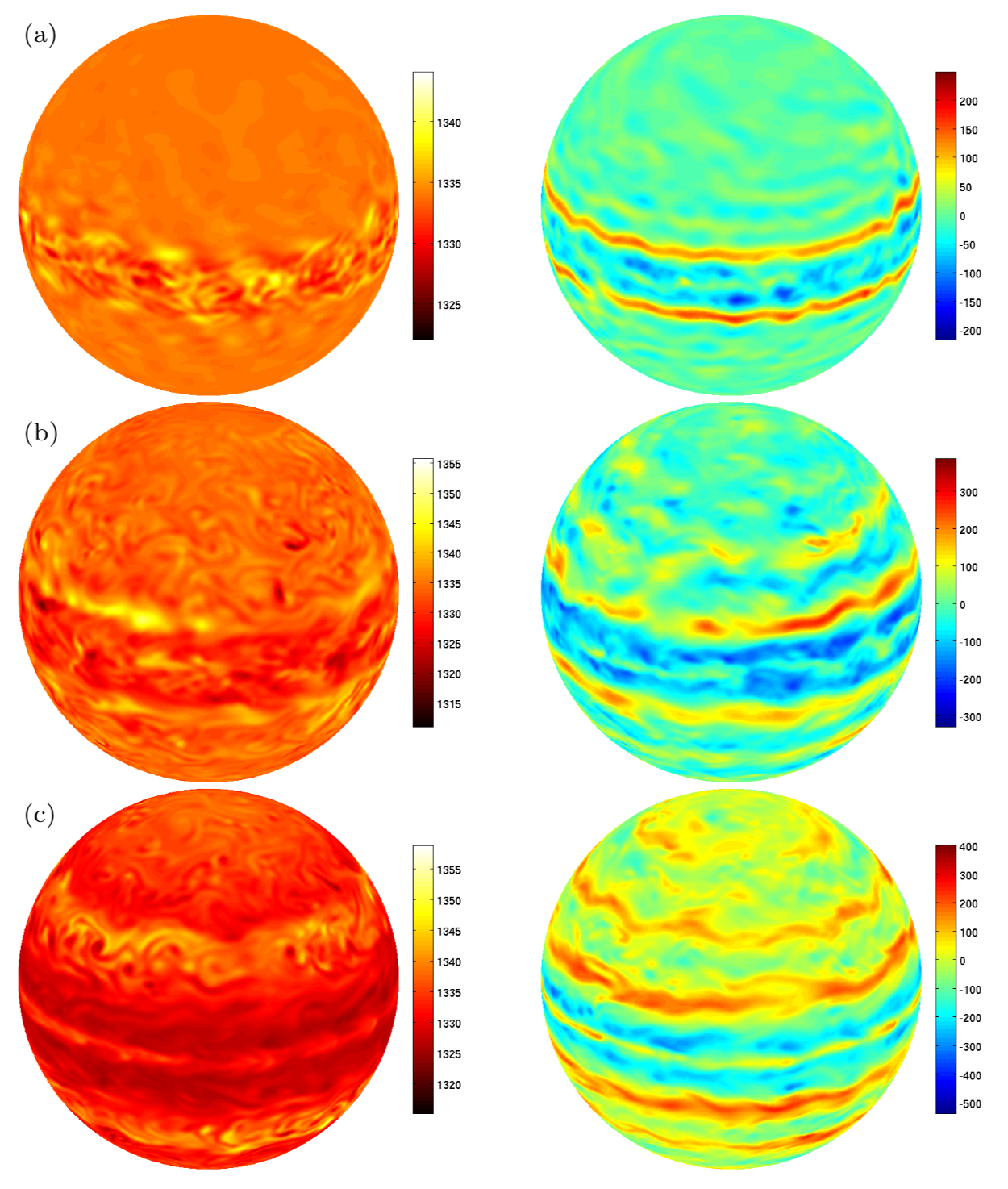}
\centering
\caption{Global models of atmospheric circulation on brown dwarfs
  forced by small-scale convective perturbations at the base of the
  atmosphere, with no external stellar irradiation, from
  \citet{showman2019}.  Left column shows temperature (K) and
  right column shows zonal wind ($\rm m\,s^{-1}$) in three simulations
  with different radiative time constants---$10^5$, $10^6$, and
  $10^7\rm\,s$ in the top, middle, and bottom rows, respectively. Rotation
  period is 5~hours, typical for brown dwarfs.}
\label{fig.26}
\end{figure*}

To further investigate whether convective perturbations can drive an
atmospheric circulation and understand how the flow organizes,
\citet{zhang&showman2014} presented global simulations of the
stratified atmosphere, using the 1.5-layer shallow-water equations.
Random, small-scale mass sources and sinks were added to represent
convective perturbations at the base of the atmosphere.  Radiation was
represented using a Newtonian cooling scheme that damps horizontal
thickness perturbations, consistent with the expectation that the
radiative equilibrium temperature structure is nearly independent of
longitude and latitude.  Zhang \& Showman found that when the
convective forcing was strong and/or the radiative damping was weak,
the flow spontaneously organized into a banded pattern comprising
zonal jets with superposed vortices.  But when the convective forcing
was weak and/or the radiative damping was strong, the atmospheric
turbulence damped before it could self-organize into jets, and the
flow instead comprised isotropic turbulence and vortices
(Figure~\ref{fig.25}).  Zonal wind speeds ranged from tens to
hundreds of $\rm m\,s^{-1}$, consistent with the analytical
predictions by \citet{showman&kaspi2013}. The transition between the
jet-dominated and vortex-dominated regimes is relevant to brown
dwarfs; for plausible forcing amplitudes, jets and banding tend to
occur when $\tau_{\rm rad}\gtrsim 10^5$--$10^6\rm\,s$. This suggests
that some brown dwarfs may be dominated by spots, whereas others may
be dominated by bands.

These models have implications for explaining surface patchiness and
lightcurve variability.  Simulated lightcurves calculated from
\citet{zhang&showman2014}'s models shows that the lightcurves can
exhibit a variety of timescales---including not only the rotational
modulation but longer, dynamical timescales as well---and allows for
changes qualitatively similar to those seen in observed lightcurves
(Figure~\ref{fig.25}).  The overall amplitude of the variability
can vary from nearly zero to $\sim$2\%, and the lightcurves
over rotational timescales can alternate between sinusoidal and multi-peaked
shapes, similar to those seen in observations (Figure~\ref{fig.21}),
depending on the evolving global pattern of the turbulence.

Continuing this line of inquiry, \citet{showman2019} presented
high-resolution, 3D numerical simulations of the atmospheric
circulation on brown dwarfs forced by small-scale convective
perturbations at the base of the stratified atmosphere.  
%This study can be considered a 3D version of the model explored by
%\citet{zhang-showman-2014} (as well as previous one-layer studies in
%the Jovian literature, such as \citet{scott-polvani-2007}).
Radiative heating/cooling was represented using a Newtonian cooling
scheme that, like in \citet{zhang&showman2014} damps thermal
perturbations and therefore removes energy from the flow.  The
convective perturbations trigger a wide population of waves, which
propagate upward into the atmosphere and induce a wave-driven
circulation through a mechanism similar to that described in
Figure~\ref{fig.24}.  Horizontal temperature
variations of tens of K on isobars and zonal winds reaching
$200$--$400\rm\,m\,s^{-1}$ occur (Figure~\ref{fig.26}), qualitatively
consistent with the predictions of \citet{showman&kaspi2013}.
\citet{showman2019} found that strong zonal jets are confined to low
latitudes when the radiative damping is strong (Figure~\ref{fig.26}a),
but the jets occur at all latitudes when the radiative damping is
weaker (Figure~\ref{fig.26}c).   

In these 3D models, the equatorial jet often exhibits an oscillation analogous to well-known stratospheric oscillations on solar system planets. In Earth’s equatorial stratosphere, vertically stacked eastward and westward jets---and associated temperature anomalies---migrate downward over time with a period of approximately two years. This phenomenon, dubbed the Quasi-Biennial Oscillation (QBO), is driven by the upward propagation of atmospheric waves—including equatorially trapped Rossby, Kelvin and inertio-gravity waves—that are absorbed in the stratosphere \citep{baldwin2001}. Similar oscillations have been observed on Jupiter and Saturn. On Jupiter, the oscillation has a period of four years and is called the Quasi-Quadrennial Oscillation (QQO). The oscillation on Saturn has a period of 15 years (half a Saturn year), and is called the Saturn Semi-Annual Oscillation (see \citealp{showman2018review} for a review). These are the first full 3D models of a giant planet to capture this type of oscillation. In the simulations, the oscillation typically has periods of several to $\sim$12 years, broadly similar to the periods of the QQO and SAO. These models highlight the likelihood that long-term oscillations may be a generic feature on warm-to-cool giant planets and brown dwarfs. A key implication is that brown dwarfs will probably be variable not only on short timescales, but also on multiannual or multidecadal timescales as well.

There is a subtle difference in the models of \cite{zhang&showman2014} and \cite{showman2019} as to whether jets exist when damping is very strong. In \cite{showman2019}, models of a given wind speed exhibit more coherent zonal banding when the forcing and damping are weak than when they are strong, because in the former case, the flow has ample time to reorganize into a strong series of zonal jets, whereas in the latter case, the forcing and damping continually attempt to disrupt the banding pattern as it forms. However, some degree of zonal banding always occurs—none of the models in Showman et al. exhibit entirely isotropic turbulence, no matter how short the radiative time constant. This occurs because, in a 3D model, the radiation damps the horizontal temperature variations, which via thermal wind suppresses the vertical shear of the zonal wind. However, the barotropic mode---that is, the height-independent component of the zonal flow---is not damped by radiative forcing. This differs from \cite{zhang&showman2014} because their 1.5-layer model---which assumes a quiescent abyssal layer---lacks a barotropic mode. In the 3D formulation, not only strong radiative damping but also strong frictional drag at the base of the atmosphere is required to suppress the jets entirely. Such frictional damping could be supplied by Lorentz forces that act to brake the deep roots of the zonal jets that penetrate into the deep interior, but this is a subtle issue that requires further investigation.

The vertical motions that cause the temperature contrasts in these
models (Figure~\ref{fig.26}) will transport condensate species upward,
which can lead to cloud formation if the condensation level occurs in
the atmosphere, as expected for L and early T dwarfs.  Very small
cloud condensate particles (radii less than perhaps $0.1\rm\,\mu$m)
exhibit very slow gravitational settling, so they may become well
mixed by the circulation into quasi-ubiquitous layers of fine
particles.  Larger particles will settle more quickly after
condensation, and since the regions of condensation are confined to
ascending air---which covers only a fraction of the globe---this will
naturally lead to cloud patchiness on large scales.  \citet{tan2018}
confirmed this picture with numerical simulations similar to those
described here but that include coupling to passive cloud tracers that
advect with the circulation and settle gravitationally but do not influence
the dynamics (i.e., no cloud radiative feedback).  As expected from
the qualitative considerations above, these models produce regionally
patchy cloud decks, which can modulate the outgoing IR flux and lead
to lightcurve variability.

\subsubsection{Atmospheric circulation II: 1D cloud-radiative feedbacks in an atmospheric column}

Recent work demonstrates that the feedback between radiation,
dynamics, and cloud formation can lead to spontaneous emergence of an
atmospheric circulation, and therefore represents a totally distinct
means of generating an atmospheric circulation on brown dwarfs than
the wave-driven circulation mechanism just described.  Here, we 
survey two distinct feedbacks that can lead to atmospheric circulation
on brown dwarfs, each of which can separately lead to a vigorous
atmospheric circulation and regional-scale cloud patchiness---even
in the absence of convective forcing that might trigger any
wave-driven circulation.

The importance of cloud radiative feedbacks on brown dwarfs is easy to appreciate. Imagine an atmospheric vertical column with an opaque cloud deck---perhaps several thousand km wide---surrounded by cloud-free air. In the cloudy region, the IR radiation to space escapes from top of the cloud, at relatively low pressure, whereas in the cloud-free region, the radiation to space occurs from much deeper pressures, perhaps a few bars. The two regions will therefore experience extremely different vertical profiles of the radiative heating/cooling, which will lead to horizontal temperature differences on isobars. These horizontal temperature contrasts will drive an overturning circulation that, in turn, can advect cloud condensate vertically, and in principle might be capable of maintaining the cloud patchiness.

To illustrate, imagine an initially quiescent, cloud-free atmosphere
in radiative equilibrium, with an effective temperature of $T_{\rm
  eff} = 1000\rm\,K$, corresponding to an IR flux to space $F=\sigma
T_{\rm eff}^4 = 6\times10^4\rm\,W\,m^{-2}$, which is likely radiated
to space from a pressure near $1\rm\,bar$ (Figure~\ref{fig.22}).
Because the atmosphere is in radiative equilibrium, this same flux is
supplied to the atmosphere from below, such that the next heating of
the column is zero.  Now for sake of argument suppose that a cloud
appears at a higher altitude, where the temperature is cooler---let us
say $900\rm\,K$. Radiation to space from the cloudy region will occur
from the top of the cloud deck, corresponding to a smaller flux
$4\times10^4\rm\,W\,m^{-2}$.  The perturbation in outgoing IR flux is
$\Delta F\approx 2\times10^4\rm\,W\,m^{-2}$.  Integrated over an
atmospheric column, this differential heating induces a rate of
temperature change over time $\Delta F g/c_p p \sim 10^{-2}\rm\,K
\,s^{-1}$, where we have adopted $g\approx 500\rm\,m\,s^{-2}$,
$c_p=1.3\times10^4\rm\,J\,kg^{-1}\,K^{-1}$, and $p\approx
1\rm\,bar$\footnote{In a hydrostatically balanced atmosphere, the mass
  per unit area in a column is $p/g$, so the radiative heating/cooling
  per unit mass (units of $\rm W\,kg^{-1}$) is $\Delta F\, g/p$, and the cooling
  in units of $\rm\,K\,s^{-1}$ is $\Delta F \,g/p c_p$.}.  Over only a few
brown dwarf rotations, such a heating differential would cause
temperature changes $\gtrsim100\rm\,K$ within the atmospheric column.

 \begin{figure*}      % use "figure*" instead of "figure" if you want your figure to span both columns
\epsscale{1.2}      % adjust this number to change the size of your figure
\includegraphics[scale=0.3]{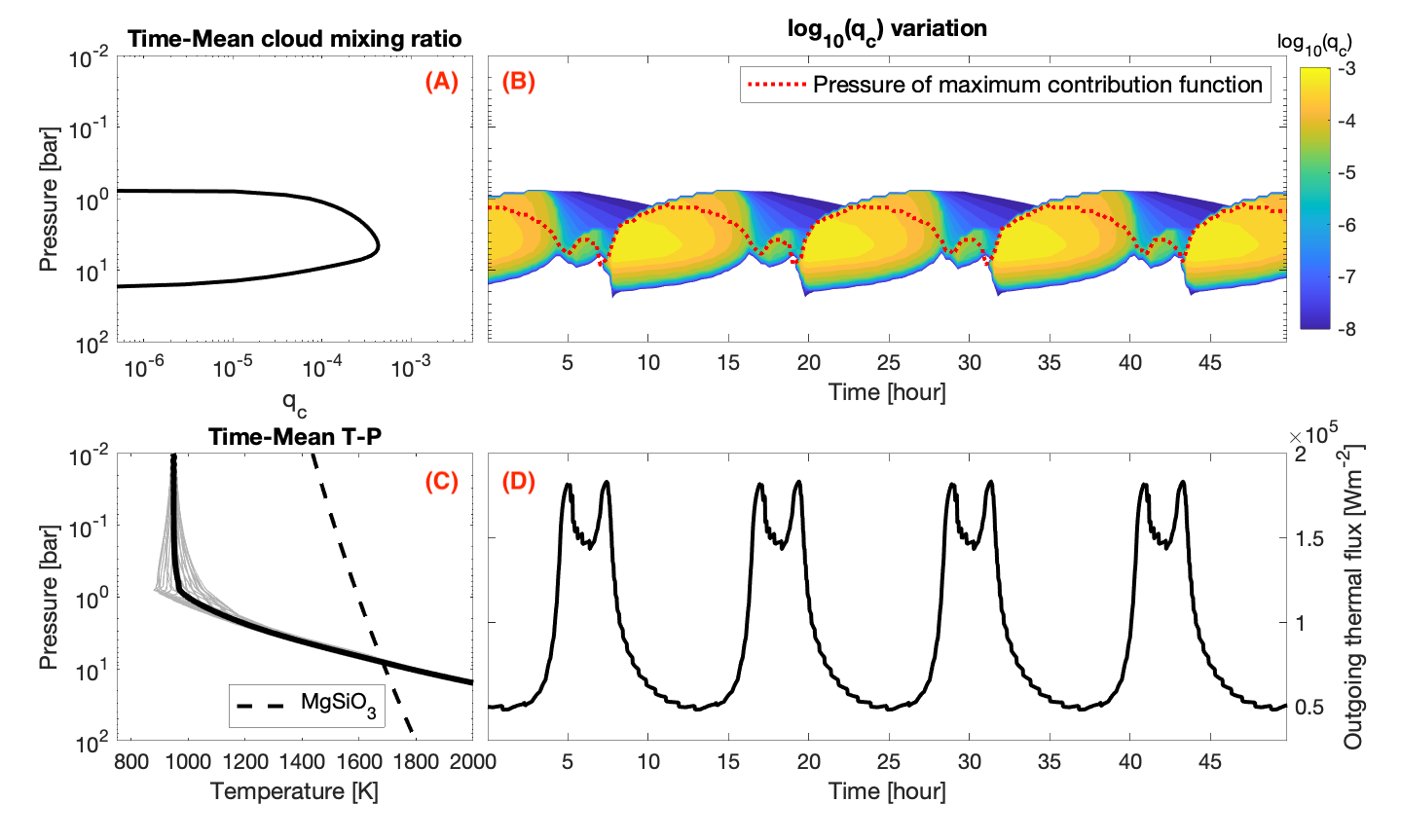}
\centering
\caption{{ Results of  1D time-dependent models of brown dwarfs showing the development of cyclical oscillations in the cloud structure and thermal flux radiated to space. From \cite{tan2019bd}. \textbf{(A)} Time-mean cloud mixing ratio ($q_c$) as a function of pressure. \textbf{(B)} Time evolution of cloud mixing ratio as a function of time and pressure in logarithmic scale. The dotted red line is the pressure at which the contribution function is at a maximum, which is where the majority of flux escape from the atmosphere.  \textbf{(C)} The time-mean temperature as a function of pressure is represented as the thick solid line;  the dashed line represents enstatite (MgSiO$_3$) condensation curve; and the grey lines represent the envelope of the variation range of the temperature-pressure profile. \textbf{(D)} The corresponding outgoing thermal flux as a function of time. }
}
\label{fig.27}
\end{figure*}

Such thermal changes due to the appearance of clouds cause two distinct processes relevant for driving an atmospheric circulation. First, they cause an adjustment of the thermal structure of the entire atmospheric column, which changes the atmospheric stratification profile and potentially alters the altitude ranges where vertical mixing and clouds occur. Second, if such a cloudy atmospheric column is surrounded by columns of relatively cloud-free air, the horizontal {\it differences} in the vertical heating/cooling profile causes horizontal temperature contrasts, and the resulting horizontal pressure-gradient forces can potentially drive an overturning circulation. The key question is whether either of these process is self-generating and self-sustaining, that is, whether an initially quiescent atmosphere will experience spontaneous emergence of atmospheric circulation due to these processes. Recent work indicates that the answer is yes.

\citet{tan2019bd} constructed a 1D, time-dependent model of the
vertical structure in a brown dwarf atmosphere to investigate the
coupling between clouds, radiation, and mixing in a vertical
atmospheric column.  This model is similar in spirit to a long history
1D brown dwarf atmosphere models \citep[e.g][]{burrows1997, burrows2006, marley2002, marley2010,
  saumon2008}, except that most of these models assume steady
state and solve for the radiative-convective-equilibrium, whereas
\citet{tan2019bd} allowed for time dependence associated with
time-varying deviations from radiative equilibrium.  They adopted a
simple cloud scheme in which clouds settle gravitationally and are
vertically mixed in regions where convection should occur.  Their
models show that the equilibrium state is generally not steady;
radiative-cloud feedbacks cause the thermal and cloud structure to
vary, with growth and collapse of cloud decks occurring episodically
on timescales of $\sim$1 to $\sim$30~hours (Figure~\ref{fig.27}).
%The typicalvariability period is $\sim$1 to $\sim$30~hours.  
As foreshadowed above, temperature varies on isobars by
$\sim$$100\rm\,K$ as the cloud deck grows and decays in a cyclical
fashion.  Because of the time-varying cloud opacity, the IR
photosphere pressure moves vertically by two scale heights, which
causes the effective temperature to vary by hundreds of K over a
cycle.  \citet{tan2019bd} showed that the existence of
variability is robust over a wide range of parameters including number
density of cloud condensation nuclei (CCN), gravity, interior entropy,
and the implementation of the vertical mixing---although the exact
oscillation period and other details depend on these parameter values.
A key point is that the oscillatory behavior is entirely spontaneous
and does not require any externally imposed perturbations.  This
contrasts with other 1D models where variability emerged only in
response to imposed thermal perturbations
\citep{robinson2014}.

The growth and collapse of cloud decks shown in Figure~\ref{fig.27}
results from several interacting feedbacks.  In the growth phase, a
positive feedback promotes upward expansion of the cloud top, as
follows: an upwardly
perturbed cloud top cools by adiabatic expansion, leading to cooler
cloud-top temperatures and smaller IR flux radiating to space from the
cloud top.  Since the temperature profile {\it above} the cloud top is
determined by the strength of this IR flux, a lessening of this OLR
causes the entire overlying $T/p$ profile (above the cloud) to cool
off.  For a given temperature on an isobar near the top of the cloud,
a decrease in the overlying temperatures acts to destabilize the air
near the cloud top, allowing the cloud to expand upward.
%During this process, the strong IR flux from the cloud top ensures
%that a spike of strong radiative cooling occurs at the cloud top,
%which drives the convection within the cloud that
But this growth process is eventually halted by a negative feedback
that kills the cloud: although the cloud itself remains
relatively well-mixed, a large upward heat flux against the
base of the cloud generates stable stratification {\it below} the
cloud, which shuts off the source of moisture (i.e. condensable
chemical species) into the cloud.  Without a source of condensable
material to replenish it, gravitational settling of particles then
inevitably depletes the cloud of particulate matter, causing the cloud
to dissipate.  The cycle can then repeat.

To reiterate, this is a solely 1D feedback that does not require large-scale dynamics. On a brown dwarf, distinct atmospheric vertical columns in different location on the globe cloud in principle independently undergo such oscillatory behavior; the interactions between such columns will not be negligible, a topic requiring a multidimensional dynaical model to investigate. It is also important to emphasis that the 1 to 30 hour period of the oscillation has nothing to do with rotation---it purely results from the interactions between radiation, convective mixing, and cloud physics---yet its similarity to the rotation timescale means that brown dwarf lightcurves can vary over short (rotational) timescales. Note that this 1D feedback occurs more easily when the cloud condensation occurs near the top of the convective zone. { This is because convection below the cloud base is needed to replenish the condensable vapor after clouds gravitationally settling out---otherwise without replenishment, the clouds will  dissipate away and the system will be eventually inactive.  } 1D model atmospheres where the cloud deck would form in the stratified atmosphere could potentially lack variability from this mechanism \citep{tan2019bd}. 

\subsubsection{Atmospheric circulation III: multi-D feedbacks between clouds, radiation, and dynamics}

Next, we turn to consider the second cloud-radiative feedback.
Imagine an initially quiescent, cloud-free atmosphere in radiative
equilibrium, and imagine seeding a small cloud into a vertical air
column that is surrounded by cloud-free air.  Being at higher
altitudes where temperatures are cooler, the IR flux to space from the
top of the cloud is less than that of the surrounding, cloud-free
regions.  If the initial air columns were in radiative
equilibrium---receiving the same heat flux from below as they radiate
to space---then this decrease in outgoing heat flux caused by the
cloud implies that, integrated vertically over an atmospheric column,
the cloudy air column experiences net heating.  This heating promotes
ascent, causing the cloud to grow to higher altitudes.  Adiabatic
expansion decreases the temperature still further, causing greater net
heating of the column and further promoting growth of the cloud.
\citet{gierasch1973} demonstrated this system is linearly
unstable, in that small perturbations will grow over time, and he
suggested its possible applicability to giant planets.  The mechanism
is thought to promote convective self-aggregration in Earth's tropics
\citep[e.g.,][]{wing2017}.  However, the mechanism has not been
previously investigated for giant planets in the fully developed,
nonlinear regime.

\begin{figure*}      % use "figure*" instead of "figure" if you want your figure to span both columns
\epsscale{1.2}      % adjust this number to change the size of your figure
\includegraphics[scale=0.6]{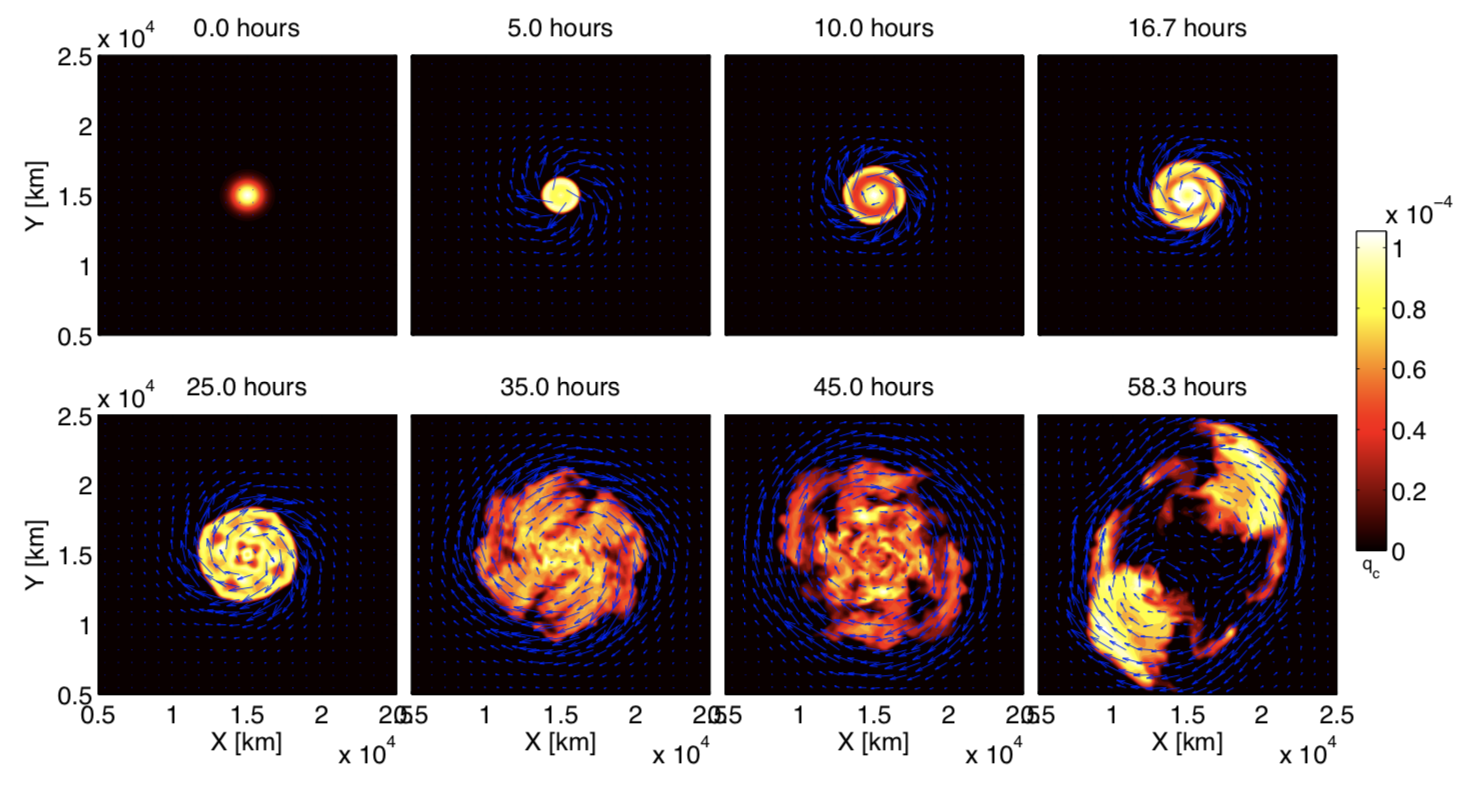}
\centering
\caption{Growth of a cloudy vortex on an $f$-plane from an initially
  cloud-free environment through a positive feedback between
  radiation, dynamics, and cloud development.  Plotted is the cloud mixing 
ratio (colorscale) and winds (vectors) at 0.4~bars.  In this model,
  $f=3\times10^{-4}\rm\,s^{-1}$. From \citet{tan2017}.}
\label{fig.28}
\end{figure*}

\cite{tan2020bd} have shown that this mechanism can cause the spontaneous emergence of a self-generating atmospheric circulation on brown dwarfs, even when the atmosphere would be quiescent in the absence of clouds. Figure \ref{fig.28} shows a numerical simulation from \cite{tan2020bd} demonstrating that this mechanism indeed constitutes a positive feedback that can generate atmospheric turbulence. \cite{tan2020bd} solved the primitive equations for the stratified atmosphere on an $f$-plane (corresponding to a 3D Cartesian domain where the Coriolis parameter $f$ is assumed constant) over a vertical domain from 10 to 0.01 bars. The initial, cloud-free atmosphere was set to be in radiative equilibrium. Through the positive feedback just described, a small cloud seeded into the domain (top left panel) leads to net heating of the column, ascending motion, cloud growth, and horizontal divergence of the flow. The Coriolis force deflects the horizontally diverging air, leading to the generation of a warm-core, anticyclonic, geostrophically balanced, cloud-covered vortex with a diameter of $\sim$20,000 km on a timescale of tens of hours.

\begin{figure*}      % use "figure*" instead of "figure" if you want your figure to span both columns
\epsscale{1.2}      % adjust this number to change the size of your figure
\includegraphics[scale=0.5]{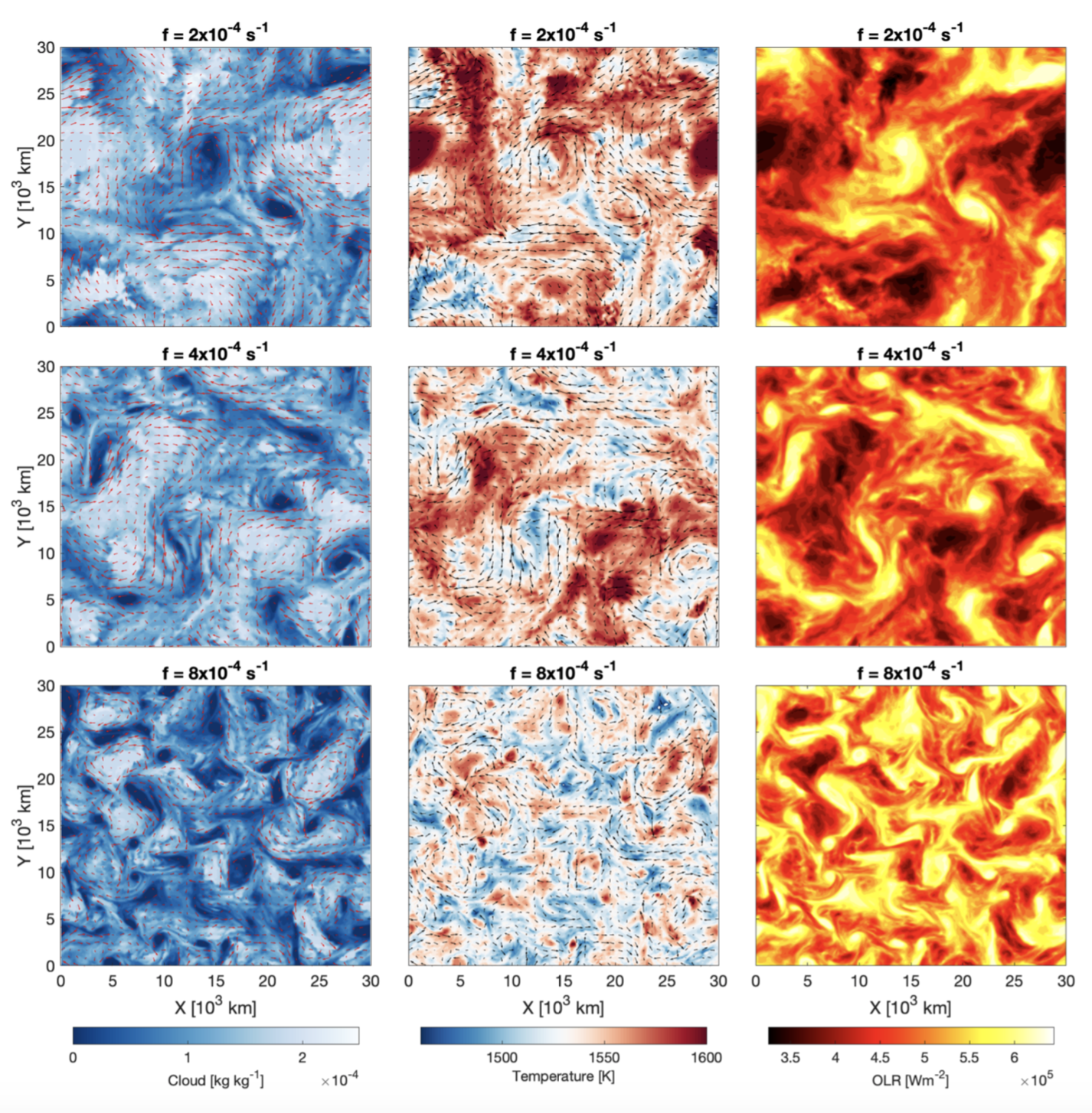}
\centering
\caption{3D simulations of cloudy brown dwarf atmospheres on the $f$-plane from simulations performed in \cite{tan2020bd}. Shown are cloud abundance at 0.2 bars (left), temperature at 0.2 bars (middle) and outgoing IR flux to space (right) for three different rotation rates in the top, middle, and bottom rows. Cloud-radiative-dynamical feedbacks lead to an active atmospheric circulation.}
\label{fig.29}
\end{figure*}

Long-time integrations demonstrate that this process is
self-generating and leads to an active atmospheric circulation that
reaches a statistical, fluctuating turbulent equilibrium. Figure \ref{fig.29}, from \cite{tan2020bd}, show snapshots of the equilibrated state for several different rotation rates. Cloudy and cloud-free vortices, filaments, and other time-variable turbulent structures emerge, with characteristic length scales that are smaller the faster the rotation rate. This trend matches expectations, since the geostrophic adjustment process should produce vortices with sizes close to the deformation radius, which decreases with increasing rotation rate. However, vortex mergers can cause an inverse energy cascade that transfers energy to larger and larger scales, so the characteristic flow length scale in the fully turbulent state (Figure \ref{fig.29}) can potentially become greater than the deformation radius, depending on the dissipation rate of the kinetic energy (implemented as a bottom frictional drag in the models). For typical brown-dwarf parameters, horizontally averaged  wind speeds are typically $300-500 \;\rm{m \;s^{-1}}$, and the local, outgoing IR flux to space can vary by nearly a factor of two.

\begin{figure*}      % use "figure*" instead of "figure" if you want your figure to span both columns
\epsscale{1.2}      % adjust this number to change the size of your figure
\includegraphics[scale=0.5]{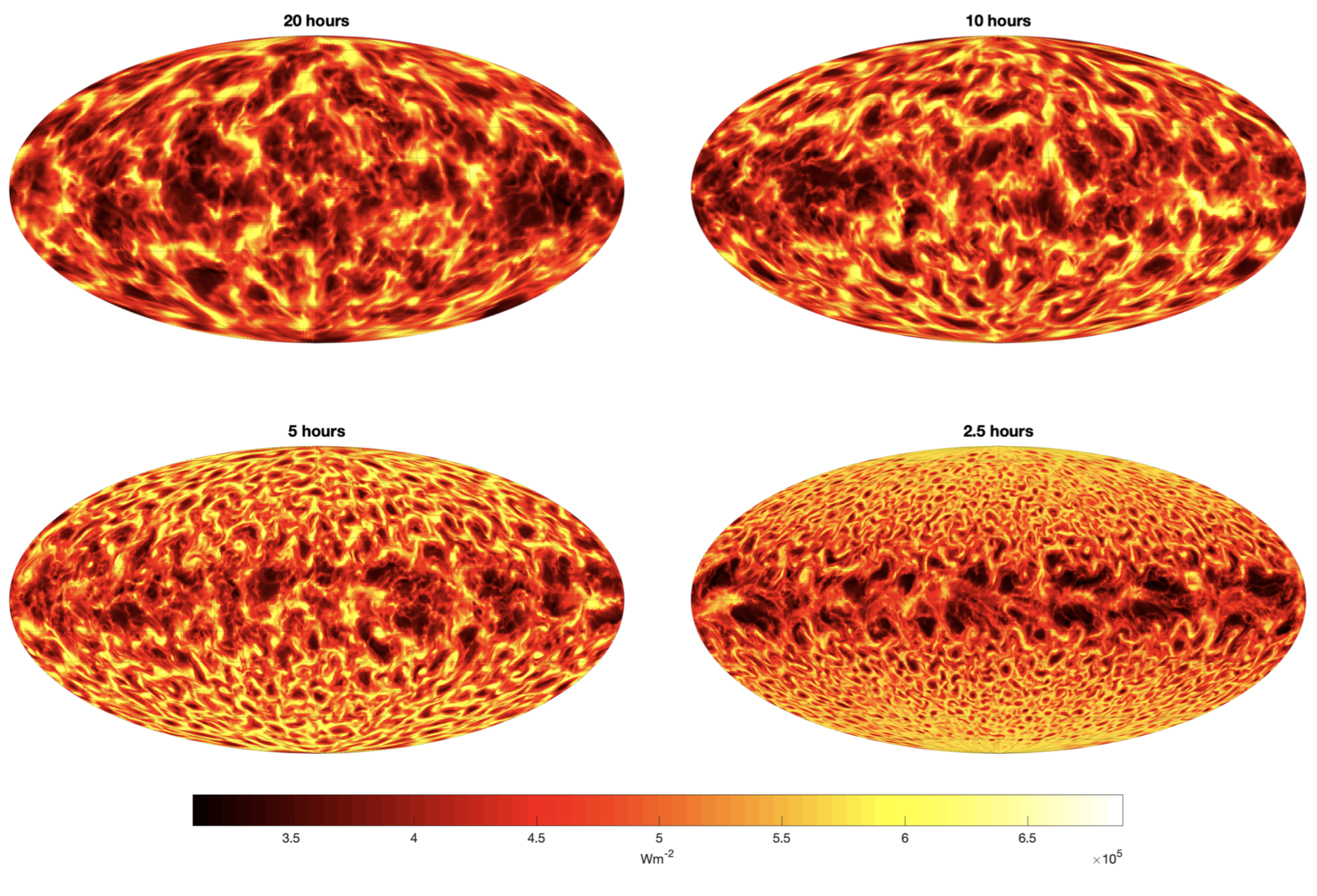}
\centering
\caption{Global circulation models of isolated brown dwarfs with a dual-band radiative transfer scheme, coupled to a cloud cycle, showing the development of highly time-variable turbulence and cloud patchiness. Shown is the total outgoing IR flux, which varies locally by a factor of two due to local variations in cloud opacity and temperature structure. The characteristic size of the turbulent structures decreases with decreasing rotation period (increasing rotation rate). From \cite{tan2018} and Tan \& Showman (in prep).}
\label{fig.30}
\end{figure*}

Global, fully spherical models of this type from \cite{tan2018} and Tan \& Showman (in prep) likewise show the development of complex, time evolving turbulence involving cloudy and cloud-free patches. Figure \ref{fig.30} shows the outgoing IR flux in the fully equilibrated state for models with rotation periods of 2.5, 5, 10, and 20 hours. Again, as expected, the dominant turbulent length scales are smaller when the rotation rate is faster (rotation period is shorter). The equatorial dynamics differ from those at higher latitudes because of the larger local Rossby numbers and the presence of equatorially trapped waves, which couple to the cloud-radiative feedbacks in complex ways. Locally, the outgoing IR flux again varies by a factor of two (Figure 30), and integrated over the disk, the lightcurve variability on rotational timescales can exceed 1\%, consistent with those commonly observed on brown dwarfs (e.g., \citealp{metchev2015}).

\subsection{Summary of brown dwarf dynamics}
Studies performed to date demonstrate that at least three independent mechanisms can lead to a vigorous circulation in the stratified atmospheres of brown dwarfs, despite the absence of any stellar irradiation that could produce day-night or equator-to-pole thermal contrasts. These mechanisms include:
\begin{itemize}
\item Cloud-free dynamics: Convection impinges on the stably stratified atmosphere, triggering waves and driving turbulence, zonal jets, and vortices. This wave-driven circulation can organize clouds into a patchy structure on large scales, even if cloud radiative feedbacks are not considered.

\item 1D cloud feedback: A cloud-radiative-mixing feed- back in 1D vertical columns drives cyclical growth and collapse of cloud decks in an atmospheric column. This causes IR flux variability of hundreds of K in effective temperature over periods of $\sim$1–30 hours.

\item Multi-D cloud feedback: A cloud-radiative-dynamical instability in 3D drives a circulation that maintains the cloud patchiness. Wind speeds reach $300-500$ ${\rm m\;s^{-1}}$, horizontal temperature differences
on isobars reaches 100-200K, and local variations in effective temperature reach hundreds of K, corresponding to local variations in the outgoing IR flux of a factor of two.
\end{itemize}

We note that the global-scale models for these three processes performed to date have essentially been idealized process studies that have aimed to investigate each of the three processes in isolation. The behavior when the three mechanisms are combined has not yet been simulated in global-scale numerical models but will likely involve even richer behavior than the models shown here that have investigated them one by one. { Future modeling efforts beyond the idealized framework summarized here include the use of GCMs with non-grey radiative transfer along with equilibrium or transport-induced disequilibrium chemistry, which are essential for straight model-observation comparisons. Other than clouds, possible transport-induced chemical patchiness and its radiative effect may also contribute to affect the global circulation as well as lightcurve variability, similar to those explored in the context of hot Jupiters (see Section 2.6).  Utilizing convection resolving models and the inclusion of both the convective and stratified zones would help to understand turbulence and wave generation in the  stratified layers and their effects on large-scale flow. Convection models may also be used to examine the relatively small-scale cloud formation and its associated dynamics, as well as  effects of chemical reactions in modulating the convective instability in brown dwarf atmospheres proposed by \cite{tremblin2017bd} and \cite{tremblin2019}.
}

\section{Irradiated brown dwarfs}

\label{irradiated-brown-dwarfs}

%There exists a population of objects more massive than
%typical hot Jupiters, and yet more strongly irradiated
%than typical brown dwarfs. 

Although most close-in, highly irradiated EGPs have masses less than a
few $M_J$, and most known brown dwarfs are isolated objects receiving
negligible stellar flux, there exists a population of massive brown
dwarfs on close-in orbits that receive intense stellar irradiation.
The irradiation levels they experience are comparable to typical hot
Jupiters, and yet because they are massive, they transport interior
heat fluxes that rival or even exceed the energy flux absorbed from
their parent star.  As such, they populate the upper right corner of
Figure~\ref{fig.2}.  While currently a less well-known subclass
than that of hot Jupiters or field brown dwarfs, they are equally
important, for they provide the link in the chain of understanding
between brown dwarfs and hot Jupiters.  One can think of them as hot
Jupiters with enormous convective heat fluxes, or as brown dwarfs that
are modified by introducing external stellar forcing.

These objects fall into two distinct populations.  The first are brown
dwarfs orbiting main-sequence stars in close-in orbits.  The rarity of
these objects relative to less massive, close-in EGPs was discovered
early, a phenomenon dubbed the ``brown dwarf desert''
\citep[e.g.,][]{marcy2000, sahlmann2011}.  Nevertheless,
about two dozen transiting, close-in brown dwarfs of this type have
been discovered.  Their orbital periods---and hence rotation periods
if synchronously rotating---range typically from 3--5~days, implying
that rotation plays a dynamical role similar to that on hot Jupiters.
The second population are brown dwarf companions to white dwarf stars,
which survived a common envelope phase during which the more massive
companion swelled during the red giant phase to envelope the
less-massive companion.  Friction caused by orbital motion of the
companion through the red giant's tenuous outer layers causes the
brown dwarf to slowly spiral in.  Once the red-giant phase ends, the
brown dwarf is left orbiting the remnant white dwarf on an extremely
tight orbit with a period of typically 1--4 hours.  Because they are
likewise expected to be synchrously rotating, their rotation periods
should be equally short, implying that rotation plays a more
controlling role in their atmospheric dynamics.  Moreover, because the
white dwarf's radiation is primarily in the ultraviolet, where typical
atmospheric gases are much more absorbing than in the visible, such
brown dwarfs may experience very different atmospheric thermal
structures and chemistry than their counterparts orbiting sunlike
stars.

\begin{figure*}      % use "figure*" instead of "figure" if you want your figure to span both columns
\epsscale{1.2}      % adjust this number to change the size of your figure
\includegraphics[scale=0.5]{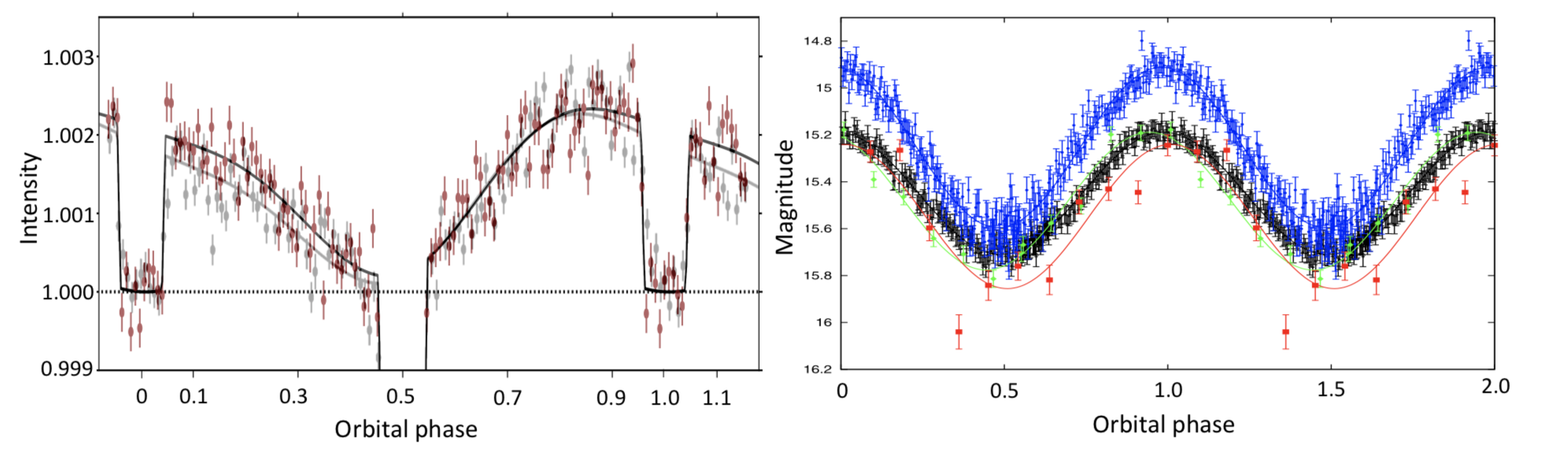}
\centering
\caption{Spitzer IR lightcurves of irradiated brown dwarfs.  Left
  panel shows $3.6\,\mu$m (grey) and $4.5\,\mu$m phase curves of the
  $27M_J$ brown dwarf KELT-1b, orbiting a main-sequence (F5) star in a
  1.2-day orbit.  The phase curves shows two secondary eclipses (at
  phase 0 and 1) and one transit (at phase 0.5).  Modified from
  \citet{beatty2019}.  Dashed line shows the bottom of the
  secondary eclipses, which represents the radiation from the star
  alone.  Right panel shows $3.6\,\mu$m (black), $4.5\,\mu$m (blue),
  $5.8\,\mu$m (green) and $8.0\,\mu$m phase curves of WD0137-349,
  which comprises a $53\,M_J$ brown dwarf orbiting a white dwarf in a
  1.9-hour orbit.  Two full orbits are shown.  In both phase curves, 0
  and 1 represent the times when the brown dwarf's dayside is aimed
  toward Earth, whereas phase 0.5 and 1.5 represent times when its
  nightside is aimed toward Earth. From \citet{casewell2015}.}
\label{fig.31}
\end{figure*}

Lightcurves and other observations of both populations have been
acquired, suggesting that these objects, like hot Jupiters, experience
large day-night temperature differences.  Examples are presented in
Figure~\ref{fig.31}, which shows IR lightcurves from
KELT-1b \citep{beatty2019}, a 27-$M_J$ brown dwarf orbiting an F
star on a 1.2-day orbit, and WD0137-349B, which is a 53-$M_J$ brown
dwarf orbiting its primary white dwarf every 1.9 hours.  Like hot
Jupiters, the IR lightcurve of KELT-1b reaches a peak flux
substantially before secondary eclipse
(Figure~\ref{fig.31}, left panel), suggesting the
possibility of a similar dynamical regime involving equatorial
superrotation and global-scale waves driven by the day-night thermal
forcing.  In contrast, the brown dwarf orbiting a white dwarf lacks
such a phase-curve offset (Figure~\ref{fig.31}, right
panel), suggesting that differing processes operate at fast rotation. { Several such brown-dwarf-white-dwarf systems have been characterized, including  NLTT 5306 \citep{steele2013},  WD0137-349 \citep{casewell2015, longstaff2017},  EPIC 21223532 \citep{casewell2018}, WD 1202-024 \citep{rappaport2017} and SDSS J141126.20+200911.1  \citep{littlefair2014,casewell2018b}.}

Note that, from the perspective of observational characterization,
brown dwarfs around main-sequence stars face similar observational
challenges as typical hot Jupiters---the planetary flux is typically
thousands of times dimmer than the parent star, so teasing out the
planetary signature from total signal is challenging.  In the case of
brown-dwarf-white-dwarf systems, however, the IR flux from the brown
dwarf often dominates, aiding observational characterization of the
brown dwarf.\footnote{The white dwarf typically dominates the {\it
    total} flux, but most of this flux emanates in the UV.  Because
  the white dwarf is physically much smaller than the brown dwarf, its
  IR flux is smaller, despite having a higher temperature.} Many brown
dwarfs of this type are cataclysmic variables
\citep[e.g.,][]{hellier2001}, in which the brown dwarf sheds mass
onto the white dwarf.  In this situation, the brown dwarf's
non-spherical shape and the presence of an accretion disc around the
white dwarf can complicate an interpretation of the observations
\citep{santisteban2016}.  Still, some observed
brown-dwarf-white-dwarf systems---including the one shown in
Figure~\ref{fig.31}---are ``detached,'' exhibiting
sufficiently great separations for these effects to be minimal.

\begin{figure*}      % use "figure*" instead of "figure" if you want your figure to span both columns
\epsscale{1.2}      % adjust this number to change the size of your figure
\includegraphics[scale=0.5]{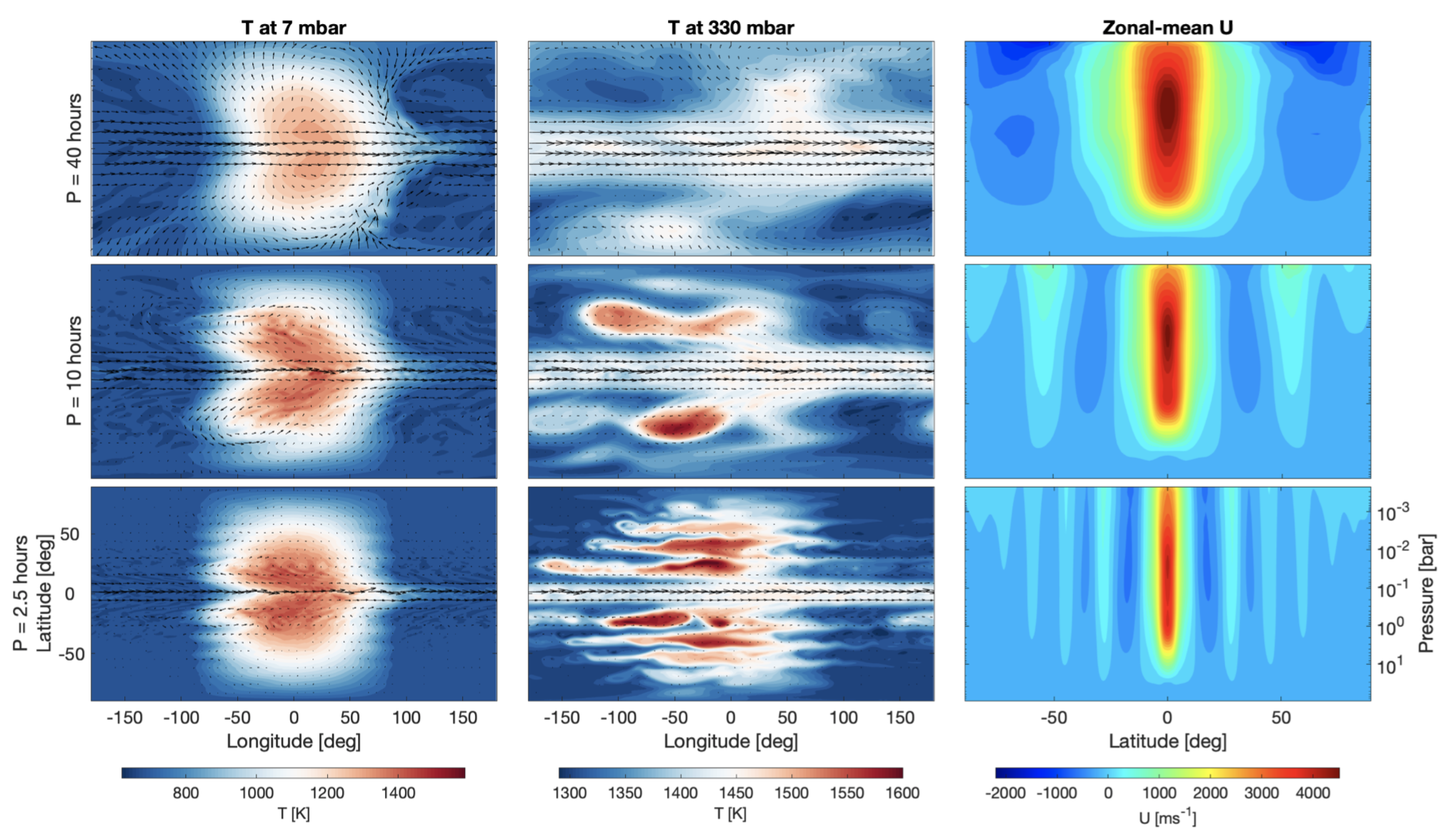}
\centering
\caption{Atmospheric circulation in synchronously rotating, strongly
  irradiated atmospheres at modest to very fast rotation rate,
  motivated by observations of brown-dwarf-white-dwarf systems with
  orbital periods of several hours.  Top, middle, and bottom rows
  represent idealized GCM simulations with rotation periods of 40, 10,
  and $2.5\rm\,hours$, respectively.  The forcing setup is identical
  in all models; only the rotation period differs.  Left and middle
  columns show temperature (colorscale) and winds (arrows) in
  snapshots at 7~mbar and 330~mbar, respectively; right column shows
  the zonal-mean zonal wind (colorscale, in $\rm m\,s^{-1}$) as a
  function of latitude and pressure.  As rotation period decreases,
  the day-night temperature difference increases, the equatorial jet
  narrows, the hemisphere-scale dayside hotspot offset decreases, and
  numerous off-equatorial zonal jets emerge in the mid-to-high
  latitudes.  From \citet{tan2020wdbd}.}
\label{fig.32}
\end{figure*}

Motivated by these systems, \citet{tan2020wdbd} performed
idealized GCM experiments of sychronously rotating, strongly
irradiated giant planets with rotation periods ranging from 80 hours
to 2.5~hours to characterize the role of fast rotation in shaping
their circulation. A subset of these models is shown in
Figure~\ref{fig.32}.  Under identical forcing conditions,
faster rotation leads to larger day-night temperature differences in
the observable atmosphere (pressures less than a few hundred mbar),
which results from the inhibition of day-night heat transport caused
by rapid planetary rotation.  Moreover, because of the decrease in the
Rossby deformation radius at fast rotation, the equatorial waveguide,
Matsuno-Gill pattern, and the equatorial superrotating jet all become
confined within $10^\circ$ of the equator.  As a result, in contrast
to hot Jupiters, the overall dayside temperature pattern does not
exhibit any prominent eastward offset when the rotation rate is very
fast (essentially, the eastward offset still occurs but becomes
confined preferentially closer and closer to the equator as rotation
rate increases).  Infrared lightcurves therefore exhibit flux peaks
that coincide with the time of secondary eclipse.  This helps to
explain the fact that many brown dwarfs on close-in orbits around
white dwarfs lack prominent offsets of the IR flux peaks, as shown in
Figure~\ref{fig.31}.  Interestingly,
Figure~\ref{fig.32} shows that deeper in the troposphere---at
pressures of a few hundred mbar---the dayside temperature develops a
``fingering'' pattern wherein meridionally narrow, zonally elongated
tongues of hot and cold air interleave.  Outside the equatorial jet, a
prominent pattern of alternating eastward and westward zonal jets
emerges.  \citet{tan2020wdbd} showed that, despite the novel
forcing, these zonal jets scale well with the Rhines scale.  Further
work is needed to understand the implications for
brown-dwarf-white-dwarf systems and the connection to zonal jets on
Jupiter and Saturn.

{ \cite{lee2020} simulated the atmospheric circulation of the brown dwarf orbiting the white dwarf WD0137-349 with an orbital period of about 2 hours. They adopted the ExoFMS GCM  which utilizes  a dual-grey radiative transfer scheme assuming constant opacity in both thermal and visible band. Their results are qualitatively similar to the rapidly rotating cases in \cite{tan2020wdbd}, showing a meridionally narrower equatorial superrotating jet compared to canonical hot Jupiter simulations, nearly geostrophic flows at mid-high latitudes and large day-night temperature difference. Although using different GCMs and radiative forcing setup, the agreement between \cite{lee2020} and \cite{tan2020wdbd} suggests that  strong rotation could robustly and profoundly influence the atmospheric circulation of tidally locked objects.
}

{ Future endeavors beyond the above idealized work are needed to obtain a completed understanding on the 3D atmospheres of this group of objects.  The spectrum  of white dwarfs peak at UV, and significant photochemistry are expected to occur in atmospheres of the companion BDs \citep{longstaff2017,casewell2018b}. Effects of photochemical products and radiative heating on the thermal structures and circulation, along with the effects of hydrogen thermal  dissociation/recombination similar to those found in ultra-hot Jupiter atmospheres \citep{bell2018,komacek2018rnaas,tan2019}, are worth to explored in a thorough manner.   Given the insufficient day-night heat transport, various mineral clouds are expected to condense in the nightsides of these objects \citep{lee2020}, and may have interesting effects on lightcurve and dynamics. Strong magnetic field ($\sim$kG) has been inferred from observations of field brown dwarfs (e.g., \citealp{kao2016,kao2018}). If brown  dwarfs around white dwarfs host such  strong fields, their strongly irradiated atmospheres may experience significant  MHD effects.  
}

\section{DISCUSSION}
Observations reveal a wealth of behavior in the atmospheres of EGPs and brown dwarfs. The intense daynight radiative forcing on hot Jupiters leads to large day-night temperature differences and planetary-scale waves that, according to most GCMs, drive a fast superrotating jet at the equator. Adjustment of the thermal structure by planetary-scale waves provides a framework for understanding the day-night temperature contrasts and their dependence on the strength of stellar irradiation and other parameters. Analytic theories and idealized numerical models have provided significant insights into these processes. Overall, a basic picture of the dynamical processes that maintain the circulation on hot Jupiters is slowly emerging. But many aspects remain poorly understood; for example, the eddies that maintain the superrotating jet are nonlinear and interact with the shear of the jet, making their structure difficult to predict in detail. The ways the circulation interacts with patchy clouds and the detailed dynamics of vertical and horizontal mixing of trace constituents remain less well understood. Predictions for various regime transitions across the hot Jupiter population—and between hot and warm Jupiters---have yet to be confronted by observational data. An open frontier that has been little studied is the extent to which (and how) the atmosphere and interior of hot Jupiters interact, including the depth to which the equatorial jet extends, the type of winds that may exist in the deep convection zone, as well as the degree of mixing, the nature of the thermal structure, and the modes of energy exchange between the atmosphere and deep interior.

Warm Jupiters are a fascinating target for future observations, and help to bridge the gap between hot Jupiters on the one hand and Jupiter and Saturn on the other. At present, few observational data exist to con- strain the dynamical behavior of warm Jupiters, but a growing body of models suggest a wide range of behav- ior that organizes into a variety of regimes depending on the rotation rate, obliquity, orbital eccentricity, and how the photosphere-level radiative time constant compares to the rotational and orbital timescales.

A wealth of data provide information about the thermal and cloud structure of isolated brown dwarfs, indicating that these worlds are covered by patchy clouds modulated by a vigorous atmospheric circulation. This regime differs from hot and warm Jupiters in being fundamentally driven by interior heat loss. In comparison to hot Jupiters, dynamical models of the global circulation on brown dwarfs are in their infancy and represents a ripe frontier for future research. Models point toward rich interactions between the convective interior and stratified atmosphere and of various subtle feedbacks between dynamics, radiation and clouds that can lead to a spontaneous, self-generating circulation comprising turbulence, vortices, and zonal jets, associated with patchy cloud structures like those observed. The continuum between hot brown dwarf on the one hand and Jupiter and Saturn on the other deserves further study, and can face observational tests from the growing quality of data on cooler T and Y dwarfs. Irradiated brown dwarfs represent an even more open frontier, and can help bridge the gap between hot Jupiters and isolated brown dwarfs. Observations are starting to clarify the thermal structure of these objects, but so far almost no modeling and theory has been performed to explore this regime.

We have argued that there is value in considering these distinct populations together as a class. The synergies that can result from comparisons between the diverse populations have yet to be well exploited but can potentially lead to major insights across these populations. The grand challenge of understanding the behavior of giant planets as an integrated population across many orders of magnitude in internal heat flux, external irradiation, rotation rate, and other parameters beckons us.

\if\bibinc n
\bibliography{draft}

\begin{thebibliography}{308}
\expandafter\ifx\csname natexlab\endcsname\relax\def\natexlab#1{#1}\fi

\bibitem[{{Ackerman} \& {Marley}(2001)}]{ackerman2001}
{Ackerman}, A.~S. \& {Marley}, M.~S. 2001, \apj, 556, 872

\bibitem[{{Ag{\'u}ndez} {et~al.}(2014){Ag{\'u}ndez}, {Parmentier}, {Venot},
  {Hersant}, \& {Selsis}}]{agundez2014}
{Ag{\'u}ndez}, M., {Parmentier}, V., {Venot}, O., {Hersant}, F., \& {Selsis},
  F. 2014, \aap, 564, A73

\bibitem[{{Allers} {et~al.}(2020){Allers}, {Vos}, {Biller}, \&
  {Williams}}]{allers2020}
{Allers}, K.~N., {Vos}, J.~M., {Biller}, B.~A., \& {Williams}, P. K.~G. 2020,
  Science, 368, 169

\bibitem[{{Amundsen} {et~al.}(2014){Amundsen}, {Baraffe}, {Tremblin},
  {Manners}, {Hayek}, {Mayne}, \& {Acreman}}]{amundsen2014}
{Amundsen}, D.~S., {Baraffe}, I., {Tremblin}, P., {Manners}, J., {Hayek}, W.,
  {Mayne}, N.~J., \& {Acreman}, D.~M. 2014, \aap, 564, A59

\bibitem[{{Amundsen} {et~al.}(2016){Amundsen}, {Mayne}, {Baraffe}, {Manners},
  {Tremblin}, {Drummond}, {Smith}, {Acreman}, \& {Homeier}}]{amundsen2016}
{Amundsen}, D.~S., {Mayne}, N.~J., {Baraffe}, I., {Manners}, J., {Tremblin},
  P., {Drummond}, B., {Smith}, C., {Acreman}, D.~M., \& {Homeier}, D. 2016,
  \aap, 595, A36

\bibitem[{{Anderson} {et~al.}(2014){Anderson}, {Collier Cameron}, {Delrez},
  {Doyle}, {Faedi}, {Fumel}, {Gillon}, {G{\'o}mez Maqueo Chew}, {Hellier},
  {Jehin}, {Lendl}, {Maxted}, {Pepe}, {Pollacco}, {Queloz}, {S{\'e}gransan},
  {Skillen}, {Smalley}, {Smith}, {Southworth}, {Triaud}, {Turner}, {Udry}, \&
  {West}}]{anderson2014}
{Anderson}, D.~R., {Collier Cameron}, A., {Delrez}, L., {Doyle}, A.~P.,
  {Faedi}, F., {Fumel}, A., {Gillon}, M., {G{\'o}mez Maqueo Chew}, Y.,
  {Hellier}, C., {Jehin}, E., {Lendl}, M., {Maxted}, P.~F.~L., {Pepe}, F.,
  {Pollacco}, D., {Queloz}, D., {S{\'e}gransan}, D., {Skillen}, I., {Smalley},
  B., {Smith}, A.~M.~S., {Southworth}, J., {Triaud}, A.~H.~M.~J., {Turner},
  O.~D., {Udry}, S., \& {West}, R.~G. 2014, \mnras, 445, 1114

\bibitem[{Andrews {et~al.}(1987)Andrews, Holton, \& Leovy}]{andrews1987}
Andrews, D.~G., Holton, J.~R., \& Leovy, C.~B. 1987, Middle atmosphere
  dynamics, Vol.~40 (Academic press)

\bibitem[{{Angerhausen} {et~al.}(2015){Angerhausen}, {DeLarme}, \&
  {Morse}}]{angerhausen2015}
{Angerhausen}, D., {DeLarme}, E., \& {Morse}, J.~A. 2015, \pasp, 127, 1113

\bibitem[{Apai {et~al.}(2017)Apai, Karalidi, Marley, Yang, Flateau, Metchev,
  Cowan, Buenzli, Burgasser, Radigan, {et~al.}}]{apai2017}
Apai, D., Karalidi, T., Marley, M., Yang, H., Flateau, D., Metchev, S., Cowan,
  N., Buenzli, E., Burgasser, A., Radigan, J., {et~al.} 2017, Science, 357, 683

\bibitem[{{Apai} {et~al.}(2013){Apai}, {Radigan}, {Buenzli}, {Burrows}, {Reid},
  \& {Jayawardhana}}]{apai2013}
{Apai}, D., {Radigan}, J., {Buenzli}, E., {Burrows}, A., {Reid}, I.~N., \&
  {Jayawardhana}, R. 2013, \apj, 768, 121

\bibitem[{{Arcangeli} {et~al.}(2019){Arcangeli}, {D{\'e}sert}, {Parmentier},
  {Stevenson}, {Bean}, {Line}, {Kreidberg}, {Fortney}, \&
  {Showman}}]{Arcangeli2019}
{Arcangeli}, J., {D{\'e}sert}, J.-M., {Parmentier}, V., {Stevenson}, K.~B.,
  {Bean}, J.~L., {Line}, M.~R., {Kreidberg}, L., {Fortney}, J.~J., \&
  {Showman}, A.~P. 2019, \aap, 625, A136

\bibitem[{{Armstrong} {et~al.}(2016){Armstrong}, {de Mooij}, {Barstow},
  {Osborn}, {Blake}, \& {Saniee}}]{armstrong2016}
{Armstrong}, D.~J., {de Mooij}, E., {Barstow}, J., {Osborn}, H.~P., {Blake},
  J., \& {Saniee}, N.~F. 2016, Nature Astronomy, 1, 0004

\bibitem[{Arnold {et~al.}(2012)Arnold, Tziperman, \& Farrell}]{arnold2012}
Arnold, N.~P., Tziperman, E., \& Farrell, B. 2012, Journal of the atmospheric
  sciences, 69, 626

\bibitem[{{Artigau}(2018)}]{artigau2018}
{Artigau}, {\'E}. 2018, ArXiv e-prints

\bibitem[{{Artigau} {et~al.}(2009){Artigau}, {Bouchard}, {Doyon}, \&
  {Lafreni{\`e}re}}]{artigau2009}
{Artigau}, {\'E}., {Bouchard}, S., {Doyon}, R., \& {Lafreni{\`e}re}, D. 2009,
  The Astrophysical Journal, 701, 1534

\bibitem[{{Bakos} {et~al.}(2018){Bakos}, {Bayliss}, {Bento}, {Bhatti}, {Brahm},
  {Csubry}, {Espinoza}, {Hartman}, {Henning}, {Jord{\'a}n}, {Mancini}, {Penev},
  {Rabus}, {Sarkis}, {Suc}, {de Val-Borro}, {Zhou}, {Butler}, {Crane},
  {Durkan}, {Shectman}, {Kim}, {L{\'a}z{\'a}r}, {Papp}, {S{\'a}ri}, {Ricker},
  {Vanderspek}, {Latham}, {Seager}, {Winn}, {Jenkins}, {Chacon},
  {F{\H{u}}r{\'e}sz}, {Goeke}, {Li}, {Quinn}, {Quintana}, {Tenenbaum}, {Teske},
  {Vezie}, {Yu}, {Stockdale}, {Evans}, \& {Relles}}]{bakos2018}
{Bakos}, G.~{\'A}., {Bayliss}, D., {Bento}, J., {Bhatti}, W., {Brahm}, R.,
  {Csubry}, Z., {Espinoza}, N., {Hartman}, J.~D., {Henning}, T., {Jord{\'a}n},
  A., {Mancini}, L., {Penev}, K., {Rabus}, M., {Sarkis}, P., {Suc}, V., {de
  Val-Borro}, M., {Zhou}, G., {Butler}, R.~P., {Crane}, J., {Durkan}, S.,
  {Shectman}, S., {Kim}, J., {L{\'a}z{\'a}r}, J., {Papp}, I., {S{\'a}ri}, P.,
  {Ricker}, G., {Vanderspek}, R., {Latham}, D.~W., {Seager}, S., {Winn}, J.~N.,
  {Jenkins}, J., {Chacon}, A.~D., {F{\H{u}}r{\'e}sz}, G., {Goeke}, B., {Li},
  J., {Quinn}, S., {Quintana}, E.~V., {Tenenbaum}, P., {Teske}, J., {Vezie},
  M., {Yu}, L., {Stockdale}, C., {Evans}, P., \& {Relles}, H.~M. 2018, arXiv
  e-prints, arXiv:1812.09406

\bibitem[{Baldwin {et~al.}(2001)Baldwin, Gray, Dunkerton, Hamilton, Haynes,
  Randel, Holton, Alexander, Hirota, Horinouchi, {et~al.}}]{baldwin2001}
Baldwin, M., Gray, L., Dunkerton, T., Hamilton, K., Haynes, P., Randel, W.,
  Holton, J., Alexander, M., Hirota, I., Horinouchi, T., {et~al.} 2001, Reviews
  of Geophysics, 39, 179

\bibitem[{{Baraffe} {et~al.}(2008){Baraffe}, {Chabrier}, \&
  {Barman}}]{baraffe2008}
{Baraffe}, I., {Chabrier}, G., \& {Barman}, T. 2008, \aap, 482, 315

\bibitem[{{Baraffe} {et~al.}(2014){Baraffe}, {Chabrier}, {Fortney}, \&
  {Sotin}}]{baraffe2014}
{Baraffe}, I., {Chabrier}, G., {Fortney}, J., \& {Sotin}, C. 2014, in
  Protostars and Planets VI, ed. H.~{Beuther}, R.~S. {Klessen}, C.~P.
  {Dullemond}, \& T.~{Henning}, 763

\bibitem[{{Barstow} {et~al.}(2017){Barstow}, {Aigrain}, {Irwin}, \&
  {Sing}}]{barstow2017}
{Barstow}, J.~K., {Aigrain}, S., {Irwin}, P.~G.~J., \& {Sing}, D.~K. 2017,
  \apj, 834, 50

\bibitem[{{Batygin} \& {Stanley}(2014)}]{Batygin2014}
{Batygin}, K. \& {Stanley}, S. 2014, \apj, 794, 10

\bibitem[{{Batygin} {et~al.}(2013){Batygin}, {Stanley}, \&
  {Stevenson}}]{Batygin2013}
{Batygin}, K., {Stanley}, S., \& {Stevenson}, D.~J. 2013, \apj, 776, 53

\bibitem[{{Batygin} \& {Stevenson}(2010)}]{Batygin2010}
{Batygin}, K. \& {Stevenson}, D.~J. 2010, \apjl, 714, L238

\bibitem[{Bayliss {et~al.}(2016)Bayliss, Hojjatpanah, Santerne, Dragomir, Zhou,
  Shporer, Col{\'o}n, Almenara, Armstrong, Barrado, {et~al.}}]{bayliss2016}
Bayliss, D., Hojjatpanah, S., Santerne, A., Dragomir, D., Zhou, G., Shporer,
  A., Col{\'o}n, K., Almenara, J., Armstrong, D., Barrado, D., {et~al.} 2016,
  The Astronomical Journal, 153, 15

\bibitem[{{Beatty} {et~al.}(2019){Beatty}, {Marley}, {Gaudi}, {Col{\'o}n},
  {Fortney}, \& {Showman}}]{beatty2019}
{Beatty}, T.~G., {Marley}, M.~S., {Gaudi}, B.~S., {Col{\'o}n}, K.~D.,
  {Fortney}, J.~J., \& {Showman}, A.~P. 2019, \aj, 158, 166

\bibitem[{{Bell} \& {Cowan}(2018)}]{bell2018}
{Bell}, T.~J. \& {Cowan}, N.~B. 2018, \apjl, 857, L20

\bibitem[{{Bending} {et~al.}(2013){Bending}, {Lewis}, \& {Kolb}}]{bending2013}
{Bending}, V.~L., {Lewis}, S.~R., \& {Kolb}, U. 2013, \mnras, 428, 2874

\bibitem[{Biller(2017)}]{biller2017}
Biller, B. 2017, Astronomical Review, 13, 1

\bibitem[{{Bowler}(2016)}]{bowler2016}
{Bowler}, B.~P. 2016, \pasp, 128, 102001

\bibitem[{{Bowler} {et~al.}(2020){Bowler}, {Zhou}, {Morley}, {Kataria},
  {Bryan}, {Benneke}, \& {Batygin}}]{bowler2020}
{Bowler}, B.~P., {Zhou}, Y., {Morley}, C.~V., {Kataria}, T., {Bryan}, M.~L.,
  {Benneke}, B., \& {Batygin}, K. 2020, \apjl, 893, L30

\bibitem[{{Brahm} {et~al.}(2016){Brahm}, {Jord{\'a}n}, {Bakos}, {Penev},
  {Espinoza}, {Rabus}, {Hartman}, {Bayliss}, {Ciceri}, {Zhou}, {Mancini},
  {Tan}, {de Val-Borro}, {Bhatti}, {Csubry}, {Bento}, {Henning}, {Schmidt},
  {Rojas}, {Suc}, {L{\'a}z{\'a}r}, {Papp}, \& {S{\'a}ri}}]{brahm2016}
{Brahm}, R., {Jord{\'a}n}, A., {Bakos}, G.~{\'A}., {Penev}, K., {Espinoza}, N.,
  {Rabus}, M., {Hartman}, J.~D., {Bayliss}, D., {Ciceri}, S., {Zhou}, G.,
  {Mancini}, L., {Tan}, T.~G., {de Val-Borro}, M., {Bhatti}, W., {Csubry}, Z.,
  {Bento}, J., {Henning}, T., {Schmidt}, B., {Rojas}, F., {Suc}, V.,
  {L{\'a}z{\'a}r}, J., {Papp}, I., \& {S{\'a}ri}, P. 2016, \aj, 151, 89

\bibitem[{{Buenzli} {et~al.}(2014){Buenzli}, {Apai}, {Radigan}, {Reid}, \&
  {Flateau}}]{buenzli2014}
{Buenzli}, E., {Apai}, D., {Radigan}, J., {Reid}, I.~N., \& {Flateau}, D. 2014,
  The Astrophysical Journal, 782, 77

\bibitem[{Buenzli {et~al.}(2015)Buenzli, Saumon, Marley, Apai, Radigan, Bedin,
  Reid, \& Morley}]{buenzli2015}
Buenzli, E., Saumon, D., Marley, M.~S., Apai, D., Radigan, J., Bedin, L.~R.,
  Reid, I.~N., \& Morley, C.~V. 2015, The Astrophysical Journal, 798, 127

\bibitem[{{Burgasser} {et~al.}(2014){Burgasser}, {Gillon}, {Faherty},
  {Radigan}, {Triaud}, {Plavchan}, {Street}, {Jehin}, {Delrez}, \&
  {Opitom}}]{burgasser2014}
{Burgasser}, A.~J., {Gillon}, M., {Faherty}, J.~K., {Radigan}, J., {Triaud}, A.
  H.~M.~J., {Plavchan}, P., {Street}, R., {Jehin}, E., {Delrez}, L., \&
  {Opitom}, C. 2014, \apj, 785, 48

\bibitem[{Burgasser {et~al.}(2002)Burgasser, Marley, Ackerman, Saumon, Lodders,
  Dahn, Harris, \& Kirkpatrick}]{burgasser2002}
Burgasser, A.~J., Marley, M.~S., Ackerman, A.~S., Saumon, D., Lodders, K.,
  Dahn, C.~C., Harris, H.~C., \& Kirkpatrick, J.~D. 2002, The Astrophysical
  Journal Letters, 571, L151

\bibitem[{{Burrows} {et~al.}(2001){Burrows}, {Hubbard}, {Lunine}, \&
  {Liebert}}]{burrows2001}
{Burrows}, A., {Hubbard}, W.~B., {Lunine}, J.~I., \& {Liebert}, J. 2001,
  Reviews of Modern Physics, 73, 719

\bibitem[{{Burrows} {et~al.}(1997){Burrows}, {Marley}, {Hubbard}, {Lunine},
  {Guillot}, {Saumon}, {Freedman}, {Sudarsky}, \& {Sharp}}]{burrows1997}
{Burrows}, A., {Marley}, M., {Hubbard}, W.~B., {Lunine}, J.~I., {Guillot}, T.,
  {Saumon}, D., {Freedman}, R., {Sudarsky}, D., \& {Sharp}, C. 1997, \apj, 491,
  856

\bibitem[{{Burrows} {et~al.}(2006){Burrows}, {Sudarsky}, \&
  {Hubeny}}]{burrows2006}
{Burrows}, A., {Sudarsky}, D., \& {Hubeny}, I. 2006, \apj, 640, 1063

\bibitem[{Burrows(2014)}]{burrows2014}
Burrows, A.~S. 2014, Nature, 513, 345

\bibitem[{Carone {et~al.}(2020)Carone, Baeyens, Molli{\`e}re, Barth, Vazan,
  Decin, Sarkis, Venot, \& Henning}]{carone2020}
Carone, L., Baeyens, R., Molli{\`e}re, P., Barth, P., Vazan, A., Decin, L.,
  Sarkis, P., Venot, O., \& Henning, T. 2020, Monthly Notices of the Royal
  Astronomical Society, 496, 3582

\bibitem[{Casewell {et~al.}(2018{\natexlab{a}})Casewell, Braker, Parsons,
  Hermes, Burleigh, Belardi, Chaushev, Finch, Roy, Littlefair,
  {et~al.}}]{casewell2018}
Casewell, S., Braker, I., Parsons, S., Hermes, J., Burleigh, M., Belardi, C.,
  Chaushev, A., Finch, N., Roy, M., Littlefair, S., {et~al.}
  2018{\natexlab{a}}, Monthly Notices of the Royal Astronomical Society, 476,
  1405

\bibitem[{Casewell {et~al.}(2018{\natexlab{b}})Casewell, Littlefair, Parsons,
  Marsh, Fortney, \& Marley}]{casewell2018b}
Casewell, S., Littlefair, S., Parsons, S., Marsh, T., Fortney, J., \& Marley,
  M. 2018{\natexlab{b}}, Monthly Notices of the Royal Astronomical Society,
  481, 5216

\bibitem[{{Casewell} {et~al.}(2015){Casewell}, {Lawrie}, {Maxted}, {Marley},
  {Fortney}, {Rimmer}, {Littlefair}, {Wynn}, {Burleigh}, \&
  {Helling}}]{casewell2015}
{Casewell}, S.~L., {Lawrie}, K.~A., {Maxted}, P.~F.~L., {Marley}, M.~S.,
  {Fortney}, J.~J., {Rimmer}, P.~B., {Littlefair}, S.~P., {Wynn}, G.,
  {Burleigh}, M.~R., \& {Helling}, C. 2015, \mnras, 447, 3218

\bibitem[{{Chabrier} {et~al.}(2000){Chabrier}, {Baraffe}, {Allard}, \&
  {Hauschildt}}]{chabrier2000}
{Chabrier}, G., {Baraffe}, I., {Allard}, F., \& {Hauschildt}, P. 2000, \apj,
  542, 464

\bibitem[{Chapman(1970)}]{chapman1970}
Chapman, S. 1970, D. Reidel Publishing Company, Dordrecht

\bibitem[{{Charnay} {et~al.}(2015){Charnay}, {Meadows}, {Misra}, {Leconte}, \&
  {Arney}}]{charnay2015}
{Charnay}, B., {Meadows}, V., {Misra}, A., {Leconte}, J., \& {Arney}, G. 2015,
  \apjl, 813, L1

\bibitem[{{Cho} {et~al.}(2003){Cho}, {Menou}, {Hansen}, \& {Seager}}]{cho2003}
{Cho}, J. Y.~K., {Menou}, K., {Hansen}, B. M.~S., \& {Seager}, S. 2003, \apjl,
  587, L117

\bibitem[{{Cho} {et~al.}(2015){Cho}, {Polichtchouk}, \&
  {Thrastarson}}]{cho2015}
{Cho}, J.~Y.~K., {Polichtchouk}, I., \& {Thrastarson}, H.~T. 2015, \mnras, 454,
  3423

\bibitem[{{Choi} {et~al.}(2009){Choi}, {Showman}, \& {Brown}}]{choi2009}
{Choi}, D.~S., {Showman}, A.~P., \& {Brown}, R.~H. 2009, Journal of Geophysical
  Research (Planets), 114, E04007

\bibitem[{{Conrath} {et~al.}(1990){Conrath}, {Gierasch}, \&
  {Leroy}}]{conrath1990}
{Conrath}, B.~J., {Gierasch}, P.~J., \& {Leroy}, S.~S. 1990, \icarus, 83, 255

\bibitem[{{Cooper} \& {Showman}(2005)}]{cooper2005}
{Cooper}, C.~S. \& {Showman}, A.~P. 2005, \apjl, 629, L45

\bibitem[{{Cooper} \& {Showman}(2006)}]{cooper2006}
---. 2006, \apj, 649, 1048

\bibitem[{{Cowan} \& {Agol}(2011{\natexlab{a}})}]{cowan2011b}
{Cowan}, N.~B. \& {Agol}, E. 2011{\natexlab{a}}, \apj, 726, 82

\bibitem[{{Cowan} \& {Agol}(2011{\natexlab{b}})}]{cowan2011}
---. 2011{\natexlab{b}}, \apj, 729, 54

\bibitem[{{Crossfield} {et~al.}(2014){Crossfield}, {Biller}, {Schlieder},
  {Deacon}, {Bonnefoy}, {Homeier}, {Allard}, {Buenzli}, {Henning}, {Brandner},
  {Goldman}, \& {Kopytova}}]{crossfield2014}
{Crossfield}, I.~J.~M., {Biller}, B., {Schlieder}, J.~E., {Deacon}, N.~R.,
  {Bonnefoy}, M., {Homeier}, D., {Allard}, F., {Buenzli}, E., {Henning}, T.,
  {Brandner}, W., {Goldman}, B., \& {Kopytova}, T. 2014, \nat, 505, 654

\bibitem[{{Cushing} {et~al.}(2011){Cushing}, {Kirkpatrick}, {Gelino},
  {Griffith}, {Skrutskie}, {Mainzer}, {Marsh}, {Beichman}, {Burgasser},
  {Prato}, {Simcoe}, {Marley}, {Saumon}, {Freedman}, {Eisenhardt}, \&
  {Wright}}]{cushing2011}
{Cushing}, M.~C., {Kirkpatrick}, J.~D., {Gelino}, C.~R., {Griffith}, R.~L.,
  {Skrutskie}, M.~F., {Mainzer}, A., {Marsh}, K.~A., {Beichman}, C.~A.,
  {Burgasser}, A.~J., {Prato}, L.~A., {Simcoe}, R.~A., {Marley}, M.~S.,
  {Saumon}, D., {Freedman}, R.~S., {Eisenhardt}, P.~R., \& {Wright}, E.~L.
  2011, The Astrophysical Journal, 743, 50

\bibitem[{{Cushing} {et~al.}(2006){Cushing}, {Roellig}, {Marley}, {Saumon},
  {Leggett}, {Kirkpatrick}, {Wilson}, {Sloan}, {Mainzer}, {Van Cleve}, \&
  {Houck}}]{cushing2006}
{Cushing}, M.~C., {Roellig}, T.~L., {Marley}, M.~S., {Saumon}, D., {Leggett},
  S.~K., {Kirkpatrick}, J.~D., {Wilson}, J.~C., {Sloan}, G.~C., {Mainzer},
  A.~K., {Van Cleve}, J.~E., \& {Houck}, J.~R. 2006, \apj, 648, 614

\bibitem[{{Dang} {et~al.}(2018){Dang}, {Cowan}, {Schwartz}, {Rauscher},
  {Zhang}, {Knutson}, {Line}, {Dobbs-Dixon}, {Deming}, {Sundararajan},
  {Fortney}, \& {Zhao}}]{Dang2018}
{Dang}, L., {Cowan}, N.~B., {Schwartz}, J.~C., {Rauscher}, E., {Zhang}, M.,
  {Knutson}, H.~A., {Line}, M., {Dobbs-Dixon}, I., {Deming}, D.,
  {Sundararajan}, S., {Fortney}, J.~J., \& {Zhao}, M. 2018, Nature Astronomy,
  2, 220

\bibitem[{{Dawson} \& {Johnson}(2018)}]{dawson2018}
{Dawson}, R.~I. \& {Johnson}, J.~A. 2018, \araa, 56, 175

\bibitem[{{de Wit} {et~al.}(2017){de Wit}, {Lewis}, {Knutson}, {Fuller},
  {Antoci}, {Fulton}, {Laughlin}, {Deming}, {Shporer}, {Batygin}, {Cowan},
  {Agol}, {Burrows}, {Fortney}, {Langton}, \& {Showman}}]{dewit2017}
{de Wit}, J., {Lewis}, N.~K., {Knutson}, H.~A., {Fuller}, J., {Antoci}, V.,
  {Fulton}, B.~J., {Laughlin}, G., {Deming}, D., {Shporer}, A., {Batygin}, K.,
  {Cowan}, N.~B., {Agol}, E., {Burrows}, A.~S., {Fortney}, J.~J., {Langton},
  J., \& {Showman}, A.~P. 2017, \apjl, 836, L17

\bibitem[{Debras {et~al.}(2019)Debras, Mayne, Baraffe, Goffrey, \&
  Thuburn}]{debras2019}
Debras, F., Mayne, N., Baraffe, I., Goffrey, T., \& Thuburn, J. 2019, Astronomy
  \& Astrophysics, 631, A36

\bibitem[{{Debras} {et~al.}(2020){Debras}, {Mayne}, {Baraffe}, {Jaupart},
  {Mourier}, {Laibe}, {Goffrey}, \& {Thuburn}}]{debras2020}
{Debras}, F., {Mayne}, N., {Baraffe}, I., {Jaupart}, E., {Mourier}, P.,
  {Laibe}, G., {Goffrey}, T., \& {Thuburn}, J. 2020, \aap, 633, A2

\bibitem[{{Del Genio} {et~al.}(2009){Del Genio}, {Achterberg}, {Baines},
  {Flasar}, {Read}, {S{\'a}nchez-Lavega}, \& {Showman}}]{delGenio2009}
{Del Genio}, A.~D., {Achterberg}, R.~K., {Baines}, K.~H., {Flasar}, F.~M.,
  {Read}, P.~L., {S{\'a}nchez-Lavega}, A., \& {Showman}, A.~P. {Saturn
  Atmospheric Structure and Dynamics}, ed. M.~K. {Dougherty}, L.~W. {Esposito},
  \& S.~M. {Krimigis}, 113

\bibitem[{Deming \& Seager(2009)}]{deming2009}
Deming, D. \& Seager, S. 2009, Nature, 462, 301

\bibitem[{Deming \& Seager(2017)}]{deming2017}
Deming, L.~D. \& Seager, S. 2017, Journal of Geophysical Research: Planets,
  122, 53

\bibitem[{{Demory} {et~al.}(2013){Demory}, {de Wit}, {Lewis}, {Fortney},
  {Zsom}, {Seager}, {Knutson}, {Heng}, {Madhusudhan}, {Gillon}, {Barclay},
  {Desert}, {Parmentier}, \& {Cowan}}]{demory2013}
{Demory}, B.-O., {de Wit}, J., {Lewis}, N., {Fortney}, J., {Zsom}, A.,
  {Seager}, S., {Knutson}, H., {Heng}, K., {Madhusudhan}, N., {Gillon}, M.,
  {Barclay}, T., {Desert}, J.-M., {Parmentier}, V., \& {Cowan}, N.~B. 2013,
  \apjl, 776, L25

\bibitem[{{Dobbs-Dixon} \& {Agol}(2013)}]{dobbsdixon2013}
{Dobbs-Dixon}, I. \& {Agol}, E. 2013, \mnras, 435, 3159

\bibitem[{{Dobbs-Dixon} {et~al.}(2012){Dobbs-Dixon}, {Agol}, \&
  {Burrows}}]{dobbsdixon2012}
{Dobbs-Dixon}, I., {Agol}, E., \& {Burrows}, A. 2012, \apj, 751, 87

\bibitem[{{Dobbs-Dixon} {et~al.}(2010){Dobbs-Dixon}, {Cumming}, \&
  {Lin}}]{DobbsDixon2010}
{Dobbs-Dixon}, I., {Cumming}, A., \& {Lin}, D.~N.~C. 2010, \apj, 710, 1395

\bibitem[{{Dobbs-Dixon} \& {Lin}(2008)}]{DobbsDixon2008}
{Dobbs-Dixon}, I. \& {Lin}, D.~N.~C. 2008, \apj, 673, 513

\bibitem[{{Drummond} {et~al.}(2020){Drummond}, {H{\'e}brard}, {Mayne}, {Venot},
  {Ridgway}, {Changeat}, {Tsai}, {Manners}, {Tremblin}, {Abraham}, {Sing}, \&
  {Kohary}}]{drummond2020}
{Drummond}, B., {H{\'e}brard}, E., {Mayne}, N.~J., {Venot}, O., {Ridgway},
  R.~J., {Changeat}, Q., {Tsai}, S.-M., {Manners}, J., {Tremblin}, P.,
  {Abraham}, N.~L., {Sing}, D., \& {Kohary}, K. 2020, \aap, 636, A68

\bibitem[{{Drummond} {et~al.}(2018{\natexlab{a}}){Drummond}, {Mayne},
  {Manners}, {Baraffe}, {Goyal}, {Tremblin}, {Sing}, \&
  {Kohary}}]{drummond2018}
{Drummond}, B., {Mayne}, N.~J., {Manners}, J., {Baraffe}, I., {Goyal}, J.,
  {Tremblin}, P., {Sing}, D.~K., \& {Kohary}, K. 2018{\natexlab{a}}, \apj, 869,
  28

\bibitem[{{Drummond} {et~al.}(2018{\natexlab{b}}){Drummond}, {Mayne},
  {Manners}, {Carter}, {Boutle}, {Baraffe}, {H{\'e}brard}, {Tremblin}, {Sing},
  {Amundsen}, \& {Acreman}}]{drummond2018b}
{Drummond}, B., {Mayne}, N.~J., {Manners}, J., {Carter}, A.~L., {Boutle},
  I.~A., {Baraffe}, I., {H{\'e}brard}, {\'E}., {Tremblin}, P., {Sing}, D.~K.,
  {Amundsen}, D.~S., \& {Acreman}, D. 2018{\natexlab{b}}, \apjl, 855, L31

\bibitem[{{Eisner} {et~al.}(2020){Eisner}, {Barrag{\'a}n}, {Aigrain},
  {Lintott}, {Miller}, {Zicher}, {Boyajian}, {Brice{\~n}o}, {Bryant},
  {Christiansen}, {Feinstein}, {Flor-Torres}, {Fridlund}, {Gand olfi},
  {Gilbert}, {Guerrero}, {Jenkins}, {Jones}, {Kristiansen}, {Vanderburg},
  {Law}, {L{\'o}pez-S{\'a}nchez}, {Mann}, {Safron}, {Schwamb}, {Stassun},
  {Osborn}, {Wang}, {Zic}, {Ziegler}, {Barnet}, {Bean}, {Bundy}, {Chetnik},
  {Dawson}, {Garstone}, {Stenner}, {Huten}, {Larish}, {Melanson}, {Mitchell},
  {Moore}, {Peltsch}, {Rogers}, {Schuster}, {Smith}, {Simister}, {Tanner},
  {Terentev}, \& {Tsymbal}}]{Eisner2020}
{Eisner}, N.~L., {Barrag{\'a}n}, O., {Aigrain}, S., {Lintott}, C., {Miller},
  G., {Zicher}, N., {Boyajian}, T.~S., {Brice{\~n}o}, C., {Bryant}, E.~M.,
  {Christiansen}, J.~L., {Feinstein}, A.~D., {Flor-Torres}, L.~M., {Fridlund},
  M., {Gand olfi}, D., {Gilbert}, J., {Guerrero}, N., {Jenkins}, J.~M.,
  {Jones}, K., {Kristiansen}, M.~H., {Vanderburg}, A., {Law}, N.,
  {L{\'o}pez-S{\'a}nchez}, A.~R., {Mann}, A.~W., {Safron}, E.~J., {Schwamb},
  M.~E., {Stassun}, K.~G., {Osborn}, H.~P., {Wang}, J., {Zic}, A., {Ziegler},
  C., {Barnet}, F., {Bean}, S.~J., {Bundy}, D.~M., {Chetnik}, Z., {Dawson},
  J.~L., {Garstone}, J., {Stenner}, A.~G., {Huten}, M., {Larish}, S.,
  {Melanson}, L.~D., {Mitchell}, T., {Moore}, C., {Peltsch}, K., {Rogers},
  D.~J., {Schuster}, C., {Smith}, D.~S., {Simister}, D.~J., {Tanner}, C.,
  {Terentev}, I., \& {Tsymbal}, A. 2020, \mnras, 494, 750

\bibitem[{{Esteves} {et~al.}(2013){Esteves}, {De Mooij}, \&
  {Jayawardhana}}]{esteves2013}
{Esteves}, L.~J., {De Mooij}, E. J.~W., \& {Jayawardhana}, R. 2013, \apj, 772,
  51

\bibitem[{{Esteves} {et~al.}(2015){Esteves}, {De Mooij}, \&
  {Jayawardhana}}]{esteves2015}
---. 2015, \apj, 804, 150

\bibitem[{{Evans} {et~al.}(2013){Evans}, {Pont}, {Sing}, {Aigrain}, {Barstow},
  {D{\'e}sert}, {Gibson}, {Heng}, {Knutson}, \& {Lecavelier des
  Etangs}}]{evans2013}
{Evans}, T.~M., {Pont}, F., {Sing}, D.~K., {Aigrain}, S., {Barstow}, J.~K.,
  {D{\'e}sert}, J.-M., {Gibson}, N., {Heng}, K., {Knutson}, H.~A., \&
  {Lecavelier des Etangs}, A. 2013, \apjl, 772, L16

\bibitem[{{Fegley} \& {Lodders}(1996)}]{fegley1996}
{Fegley}, Jr., B. \& {Lodders}, K. 1996, \apjl, 472, L37

\bibitem[{{Fleury} {et~al.}(2019){Fleury}, {Gudipati}, {Henderson}, \&
  {Swain}}]{fleury2019}
{Fleury}, B., {Gudipati}, M.~S., {Henderson}, B.~L., \& {Swain}, M. 2019, \apj,
  871, 158

\bibitem[{{Flowers} {et~al.}(2019){Flowers}, {Brogi}, {Rauscher}, {Kempton}, \&
  {Chiavassa}}]{flowers2019}
{Flowers}, E., {Brogi}, M., {Rauscher}, E., {Kempton}, E. M.~R., \&
  {Chiavassa}, A. 2019, \aj, 157, 209

\bibitem[{{Fortney} {et~al.}(2010){Fortney}, {Baraffe}, \&
  {Militzer}}]{fortney2010}
{Fortney}, J.~J., {Baraffe}, I., \& {Militzer}, B. {Giant Planet Interior
  Structure and Thermal Evolution}, ed. S.~{Seager}, 397--418

\bibitem[{{Fortney} {et~al.}(2008){Fortney}, {Lodders}, {Marley}, \&
  {Freedman}}]{fortney2008}
{Fortney}, J.~J., {Lodders}, K., {Marley}, M.~S., \& {Freedman}, R.~S. 2008,
  \apj, 678, 1419

\bibitem[{{Fortney} {et~al.}(2007){Fortney}, {Marley}, \&
  {Barnes}}]{fortney2007}
{Fortney}, J.~J., {Marley}, M.~S., \& {Barnes}, J.~W. 2007, \apj, 659, 1661

\bibitem[{Fortney \& Nettelmann(2010)}]{fortney2010interior}
Fortney, J.~J. \& Nettelmann, N. 2010, Space Science Reviews, 152, 423

\bibitem[{{Freytag} {et~al.}(2010){Freytag}, {Allard}, {Ludwig}, {Homeier}, \&
  {Steffen}}]{freytag2010}
{Freytag}, B., {Allard}, F., {Ludwig}, H.-G., {Homeier}, D., \& {Steffen}, M.
  2010, \aap, 513, A19

\bibitem[{{Fromang} {et~al.}(2016){Fromang}, {Leconte}, \&
  {Heng}}]{fromang2016}
{Fromang}, S., {Leconte}, J., \& {Heng}, K. 2016, \aap, 591, A144

\bibitem[{Gelino {et~al.}(2002)Gelino, Marley, Holtzman, Ackerman, \&
  Lodders}]{gelino2002}
Gelino, C.~R., Marley, M.~S., Holtzman, J.~A., Ackerman, A.~S., \& Lodders, K.
  2002, The Astrophysical Journal, 577, 433

\bibitem[{Gierasch {et~al.}(1973)Gierasch, Ingersoll, \&
  Williams}]{gierasch1973}
Gierasch, P.~J., Ingersoll, A.~P., \& Williams, R.~T. 1973, Icarus, 19, 473

\bibitem[{Gill(1980)}]{gill1980}
Gill, A.~E. 1980, Quarterly Journal of the Royal Meteorological Society, 106,
  447

\bibitem[{{Ginzburg} \& {Sari}(2016)}]{ginzburg2016}
{Ginzburg}, S. \& {Sari}, R. 2016, \apj, 819, 116

\bibitem[{{Guillot}(1999)}]{guillot1999}
{Guillot}, T. 1999, Science, 286, 72

\bibitem[{{Guillot}(2010)}]{guillot2010}
---. 2010, \aap, 520, A27

\bibitem[{{Guillot} {et~al.}(1996){Guillot}, {Burrows}, {Hubbard}, {Lunine}, \&
  {Saumon}}]{guillot1996}
{Guillot}, T., {Burrows}, A., {Hubbard}, W.~B., {Lunine}, J.~I., \& {Saumon},
  D. 1996, \apjl, 459, L35

\bibitem[{Guillot {et~al.}(2018)Guillot, Miguel, Militzer, Hubbard, Kaspi,
  Galanti, Cao, Helled, Wahl, Iess, {et~al.}}]{guillot2018}
Guillot, T., Miguel, Y., Militzer, B., Hubbard, W., Kaspi, Y., Galanti, E.,
  Cao, H., Helled, R., Wahl, S., Iess, L., {et~al.} 2018, Nature, 555, 227

\bibitem[{{Guillot} \& {Showman}(2002)}]{Guillot2002}
{Guillot}, T. \& {Showman}, A.~P. 2002, \aap, 385, 156

\bibitem[{{Hammond} \& {Pierrehumbert}(2018)}]{hammond2018}
{Hammond}, M. \& {Pierrehumbert}, R.~T. 2018, \apj, 869, 65

\bibitem[{{Harada} {et~al.}(2019){Harada}, {Kempton}, {Rauscher}, {Roman}, \&
  {Brinjikji}}]{harada2019}
{Harada}, C.~K., {Kempton}, E. M.~R., {Rauscher}, E., {Roman}, M., \&
  {Brinjikji}, M. 2019, arXiv e-prints, arXiv:1912.02268

\bibitem[{{Hellier}(2001)}]{hellier2001}
{Hellier}, C. 2001, {Cataclysmic Variable Stars}

\bibitem[{Helling \& Casewell(2014)}]{helling2014}
Helling, C. \& Casewell, S. 2014, The Astronomy and Astrophysics Review, 22, 1

\bibitem[{{Helling} {et~al.}(2019{\natexlab{a}}){Helling}, {Gourbin}, {Woitke},
  \& {Parmentier}}]{Helling2018}
{Helling}, C., {Gourbin}, P., {Woitke}, P., \& {Parmentier}, V.
  2019{\natexlab{a}}, \aap, 626, A133

\bibitem[{{Helling} {et~al.}(2019{\natexlab{b}}){Helling}, {Iro}, {Corrales},
  {Samra}, {Ohno}, {Alam}, {Steinrueck}, {Lew}, {Molaverdikhani}, {MacDonald},
  {Herbort}, {Woitke}, \& {Parmentier}}]{helling2019}
{Helling}, C., {Iro}, N., {Corrales}, L., {Samra}, D., {Ohno}, K., {Alam},
  M.~K., {Steinrueck}, M., {Lew}, B., {Molaverdikhani}, K., {MacDonald}, R.~J.,
  {Herbort}, O., {Woitke}, P., \& {Parmentier}, V. 2019{\natexlab{b}}, \aap,
  631, A79

\bibitem[{{Heng}(2016)}]{heng2016}
{Heng}, K. 2016, \apjl, 826, L16

\bibitem[{{Heng} \& {Demory}(2013)}]{heng2013}
{Heng}, K. \& {Demory}, B.-O. 2013, \apj, 777, 100

\bibitem[{{Heng} {et~al.}(2011{\natexlab{a}}){Heng}, {Frierson}, \&
  {Phillipps}}]{heng2011b}
{Heng}, K., {Frierson}, D.~M.~W., \& {Phillipps}, P.~J. 2011{\natexlab{a}},
  \mnras, 418, 2669

\bibitem[{{Heng} {et~al.}(2011{\natexlab{b}}){Heng}, {Menou}, \&
  {Phillipps}}]{heng2011}
{Heng}, K., {Menou}, K., \& {Phillipps}, P.~J. 2011{\natexlab{b}}, \mnras, 413,
  2380

\bibitem[{{Heng} \& {Showman}(2015)}]{heng2015}
{Heng}, K. \& {Showman}, A.~P. 2015, Annual Review of Earth and Planetary
  Sciences, 43, 509

\bibitem[{{Heng} \& {Workman}(2014)}]{heng2014}
{Heng}, K. \& {Workman}, J. 2014, \apjs, 213, 27

\bibitem[{Hide(1969)}]{hide1969}
Hide, R. 1969, Journal of the Atmospheric Sciences, 26, 841

\bibitem[{Holton(1986)}]{holton1986}
Holton, J.~R. 1986, Journal of Geophysical Research: Atmospheres, 91, 2681

\bibitem[{Holton \& Hakim(2012)}]{holton2012}
Holton, J.~R. \& Hakim, G.~J. 2012, An introduction to dynamic meteorology,
  Vol.~88 (Academic press)

\bibitem[{{Howard} {et~al.}(2012){Howard}, {Bakos}, {Hartman}, {Torres},
  {Shporer}, {Mazeh}, {Kov{\'a}cs}, {Latham}, {Noyes}, {Fischer}, {Johnson},
  {Marcy}, {Esquerdo}, {B{\'e}ky}, {Butler}, {Sasselov}, {Stefanik},
  {Perumpilly}, {L{\'a}z{\'a}r}, {Papp}, \& {S{\'a}ri}}]{howard2012}
{Howard}, A.~W., {Bakos}, G.~{\'A}., {Hartman}, J., {Torres}, G., {Shporer},
  A., {Mazeh}, T., {Kov{\'a}cs}, G., {Latham}, D.~W., {Noyes}, R.~W.,
  {Fischer}, D.~A., {Johnson}, J.~A., {Marcy}, G.~W., {Esquerdo}, G.~A.,
  {B{\'e}ky}, B., {Butler}, R.~P., {Sasselov}, D.~D., {Stefanik}, R.~P.,
  {Perumpilly}, G., {L{\'a}z{\'a}r}, J., {Papp}, I., \& {S{\'a}ri}, P. 2012,
  \apj, 749, 134

\bibitem[{Hubbard {et~al.}(1991)Hubbard, Nellis, Mitchell, Holmes, Limaye, \&
  McCandless}]{hubbard1991}
Hubbard, W.~B., Nellis, W., Mitchell, A., Holmes, N., Limaye, S., \&
  McCandless, P. 1991, Science, 253, 648

\bibitem[{Iess {et~al.}(2019)Iess, Militzer, Kaspi, Nicholson, Durante,
  Racioppa, Anabtawi, Galanti, Hubbard, Mariani, {et~al.}}]{iess2019}
Iess, L., Militzer, B., Kaspi, Y., Nicholson, P., Durante, D., Racioppa, P.,
  Anabtawi, A., Galanti, E., Hubbard, W., Mariani, M., {et~al.} 2019, Science,
  364

\bibitem[{{Iyer} {et~al.}(2016){Iyer}, {Swain}, {Zellem}, {Line}, {Roudier},
  {Rocha}, \& {Livingston}}]{iyer2016}
{Iyer}, A.~R., {Swain}, M.~R., {Zellem}, R.~T., {Line}, M.~R., {Roudier}, G.,
  {Rocha}, G., \& {Livingston}, J.~H. 2016, \apj, 823, 109

\bibitem[{{Jackson} {et~al.}(2019){Jackson}, {Adams}, {Sandidge}, {Kreyche}, \&
  {Briggs}}]{jackson2019}
{Jackson}, B., {Adams}, E., {Sandidge}, W., {Kreyche}, S., \& {Briggs}, J.
  2019, \aj, 157, 239

\bibitem[{Kao {et~al.}(2016)Kao, Hallinan, Pineda, Escala, Burgasser, Bourke,
  \& Stevenson}]{kao2016}
Kao, M.~M., Hallinan, G., Pineda, J.~S., Escala, I., Burgasser, A., Bourke, S.,
  \& Stevenson, D. 2016, The Astrophysical Journal, 818, 24

\bibitem[{Kao {et~al.}(2018)Kao, Hallinan, Pineda, Stevenson, \&
  Burgasser}]{kao2018}
Kao, M.~M., Hallinan, G., Pineda, J.~S., Stevenson, D., \& Burgasser, A. 2018,
  The Astrophysical Journal Supplement Series, 237, 25

\bibitem[{Karalidi {et~al.}(2016)Karalidi, Apai, Marley, \&
  Buenzli}]{karalidi2016}
Karalidi, T., Apai, D., Marley, M.~S., \& Buenzli, E. 2016, The Astrophysical
  Journal, 825, 90

\bibitem[{Kaspi {et~al.}(2009)Kaspi, Flierl, \& Showman}]{kaspi2009}
Kaspi, Y., Flierl, G.~R., \& Showman, A.~P. 2009, Icarus, 202, 525

\bibitem[{Kaspi {et~al.}(2018)Kaspi, Galanti, Hubbard, Stevenson, Bolton, Iess,
  Guillot, Bloxham, Connerney, Cao, {et~al.}}]{kaspi2018}
Kaspi, Y., Galanti, E., Hubbard, W.~B., Stevenson, D., Bolton, S., Iess, L.,
  Guillot, T., Bloxham, J., Connerney, J., Cao, H., {et~al.} 2018, Nature, 555,
  223

\bibitem[{{Kataria} {et~al.}(2014){Kataria}, {Showman}, {Fortney}, {Marley}, \&
  {Freedman}}]{kataria2014}
{Kataria}, T., {Showman}, A.~P., {Fortney}, J.~J., {Marley}, M.~S., \&
  {Freedman}, R.~S. 2014, \apj, 785, 92

\bibitem[{{Kataria} {et~al.}(2015){Kataria}, {Showman}, {Fortney}, {Stevenson},
  {Line}, {Kreidberg}, {Bean}, \& {D{\'e}sert}}]{kataria2015}
{Kataria}, T., {Showman}, A.~P., {Fortney}, J.~J., {Stevenson}, K.~B., {Line},
  M.~R., {Kreidberg}, L., {Bean}, J.~L., \& {D{\'e}sert}, J.-M. 2015, \apj,
  801, 86

\bibitem[{{Kataria} {et~al.}(2013){Kataria}, {Showman}, {Lewis}, {Fortney},
  {Marley}, \& {Freedman}}]{kataria2013}
{Kataria}, T., {Showman}, A.~P., {Lewis}, N.~K., {Fortney}, J.~J., {Marley},
  M.~S., \& {Freedman}, R.~S. 2013, \apj, 767, 76

\bibitem[{{Kataria} {et~al.}(2016){Kataria}, {Sing}, {Lewis}, {Visscher},
  {Showman}, {Fortney}, \& {Marley}}]{kataria2016}
{Kataria}, T., {Sing}, D.~K., {Lewis}, N.~K., {Visscher}, C., {Showman}, A.~P.,
  {Fortney}, J.~J., \& {Marley}, M.~S. 2016, \apj, 821, 9

\bibitem[{{Kempton} {et~al.}(2017){Kempton}, {Bean}, \&
  {Parmentier}}]{kempton2017}
{Kempton}, E. M.~R., {Bean}, J.~L., \& {Parmentier}, V. 2017, \apjl, 845, L20

\bibitem[{Kiladis {et~al.}(2009)Kiladis, Wheeler, Haertel, Straub, \&
  Roundy}]{kiladis2009}
Kiladis, G.~N., Wheeler, M.~C., Haertel, P.~T., Straub, K.~H., \& Roundy, P.~E.
  2009, Reviews of Geophysics, 47

\bibitem[{{Kipping} {et~al.}(2019){Kipping}, {Nesvorn{\'y}}, {Hartman},
  {Torres}, {Bakos}, {Jansen}, \& {Teachey}}]{kipping2019}
{Kipping}, D., {Nesvorn{\'y}}, D., {Hartman}, J., {Torres}, G., {Bakos}, G.,
  {Jansen}, T., \& {Teachey}, A. 2019, \mnras, 486, 4980

\bibitem[{{Kirkpatrick}(2005)}]{kirkpatrick2005}
{Kirkpatrick}, J.~D. 2005, \araa, 43, 195

\bibitem[{{Kirkpatrick} {et~al.}(2012){Kirkpatrick}, {Gelino}, {Cushing},
  {Mace}, {Griffith}, {Skrutskie}, {Marsh}, {Wright}, {Eisenhardt}, {McLean},
  {Mainzer}, {Burgasser}, {Tinney}, {Parker}, \& {Salter}}]{kirkpatrick2012}
{Kirkpatrick}, J.~D., {Gelino}, C.~R., {Cushing}, M.~C., {Mace}, G.~N.,
  {Griffith}, R.~L., {Skrutskie}, M.~F., {Marsh}, K.~A., {Wright}, E.~L.,
  {Eisenhardt}, P.~R., {McLean}, I.~S., {Mainzer}, A.~K., {Burgasser}, A.~J.,
  {Tinney}, C.~G., {Parker}, S., \& {Salter}, G. 2012, The Astrophysical
  Journal, 753, 156

\bibitem[{Kitzmann {et~al.}(2018)Kitzmann, Heng, Rimmer, Hoeijmakers, Tsai,
  Malik, Lendl, Deitrick, \& Demory}]{kitzmann2018}
Kitzmann, D., Heng, K., Rimmer, P.~B., Hoeijmakers, H.~J., Tsai, S.-M., Malik,
  M., Lendl, M., Deitrick, R., \& Demory, B.-O. 2018, The Astrophysical
  Journal, 863, 183

\bibitem[{Knapp {et~al.}(2004)Knapp, Leggett, Fan, Marley, Geballe, Golimowski,
  Finkbeiner, Gunn, Hennawi, Ivezi{\'c}, {et~al.}}]{knapp2004}
Knapp, G., Leggett, S.~K., Fan, X., Marley, M., Geballe, T., Golimowski, D.,
  Finkbeiner, D., Gunn, J., Hennawi, J., Ivezi{\'c}, Z., {et~al.} 2004, The
  Astronomical Journal, 127, 3553

\bibitem[{{Knutson} {et~al.}(2007){Knutson}, {Charbonneau}, {Allen}, {Fortney},
  {Agol}, {Cowan}, {Showman}, {Cooper}, \& {Megeath}}]{knutson2007}
{Knutson}, H.~A., {Charbonneau}, D., {Allen}, L.~E., {Fortney}, J.~J., {Agol},
  E., {Cowan}, N.~B., {Showman}, A.~P., {Cooper}, C.~S., \& {Megeath}, S.~T.
  2007, \nat, 447, 183

\bibitem[{{Knutson} {et~al.}(2012){Knutson}, {Lewis}, {Fortney}, {Burrows},
  {Showman}, {Cowan}, {Agol}, {Aigrain}, {Charbonneau}, {Deming}, {D{\'e}sert},
  {Henry}, {Langton}, \& {Laughlin}}]{knutson2012}
{Knutson}, H.~A., {Lewis}, N., {Fortney}, J.~J., {Burrows}, A., {Showman},
  A.~P., {Cowan}, N.~B., {Agol}, E., {Aigrain}, S., {Charbonneau}, D.,
  {Deming}, D., {D{\'e}sert}, J.-M., {Henry}, G.~W., {Langton}, J., \&
  {Laughlin}, G. 2012, \apj, 754, 22

\bibitem[{{Koll} \& {Komacek}(2018)}]{koll2018}
{Koll}, D.~D.~B. \& {Komacek}, T.~D. 2018, \apj, 853, 133

\bibitem[{Komacek \& Showman(2016)}]{komacek2016}
Komacek, T.~D. \& Showman, A.~P. 2016, The Astrophysical Journal, 821, 16

\bibitem[{{Komacek} \& {Showman}(2020)}]{komacek2020}
{Komacek}, T.~D. \& {Showman}, A.~P. 2020, \apj, 888, 2

\bibitem[{{Komacek} {et~al.}(2019){Komacek}, {Showman}, \&
  {Parmentier}}]{komacek2019}
{Komacek}, T.~D., {Showman}, A.~P., \& {Parmentier}, V. 2019, \apj, 881, 152

\bibitem[{{Komacek} {et~al.}(2017){Komacek}, {Showman}, \& {Tan}}]{komacek2017}
{Komacek}, T.~D., {Showman}, A.~P., \& {Tan}, X. 2017, \apj, 835, 198

\bibitem[{{Komacek} \& {Tan}(2018)}]{komacek2018rnaas}
{Komacek}, T.~D. \& {Tan}, X. 2018, Research Notes of the American Astronomical
  Society, 2, 36

\bibitem[{Komacek \& Youdin(2017)}]{komacek2017structure}
Komacek, T.~D. \& Youdin, A.~N. 2017, The Astrophysical Journal, 844, 94

\bibitem[{{Kreidberg} {et~al.}(2018){Kreidberg}, {Line}, {Parmentier},
  {Stevenson}, {Louden}, {Bonnefoy}, {Faherty}, {Henry}, {Williamson},
  {Stassun}, {Beatty}, {Bean}, {Fortney}, {Showman}, {D{\'e}sert}, \&
  {Arcangeli}}]{kreidberg2018}
{Kreidberg}, L., {Line}, M.~R., {Parmentier}, V., {Stevenson}, K.~B., {Louden},
  T., {Bonnefoy}, M., {Faherty}, J.~K., {Henry}, G.~W., {Williamson}, M.~H.,
  {Stassun}, K., {Beatty}, T.~G., {Bean}, J.~L., {Fortney}, J.~J., {Showman},
  A.~P., {D{\'e}sert}, J.-M., \& {Arcangeli}, J. 2018, The Astronomical
  Journal, 156, 17

\bibitem[{{Langton} \& {Laughlin}(2007)}]{langton2007}
{Langton}, J. \& {Laughlin}, G. 2007, \apjl, 657, L113

\bibitem[{{Langton} \& {Laughlin}(2008)}]{langton2008}
---. 2008, \apj, 674, 1106

\bibitem[{{Laughlin} {et~al.}(2009){Laughlin}, {Deming}, {Langton}, {Kasen},
  {Vogt}, {Butler}, {Rivera}, \& {Meschiari}}]{laughlin2009}
{Laughlin}, G., {Deming}, D., {Langton}, J., {Kasen}, D., {Vogt}, S., {Butler},
  P., {Rivera}, E., \& {Meschiari}, S. 2009, \nat, 457, 562

\bibitem[{{Lavvas} \& {Koskinen}(2017)}]{lavvas2017}
{Lavvas}, P. \& {Koskinen}, T. 2017, \apj, 847, 32

\bibitem[{{Leconte} {et~al.}(2017){Leconte}, {Selsis}, {Hersant}, \&
  {Guillot}}]{Leconte2017}
{Leconte}, J., {Selsis}, F., {Hersant}, F., \& {Guillot}, T. 2017, \aap, 598,
  A98

\bibitem[{{Lee} {et~al.}(2016){Lee}, {Dobbs-Dixon}, {Helling}, {Bognar}, \&
  {Woitke}}]{lee2016}
{Lee}, G., {Dobbs-Dixon}, I., {Helling}, C., {Bognar}, K., \& {Woitke}, P.
  2016, \aap, 594, A48

\bibitem[{{Lee} {et~al.}(2020){Lee}, {Casewell}, {Chubb}, {Hammond}, {Tan},
  {Tsai}, \& {Pierrehumbert}}]{lee2020}
{Lee}, G. K.~H., {Casewell}, S.~L., {Chubb}, K.~L., {Hammond}, M., {Tan}, X.,
  {Tsai}, S.-M., \& {Pierrehumbert}, R.~T. 2020, \mnras, 496, 4674

\bibitem[{{Leggett} {et~al.}(2007){Leggett}, {Saumon}, {Marley}, {Geballe},
  {Golimowski}, {Stephens}, \& {Fan}}]{leggett2007}
{Leggett}, S.~K., {Saumon}, D., {Marley}, M.~S., {Geballe}, T.~R.,
  {Golimowski}, D.~A., {Stephens}, D., \& {Fan}, X. 2007, \apj, 655, 1079

\bibitem[{{Leggett} {et~al.}(2017){Leggett}, {Tremblin}, {Esplin}, {Luhman}, \&
  {Morley}}]{leggett2017}
{Leggett}, S.~K., {Tremblin}, P., {Esplin}, T.~L., {Luhman}, K.~L., \&
  {Morley}, C.~V. 2017, \apj, 842, 118

\bibitem[{Lew {et~al.}(2016)Lew, Apai, Zhou, Schneider, Burgasser, Karalidi,
  Yang, Marley, Cowan, Bedin, {et~al.}}]{lew2016}
Lew, B.~W., Apai, D., Zhou, Y., Schneider, G., Burgasser, A.~J., Karalidi, T.,
  Yang, H., Marley, M.~S., Cowan, N.~B., Bedin, L.~R., {et~al.} 2016, The
  Astrophysical Journal Letters, 829, L32

\bibitem[{{Lewis} {et~al.}(2013){Lewis}, {Knutson}, {Showman}, {Cowan},
  {Laughlin}, {Burrows}, {Deming}, {Crepp}, {Mighell}, {Agol}, {Bakos},
  {Charbonneau}, {D{\'e}sert}, {Fischer}, {Fortney}, {Hartman}, {Hinkley},
  {Howard}, {Johnson}, {Kao}, {Langton}, \& {Marcy}}]{lewis2013}
{Lewis}, N.~K., {Knutson}, H.~A., {Showman}, A.~P., {Cowan}, N.~B., {Laughlin},
  G., {Burrows}, A., {Deming}, D., {Crepp}, J.~R., {Mighell}, K.~J., {Agol},
  E., {Bakos}, G.~{\'A}., {Charbonneau}, D., {D{\'e}sert}, J.-M., {Fischer},
  D.~A., {Fortney}, J.~J., {Hartman}, J.~D., {Hinkley}, S., {Howard}, A.~W.,
  {Johnson}, J.~A., {Kao}, M., {Langton}, J., \& {Marcy}, G.~W. 2013, \apj,
  766, 95

\bibitem[{{Lewis} {et~al.}(2017){Lewis}, {Parmentier}, {Kataria}, {de Wit},
  {Showman}, {Fortney}, \& {Marley}}]{lewis2017}
{Lewis}, N.~K., {Parmentier}, V., {Kataria}, T., {de Wit}, J., {Showman},
  A.~P., {Fortney}, J.~J., \& {Marley}, M.~S. 2017, arXiv e-prints,
  arXiv:1706.00466

\bibitem[{{Lewis} {et~al.}(2014){Lewis}, {Showman}, {Fortney}, {Knutson}, \&
  {Marley}}]{lewis2014}
{Lewis}, N.~K., {Showman}, A.~P., {Fortney}, J.~J., {Knutson}, H.~A., \&
  {Marley}, M.~S. 2014, \apj, 795, 150

\bibitem[{{Lewis} {et~al.}(2010){Lewis}, {Showman}, {Fortney}, {Marley},
  {Freedman}, \& {Lodders}}]{lewis2010}
{Lewis}, N.~K., {Showman}, A.~P., {Fortney}, J.~J., {Marley}, M.~S.,
  {Freedman}, R.~S., \& {Lodders}, K. 2010, \apj, 720, 344

\bibitem[{{Li} {et~al.}(2004){Li}, {Ingersoll}, {Vasavada}, {Porco}, {Del
  Genio}, \& {Ewald}}]{li2004}
{Li}, L., {Ingersoll}, A.~P., {Vasavada}, A.~R., {Porco}, C.~C., {Del Genio},
  A.~D., \& {Ewald}, S.~P. 2004, \icarus, 172, 9

\bibitem[{Li {et~al.}(2018)Li, Jiang, West, Gierasch, Perez-Hoyos,
  Sanchez-Lavega, Fletcher, Fortney, Knowles, Porco, {et~al.}}]{li2018}
Li, L., Jiang, X., West, R., Gierasch, P., Perez-Hoyos, S., Sanchez-Lavega, A.,
  Fletcher, L., Fortney, J., Knowles, B., Porco, C., {et~al.} 2018, Nature
  communications, 9, 1

\bibitem[{Line {et~al.}(2016)Line, Marley, Liu, Morley, Burningham, Hinkel,
  Teske, \& Fortney}]{line2016}
Line, M.~R., Marley, M.~S., Liu, M.~L., Morley, C.~V., Burningham, B., Hinkel,
  N.~R., Teske, J., \& Fortney, J.~J. 2016, arXiv preprint arXiv:1612.02809

\bibitem[{Lines {et~al.}(2018)Lines, Mayne, Boutle, Manners, Lee, Helling,
  Drummond, Amundsen, Goyal, Acreman, {et~al.}}]{lines2018}
Lines, S., Mayne, N., Boutle, I.~A., Manners, J., Lee, G. K.~H., Helling, C.,
  Drummond, B., Amundsen, D.~S., Goyal, J., Acreman, D.~M., {et~al.} 2018,
  Astronomy \& Astrophysics

\bibitem[{{Lines} {et~al.}(2019){Lines}, {Mayne}, {Manners}, {Boutle},
  {Drummond}, {Mikal-Evans}, {Kohary}, \& {Sing}}]{lines2019}
{Lines}, S., {Mayne}, N.~J., {Manners}, J., {Boutle}, I.~A., {Drummond}, B.,
  {Mikal-Evans}, T., {Kohary}, K., \& {Sing}, D.~K. 2019, \mnras, 488, 1332

\bibitem[{Littlefair {et~al.}(2014)Littlefair, Casewell, Parsons, Dhillon,
  Marsh, G{\"a}nsicke, Bloemen, Catalan, Irawati, Hardy,
  {et~al.}}]{littlefair2014}
Littlefair, S., Casewell, S., Parsons, S., Dhillon, V., Marsh, T.,
  G{\"a}nsicke, B., Bloemen, S., Catalan, S., Irawati, P., Hardy, L., {et~al.}
  2014, Monthly Notices of the Royal Astronomical Society, 445, 2106

\bibitem[{{Liu} \& {Showman}(2013)}]{liu2013}
{Liu}, B. \& {Showman}, A.~P. 2013, \apj, 770, 42

\bibitem[{Liu {et~al.}(2008)Liu, Goldreich, \& Stevenson}]{liu2008}
Liu, J., Goldreich, P.~M., \& Stevenson, D.~J. 2008, Icarus, 196, 653

\bibitem[{Longstaff {et~al.}(2017)Longstaff, Casewell, Wynn, Maxted, \&
  Helling}]{longstaff2017}
Longstaff, E., Casewell, S., Wynn, G., Maxted, P., \& Helling, C. 2017, Monthly
  Notices of the Royal Astronomical Society, 471, 1728

\bibitem[{{Louden} \& {Wheatley}(2019)}]{louden2019ESS}
{Louden}, T. \& {Wheatley}, P. 2019, in AAS/Division for Extreme Solar Systems
  Abstracts, Vol.~51, AAS/Division for Extreme Solar Systems Abstracts, 326.43

\bibitem[{{Louden} \& {Wheatley}(2015)}]{louden2015}
{Louden}, T. \& {Wheatley}, P.~J. 2015, \apjl, 814, L24

\bibitem[{{Mac Low} \& {Ingersoll}(1986)}]{maclow1986}
{Mac Low}, M.~M. \& {Ingersoll}, A.~P. 1986, \icarus, 65, 353

\bibitem[{{Madhusudhan}(2019)}]{madhusudhan2019}
{Madhusudhan}, N. 2019, \araa, 57, 617

\bibitem[{{Madhusudhan} {et~al.}(2014){Madhusudhan}, {Knutson}, {Fortney}, \&
  {Barman}}]{madhusudhan2014}
{Madhusudhan}, N., {Knutson}, H., {Fortney}, J.~J., \& {Barman}, T. 2014, in
  Protostars and Planets VI, ed. H.~{Beuther}, R.~S. {Klessen}, C.~P.
  {Dullemond}, \& T.~{Henning}, 739

\bibitem[{{Mankovich} {et~al.}(2016){Mankovich}, {Fortney}, \&
  {Moore}}]{Mankovich2016}
{Mankovich}, C., {Fortney}, J.~J., \& {Moore}, K.~L. 2016, \apj, 832, 113

\bibitem[{{Mansfield} {et~al.}(2020){Mansfield}, {Bean}, {Stevenson},
  {Komacek}, {Bell}, {Tan}, {Malik}, {Beatty}, {Wong}, {Cowan}, {Dang},
  {D{\'e}sert}, {Fortney}, {Gaudi}, {Keating}, {Kempton}, {Kreidberg}, {Line},
  {Parmentier}, {Stassun}, {Swain}, \& {Zellem}}]{mansfield2020}
{Mansfield}, M., {Bean}, J.~L., {Stevenson}, K.~B., {Komacek}, T.~D., {Bell},
  T.~J., {Tan}, X., {Malik}, M., {Beatty}, T.~G., {Wong}, I., {Cowan}, N.~B.,
  {Dang}, L., {D{\'e}sert}, J.-M., {Fortney}, J.~J., {Gaudi}, B.~S., {Keating},
  D., {Kempton}, E. M.~R., {Kreidberg}, L., {Line}, M.~R., {Parmentier}, V.,
  {Stassun}, K.~G., {Swain}, M.~R., \& {Zellem}, R.~T. 2020, \apjl, 888, L15

\bibitem[{{Marcy} \& {Butler}(2000)}]{marcy2000}
{Marcy}, G.~W. \& {Butler}, R.~P. 2000, \pasp, 112, 137

\bibitem[{Marley \& Robinson(2015)}]{marley2015}
Marley, M.~S. \& Robinson, T.~D. 2015, Annual Review of Astronomy and
  Astrophysics, 53, 279

\bibitem[{{Marley} {et~al.}(2010){Marley}, {Saumon}, \&
  {Goldblatt}}]{marley2010}
{Marley}, M.~S., {Saumon}, D., \& {Goldblatt}, C. 2010, \apjl, 723, L117

\bibitem[{{Marley} {et~al.}(2002){Marley}, {Seager}, {Saumon}, {Lodders},
  {Ackerman}, {Freedman}, \& {Fan}}]{marley2002}
{Marley}, M.~S., {Seager}, S., {Saumon}, D., {Lodders}, K., {Ackerman}, A.~S.,
  {Freedman}, R.~S., \& {Fan}, X. 2002, \apj, 568, 335

\bibitem[{Matsuno(1966)}]{matsuno1966}
Matsuno, T. 1966, Journal of the Meteorological Society of Japan. Ser. II, 44,
  25

\bibitem[{{Mayne} {et~al.}(2014){Mayne}, {Baraffe}, {Acreman}, {Smith},
  {Browning}, {Sk{\aa}lid Amundsen}, {Wood}, {Thuburn}, \&
  {Jackson}}]{mayne2014}
{Mayne}, N.~J., {Baraffe}, I., {Acreman}, D.~M., {Smith}, C., {Browning},
  M.~K., {Sk{\aa}lid Amundsen}, D., {Wood}, N., {Thuburn}, J., \& {Jackson},
  D.~R. 2014, \aap, 561, A1

\bibitem[{{Mayne} {et~al.}(2017){Mayne}, {Debras}, {Baraffe}, {Thuburn},
  {Amundsen}, {Acreman}, {Smith}, {Browning}, {Manners}, \& {Wood}}]{mayne2017}
{Mayne}, N.~J., {Debras}, F., {Baraffe}, I., {Thuburn}, J., {Amundsen}, D.~S.,
  {Acreman}, D.~M., {Smith}, C., {Browning}, M.~K., {Manners}, J., \& {Wood},
  N. 2017, \aap, 604, A79

\bibitem[{{Mayne} {et~al.}(2019){Mayne}, {Drummond}, {Debras}, {Jaupart},
  {Manners}, {Boutle}, {Baraffe}, \& {Kohary}}]{mayne2019}
{Mayne}, N.~J., {Drummond}, B., {Debras}, F., {Jaupart}, E., {Manners}, J.,
  {Boutle}, I.~A., {Baraffe}, I., \& {Kohary}, K. 2019, \apj, 871, 56

\bibitem[{{Mayor} \& {Queloz}(1995)}]{mayor1995}
{Mayor}, M. \& {Queloz}, D. 1995, \nat, 378, 355

\bibitem[{{Mendon{\c c}a} {et~al.}(2016){Mendon{\c c}a}, {Grimm}, {Grosheintz},
  \& {Heng}}]{mendonca2016}
{Mendon{\c c}a}, J.~M., {Grimm}, S.~L., {Grosheintz}, L., \& {Heng}, K. 2016,
  \apj, 829, 115

\bibitem[{{Mendon{\c{c}}a}(2020)}]{mendonca2020}
{Mendon{\c{c}}a}, J.~M. 2020, \mnras, 491, 1456

\bibitem[{{Mendon{\c{c}}a} {et~al.}(2018){Mendon{\c{c}}a}, {Malik}, {Demory},
  \& {Heng}}]{mendonca2018}
{Mendon{\c{c}}a}, J.~M., {Malik}, M., {Demory}, B.-O., \& {Heng}, K. 2018, \aj,
  155, 150

\bibitem[{{Menou}(2012{\natexlab{a}})}]{Menou2012}
{Menou}, K. 2012{\natexlab{a}}, The Astrophysical Journal Letter, 744, L16

\bibitem[{{Menou}(2012{\natexlab{b}})}]{Menou2012b}
---. 2012{\natexlab{b}}, \apj, 745, 138

\bibitem[{{Menou}(2012{\natexlab{c}})}]{Menou2012c}
---. 2012{\natexlab{c}}, \apjl, 754, L9

\bibitem[{{Menou}(2019)}]{menou2019}
---. 2019, \mnras, 485, L98

\bibitem[{{Menou} {et~al.}(2003){Menou}, {Cho}, {Seager}, \&
  {Hansen}}]{menou2003}
{Menou}, K., {Cho}, J. Y.~K., {Seager}, S., \& {Hansen}, B. M.~S. 2003, \apjl,
  587, L113

\bibitem[{{Menou} \& {Rauscher}(2009)}]{menou2009}
{Menou}, K. \& {Rauscher}, E. 2009, \apj, 700, 887

\bibitem[{{Metchev} {et~al.}(2015){Metchev}, {Heinze}, {Apai}, {Flateau},
  {Radigan}, {Burgasser}, {Marley}, {Artigau}, {Plavchan}, \&
  {Goldman}}]{metchev2015}
{Metchev}, S.~A., {Heinze}, A., {Apai}, D., {Flateau}, D., {Radigan}, J.,
  {Burgasser}, A., {Marley}, M.~S., {Artigau}, {\'E}., {Plavchan}, P., \&
  {Goldman}, B. 2015, \apj, 799, 154

\bibitem[{{Miles} {et~al.}(2020){Miles}, {Skemer}, {Morley}, {Marley},
  {Fortney}, {Allers}, {Faherty}, {Geballe}, {Visscher}, {Schneider}, {Lupu},
  {Freedman}, \& {Bjoraker}}]{miles2020}
{Miles}, B.~E., {Skemer}, A. J.~I., {Morley}, C.~V., {Marley}, M.~S.,
  {Fortney}, J.~J., {Allers}, K.~N., {Faherty}, J.~K., {Geballe}, T.~R.,
  {Visscher}, C., {Schneider}, A.~C., {Lupu}, R., {Freedman}, R.~S., \&
  {Bjoraker}, G.~L. 2020, \aj, 160, 63

\bibitem[{{Miller-Ricci Kempton} \& {Rauscher}(2012)}]{kempton2012}
{Miller-Ricci Kempton}, E. \& {Rauscher}, E. 2012, \apj, 751, 117

\bibitem[{{Moreno} \& {Sedano}(1997)}]{moreno1997}
{Moreno}, F. \& {Sedano}, J. 1997, \icarus, 130, 36

\bibitem[{Morley {et~al.}(2012)Morley, Fortney, Marley, Visscher, Saumon, \&
  Leggett}]{morley2012}
Morley, C.~V., Fortney, J.~J., Marley, M.~S., Visscher, C., Saumon, D., \&
  Leggett, S. 2012, The Astrophysical Journal, 756, 172

\bibitem[{{Moses} {et~al.}(2011){Moses}, {Visscher}, {Fortney}, {Showman},
  {Lewis}, {Griffith}, {Klippenstein}, {Shabram}, {Friedson}, {Marley}, \&
  {Freedman}}]{moses2011}
{Moses}, J.~I., {Visscher}, C., {Fortney}, J.~J., {Showman}, A.~P., {Lewis},
  N.~K., {Griffith}, C.~A., {Klippenstein}, S.~J., {Shabram}, M., {Friedson},
  A.~J., {Marley}, M.~S., \& {Freedman}, R.~S. 2011, \apj, 737, 15

\bibitem[{{Nakajima} {et~al.}(1995){Nakajima}, {Oppenheimer}, {Kulkarni},
  {Golimowski}, {Matthews}, \& {Durrance}}]{nakajima1995}
{Nakajima}, T., {Oppenheimer}, B.~R., {Kulkarni}, S.~R., {Golimowski}, D.~A.,
  {Matthews}, K., \& {Durrance}, S.~T. 1995, \nat, 378, 463

\bibitem[{{Ohno} \& {Zhang}(2019{\natexlab{a}})}]{ohno2019a}
{Ohno}, K. \& {Zhang}, X. 2019{\natexlab{a}}, \apj, 874, 1

\bibitem[{{Ohno} \& {Zhang}(2019{\natexlab{b}})}]{ohno2019b}
---. 2019{\natexlab{b}}, \apj, 874, 2

\bibitem[{{Oppenheimer} {et~al.}(1995){Oppenheimer}, {Kulkarni}, {Matthews}, \&
  {Nakajima}}]{oppenheimer1995}
{Oppenheimer}, B.~R., {Kulkarni}, S.~R., {Matthews}, K., \& {Nakajima}, T.
  1995, Science, 270, 1478

\bibitem[{{Parmentier} \& {Crossfield}(2018)}]{parmentier2018review}
{Parmentier}, V. \& {Crossfield}, I. J.~M. {Exoplanet Phase Curves:
  Observations and Theory}, 116

\bibitem[{{Parmentier} {et~al.}(2016){Parmentier}, {Fortney}, {Showman},
  {Morley}, \& {Marley}}]{parmentier2016}
{Parmentier}, V., {Fortney}, J.~J., {Showman}, A.~P., {Morley}, C., \&
  {Marley}, M.~S. 2016, \apj, 828, 22

\bibitem[{{Parmentier} \& {Guillot}(2014)}]{Parmentier2014}
{Parmentier}, V. \& {Guillot}, T. 2014, \aap, 562, A133

\bibitem[{{Parmentier} {et~al.}(2018){Parmentier}, {Line}, {Bean}, {Mansfield},
  {Kreidberg}, {Lupu}, {Visscher}, {D{\'e}sert}, {Fortney}, {Deleuil},
  {Arcangeli}, {Showman}, \& {Marley}}]{parmentier2018}
{Parmentier}, V., {Line}, M.~R., {Bean}, J.~L., {Mansfield}, M., {Kreidberg},
  L., {Lupu}, R., {Visscher}, C., {D{\'e}sert}, J.-M., {Fortney}, J.~J.,
  {Deleuil}, M., {Arcangeli}, J., {Showman}, A.~P., \& {Marley}, M.~S. 2018,
  \aap, 617, A110

\bibitem[{{Parmentier} {et~al.}(2015){Parmentier}, {Showman}, \& {de
  Wit}}]{parmentier2015}
{Parmentier}, V., {Showman}, A.~P., \& {de Wit}, J. 2015, Experimental
  Astronomy, 40, 481

\bibitem[{{Parmentier} {et~al.}(2020){Parmentier}, {Showman}, \&
  {Fortney}}]{Parmentier2020}
{Parmentier}, V., {Showman}, A.~P., \& {Fortney}. 2020, submitted

\bibitem[{Parmentier {et~al.}(2013)Parmentier, Showman, \&
  Lian}]{parmentier2013}
Parmentier, V., Showman, A.~P., \& Lian, Y. 2013, Astronomy \& Astrophysics,
  558, A91

\bibitem[{Pearl \& Conrath(1991)}]{pearl1991}
Pearl, J. \& Conrath, B. 1991, Journal of Geophysical Research: Space Physics,
  96, 18921

\bibitem[{Pedlosky(2013)}]{pedlosky_book}
Pedlosky, J. 2013, Geophysical fluid dynamics (Springer Science \& Business
  Media)

\bibitem[{{Penn} \& {Vallis}(2017)}]{penn2017}
{Penn}, J. \& {Vallis}, G.~K. 2017, \apj, 842, 101

\bibitem[{{Perez-Becker} \& {Showman}(2013)}]{perezbecker2013}
{Perez-Becker}, D. \& {Showman}, A.~P. 2013, \apj, 776, 134

\bibitem[{{Perna} {et~al.}(2012){Perna}, {Heng}, \& {Pont}}]{perna2012}
{Perna}, R., {Heng}, K., \& {Pont}, F. 2012, \apj, 751, 59

\bibitem[{Phlips \& Gill(1987)}]{phlips1987}
Phlips, P. \& Gill, A. 1987, Quarterly Journal of the Royal Meteorological
  Society, 113, 213

\bibitem[{Pierrehumbert \& Hammond(2019)}]{pierrehumbert2019}
Pierrehumbert, R.~T. \& Hammond, M. 2019, Annual Review of Fluid Mechanics

\bibitem[{{Pont} {et~al.}(2013){Pont}, {Sing}, {Gibson}, {Aigrain}, {Henry}, \&
  {Husnoo}}]{pont2013}
{Pont}, F., {Sing}, D.~K., {Gibson}, N.~P., {Aigrain}, S., {Henry}, G., \&
  {Husnoo}, N. 2013, \mnras, 432, 2917

\bibitem[{{Powell} {et~al.}(2019){Powell}, {Louden}, {Kreidberg}, {Zhang},
  {Gao}, \& {Parmentier}}]{powell2019}
{Powell}, D., {Louden}, T., {Kreidberg}, L., {Zhang}, X., {Gao}, P., \&
  {Parmentier}, V. 2019, \apj, 887, 170

\bibitem[{{Powell} {et~al.}(2018){Powell}, {Zhang}, {Gao}, \&
  {Parmentier}}]{powell2018}
{Powell}, D., {Zhang}, X., {Gao}, P., \& {Parmentier}, V. 2018, \apj, 860, 18

\bibitem[{{Radigan} {et~al.}(2012){Radigan}, {Jayawardhana}, {Lafreni{\`e}re},
  {Artigau}, {Marley}, \& {Saumon}}]{radigan2012}
{Radigan}, J., {Jayawardhana}, R., {Lafreni{\`e}re}, D., {Artigau}, {\'E}.,
  {Marley}, M., \& {Saumon}, D. 2012, \apj, 750, 105

\bibitem[{Rappaport {et~al.}(2017)Rappaport, Vanderburg, Nelson, Gary, Kaye,
  Kalomeni, Howell, Thorstensen, Lachapelle, Lundy, {et~al.}}]{rappaport2017}
Rappaport, S., Vanderburg, A., Nelson, L., Gary, B., Kaye, T., Kalomeni, B.,
  Howell, S., Thorstensen, J., Lachapelle, F.-R., Lundy, M., {et~al.} 2017,
  Monthly Notices of the Royal Astronomical Society, 471, 948

\bibitem[{{Rauscher}(2017)}]{rauscher2017}
{Rauscher}, E. 2017, \apj, 846, 69

\bibitem[{{Rauscher} \& {Kempton}(2014)}]{rauscher2014}
{Rauscher}, E. \& {Kempton}, E. M.~R. 2014, \apj, 790, 79

\bibitem[{{Rauscher} \& {Menou}(2010)}]{rauscher2010}
{Rauscher}, E. \& {Menou}, K. 2010, \apj, 714, 1334

\bibitem[{Rauscher \& Menou(2012)}]{rauscher2012}
Rauscher, E. \& Menou, K. 2012, The Astrophysical Journal, 750, 96

\bibitem[{{Rauscher} \& {Menou}(2012)}]{rauscher2012b}
{Rauscher}, E. \& {Menou}, K. 2012, \apj, 745, 78

\bibitem[{{Rauscher} \& {Menou}(2013)}]{rauscher2013}
---. 2013, \apj, 764, 103

\bibitem[{Rauscher \& Showman(2014)}]{rauscher2014influence}
Rauscher, E. \& Showman, A.~P. 2014, The Astrophysical Journal, 784, 160

\bibitem[{{Reiners} \& {Basri}(2008)}]{reiners2008}
{Reiners}, A. \& {Basri}, G. 2008, \apj, 684, 1390

\bibitem[{Robinson \& Marley(2014)}]{robinson2014}
Robinson, T.~D. \& Marley, M.~S. 2014, The Astrophysical Journal, 785, 158

\bibitem[{Rogers(2017)}]{rogers2017}
Rogers, T. 2017, Nature Astronomy, 1, 0131

\bibitem[{Rogers \& Komacek(2014)}]{rogers2014komacek}
Rogers, T.~M. \& Komacek, T.~D. 2014, The Astrophysical Journal, 794, 132

\bibitem[{{Rogers} \& {McElwaine}(2017)}]{Rogers2017McElwaine}
{Rogers}, T.~M. \& {McElwaine}, J.~N. 2017, \apjl, 841, L26

\bibitem[{{Rogers} \& {Showman}(2014)}]{rogers2014}
{Rogers}, T.~M. \& {Showman}, A.~P. 2014, \apjl, 782, L4

\bibitem[{{Roman} \& {Rauscher}(2017)}]{roman2017}
{Roman}, M. \& {Rauscher}, E. 2017, \apj, 850, 17

\bibitem[{{Roman} \& {Rauscher}(2019)}]{roman2019}
---. 2019, \apj, 872, 1

\bibitem[{{Sahlmann} {et~al.}(2011){Sahlmann}, {S{\'e}gransan}, {Queloz},
  {Udry}, {Santos}, {Marmier}, {Mayor}, {Naef}, {Pepe}, \&
  {Zucker}}]{sahlmann2011}
{Sahlmann}, J., {S{\'e}gransan}, D., {Queloz}, D., {Udry}, S., {Santos}, N.~C.,
  {Marmier}, M., {Mayor}, M., {Naef}, D., {Pepe}, F., \& {Zucker}, S. 2011,
  \aap, 525, A95

\bibitem[{{Sainsbury-Martinez} {et~al.}(2019){Sainsbury-Martinez}, {Wang},
  {Fromang}, {Tremblin}, {Dubos}, {Meurdesoif}, {Spiga}, {Leconte}, {Baraffe},
  {Chabrier}, {Mayne}, {Drummond}, \& {Debras}}]{Sainsbury-Martinez2019}
{Sainsbury-Martinez}, F., {Wang}, P., {Fromang}, S., {Tremblin}, P., {Dubos},
  T., {Meurdesoif}, Y., {Spiga}, A., {Leconte}, J., {Baraffe}, I., {Chabrier},
  G., {Mayne}, N., {Drummond}, B., \& {Debras}, F. 2019, \aap, 632, A114

\bibitem[{Salby \& Garcia(1987)}]{salby1987}
Salby, M.~L. \& Garcia, R.~R. 1987, Journal of the atmospheric sciences, 44,
  458

\bibitem[{Santisteban {et~al.}(2016)Santisteban, Knigge, Littlefair, Breton,
  Dhillon, G{\"a}nsicke, Marsh, Pretorius, Southworth, \&
  Hauschildt}]{santisteban2016}
Santisteban, J. V.~H., Knigge, C., Littlefair, S.~P., Breton, R.~P., Dhillon,
  V.~S., G{\"a}nsicke, B.~T., Marsh, T.~R., Pretorius, M.~L., Southworth, J.,
  \& Hauschildt, P.~H. 2016, Nature, 533, 366

\bibitem[{{Saumon} {et~al.}(2000){Saumon}, {Geballe}, {Leggett}, {Marley},
  {Freedman}, {Lodders}, {Fegley}, \& {Sengupta}}]{saumon2000}
{Saumon}, D., {Geballe}, T.~R., {Leggett}, S.~K., {Marley}, M.~S., {Freedman},
  R.~S., {Lodders}, K., {Fegley}, B., J., \& {Sengupta}, S.~K. 2000, \apj, 541,
  374

\bibitem[{Saumon \& Marley(2008)}]{saumon2008}
Saumon, D. \& Marley, M.~S. 2008, The Astrophysical Journal, 689, 1327

\bibitem[{{Saumon} {et~al.}(2006){Saumon}, {Marley}, {Cushing}, {Leggett},
  {Roellig}, {Lodders}, \& {Freedman}}]{saumon2006}
{Saumon}, D., {Marley}, M.~S., {Cushing}, M.~C., {Leggett}, S.~K., {Roellig},
  T.~L., {Lodders}, K., \& {Freedman}, R.~S. 2006, \apj, 647, 552

\bibitem[{{Schneider} \& {Liu}(2009)}]{schneider2009}
{Schneider}, T. \& {Liu}, J. 2009, Journal of Atmospheric Sciences, 66, 579

\bibitem[{{Schwartz} \& {Cowan}(2015)}]{schwartz2015}
{Schwartz}, J.~C. \& {Cowan}, N.~B. 2015, \mnras, 449, 4192

\bibitem[{Seager \& Deming(2010)}]{seager2010}
Seager, S. \& Deming, D. 2010, Annual Review of Astronomy and Astrophysics, 48,
  631

\bibitem[{{Showman} {et~al.}(2010){Showman}, {Cho}, \& {Menou}}]{showman2010}
{Showman}, A.~P., {Cho}, J.~Y.-K., \& {Menou}, K. {Atmospheric Circulation of
  Exoplanets}, ed. S.~{Seager}, 471--516

\bibitem[{{Showman} {et~al.}(2008{\natexlab{a}}){Showman}, {Cooper}, {Fortney},
  \& {Marley}}]{showman2008}
{Showman}, A.~P., {Cooper}, C.~S., {Fortney}, J.~J., \& {Marley}, M.~S.
  2008{\natexlab{a}}, \apj, 682, 559

\bibitem[{{Showman} {et~al.}(2013){Showman}, {Fortney}, {Lewis}, \&
  {Shabram}}]{showman2013b}
{Showman}, A.~P., {Fortney}, J.~J., {Lewis}, N.~K., \& {Shabram}, M. 2013,
  \apj, 762, 24

\bibitem[{{Showman} {et~al.}(2009){Showman}, {Fortney}, {Lian}, {Marley},
  {Freedman}, {Knutson}, \& {Charbonneau}}]{showman2009}
{Showman}, A.~P., {Fortney}, J.~J., {Lian}, Y., {Marley}, M.~S., {Freedman},
  R.~S., {Knutson}, H.~A., \& {Charbonneau}, D. 2009, \apj, 699, 564

\bibitem[{{Showman} \& {Guillot}(2002)}]{showman2002}
{Showman}, A.~P. \& {Guillot}, T. 2002, \aap, 385, 166

\bibitem[{{Showman} \& {Ingersoll}(1998)}]{showman&ingersoll1998}
{Showman}, A.~P. \& {Ingersoll}, A.~P. 1998, \icarus, 132, 205

\bibitem[{Showman {et~al.}(2018)Showman, Ingersoll, Achterberg, \&
  Kaspi}]{showman2018review}
Showman, A.~P., Ingersoll, A.~P., Achterberg, R., \& Kaspi, Y. 2018, Saturn in
  the 21st Century, 20, 295

\bibitem[{{Showman} \& {Kaspi}(2013)}]{showman&kaspi2013}
{Showman}, A.~P. \& {Kaspi}, Y. 2013, \apj, 776, 85

\bibitem[{Showman {et~al.}(2011)Showman, Kaspi, \& Flierl}]{showman2011scaling}
Showman, A.~P., Kaspi, Y., \& Flierl, G.~R. 2011, Icarus, 211, 1258

\bibitem[{Showman {et~al.}(2015)Showman, Lewis, \& Fortney}]{showman2015}
Showman, A.~P., Lewis, N.~K., \& Fortney, J.~J. 2015, The Astrophysical
  Journal, 801, 95

\bibitem[{{Showman} {et~al.}(2008{\natexlab{b}}){Showman}, {Menou}, \&
  {Cho}}]{showman2008b}
{Showman}, A.~P., {Menou}, K., \& {Cho}, J.~Y.~K. Astronomical Society of the
  Pacific Conference Series, Vol. 398, {Atmospheric Circulation of Hot
  Jupiters: A Review of Current Understanding}, ed. D.~{Fischer}, F.~A.
  {Rasio}, S.~E. {Thorsett}, \& A.~{Wolszczan}, 419

\bibitem[{Showman \& Polvani(2010)}]{showman2010b}
Showman, A.~P. \& Polvani, L.~M. 2010, Geophysical research letters, 37

\bibitem[{Showman \& Polvani(2011)}]{showman2011}
---. 2011, The Astrophysical Journal, 738, 71

\bibitem[{{Showman} {et~al.}(2019){Showman}, {Tan}, \& {Zhang}}]{showman2019}
{Showman}, A.~P., {Tan}, X., \& {Zhang}, X. 2019, \apj, 883, 4

\bibitem[{Showman {et~al.}(2013)Showman, Wordsworth, Merlis, \&
  Kaspi}]{showman2013}
Showman, A.~P., Wordsworth, R.~D., Merlis, T.~M., \& Kaspi, Y. 2013,
  Comparative Climatology of Terrestrial Planets, 1, 277

\bibitem[{{Shporer}(2017)}]{shporer2017}
{Shporer}, A. 2017, \pasp, 129, 072001

\bibitem[{{Shporer} \& {Hu}(2015)}]{shporer2015}
{Shporer}, A. \& {Hu}, R. 2015, \aj, 150, 112

\bibitem[{{Sing} {et~al.}(2016){Sing}, {Fortney}, {Nikolov}, {Wakeford},
  {Kataria}, {Evans}, {Aigrain}, {Ballester}, {Burrows}, {Deming},
  {D{\'e}sert}, {Gibson}, {Henry}, {Huitson}, {Knutson}, {Lecavelier Des
  Etangs}, {Pont}, {Showman}, {Vidal-Madjar}, {Williamson}, \&
  {Wilson}}]{sing2016}
{Sing}, D.~K., {Fortney}, J.~J., {Nikolov}, N., {Wakeford}, H.~R., {Kataria},
  T., {Evans}, T.~M., {Aigrain}, S., {Ballester}, G.~E., {Burrows}, A.~S.,
  {Deming}, D., {D{\'e}sert}, J.-M., {Gibson}, N.~P., {Henry}, G.~W.,
  {Huitson}, C.~M., {Knutson}, H.~A., {Lecavelier Des Etangs}, A., {Pont}, F.,
  {Showman}, A.~P., {Vidal-Madjar}, A., {Williamson}, M.~H., \& {Wilson}, P.~A.
  2016, \nat, 529, 59

\bibitem[{{Sing} {et~al.}(2011){Sing}, {Pont}, {Aigrain}, {Charbonneau},
  {D{\'e}sert}, {Gibson}, {Gilliland}, {Hayek}, {Henry}, {Knutson}, {Lecavelier
  Des Etangs}, {Mazeh}, \& {Shporer}}]{sing2011}
{Sing}, D.~K., {Pont}, F., {Aigrain}, S., {Charbonneau}, D., {D{\'e}sert},
  J.~M., {Gibson}, N., {Gilliland}, R., {Hayek}, W., {Henry}, G., {Knutson},
  H., {Lecavelier Des Etangs}, A., {Mazeh}, T., \& {Shporer}, A. 2011, \mnras,
  416, 1443

\bibitem[{{Snellen} {et~al.}(2014){Snellen}, {Brandl}, {de Kok}, {Brogi},
  {Birkby}, \& {Schwarz}}]{snellen2014}
{Snellen}, I.~A.~G., {Brandl}, B.~R., {de Kok}, R.~J., {Brogi}, M., {Birkby},
  J., \& {Schwarz}, H. 2014, \nat, 509, 63

\bibitem[{{Snellen} {et~al.}(2010){Snellen}, {de Kok}, {de Mooij}, \&
  {Albrecht}}]{snellen2010}
{Snellen}, I. A.~G., {de Kok}, R.~J., {de Mooij}, E. J.~W., \& {Albrecht}, S.
  2010, \nat, 465, 1049

\bibitem[{Steele {et~al.}(2013)Steele, Saglia, Burleigh, Marsh, G{\"a}nsicke,
  Lawrie, Cappetta, Girven, \& Napiwotzki}]{steele2013}
Steele, P., Saglia, R., Burleigh, M.~R., Marsh, T., G{\"a}nsicke, B., Lawrie,
  K., Cappetta, M., Girven, J., \& Napiwotzki, R. 2013, Monthly Notices of the
  Royal Astronomical Society, 429, 3492

\bibitem[{{Steinrueck} {et~al.}(2019){Steinrueck}, {Parmentier}, {Showman},
  {Lothringer}, \& {Lupu}}]{steinrueck2019}
{Steinrueck}, M.~E., {Parmentier}, V., {Showman}, A.~P., {Lothringer}, J.~D.,
  \& {Lupu}, R.~E. 2019, \apj, 880, 14

\bibitem[{{Stephens} {et~al.}(2009){Stephens}, {Leggett}, {Cushing}, {Marley},
  {Saumon}, {Geballe}, {Golimowski}, {Fan}, \& {Noll}}]{stephens2009}
{Stephens}, D.~C., {Leggett}, S.~K., {Cushing}, M.~C., {Marley}, M.~S.,
  {Saumon}, D., {Geballe}, T.~R., {Golimowski}, D.~A., {Fan}, X., \& {Noll},
  K.~S. 2009, \apj, 702, 154

\bibitem[{{Stevenson}(1991)}]{stevenson1991}
{Stevenson}, D.~J. 1991, \araa, 29, 163

\bibitem[{{Stevenson}(2016)}]{stevenson2016}
{Stevenson}, K.~B. 2016, \apjl, 817, L16

\bibitem[{{Stevenson} {et~al.}(2017){Stevenson}, {Line}, {Bean}, {D{\'e}sert},
  {Fortney}, {Showman}, {Kataria}, {Kreidberg}, \& {Feng}}]{stevenson2017}
{Stevenson}, K.~B., {Line}, M.~R., {Bean}, J.~L., {D{\'e}sert}, J.-M.,
  {Fortney}, J.~J., {Showman}, A.~P., {Kataria}, T., {Kreidberg}, L., \&
  {Feng}, Y.~K. 2017, \aj, 153, 68

\bibitem[{Tan(2018)}]{tan2018}
Tan, X. 2018, PhD thesis, Ph. D. Thesis, University of Arizona

\bibitem[{{Tan} \& {Komacek}(2019)}]{tan2019}
{Tan}, X. \& {Komacek}, T.~D. 2019, \apj, 886, 26

\bibitem[{{Tan} \& {Showman}(2017)}]{tan2017}
{Tan}, X. \& {Showman}, A.~P. 2017, \apj, 835, 186

\bibitem[{{Tan} \& {Showman}(2019)}]{tan2019bd}
---. 2019, \apj, 874, 111

\bibitem[{{Tan} \& {Showman}(2020{\natexlab{a}})}]{tan2020bd}
---. 2020{\natexlab{a}}, arXiv e-prints, arXiv:2005.12152

\bibitem[{{Tan} \& {Showman}(2020{\natexlab{b}})}]{tan2020wdbd}
---. 2020{\natexlab{b}}, arXiv e-prints, arXiv:2001.06269

\bibitem[{{Thorngren} {et~al.}(2019){Thorngren}, {Gao}, \&
  {Fortney}}]{Thorngren2019}
{Thorngren}, D., {Gao}, P., \& {Fortney}, J.~J. 2019, \apjl, 884, L6

\bibitem[{Tremblin {et~al.}(2017)Tremblin, Chabrier, Baraffe, Liu, Magnier,
  Lagage, De~Oliveira, Burgasser, Amundsen, \& Drummond}]{tremblin2017bd}
Tremblin, P., Chabrier, G., Baraffe, I., Liu, M.~C., Magnier, E., Lagage,
  P.-O., De~Oliveira, C.~A., Burgasser, A., Amundsen, D., \& Drummond, B. 2017,
  The Astrophysical Journal, 850, 46

\bibitem[{{Tremblin} {et~al.}(2017){Tremblin}, {Chabrier}, {Mayne}, {Amundsen},
  {Baraffe}, {Debras}, {Drummond}, {Manners}, \& {Fromang}}]{tremblin2017}
{Tremblin}, P., {Chabrier}, G., {Mayne}, N.~J., {Amundsen}, D.~S., {Baraffe},
  I., {Debras}, F., {Drummond}, B., {Manners}, J., \& {Fromang}, S. 2017, \apj,
  841, 30

\bibitem[{Tremblin {et~al.}(2019)Tremblin, Padioleau, Phillips, Chabrier,
  Baraffe, Fromang, Audit, Bloch, Burgasser, Drummond, {et~al.}}]{tremblin2019}
Tremblin, P., Padioleau, T., Phillips, M., Chabrier, G., Baraffe, I., Fromang,
  S., Audit, E., Bloch, H., Burgasser, A., Drummond, B., {et~al.} 2019, The
  Astrophysical Journal, 876, 144

\bibitem[{{Tsai} {et~al.}(2014){Tsai}, {Dobbs-Dixon}, \& {Gu}}]{tsai2014}
{Tsai}, S.-M., {Dobbs-Dixon}, I., \& {Gu}, P.-G. 2014, \apj, 793, 141

\bibitem[{{Tsai} {et~al.}(2018){Tsai}, {Kitzmann}, {Lyons}, {Mendon{\c{c}}a},
  {Grimm}, \& {Heng}}]{tsai2018}
{Tsai}, S.-M., {Kitzmann}, D., {Lyons}, J.~R., {Mendon{\c{c}}a}, J., {Grimm},
  S.~L., \& {Heng}, K. 2018, \apj, 862, 31

\bibitem[{{Tsuji}(2002)}]{tsuji2002}
{Tsuji}, T. 2002, \apj, 575, 264

\bibitem[{Vallis(2006)}]{vallis2006}
Vallis, G.~K. 2006, Atmospheric and oceanic fluid dynamics: fundamentals and
  large-scale circulation (Cambridge University Press)

\bibitem[{{Vasavada} {et~al.}(2006){Vasavada}, {H{\"o}rst}, {Kennedy},
  {Ingersoll}, {Porco}, {Del Genio}, \& {West}}]{vasavada2006}
{Vasavada}, A.~R., {H{\"o}rst}, S.~M., {Kennedy}, M.~R., {Ingersoll}, A.~P.,
  {Porco}, C.~C., {Del Genio}, A.~D., \& {West}, R.~A. 2006, Journal of
  Geophysical Research (Planets), 111, E05004

\bibitem[{{Venot} {et~al.}(2019){Venot}, {Bounaceur}, {Dobrijevic},
  {H{\'e}brard}, {Cavali{\'e}}, {Tremblin}, {Drummond}, \&
  {Charnay}}]{venot2019}
{Venot}, O., {Bounaceur}, R., {Dobrijevic}, M., {H{\'e}brard}, E.,
  {Cavali{\'e}}, T., {Tremblin}, P., {Drummond}, B., \& {Charnay}, B. 2019,
  \aap, 624, A58

\bibitem[{{Visscher}(2012)}]{visscher2012}
{Visscher}, C. 2012, \apj, 757, 5

\bibitem[{{Vos} {et~al.}(2020){Vos}, {Biller}, {Allers}, {Faherty}, {Liu},
  {Metchev}, {Eriksson}, {Manjavacas}, {Dupuy}, {Janson}, {Radigan-Hoffmanf},
  {Crossfield}, {Bonnefoy}, {Best}, {Homeier}, {Schlieder}, {Brandner},
  {Henning}, {Bonavita}, \& {Buenzli}}]{vos2020}
{Vos}, J.~M., {Biller}, B.~A., {Allers}, K.~N., {Faherty}, J.~K., {Liu}, M.~C.,
  {Metchev}, S., {Eriksson}, S., {Manjavacas}, E., {Dupuy}, T.~J., {Janson},
  M., {Radigan-Hoffmanf}, J., {Crossfield}, I., {Bonnefoy}, M., {Best}, W.
  M.~J., {Homeier}, D., {Schlieder}, J.~E., {Brandner}, W., {Henning}, T.,
  {Bonavita}, M., \& {Buenzli}, E. 2020, \aj, 160, 38

\bibitem[{Vos {et~al.}(2019)Vos, Biller, Bonavita, Eriksson, Liu, Best,
  Metchev, Radigan, Allers, Janson, {et~al.}}]{vos2019}
Vos, J.~M., Biller, B.~A., Bonavita, M., Eriksson, S., Liu, M.~C., Best, W.~M.,
  Metchev, S., Radigan, J., Allers, K.~N., Janson, M., {et~al.} 2019, Monthly
  Notices of the Royal Astronomical Society, 483, 480

\bibitem[{{Wakeford} {et~al.}(2017){Wakeford}, {Visscher}, {Lewis}, {Kataria},
  {Marley}, {Fortney}, \& {Mand ell}}]{wakeford2017}
{Wakeford}, H.~R., {Visscher}, C., {Lewis}, N.~K., {Kataria}, T., {Marley},
  M.~S., {Fortney}, J.~J., \& {Mand ell}, A.~M. 2017, \mnras, 464, 4247

\bibitem[{Wang \& Wordsworth(2020)}]{wang2020}
Wang, H. \& Wordsworth, R. 2020, The Astrophysical Journal, 891, 7

\bibitem[{{West} {et~al.}(1992){West}, {Friedson}, \& {Appleby}}]{west1992}
{West}, R.~A., {Friedson}, A.~J., \& {Appleby}, J.~F. 1992, \icarus, 100, 245

\bibitem[{Wheeler \& Kiladis(1999)}]{wheeler1999}
Wheeler, M. \& Kiladis, G.~N. 1999, Journal of the Atmospheric Sciences, 56,
  374

\bibitem[{Williams(1988)}]{williams1988}
Williams, G.~P. 1988, Climate Dynamics, 2, 205

\bibitem[{{Wilson} {et~al.}(2014){Wilson}, {Rajan}, \& {Patience}}]{wilson2014}
{Wilson}, P.~A., {Rajan}, A., \& {Patience}, J. 2014, \aap, 566, A111

\bibitem[{Wing {et~al.}(2017)Wing, Emanuel, Holloway, \& Muller}]{wing2017}
Wing, A.~A., Emanuel, K., Holloway, C.~E., \& Muller, C. 2017, in Shallow
  Clouds, Water Vapor, Circulation, and Climate Sensitivity (Springer), 1--25

\bibitem[{{Wong} {et~al.}(2016){Wong}, {Knutson}, {Kataria}, {Lewis},
  {Burrows}, {Fortney}, {Schwartz}, {Shporer}, {Agol}, {Cowan}, {Deming},
  {D{\'e}sert}, {Fulton}, {Howard}, {Langton}, {Laughlin}, {Showman}, \&
  {Todorov}}]{wong2016}
{Wong}, I., {Knutson}, H.~A., {Kataria}, T., {Lewis}, N.~K., {Burrows}, A.,
  {Fortney}, J.~J., {Schwartz}, J., {Shporer}, A., {Agol}, E., {Cowan}, N.~B.,
  {Deming}, D., {D{\'e}sert}, J.-M., {Fulton}, B.~J., {Howard}, A.~W.,
  {Langton}, J., {Laughlin}, G., {Showman}, A.~P., \& {Todorov}, K. 2016, \apj,
  823, 122

\bibitem[{{Wong} {et~al.}(2019){Wong}, {Shporer}, {Kitzmann}, {Morris}, {Heng},
  {Hoeijmakers}, {Demory}, {Mansfield}, {Bean}, {Daylan}, {Fetherolf},
  {Rodriguez}, {Benneke}, {Ricker}, {Latham}, {Vanderspek}, {Seager}, {Winn},
  {Jenkins}, {Burke}, {Christiansen}, {Essack}, {Rose}, {Smith}, {Tenenbaum},
  \& {Yahalomi}}]{wong2019}
{Wong}, I., {Shporer}, A., {Kitzmann}, D., {Morris}, B.~M., {Heng}, K.,
  {Hoeijmakers}, H.~J., {Demory}, B.-O., {Mansfield}, M., {Bean}, J.~L.,
  {Daylan}, T., {Fetherolf}, T., {Rodriguez}, J.~E., {Benneke}, B., {Ricker},
  G.~R., {Latham}, D.~W., {Vanderspek}, R., {Seager}, S., {Winn}, J.~N.,
  {Jenkins}, J.~M., {Burke}, C.~J., {Christiansen}, J., {Essack}, Z., {Rose},
  M.~E., {Smith}, J.~C., {Tenenbaum}, P., \& {Yahalomi}, D. 2019, arXiv
  e-prints, arXiv:1910.01607

\bibitem[{{Yang} {et~al.}(2016){Yang}, {Apai}, {Marley}, {Karalidi}, {Flateau},
  {Showman}, {Metchev}, {Buenzli}, {Radigan}, {Artigau}, {Lowrance}, \&
  {Burgasser}}]{yang2016}
{Yang}, H., {Apai}, D., {Marley}, M.~S., {Karalidi}, T., {Flateau}, D.,
  {Showman}, A.~P., {Metchev}, S., {Buenzli}, E., {Radigan}, J., {Artigau},
  {\'E}., {Lowrance}, P.~J., \& {Burgasser}, A.~J. 2016, \apj, 826, 8

\bibitem[{{Youdin} \& {Mitchell}(2010)}]{Youdin2010}
{Youdin}, A.~N. \& {Mitchell}, J.~L. 2010, \apj, 721, 1113

\bibitem[{{Zellem} {et~al.}(2014){Zellem}, {Lewis}, {Knutson}, {Griffith},
  {Showman}, {Fortney}, {Cowan}, {Agol}, {Burrows}, {Charbonneau}, {Deming},
  {Laughlin}, \& {Langton}}]{zellem2014}
{Zellem}, R.~T., {Lewis}, N.~K., {Knutson}, H.~A., {Griffith}, C.~A.,
  {Showman}, A.~P., {Fortney}, J.~J., {Cowan}, N.~B., {Agol}, E., {Burrows},
  A., {Charbonneau}, D., {Deming}, D., {Laughlin}, G., \& {Langton}, J. 2014,
  \apj, 790, 53

\bibitem[{Zhang(2020)}]{zhang2020}
Zhang, X. 2020, arXiv preprint arXiv:2006.13384

\bibitem[{{Zhang} \& {Showman}(2014)}]{zhang&showman2014}
{Zhang}, X. \& {Showman}, A.~P. 2014, \apjl, 788, L6

\bibitem[{{Zhang} \& {Showman}(2017)}]{zhang2017}
---. 2017, \apj, 836, 73

\bibitem[{{Zhang} \& {Showman}(2018{\natexlab{a}})}]{zhang2018a}
---. 2018{\natexlab{a}}, \apj, 866, 1

\bibitem[{{Zhang} \& {Showman}(2018{\natexlab{b}})}]{zhang2018b}
---. 2018{\natexlab{b}}, \apj, 866, 2

\bibitem[{{Zhou} {et~al.}(2016){Zhou}, {Apai}, {Schneider}, {Marley}, \&
  {Showman}}]{zhou2016}
{Zhou}, Y., {Apai}, D., {Schneider}, G.~H., {Marley}, M.~S., \& {Showman},
  A.~P. 2016, \apj, 818, 176

\bibitem[{{Zhou} {et~al.}(2020){Zhou}, {Bowler}, {Morley}, {Apai}, {Kataria},
  {Bryan}, \& {Benneke}}]{zhou2020}
{Zhou}, Y., {Bowler}, B.~P., {Morley}, C.~V., {Apai}, D., {Kataria}, T.,
  {Bryan}, M.~L., \& {Benneke}, B. 2020, \aj, 160, 77

\end{thebibliography}


\begin{thebibliography}
\end{thebibliography}
\fi

\if\bibinc y

\fi

\end{document}